# TOWARDS THE EUROPEAN STRATEGY FOR PARTICLE PHYSICS: THE BRIEFING BOOK

T. Åkesson[a], R. Aleksan[b], B. Allanach[c], S. Bertolucci[d], A. Blondel[e], J. Butterworth[f], M. Cavalli-Sforza[g], A. Cervera[h], S. Davidson[i], M. de Naurois[j], K. Desch[k], U. Egede[l], N. Glover[m], R. Heuer[n], A. Hoecker[o], P. Huber[p], K. Jungmann[q], R. Landua[o], J-M. Le Goff[o], F. Linde[r], A. Lombardi[o], M. Mangano[o], M. Mezzetto[s], G. Onderwater[q], N. Palanque-Delabrouille[t], K. Peach[u], A. Polosa[v], E. Rondio[w], B. Webber[c], G. Weiglein[m], J. Womersley[x], K. Wurr[n]

## ABSTRACT

This document was prepared as part of the briefing material for the Workshop of the CERN Council Strategy Group, held in DESY Zeuthen from 2nd to 6th May 2006. It gives an overview of the physics issues and of the technological challenges that will shape the future of the field, and incorporates material presented and discussed during the Symposium on the European Strategy for Particle Physics, held in Orsay from 30th January to 2nd February 2006, reflecting the various opinions of the European community as recorded in written submissions to the Strategy Group and in the discussions at the Symposium.

[a] *Lund University, Sweden*
[b] *CPPM/IN2P3-CNRS and DAPNIA/CEA, France*
[c] *Cambridge University and DAMTP, UK*
[d] *INFN and Laboratori Nazionali di Frascati, Italy*
[e] *University of Geneva, Switzerland*
[f] *University College London, UK*
[g] *IFAE, Universitat Autònoma de Barcelona, Spain*
[h] *University of Valencia, Spain*
[i] *University of Lyon, France*
[j] *LPNHE-IN2P3-CNRS and University of Paris VI&VII, France*
[k] *Freiburg University, Germany*
[l] *Imperial College London, UK*
[m] *IPPP, Durham University, UK*
[n] *University of Hamburg and DESY, Germany*
[o] *CERN, Geneva, Switzerland*
[p] *University of Wisconsin, Madison, USA*
[q] *KVI, Groningen, The Netherlands*
[r] *NIKHEF, Amsterdam, The Netherlands*
[s] *INFN and University of Padova, Italy*
[t] *DAPNIA, Saclay, France*
[u] *John Adams Institute, University of Oxford and Royal Holloway University of London, UK*
[v] *University of Rome, La Sapienza, Italy*
[w] *Soltan Institute for Nuclear Studies, Warsaw, Poland*
[x] *CCLRC, , Rutherford Appleton Laboratory , Chilton, Didcot, UK*

# TABLE OF CONTENTS

# I PREFACE

On the 16th of June 2005, CERN Council launched a process to define a European Strategy for Particle Physics. To this end it established an *ad hoc* scientific advisory group to produce a draft strategy to be considered by a special Council meeting.

To be able to formulate such a strategy it was essential to assemble a broad scientific overview of the field, as well as information on other aspects such as organization and knowledge transfer. Input was therefore called from the international community, which responded with a large number of thoughtful and informative written submissions; in addition, a symposium[1] was arranged in Orsay at the end of January 2006 as a step towards producing this scientific overview. This symposium also had a strong emphasis on discussions about the different themes.

The information collected during the symposium and through the written submissions was elaborated and printed as a Briefing Book, submitted to the scientific advisory group. This report here is the scientific overview in that Briefing Book; it includes the summaries of the discussions that took place during the symposium, as well as references to the submitted material. The full list of submissions is collected here in the Appendix. The relative references to these documents, labeled in the text as [BB2-…], are available through the Briefing Book, vol 2, link on the Strategy Group web page[2].

The process terminated in Lisbon the 14th of July 2006 at a special meeting of the CERN Council, where it unanimously adopted the European Strategy for Particle Physics[4]. This could well be the start of a new chapter in European scientific collaboration.

Several people contributed to the realization of this work. In particular, we would like here to acknowledge the help from the local organization of the Symposium in Orsay, and the support given by CERN staff, especially Brigitte Beauseroy, Isabelle Billod, Sylvia Martakis and Suzy Vascotto; without them, this could not have been done.

In the preparation of this document, we benefited from the contribution of several colleagues. For Chapter III, we acknowledge contributions from W. Buchmueller, A. De Roeck, E. Elsen, F. Gianotti, K. Jakobs, K. Moenig, and P. Zerwas in preparation of the Open Symposium and the comments of F. Gianotti and A. De Roeck on parts of the manuscript. For Chapter IV we thank B. Cros, J.-P. Delahaye, R. Garoby and M. Vretenar. For Chapter VI we acknowledge contributions by A. Baldini, A. Ceccucci, and G. Isidori, and by S. T'Jampens and M. Pierini from the CKMfitter and UTfit groups, respectively, for their help in preparing many of the plots shown. For the preparation of the Open Symposium and this document, the authors thank A. Baldini, I. Bigi, A. Ceccucci, O. Deschamps, T. Gershon, U. Haisch, M. Hazumi, T. Hurth, G. Isidori, H. Lacker, O. Pène, P. Roudeau, M. Różańska, O. Schneider, M. Smizanska, A. Stocchi and A. Variola for helpful conversations and comments. For Chapter VII, we thank H. Abele, A. Caldwell, and R.G.E. Timmermans. For Chapter VIII, we are grateful to C. Spiering for his valuable comments to the text. For Chapter XI we thank B. Bressan and M. Streit-Bianchi for their contributions to the text, and H.F. Hoffmann, G. Mikenberg, D-O. Riska and all authors mentioned in the references for their input.

---

[1] http://events.lal.in2p3.fr/conferences/Symposium06/
[2] http//cern.ch/council-strategygroup .
[4] http://cern.ch/council-strategygroup/Lisbon.html

# II PARTICLE PHYSICS: TOWARDS A NEW ERA OF FUNDAMENTAL DISCOVERIES

## II-1 The Standard Model of particle physics

During the past few decades, particle physics has made unprecedented progress in elucidating the laws of nature at the most fundamental level. We moved from the formulation and consolidation of a quantitative theory of quantum electrodynamics for elementary particles, towards the development of a framework capable of describing the whole variety of observed particles and interactions in terms of a few fundamental interactions and elementary objects. This framework, based on the formalism of relativistic quantum field theory and gauge symmetry as a dynamical principle, is known as the Standard Model (SM). Thanks to an impressive series of experimental confirmations, it has grown into a complete and accurate description of the microscopic phenomena that are the basis of our macroscopic world. Only gravity, which cannot be formulated as a simple gauge theory, still lacks a fully satisfactory understanding at the quantum level, and remains stubbornly outside the SM.

With one notable exception, all of the elementary particles and interactions whose existence had been required by the SM have been observed. The dynamical properties of the fundamental interactions, as predicted by the SM, have been confirmed to a high level of precision, up to the accuracy allowed by the difficulty of the measurements and of the theoretical calculations. The laws of electromagnetism have been tested to one part in a hundred billion, making it by far the most solid and verified field of science. The unification of electromagnetism and weak interactions has been proved and tested to one part per mille, confirming an intellectual achievement comparable to Maxwell's unification of electricity and magnetism 140 years ago. The interactions responsible for holding nuclei together, and for the multitude of unstable particles that are produced when large concentrations of energy turn into matter, have been identified and their effects quantitatively predicted at the per cent level. These successes have been made possible by a remarkable sequence of ambitious experimental programmes, starting with the detection of the charm quark and of neutral currents in the 1970's, and arriving at the simultaneous discovery of the top quark and the indirect extraction of its properties from precision electroweak measurements.

There remains one missing element in the SM: the Higgs boson. This unique elementary spin-0 particle is invoked to explain the generation of masses of the electroweak spin-1 bosons and of the fermions. It is possible to formulate alternatives to the SM that are consistent with the available data and mimic its role by other means. Therefore finding the Higgs boson or refuting this concept constitutes a primary goal of investigations at the Tevatron and the LHC. Its observation would set the seal on the SM as the most ambitious and successful attempt to unveil the laws that govern the behaviour of the Universe, rewarding the efforts of generations of natural philosophers and scientists; its refutation would constitute a revolution with long-lasting consequences.

The scientific value of the SM rests not only on its ability to describe the fundamental properties of the elementary components of the Universe. It also follows from its success, when used together with astrophysical and cosmological models based on general relativity, in describing the properties of the Universe on cosmological scales. For example, the weak interactions described by the SM and the existence of three families of light neutrino allow us to predict the detailed composition of the nuclear species produced during the early stage of the Universe, within the first few hundred seconds after the Big Bang. The agreement of these predictions with the observations provides a strong validation of the over-all theoretical framework used to describe the early Universe, a validation that has opened the way to quantitative analyses of the rich data sets that modern observational cosmology is collecting. These studies aim at linking the origin of other features of the early Universe with the detailed pattern of particles and their interactions.

The more our confidence in the SM grows, the stronger the need becomes to explore its conceptual foundations, the origin of its postulates, and its possible flaws. These three topics are intimately linked, and their exploration will redefine the frontiers of our knowledge.

Given the immense body of phenomena accurately described by the SM, it is natural to ask: Does the SM provide an answer to every question we can pose about the fundamental properties of the Universe, or should we consider it just as an effective theory, doomed to fail when probed more deeply? In the rest of this chapter we shall address this question from both the experimental and theoretical points of view.

## II-1.1 Observational shortcomings of the SM

There are today three compelling and firmly established observational facts that the SM fails to account for: neutrino masses ($\nu$M), the existence of dark matter (DM), and the size of the baryon asymmetry of the Universe (BAU). For each of these observables, the SM makes very specific statements, failing however to reproduce the experimental evidence, as briefly discussed here.

Arguably the most important experimental particle physics result in the last ten years has been the confirmation that neutrinos have mass. Since the early 1970's, experiments have succeeded in detecting electron neutrinos produced in the Sun, thus confirming our concept of how the Sun functions, but the observed flux of electron neutrinos on the Earth has always been consistently lower, by a factor 2-3, than expectations. This long-standing solar neutrino puzzle was enhanced in the 1990's by the observation of a similar deficit in atmospheric muon neutrinos. It was finally solved at the turn of this century, by the simultaneous measurement of charged- and neutral-current reactions of the neutrinos from the Sun and by experiments performed with man-made neutrinos, either from reactors or from an accelerator. The picture is currently consistent with the three neutrino families undergoing oscillations, a coherent quantum phenomenon on the scale of hundreds to millions of kilometres. This can only happen if neutrinos have masses and mix. Neutrinos of the SM are massless and the incorporation of neutrino masses requires either a new ad-hoc conservation law or new phenomena beyond the present framework.

There is no object predicted by the SM, whether elementary or composite, that can account for the amount of DM required by the recent cosmological and astrophysical observations. The successful description of nucleosynthesis alluded to earlier fixes the total amount of baryonic matter present in the Universe, independently of its state of aggregation. This rules out dark bodies such as black holes, planets or brown dwarfs as DM candidates. A minimal extension of the SM granting masses to the known neutrinos is not sufficient either, since the dynamics of light neutrinos in the early Universe cannot explain the formation of large-scale structures (galaxies and galaxy clusters).

A mechanism for the BAU is present in the SM: it is based on the CP violation in the quark sector, and on the departure from equilibrium realized at the time of the electroweak phase transition, when the temperature of the Universe fell below the Fermi scale, leading to the phase in which SU(2)xU(1) symmetry is broken. While conceptually valid, this mechanism fails quantitatively, owing to the observed values of the parameters that control the size of the resulting BAU: the size of CP violation (too small in the SM), and the Higgs mass (LEP's limits giving a mass too large for a first-order phase transition strong enough to allow the survival of the BAU at lower temperatures). In addition, even if we were ready to accept a very unnatural and highly fine-tuned primordial asymmetry between matter and antimatter, this would be washed out during the early, hot phase of evolution.

To summarize: it is precisely our confidence in the SM and our ability to calculate its consequences that lead us without a shadow of doubt to the conclusion that the SM is incomplete, and new phenomena must be anticipated.

## II-1.2 Conceptual limitations of the SM

Like any mathematical construction, the SM relies on a set of axioms (albeit based on experimental inputs), which are part of its definition rather than a consequence of its predictions. For example, the fermion masses, as well as their mixing angles and the CP phase, assume a great variety of numerical values that are *a priori* arbitrary, and must be determined experimentally. Similarly, the relative strengths of the three fundamental interactions, electromagnetic, weak and strong, are free parameters of the SM, fixed by matching to experimental data. The question naturally arises as to whether some, if not all, of these parameters, arbitrary within the SM, may have a dynamical origin in a more fundamental theory.

On a deeper level, the SM cannot provide any answer to questions about its very structure: why are there three families of quarks and of leptons? Why SU(3)xSU(2)xU(1) as a gauge group? Are there additional gauge interactions? Why should the electroweak symmetry be broken? Why the asymmetry between left and right and under time reversal? These questions could find answers, or be reformulated in dynamical terms, in field-theoretical extensions of the SM, such as grand unified theories (GUTs), where one assumes that the SM gauge group results from the breaking of a larger, unified symmetry at scales of the order of $10^{15}$ GeV. In such extensions, relationships are commonly found between the gauge couplings and between the different particle masses and mixing angles; furthermore the lepton and/or baryon number are not absolutely conserved, and the smallness of neutrino masses arises in a natural way, while the decay of the proton could be observable within the scope of conceivable detectors.

At a yet deeper level, we encounter issues that touch more profoundly on our notion of the Universe: Why do we live in 3+1 dimensions? What is the origin of the by-now established cosmic inflation, and of the observed small value of the cosmological constant? What was the origin of the Big Bang? What is the quantum structure of space-time at the shortest-distance scales? It is likely that the answers to these questions will require a radical departure from our field-theoretical framework of particle physics, with far-reaching intellectual and experimental consequences.

Theories extending beyond the SM (BSM), capable of addressing at least some of these questions and at the same time containing the solution to the above-mentioned flaws, are natural candidates for theoretical and experimental study. Their exploration is the continuation of the long quest for the ultimate understanding of nature, and is therefore a priority for the scientific community. In more pragmatic terms, the discoveries made so far in this quest have contributed to shaping our lives in the most dramatic way, and it is plausible to expect that the same will eventually follow from the future revolutions in particle physics as well.

## II-2 Looking forward and back

The coincidence of an excellent, but incomplete, theory (the SM), very concrete experimental expectations (Is there a Higgs boson?) and puzzles (What is the origin of neutrino masses, of DM and of the BAU?), together with very deep, fundamental open questions, sets the stage for an exciting new era in physics. It is not an exaggeration to compare the scientific phase we live in with the situation facing physicists at the dawn of the 20th century. Their understanding of individual classes of natural phenomena was accurate and compelling. Electromagnetism was a complete, elegant and predictive theory of electric, magnetic and optical phenomena. Likewise, mechanics had long been established as a solid basis for the formulation of dynamical principles, explaining the motion of earthly and celestial objects. Chemistry and thermodynamics were addressing the remaining realms of physical processes.

In spite of their respective successes, a few discrepancies with data and a few conceptual problems and inconsistencies between the different theories were noted here and there. For example, electromagnetism was not compatible with the Galilean transformation laws. It was the matching of the conceptual problems with the observed discrepancies that led to the major revolutions in physics of the last century, relativity and quantum mechanics. Some of the issues that led to those developments bear a close resemblance to the questions faced today by particle physics. It may provide a source of inspiration and motivation to refer occasionally to these analogies when analysing the possible paths of evolution for our field today.

### II-2.1 The structure of space-time

The first major revolution of the 20th century was the new vision of space and time. It resulted from the attempt to reconcile the symmetry properties of classical mechanics and electromagnetism. Today we face a similar need to reconcile two major building blocks of our description of nature, both individually successful in describing their respective fields of application: quantum mechanics and gravity. The most promising theories in this direction require equally revolutionary modifications of our concept of space-time: supersymmetry and extra dimensions.

According to supersymmetry, the standard commuting coordinates of space-time are accompanied by one or more directions parametrized not by bosonic, but by fermionic (anticommuting) coordinates. However intangible, they are there. Shifts from ordinary space-time towards these fermionic directions change the spin of a particle by half a unit. The product of two such shifts leads to a displacement in ordinary space-time.

Whereas the combination of quantum mechanics and special relativity required the doubling of the particle spectrum – to each particle there corresponds an antiparticle – supersymmetry requires the introduction of a superpartner for each SM state. The discovery of supersymmetric particles, which could be experimentally as close as the turn-on of the LHC, would therefore force a new revision of our idea of space-time. As a by-product, we would also find a deep origin for one of the requirements for the existence of stable atomic matter: fermionic particles, which are required in supersymmetry.

The changes to our picture of space-time would be even more far-reaching if one were led to consider the extension of supersymmetry into a superstring theory, where additional, and possibly detectable, spatial dimensions are required. Even in the absence of supersymmetry, the existence of extra dimensions is a possibility that is not ruled out by available data and needs to be investigated experimentally.

The non-invariance of the fundamental interactions under the discrete symmetries of space-time, parity (P) and time reversal (T) has played a central role in the development of the SM, and has consequences that are even crucial for the existence of life. The invariance of the combined set of discrete symmetries, P, T and charge conjugation (C), is known to be an exact property of local quantum field theories respecting Lorentz invariance, and therefore of any extension of the SM based on standard field theory. The discovery of signals of CPT violation (such as a non-zero mass difference between a particle and its antiparticle) would therefore point to short-distance modifications of the structure of space-time even more radical than the mere existence of extra dimensions, and would be likely to provide a unique experimental input into the understanding of quantum gravity.

## II-2.2 Electroweak interactions and symmetry breaking

As mentioned before, recent experiments have tested the electroweak (EW) sector of the SM with unprecedented accuracy. The flavour-universality of charged and neutral weak interactions has been tested to better than 1% in both the quark and lepton sectors. The non-abelian gauge nature of the couplings among massive vector bosons W and Z has been verified at LEP2. The effects of quantum corrections to the EW couplings of fermions have been observed. Their consistency with the predictions of the SM has been successfully demonstrated by the discovery of the top quark at the Tevatron, and by the agreement of its measured mass with that required to fit all EW precision data. However, with the crucial remaining ingredient of the SM, the Higgs boson, still missing, the mechanism of electroweak symmetry breaking (EWSB) remains to be established, and is therefore today the most burning question of particle physics.

EWSB directly affects the gauge sector of the SM, but is also responsible for the generation of particle masses, and indirectly for the differentiation between flavours. It therefore provides an important link between the two main elements of the SM, the gauge and flavour structures. This becomes more apparent in several BSM theories, where, for example, radiative or dynamic EWSB is triggered by the large value of the top mass.

The SM defines unambiguously the mechanism of EWSB and its consequences, and all experimental data are consistent with the existence of a Higgs boson, suggesting a mass larger than 114 GeV and, in the absence of new physics, smaller than about 200 GeV. There are nevertheless good reasons for theorists to suspect that BSM physics should play a key role in the dynamics of EWSB. The radiative contribution to the Higgs mass grows linearly with the scale at which the integration over short-distance quantum modes is cut off, leading to the following numerical result:

$$\delta m^2{}_H \cong (115 \,\text{GeV})^2 \, (\Lambda/400\,\text{GeV})^2 \,,$$

where $\Lambda$ is the cut-off scale. This contribution is dominated by the effect of virtual top antitop quark pairs, which interact very strongly with the Higgs boson because of the large top mass. As the cut-off is pushed to infinity, a huge and negative bare Higgs mass squared needs to be introduced by hand, to cancel this divergent radiative contribution and leave a finite value equal to the physical mass. While this regularization procedure is consistent with the renormalizability of the SM, extremely accurate fine-tuning is required to keep $m_H$ in the range of few hundred GeV, if we want to allow the cut-off to become as large as the only natural upper scale of the SM, namely the Planck mass.

This problem is known as the hierarchy problem of the SM. It might appear an academic issue, but it is worth recalling that the consideration of a similar problem of the last century, the self-energy of the electron, played a role in the development of QED. In that case, the electron mass receives a contribution from the electric field, proportional to the inverse of the electron radius, and linearly divergent if we assume a point-like electron. With the Higgs boson, the role of the electromagnetic field is replaced by the interaction with the field generated predominantly by virtual pairs of top antitop quarks.

In the case of the electron, the problem is solved by the inclusion of the positron. New contributions to the electron self-energy due to the positron cancel the classical ones, and reduce the linear divergence to a logarithmic one, which does not require fine-tuning. One can think of the positron as the new physics which intervenes to regulate the bad ultraviolet behaviour of the effective, non-relativistic theory of the electron. Its mass is of the order of the scale (the electron mass) above which the mass renormalization requires strong fine-tuning.

For the hierarchy problem of the SM, a similar solution is possible via the introduction of new states whose contributions to the Higgs self-energy cancel the leading linear divergence. As in the case of the positron, we expect their mass to be of the order of the scale at which the radiative corrections start to exceed the Higgs mass itself, namely a few hundred GeV. The excellent agreement of the SM with precision EW measurements, however, sets very stringent constraints on the possible existence of new particles with masses of the order of few hundred GeV. As a result, the search for extensions of the SM that can alleviate the divergence of the Higgs self-energy is extremely constrained.

A few models satisfying these constraints have been introduced in the past few years. They provide a rich terrain for exploration at the future experimental facilities. Among these models, we find supersymmetry, dynamical symmetry breaking induced by new strong interactions, little-Higgs theories, and theories based on the existence of extra spatial dimensions. In most of these cases, new particles with masses in the TeV range are predicted. In supersymmetry, for example, the spin-0 partner of the top quark (the stop) plays the role of the positron in QED: its coupling to the Higgs boson generates contributions that cancel the linear divergence of the Higgs self-energy due to the top quark. In little-Higgs theories this role is played by a heavier partner of the top quark, with a mass of the order of a TeV. In this case, extra massive gauge bosons are also present, with masses in the range of 1 to a few TeV. In some extra-dimensional theories, the Higgs boson itself is a 4-dimensional scalar leftover of a gauge boson in higher dimensions, and its mass is protected by gauge symmetries. Here new states appear in the form of Kaluza Klein excitations of SM particles.

Whatever the correct theory may turn out to be, the hope for the manifestation of new phenomena at the TeV scale is very strong. While the Higgs boson itself is expected to be much lighter than this, the same is not true of the other particles that would complete the EWSB sector in BSM theories. The continued exploration of physics at the TeV scale and above remains therefore our best possible tool to shed light on the EWSB phenomenon and to identify the new theoretical paradigms that will guide us toward the solution of some of the fundamental problems outlined above.

## II-2.3 The flavour problem and CP violation

We tend to associate the origin of the SM with the gauge principle and with the consolidation of Yang Mills interactions as unitary and renormalizable quantum field theories. We often forget that flavour phenomena have contributed as much as the gauge principle in shaping the overall structure of the SM.

It is the existence of flavours (in both the lepton and quark sectors) that gives the SM its family and generation structure. The arrangement of the quarks in EW doublets is needed for the suppression of flavour-changing neutral currents (FCNCs), which led to the GIM mechanism and to the prediction of the charm quark. The experimental study of kaon decays led to the discovery of CP violation, and to the three-generation quark model. Just as $K^0$ mixing played a role in setting the mass range for charm, $B_d$ mixing was the first experimental phenomenon to correctly anticipate the large value of the top quark mass. Furthermore, the observation of neutrino masses has provided the first concrete and incontrovertible evidence that the SM is incomplete. At the very least, this calls for an extension of the SM describing sterile right-handed neutrinos; more ambitiously, as reviewed below, neutrino masses may become a window on physics at the grand unification scale.

In the quark sector the description of flavour phenomena provided by the SM is as successful as the SM predictions in the gauge sector. With the large number of precise measurements of many different B-meson decay modes obtained in the B-factories, the CKM picture of mixing and CP violation is now verified at the few per cent level. The lengths of the sides of the unitarity triangle are known today with good accuracy from the measurement of $|V_{cb}/V_{ub}|$, $\Delta m(B_d)$, and the recent determination of $\Delta m(B_s)$. While these three quantities are CP-conserving, the extracted values of the triangle sides already imply non-zero angles, and therefore CP violation. Quantitatively, the directly measured CP violation in several channels, in both the K and B sectors, is perfectly consistent with the SM, and in particular with one single complex phase as the dominant – if not the only – source of CP violation in the quark sector. As a result, alternatives to this picture are strongly constrained. As already pointed out, the smallness of FCNCs and the patterns of CP have been built into the structure of the SM from the outset. In the quark sector, they result from the unitarity of the mixing matrix and from the small mixing between heavy and light generations. In the lepton sector, it is the smallness of the neutrino masses that suppresses possible evidence of mixing and CP violation for charged leptons.

There is absolutely no guarantee that the above properties survive in extensions of the SM. For example, in supersymmetry $B^0$ and $K^0$ mixings are greatly enhanced if the squark mixing matrix is not aligned with that of the quarks. In addition, a large number of new CP-violating phases will typically be present, in both flavour-changing and flavour-conserving couplings of squarks, gluinos, and possibly Higgs particles. This is *a priori* a welcome feature of BSM models, as it provides the opportunity to generate an amount of CP violation large enough to reproduce the BAU.

On the other hand, in a model where squark flavours are maximally mixed, superpartner masses should be larger than several TeV; this would sufficiently suppress these contributions without clashing with the data on mixing, and on CP violation in flavour-changing transitions and electric dipole moments. The day that supersymmetry (or some other form of new physics) is discovered at mass scales below or around a TeV, say at the LHC, understanding how the suppression of these processes is achieved will be a major step towards the identification of the new phenomena. The measurement of very rare FCNC decays such as $K^0_L \to \pi^0 \nu\nu$, $K^+ \to \pi^+ \nu\nu$ or $B_{d,s} \to \ell^+\ell^-$, the detection of new CP-violating phases in heavy-flavour decays, and of electric dipole moments of neutrons, electrons and muons, will then provide precious information on the effect of these new phenomena on low-energy physics, and perhaps give important constraints on yet unobserved heavy particles. Of particular interest will be the interplay with flavour-violating processes in the lepton sector, as discussed in the next subsection.

## II-2.4 Neutrinos and lepton-flavour violation

The observation of neutrino oscillations, and the consequent evidence that neutrinos have mass, is the first direct signal of physics beyond the SM. Neutrino masses could in principle be incorporated in a trivial extension of the SM, by adding a right-handed neutrino state $N_\ell$ for each known neutrino flavour. An SU(2)xU(1)-invariant coupling between the Higgs field, the left-handed lepton doublet, and $N_\ell$ can then be added to the SM lagrangian, giving mass to the neutrino after EWSB. This is a coupling of the same type as that giving mass to the up-type quarks, since the left-handed neutrino has weak isospin +1/2. The consequences of this scenario are twofold: first, $N_\ell$ is completely neutral, since it must have zero weak isospin and zero hypercharge. Therefore it is totally decoupled from any gauge interaction. Secondly, the Yukawa coupling for its interaction with the Higgs field should be exceptionally small, of the order of $m_\nu/m_f \cong 10^{-12}$.

Such a scenario, while phenomenologically acceptable, creates more problems than it solves. What is the role in nature of such an idle object as $N_\ell$? What is the origin of such a minuscule Yukawa coupling? Such a solution would lead to no progress in our quest for a deeper understanding of the origin of mass and of the flavour structure of the SM. In contrast, it is possible to identify frameworks in which neutrino masses are naturally linked to new phenomena occurring at very high energy scales, phenomena which, in turn, have the potential to shed light on some of the other big questions of particle physics.

The simplest and most promising alternative to the trivial extension of the SM discussed above is the so-called seesaw mechanism. In this picture, a mass term for the right-handed neutrino $N_\ell$, exists and can be arbitrarily large. It is SU(2)xU(1)-invariant and not the result of EWSB. The mixing with the left-handed neutrino $\nu_L$ induced by the Yukawa coupling, after diagonalization, leads to a value of $m_\nu$ of the order of $m^2/M_N$, where $m$ is the left-handed neutrino mass acquired via the Higgs mechanism. Assuming a natural value for $m$ of the order of the charged-fermion masses, thus restoring the symmetry between the Higgs couplings to charged and to neutral fermions, leads to values for $M_N$ around $10^{15}$ GeV. Within the seesaw mechanism, therefore, one is led to infer a possible connection between neutrino masses and physics at the GUT scale. This connection is strengthened by the fact that the state $N_\ell$ finds a natural place within the particle classification of several GUT models, such as those based on an SO(10) unified symmetry. It is remarkable that the lightest massive particles known in nature might derive their mass from phenomena taking place at the highest energies, and that their exploration could help in extracting indirect information on energy scales so remote from our laboratory experience.

Furthermore, in this context, massive neutrinos could exhibit a new property of nature. While the electric charge forces all other known fermions to be distinct from their antiparticles, the chargeless neutrinos could either be of Dirac type (with the antineutrino different from the neutrino) or of Majorana type, in which case the left-handed neutrino and antineutrino are exactly the same object. This latter possibility would induce a much richer phenomenology, such as neutrinoless double-beta nuclear decays, which would provide the first experimental evidence of a small fermion-number violation.

If we accept the role of GUT models in particle physics, several additional consequences arise. To start with, quantitative studies of the unification of strong and EW coupling at the GUT scale strongly imply the supersymmetric nature of these theories. Assuming a GUT with supersymmetry and neutrino masses, remarkable relations appear between the properties of neutrinos and the flavour structure of quarks and charged leptons. For example, neutrinos and up-type quarks, being the isospin +1/2 members of weak doublets, must have the same Yukawa coupling at the GUT scale. This results in a prediction for the hierarchy of neutrino masses similar to that of the up, charm and top quarks. In addition, the large mixing among neutrinos of different families leads unavoidably, via radiative corrections, to potentially large mixings between the supersymmetric scalar partners of the charged leptons. This generates lepton-flavour-violation phenomena such as decays of a muon into an electron-photon pair, with rates that can be accessible to the forthcoming generation of experiments. In addition, a large neutrino mixing induced by the mixing of right-handed neutrinos could imply large mixings among the right-handed components of strange and bottom squarks, leading to observable consequences in the phenomenology of B mesons.

While these are specific examples in the context of supersymmetric theories, it is a general fact that models inspired by the desire to provide a natural explanation of the small neutrino masses and large mixings ultimately lead to an immense range of profound implications, not least the possibility that CP violation in neutrino interactions could provide an explanation for the BAU. To assess the viability of this hypothesis, and to establish firmer connections between neutrinos and the other sectors of the SM or its extensions, a more complete knowledge of neutrino properties is required, starting from the determination of their absolute mass scale, a more accurate measurement of the mixing angles, and the detection of possible CP-violating phases. The last fermions to manifest a non-trivial flavour structure could become the first to point towards an explanation of some of the leading mysteries of particle physics.

## II-2.5 Cosmic connections

The connections between particle physics, astrophysics and cosmology are many and keep multiplying. The properties of elementary particles and fields control the past evolution of the Universe, its present condition and its future destiny. Consequently, theories and observations in the two fields often have implications for each other.

A wide range of cosmological data suggests that the Universe currently consists of roughly one-third cold dark matter (CDM), two-thirds dark energy (DE, a component that exerts negative pressure, tending to accelerate the expansion of the Universe), and only a few per cent of familiar baryonic matter. The situation with regard to DM is reminiscent of the problem of nuclear beta-decay in the 1920s. The rotation curves of galaxies, like the kinematics of beta-decay, defy the laws of mechanics unless an invisible component is participating in the process. This cosmic component, like the neutrino, is elusive but should be possible to detect directly, with dedicated experiments. Direct searches for cosmic DM are indeed already under way, with ever-increasing sensitivity. Again like the neutrino, the DM particle can also be hunted in high-energy collisions. In supersymmetric models it is the superpartner of the gauge bosons, as the neutrino was the gauge partner of the leptons, while in models with extra spatial dimensions it is again a gauge-boson partner. In both cases, the DM particle should be produced copiously, either directly or indirectly, in particle collisions at sufficiently high energies. The elucidation of the nature and properties of dark matter by collider experiments would surely be an outstanding example of the cross-fertilization of particle physics and cosmology.

An alternative possibility, also considered for beta-decay, is that the laws of mechanics have to be modified (MOND: modified Newtonian dynamics) and no new DM particle exists. However, in that case one has to find a way to fit the modification into a consistent over-all framework, and to explain a range of other apparently related observations (e.g. gravitational lensing and large-scale structure formation).

It is possible that the evidence for DE is also pointing to a small modification of existing laws, but it seems likely that here again a new type of field/particle is involved. A scalar field with the appropriate self-interaction ('quintessence') would naturally lead to a time-dependent DE density, an issue that will be addressed by future supernova surveys. There is the possibility that the same field could have driven the inflation of the Universe in an earlier epoch. A time-independent DE density would suggest a connection with the cosmological-constant problem, a deep mystery at the heart of quantum field theory. It is hard in any case to understand why the vacuum energy density of the known quantum fields should not completely overwhelm the observed DE density of about $10^{-47}$ GeV$^4$. In supersymmetric theories some cancellation between contributions occurs naturally, but the remainder is still too large by at least 60 orders of magnitude. There is hope that light will be shed on this contradiction if new phenomena discovered in collider experiments point the way beyond quantum field theory.

In some models with extra spatial dimensions, the scales of string dynamics and strong gravity are indeed within the range of the coming generation of colliders. That would imply an unimaginably rich prospect of new phenomena such as stringball and black-hole production. The parameters of such models are constrained by astrophysical data such as the neutrino pulse length from supernova 1987A and the diffuse cosmic gamma-ray background. These already rule out models with one or two 'large' extra dimensions accessible at the LHC, and restrict the Planck scale to be greater than 7 TeV for three extra dimensions. These models also have implications for the early Universe which have yet to be fully explored and may well yield stronger constraints.

The cosmic abundances of the lightest elements, formed in the first few minutes after the Big Bang, already place interesting constraints on particle physics. For the most part they are in agreement with expectations based on the SM. The anomalously low abundance of $^7$Li, however, may be an indication of physics beyond the SM.

Another field of strong overlap between particle physics and astrophysics is the study of high-energy cosmic rays. If cosmic rays are reaching the Earth with energies higher than that at which the microwave background becomes opaque to extragalactic Standard-Model particles (the GZK cut-off), then either they are exotic particles themselves or else they come from the decay of exotic massive local objects. Even if the GZK cut-off is satisfied, the composition and production mechanism of the highest-energy cosmic rays will pose a challenge to both particle physics and astrophysics. Information from collider experiments is also indispensable for the reliable deduction of cosmic-ray energies from their interactions in the atmosphere.

## II-2.6 A deeper understanding of SM dynamics

A continuing goal of particle physics is to probe more deeply the dynamical structure of the Standard Model. This is especially true of the electroweak sector since the mechanism of electroweak symmetry breaking remains to be fully explored. In particular the Higgs field, which fulfils the multiple roles of symmetry breaking (thereby giving mass and polarization states to the vector bosons) and providing mass to fermions, is still an unconfirmed hypothesis. The Higgs boson must be discovered and its interactions verified, including its self-interactions, which reflect directly the parameters of the Higgs potential. This is probably the area of Standard-Model physics that is most likely to yield clues to physics beyond the SM.

In the strong-interaction sector of the SM, our understanding of dynamics has made huge advances in the forty years since the discovery of the point-like substructure of hadrons. A vast range of experimental evidence, from scaling violation to jet production, has established QCD as the best description of these phenomena. However, there remains much scope for further understanding of QCD, both in its own right and as a testing ground for concepts and techniques in strongly coupled quantum field theory. The deduction of hadronic spectra and properties from first principles remains the central objective of lattice QCD. The spin structure of the nucleon, and of hadronic interactions more generally, is being probed with increasing precision and remains to be fully understood. The study of low-energy hadronic phenomena sheds light on aspects of non-perturbative physics that could uncover new effective degrees of freedom (e.g. diquarks) and also provide inspiration for handling fundamental problems in other areas, for example if BSM physics involves a strong-coupling theory such as technicolour. Heavy Ion collisions provide a means of studying QCD dynamics at high temperatures and densities, a rich field of study in its own right, and essential for the understanding of neutron stars and the early Universe.

Since almost all physics to be explored at the high-energy frontier involves hadrons in one way or another, a better understanding of QCD is also necessary across the whole range of frontier exploration. Improved parton distributions and jet fragmentation functions are needed for signal and background predictions at the LHC. The tools of perturbative QCD must be refined and validated to predict the backgrounds to new phenomena accurately. The various QCD tools used in the context of heavy-flavour physics also require development and validation, so that improved decay form factors and hadronic matrix elements will allow more accurate extraction of electroweak and new-physics parameters from B, D and K meson decays. The goal here is to match the accuracy of the theoretical tools to the projected accuracy of the measurements expected at the future flavour factories. The same considerations apply to other low-energy probes of new physics, such as the magnetic moment of the muon and nuclear electric dipole moments.

One outstanding mystery of QCD is why strong interactions do not violate CP symmetry, at least at a level comparable to the weak interaction, when CP-violating interactions are not forbidden by SU(3) gauge invariance. The most popular cure for this 'strong CP problem' requires the existence of a new, ultralight spin-0 particle, the axion, which has yet to be found experimentally. If they exist, axions should have been created in abundance in the early Universe and could even constitute the dominant form of dark matter.

# II-3 Preparing for future discoveries

To the vast array of conceptual themes characterizing the current status of particle physics, there corresponds a similarly varied panorama of experimental initiatives, which constitute an indispensable component of future progress. Three broad areas have emerged, namely physics at accelerators, without accelerators, and the interdisciplinary field of particle astrophysics. These will be briefly reviewed here, together with R&D on the indispensable tools underpinning progress in particle physics: accelerators and detectors.

## II-3.1 Physics at accelerators

Most of the contributions to this Briefing Book focus on this area, and therefore only a very sketchy list of accelerator-based experimental programs and facilities, either existing or under study, is given in this section. Here, the goal is to provide an overview of the initiatives already being undertaken or considered.

Accelerators on the high-energy frontier are still the indispensable means to tackle many of the most exciting questions in particle physics. From 2007 onwards, the LHC and its general-purpose detectors at CERN will begin exploring a new, large phase space. We anticipate that this will lead to the discovery of the Higgs boson, to the first direct exploration of the nature of the EWSB mechanism, and that it will also allow us to discover the first signals of what lies beyond the Standard Model.

Some of the possibilities for BSM physics have been mentioned in this introduction, but it must be stressed here that no-one can confidently predict now which specific model will emerge, nor the precise value of the energy scale at which it will become manifest. Different scenarios can therefore be foreseen, where the more complete understanding of the new physics will require one or more of several alternative accelerator facilities.

Depending on the nature of the discoveries made at the LHC, higher-statistics studies of these phenomena would naturally call for an increase in the LHC luminosity. This should take place roughly three-to-five years after the LHC has reached its nominal luminosity of $10^{34}$ cm$^{-2}$s$^{-1}$. This upgrade – referred to as Super-LHC (SLHC) – should bring the luminosity to about $10^{35}$ cm$^{-2}$s$^{-1}$, allowing for the rarest phenomena to be studied in greater depth, and extending the mass range over which new physics can be detected by about 20-30%.

The accurate measurement of Higgs-boson properties, as well as the study of most new phenomena discovered at the LHC, will need the cleaner environment of electron-positron collisions to be addressed precisely and completely. In the case of the Higgs boson, and of new particles below 400-500 GeV, this programme could be carried out by the International Linear Collider, whose design is already being addressed by a world-wide collaboration of physicists in the context of the Global Design Effort. More generally, it can be stated that for essentially every BSM-physics scenario involving particles in the ILC energy range, detailed research programmes have been formulated, and they lead to remarkably definite conclusions about many features of the new physics – be it in the Higgs, SUSY, or other domains. Thus an electron-positron collider of the appropriate energy reach appears to be an indispensable major initiative to complement the LHC.

New processes observed at the LHC in the highest accessible mass region, or indications of the proximity of new mass thresholds, would lead to the consideration of substantial increases in the energy of either the LHC or the ILC. Achieving multi-TeV collisions between elementary objects (partons or leptons) can be envisaged today with two approaches: a 3-4 TeV electron-positron collider, applying the promising CLIC two-beam acceleration concept, or an energy-doubled LHC, using magnets made of new superconducting materials, which are currently being investigated in several laboratories. In the longer term, physicists have been considering muon colliders and very-large hadron colliders. The former have been proposed as unique tools to study multiple Higgs scenarios and, for the more distant future, as the ultimate probe of energies in the 10 TeV range and possibly above. The latter, operating with beam energies in the 100 TeV range, would push the energy scale of exploration up by another factor of 10.

The discoveries of the last decade in neutrino physics – oscillations, neutrino masses, the surprisingly large values of the mixing amplitudes, and the corresponding possibility of observing CP violation in the lepton sector – point to a possible new window on mass-generation mechanisms and more generally on BSM physics. This makes the development of more advanced neutrino facilities imperative. In Europe, two main paths are being explored at present: the first consists of a high-power, low-energy (0.1-1 GeV) neutrino beam from pion decay (known as a *superbeam*), combined with *beta beams*, providing a pure beam of electron neutrinos with a similar spectrum; the second (a *neutrino factory)* would use a stored muon beam to provide high-energy neutrino beams containing electron and muon neutrinos with opposite leptonic charges. The investments needed may be on a somewhat smaller scale than that of the highest-energy colliders. This should make it possible to develop this line in a regional or a global framework.

Flavour physics in the quark sector remains an important research direction. After establishing the existence of CP violation in b-quark decays, *B*-factories have not finished their mission, and the LHC experiments will soon complement and extend these studies. Working at the intensity frontier and with ever-increasing luminosities, currently operating facilities may well succeed in finding decays with rates above the SM predictions. Alternatively, superfactories with luminosities in the $10^{36}$ cm$^{-2}$s$^{-1}$ range, or above, may be needed. A parallel and closely related direction is that of kaon rare decays, pursued either with lower-energy, high-luminosity colliders or with intense beams from stationary-target accelerators.

Existing and future accelerator laboratories also provide the venue for experimental programmes that push the envelope on crucial parameters of the SM framework and thus may lead to – or indicate the path to – fundamental discoveries. Examples of such programmes are muon (g-2) experiments, searches for lepton-flavour-violating processes such as $\mu \to e\gamma$, and precision experiments with very cold antiprotons.

## II-3.2 Particle physics without accelerators

As the energy range of interest to particle physics has expanded from the limits set by the next generation of accelerators all the way up to the Planck scale, the discipline has gone beyond the limits set by current accelerator technology, in order to investigate phenomena in which much higher energies may come into play.

Neutrinos are one instance of these developments because, as already discussed, the discovery of neutrino masses, the long-standing issue of the Dirac or Majorana nature of neutrinos, the seesaw concept and related theoretical ideas, all conspire to add intense interest to a multifaceted programme on the physics of neutrinos, in which non-accelerator experiments will play a crucial role and are already requiring substantial resources.

It is anticipated that continued exploitation and extension of existing experiments using solar or atmospheric neutrinos and neutrino-oscillation experiments at nuclear reactors will provide complementary means of measuring the $\theta_{e3}$ mixing angle on an early time-scale.

While running and planned experiments are expected to produce rapid progress in measuring the neutrino mass differences, the overall neutrino-mass scale remains poorly known, with the only direct limits coming from beta-decay end-point experiments. The remarkable precision reached in this type of measurement may make it possible to push mass limits down to levels similar to the mass difference scale. Cosmological arguments are also providing ever-tighter constraints on the sum of the neutrino masses.

Finally, neutrinoless double-beta decay, if detected, would constitute one of the most far-reaching discoveries, with implications way beyond neutrino physics *per se,* because of the indications it would provide about the origin of masses. This justifies the considerable efforts being devoted to the development of novel detection techniques for this purpose.

Another aspect of physics beyond the accelerator energy domain can be explored by searching for nucleon decay, the tell-tale signal of grand unification and the only currently foreseen probe of energy scale beyond $10^{15}$ GeV. The existing limits suggest that future facilities should contain of the order of $10^{35}$ nucleons, roughly corresponding to the megaton scale. Even in this case, the signal event rates expected in most theoretical scenarios are only a few events per decade. Hence the detectors must be located underground (to be shielded from cosmic rays), and taking all necessary measures to reduce natural and instrumental backgrounds to the required, very low level.

Direct detection of massive relic dark-matter particles is another very active line of particle physics that does not require accelerators and must be pursued underground. In this case, the severity of the background-suppression requirements is enhanced by the need to detect the very small signals given by nuclear recoil. Again, these delicate observations have triggered a variety of imaginative detection techniques, cryogenic detectors among them.

The requirement of underground laboratories is shared with some of the lines of research on neutrinos, both with and without particle beams. The possibility of setting up larger underground labs, located so as to permit investigation of oscillations of neutrinos produced at accelerators, detection of supernovae or cosmological neutrinos, and rare processes such as proton decay, is under examination in the particle-physics community.

The axion – mentioned earlier in connection with the strong CP problem – is another DM candidate that has been searched for in several dedicated experiments. Recently, such searches have extended to light scalar particles with weak couplings to the photon. They are mentioned here because intriguing hints of possible signals are triggering proposals for small but novel initiatives.

## II-3.3 Astroparticle physics and cosmology

Yet another way to explore the energy range beyond accelerators is to use the cosmos either as an accelerator or as a source of particles that cannot be produced on Earth. This approach is reminiscent of the days – before the 1950's – when cosmic rays were the main source of yet-undiscovered particles. Over the last two or three decades, this area of research has grown and currently involves a significant fraction of the particle physics community. We have witnessed the birth of the interdisciplinary field of *astroparticle* *physics,* in which the themes span astrophysics and particle physics, with an important interface with cosmology. As is typical of new interdisciplinary fields, there is frequent collaboration with specialists from less closely related disciplines, such as geophysics, oceanography, etc.

Traditionally, the efforts of astroparticle physicists have been focused on three lines, characterized by the particles being detected: the (charged) majority component of cosmic rays, gamma rays and neutrinos.

The long-standing but still fascinating issues of the acceleration and composition of the highest-energy cosmic rays has already been briefly mentioned above. The experimental fact driving the field is that the Universe accelerates protons to energies up to, and perhaps beyond, $10^{20}$ eV. Whether such extreme energies are reached by gradual acceleration over astronomically large distances ('bottom-up' processes) or produced by 'top-down' mechanisms, involving as-yet undiscovered particles and energy scales, is one of the fundamental questions. Very large-area facilities, of which the Auger project is the latest and most ambitious, lead the progress on the front of the highest energies and largest observation areas.

The highly complementary but observationally very different fields of high-energy gamma ray and neutrino astrophysics also have a very rich scientific programme. The common feature of these neutral, long-lived particles is that they point back to their sources. Neutrinos in particular – despite the great efforts required for their detection – may carry information from the core of some of the most active regions of the Universe.

In gamma-ray astrophysics, the detectors span a very broad energy range, from about 1 keV to tens of TeV, being located on satellites (to detect keV to $\approx$ 10 GeV gammas) or on the ground (from $\approx$ 50 GeV upwards, mostly with Imaging Atmospheric Cerenkov Telescopes (IACTs)). Greatly enhanced sensitivities, achieved with successive generations of IACTs in about two decades, have led to a rapidly increasing number of observed gamma-ray sources in the 100-GeV range; furthermore, great improvements in angular resolution have opened an age of morphological studies that have the potential to elucidate crucial questions concerning cosmic-ray origin. The available fluxes limit the accessible gamma-ray energies to the tens of TeV; however, a more fundamental limit is imposed by the absorption of such gamma rays on the relic cosmological electromagnetic radiation field (ranging from the near infrared to the CMB). This process creates a 'gamma-ray horizon' that limits the possible observation distance as a function of the gamma energy.

In contrast, astrophysically produced neutrinos propagate over cosmological distances independently of energy, and can probe deeper into sources than gamma rays, thanks to the relative transparency of the originating media to these particles. The observational challenges of high-energy neutrino astrophysics are enormous, but they are being met by kilometre-scale detectors, of which one (IceCube) is already under construction at the South Pole. A similar underwater detector (KM3NET, located in the Mediterranean sea) is under study for the Northern Hemisphere. With such facilities, the observable energy range goes from 0.1 TeV to at least a PeV. While no claim of observation of a specific source has been presented yet, such developments appear likely in the near future.

All the high-energy particle sources of the Universe – supernova remnants (SNR), active galactic nuclei (AGN), and gamma-ray bursters (GRB) have been found to be sources of very-high-energy gammas, and are targets of opportunity for neutrino astrophysics.

The astrophysical issues under investigation include the mechanisms of acceleration and particle production in such diverse environments; as of today, this looks like a realistic goal for the next decade. From IACTs, recent evidence points to the possible detection of hadron acceleration within our galaxy; besides the importance of such a contribution to the cosmic-ray acceleration problem, this observation would lend support to arguments suggesting roughly equivalent fluxes of TeV gammas and neutrinos at the source. The resulting rates of neutrino interactions define the volume of the planned high-energy neutrino detectors.

Gamma-ray and neutrino astrophysics share a potential for fundamental particle physics discoveries – here is only a small sample of the exciting discoveries that may take place over the next several years:

- Dark-matter particles, gravitationally bound to massive centres (galaxies, the Sun…) may pair-annihilate and produce characteristic gamma-ray spectra, or neutrino signals.

- Violations of Lorentz invariance may be exhibited by gammas or neutrinos produced in extremely intense gravitational fields.

- The top-down particle production phenomena that may produce the highest-energy cosmic rays would inevitably produce neutrinos of comparable energies. Although speculative, the importance of such a possible path to the highest particle-physics mass scales should not be underestimated.

Last but not at all least, research on dark energy, already mentioned in Section 2.5, has recently mobilized significant resources in the theoretical and experimental particle-physics communities. The observational techniques, whether space- or ground-based, are typical of astronomy (optical telescopes), but the instrumental, data analysis and modelling talents of particle physics have already been usefully applied. The implications for particle physics are likely to be profound, despite (or because of) the difficulty of accommodating this discovery into current theoretical frameworks.

## II-3.4 Accelerator and detector R&D, and computing for experiments

The ambitious research facilities currently being completed, like the LHC, or at an advanced stage of planning, like the ILC, are the result of an enormous amount of original work conducted over at least the past two decades. Since accelerators reaching higher energies and intensities will remain the irreplaceable driving force of further progress in particle physics for the foreseeable future, it is crucially important that, over the next decade, accelerator science and the related technological research be supported at a level that will allow progress beyond the LHC and the ILC.

Research directions leading to higher accelerating gradients, higher sustained magnetic fields and greater efficiency in power usage, already have a long history; but they need to be pushed further in order to assure to particle physics a future beyond the currently envisaged facilities. More project-specific technological issues – related, for instance, to vacuum, radiation hardness, and target performance – naturally arise concurrently, and need not be emphasized here. To address the far-future challenges, new acceleration techniques will be needed and the corresponding R&D programmes should be promoted.

Naturally, new energy/intensity domains typically require novel detectors. As in the case of accelerator research, the construction of the major LHC detectors was based on a large detector R&D programme, comprising more than thirty R&D initiatives, which spanned the range from very fundamental research on new concepts to more technically oriented, but equally important, applications directed to the establishment of economical detector construction techniques. Similar efforts must start again, notwithstanding the financial strictures upon most laboratories.

Looking into the near future, the current generation of highly sophisticated general-purpose experiments that will operate at the LHC from 2007 onwards will need substantial upgrades, involving significant detector R&D, in order to cope with SLHC luminosities. On the other hand, the ILC environment would provide an ideal testing ground for detectors reaching new goals of spatial resolution, precision of calorimetric measurements, and reliable integrability into very large systems.

Far-reaching detector developments, going beyond short-term applications to the next generation of experiments, should also be encouraged. Cross-disciplinary fertilization with new fields such as nanotechnology would be particularly welcome.

The volume of data produced by the LHC, and the associated need for Monte Carlo simulations, will place unprecedented demands upon the computing infrastructure. The increase in the price-performance characteristics of CPU, memory, disk, network bandwidth and persistent storage (tape), commonly known as "Moore's Law", will probably track the increase in instantaneous and integrated luminosity for much of the period. However, the significant increase in the complexity of the events with the large increase in the number of parasitic interactions at the SLHC, and the corresponding search through these larger volumes of data for exceedingly rare or topologically complicated events will almost certainly require a further significant increase in computing capacity. The requirements of other experiments, while substantial, will be significantly less. There is also likely to be a large demand for computer-intensive simulations for accelerator design and optimisation.

The development of the Grid Computing paradigm enabled computing resources that might otherwise have been difficult to utilise efficiently to be made available for the processing and analysis of the LHC data, and has been of significant benefit to other ongoing experiments such as BaBar, CDF, D0, H1 and ZEUS. Further work is required to improve the reliability and performance of the Grid, and to reduce the overall "cost of ownership". However, the Grid model should ensure that, at least for the foreseeable future, the amount of computing available for the experimental programme be limited by the resources rather than by the technology. Because of this, there will be a continued need for R&D into new computing methodologies and paradigms to improve performance.

# III THE PHYSICS OF THE HIGH ENERGY FRONTIER

## III-1 Introduction

The current understanding of the innermost structure of the Universe will be boosted by a wealth of new experimental information, which we expect to obtain in the near future within a coherent programme of very different experimental approaches. These range from astrophysical observations, physics with particles from cosmic rays, neutrino physics (from space, the atmosphere, from reactors and accelerators), precision experiments with low-energy high-intensity particle beams, to experiments with colliding beams at the highest energies. The latter play a central role because new fundamental particles and interactions can be discovered and studied under controllable experimental conditions and a multitude of observables is accessible in one experiment.

With the accelerators at the high energy frontier that are currently under construction or in the planning phase, particle physics is about to enter a new territory, the TeV scale, where ground-breaking discoveries are expected. The exploration of this new territory will allow us to examine the very fabric of matter, space and time. The experimental information obtained from exploring the TeV scale will be indispensable, in particular, for deciphering the mechanism that gives rise to the breaking of the electroweak symmetry, and thus establishing the origin of the masses of particles.

Furthermore it is very likely that new physics at the TeV scale is responsible for stabilizing the huge hierarchy between the electroweak and the Planck scale. The determination of the nature of the new physics may eventually lead to an understanding of the ultimate unification of forces. We also expect a deeper insight on whether space and time are embedded into a wider framework of supersymmetric (or non-supersymmetric) coordinates, and whether dark matter can be produced on Earth.

### III-1.1 Accelerators for exploring the TeV scale

From 2007 onwards, the Large Hadron Collider (LHC) and its general-purpose detectors ATLAS and CMS, will begin exploring physics at the TeV scale. The LHC will deliver proton-proton collisions at an energy of 14 TeV and a nominal luminosity of $10^{34}$ cm$^{-2}$ s$^{-1}$. ATLAS and CMS will be able to discover a SM-like or supersymmetric Higgs boson over the whole theoretically possible mass range. The LHC experiments have a broad discovery sensitivity to high-$p_T$ phenomena arising from beyond-SM physics scenarios. In particular, supersymmetry can be discovered if the SUSY particles are not unnaturally heavy. Beyond discovery, LHC can perform initial measurements of several properties of the new particles.

Higher-statistics studies of the phenomena observed at the LHC may call for an increase in the LHC luminosity. A possible upgrade, referred to as SuperLHC (SLHC), is discussed which should allow to increase the luminosity to about $10^{35}$ cm$^{-2}$s$^{-1}$.

The cleaner environment of electron-positron collisions will be required for the accurate measurement of Higgs boson properties as well as for a more precise and complete study of most new phenomena discovered at the LHC.

In the case of the Higgs boson, and of new particles below 400-500 GeV, this programme could be carried out by the International Linear Collider (ILC), whose research programme has been studied in a world-wide effort. It has been demonstrated for essentially every physics scenario beyond the Standard Model involving new particles in the ILC energy range that the ILC results, together with the results from the LHC, can reveal the detailed structure of the underlying physics. Thus an electron-positron collider of the appropriate energy reach appears to be an indispensable major initiative. The consensus that a linear collider of up to at least 400 (500) GeV, upgradeable to about a TeV, should be the next major project at the high energy frontier as well as the need for its timely realization, has been clearly expressed in statements by ECFA, ACFA, HEPAP, ICFA, GSF, and other organizations (see the corresponding documents in BB2).

New processes observed at the LHC in the highest accessible mass region, or indications for new mass thresholds from ILC precision measurements, would lead to the consideration of substantial increases in energy. Achieving multi-TeV collisions between partons can be envisaged today with two approaches: a 3-4 TeV electron-positron collider, applying the CLIC two-beam acceleration concept, or an energy-doubled LHC (DLHC), using magnets made of new superconducting materials, which are currently being investigated in several laboratories. In the longer term, a muon collider, emerging as an upgrade path of a future neutrino factory, has been considered. A muon collider represents a unique tool to study multiple Higgs s-channel production and, for the more distant future, may ultimately reach energies in the 10 TeV range. As another approach, a very large hadron collider (VLHC) is considered for reaching very high energies possibly beyond 100 TeV.

As an extension to the LHC, a Large Hadron Electron Collider (LHeC) has been suggested [BB2-2.6.03], where a 70 GeV electron or positron beam is brought to collision with one of the LHC hadron beams. Such a machine, yielding a centre-of-mass energy of about 1.4 TeV and a luminosity of $10^{33} \text{cm}^{-2}\text{s}^{-1}$, would provide sensitivity to new states in the lepton-quark sector.

The status of the above machines is discussed in more detail in Chapter IV.

## III-1.2 Physics at the TeV scale

The first and most important goal at the TeV scale is to reveal the mystery of electroweak symmetry breaking (EWSB). In the SM and many of its possible extensions, EWSB proceeds via the Higgs mechanism signalled experimentally by the presence of one or more fundamental scalars, the Higgs boson(s).

If a state resembling a Higgs boson is detected, it is crucial to test experimentally its nature. To this end the couplings of the new state to as many particles as possible must be precisely determined, which requires observation of the candidate Higgs boson in several different production and decay channels. Furthermore the spin and CP-properties of the new state need to be measured, and it must be clarified whether there is more than one Higgs state. If no light Higgs boson exists, quasi-elastic scattering processes of W and Z bosons at high energies provide a direct probe of the dynamics of electroweak symmetry breaking. This requires a detailed experimental analysis of 6-fermion processes.

If other new states are observed at the TeV scale, it will be of paramount importance to determine their properties precisely. In order to establish SUSY experimentally, for example, it will be necessary to demonstrate that every particle has a superpartner, that their spins differ by 1/2, that their gauge quantum numbers are the same, that their couplings are identical, and that certain mass relations hold. This will require a large amount of experimental information, in particular precise measurements of masses, branching ratios, cross sections, angular distributions, etc. A precise knowledge of as many supersymmetric parameters as possible will be necessary to disentangle the underlying pattern of SUSY breaking and to verify a possible supersymmetric nature of dark matter.

Other manifestations of new physics, such as extra spatial dimensions or an extended gauge structure, can give rise to a large variety of possible signals. Different scenarios may have a somewhat similar phenomenology, but a completely different physical origin. A comprehensive programme of precision measurements of the properties of new phenomena will therefore be indispensable in order to unambiguously identify the nature of new physics.

In the following, the physics possibilities at the different types of colliders will be discussed, focusing on three scenarios of possible results observed in the initial LHC runs: (i) the detection of at least one state with properties that are compatible with those of a Higgs boson; (ii) no experimental evidence for a Higgs boson; (iii) the detection of new states of physics beyond the SM.

## III-2 Physics at TeV scale colliders

While the discovery of new particles often requires access to the highest possible energies, disentangling the underlying structure calls for highest possible precision of the measurements. Quantum corrections are influenced by the whole structure of the model. Thus, the fingerprints of new physics often manifest themselves in tiny deviations. While in hadron collisions it is technically feasible to reach the highest centre-of-mass energies, in lepton collisions (in particular electron-positron collisions) the highest precision of measurements can be achieved. This complementarity has often led to a concurrent operation of hadron and lepton colliders in the past and has undoubtedly created a high degree of synergy of the physics programmes of those colliders. As an example, the Z boson was discovered at a proton-antiproton collider, the CERN SppS. Its detailed properties, on the other hand, have only been measured with high precision at electron-positron colliders, LEP at CERN and SLC at SLAC. Contrarily, the gluon was discovered at an electron-positron collider, PETRA at DESY, rather than at a hadron collider. All these measurements were crucial in establishing the SM.

Within the last decade, the results obtained at LEP and the SLC had a significant impact on the physics programme of the Tevatron proton-antiproton collider and vice versa. The top quark was discovered at the Tevatron, with a mass close to that inferred from the electroweak precision measurements at LEP and the SLC. The measurement of the top-quark mass at the Tevatron was crucial for deriving indirect constraints from LEP/SLC data on the Higgs-boson mass in the SM, while experimental bounds from the direct search were established at LEP. The experimental results obtained at LEP have been important for the physics programme of the currently ongoing Run II of the Tevatron.

The need for instruments that are optimized in different ways is typical in all branches of natural science, for example the multi-messenger approach in astroparticle physics and astronomy and the use of neutrons and photons as probes in material science. The LHC and the ILC can probe the new TeV energy regime in very different ways, as a consequence of their distinct features.

## III-2.1 The LHC[1,2]

The start of the LHC will be an exciting time for particle physics, opening a window to new physics (see contributions [BB2-2.1.06, 12, 23]). One of the great assets of the LHC is its large mass reach for direct discoveries, which extends up to typically 6-7 TeV for singly-produced particles with QCD-like couplings (e.g. excited quarks) and 2-3 TeV for pair-produced strongly interacting particles. The reach for singly produced electroweak resonances (e.g. a heavy partner of the Z boson) is about 5 TeV. The hadronic environment at the LHC, on the other hand, will be experimentally challenging. Kinematic reconstructions are normally restricted to the transverse direction. Since the initial-state particles carry colour charge, QCD cross sections at the LHC are huge, giving rise to backgrounds that are many orders of magnitude larger than important signal processes of an electroweak nature. Furthermore, operation at high luminosity entails an experimentally difficult environment such as pile-up events.

If a SM-like Higgs boson exists in nature, it will be detected at the LHC. The measurements at the LHC will allow us to determine the mass of the new particle, and its observation in different channels will provide valuable information on the couplings of the new state, and initial studies of further properties can be performed. Revealing that the new state is indeed a Higgs boson and distinguishing the Higgs boson of the SM from the states of extended Higgs theories will, however, be non-trivial. Since many extended Higgs theories over a wide part of their parameter space have a lightest Higgs scalar with properties nearly identical to those of the SM Higgs boson, a comprehensive programme of precision Higgs measurements will be necessary.

On the other hand, physics beyond the SM can give rise to Higgs phenomenology that differs drastically from the SM case. The Minimal Supersymmetric Standard Model (MSSM), for example, predicts five physical Higgs states instead of the single Higgs boson of the SM. The LHC will be able to observe all five of these states over a significant part of the MSSM parameter space. There exists also an important parameter region, however, where the LHC will detect only one of the MSSM Higgs bosons having SM-like properties. The LHC may also observe a single scalar state with a non-SM-like production or decay rate. In this case it would be difficult to tell from LHC data alone whether this is due to the presence of an extended Higgs sector, such as that predicted by the MSSM or by its most attractive extension, the Next-to-Minimal Supersymmetric Model (NMSSM), which has two more neutral Higgs bosons, or whether the observed state is an admixture of a Higgs boson with a so-called radion from extra dimensions. Similarly, the interpretation of the data will be quite difficult if an intermediate-mass scalar, with a mass above the SM bound from electroweak precision tests (for instance about 400 GeV), is observed alone. It will then be a challenge to determine whether the observed particle is the radion (with the Higgs particle left undetected), a heavy Higgs boson within a multi doublet Higgs sector (with additional contributions to precision electroweak observables that compensate for the non-standard properties of the observed scalar) or something else.

If no state compatible with the properties of a Higgs boson is detected at the LHC, quasi-elastic scattering processes of W and Z bosons at high energies need to be studied in order to investigate whether there are signs of a new kind of strong interaction in the gauge boson sector. The corresponding dynamics of strong electroweak symmetry breaking manifests itself as a deviation in the scattering amplitudes of longitudinally polarized vector bosons, possibly as resonances in the high-energy region. Collecting evidence for strong electroweak symmetry breaking will not be easy at the LHC, especially in the non-resonant case. The best non-resonant channel, namely $W^+_L W^+_L \to l^+\nu l^+\nu$, is predicted to yield signal significances below $5\sigma$ in many models.

Besides the mechanism of strong electroweak symmetry breaking, recently Higgs-less models have been proposed in the context of higher-dimensional theories. In such a scenario, boundary conditions on a brane in a warped 5$^{th}$ dimension are responsible for electroweak symmetry breaking. The mechanism for maintaining the unitarity of WW scattering may in this case be associated with Kaluza-Klein (KK) excitations of the W and Z, not much above the TeV scale, so that the detection of this kind of states at the LHC can give insight of the dynamics of electroweak symmetry breaking. New s-channel resonances coupling to both quarks and charged leptons such as KK-excitations of the Z, spin-2 resonances from warped extra dimensions, or Z' bosons from extended gauge groups, can be detected at the LHC experiments up to masses of several TeV, e.g. 5.3 TeV for a sequential Z' with SM-like couplings.

The physics of the top quark plays an important role as a possible window to new physics. The top quark is the heaviest elementary particle found so far. Since it decays much faster than the typical time for formation of top hadrons, it provides a clean source of fundamental information. The ~1Hz production rate of top quarks at the LHC (inclusive production at a luminosity of $10^{33}$ cm$^{-2}$s$^{-1}$) will provide identified samples of several million top events, leading to a determination of its mass with an expected systematic accuracy of 1 GeV, measurements of its couplings to the W boson at a few percent level, and the detection of possibly enhanced (non-standard) FCNC decays with BR up to $10^{-5}$. The top quark, furthermore, will be used as a tag of more exotic phenomena, such as production of stop squarks. The study of its couplings to the Higgs boson, finally, will be a key element in the study of the EWSB mechanism.

The production of supersymmetric particles at the LHC will be dominated by the production of coloured particles, i.e. gluinos and squarks. Searches for the signature of multi jets accompanied by large missing transverse energy at the LHC will provide sensitivity for discovering SUSY if gluino or squark masses are below 2.5-3 TeV, thus covering the largest part of the viable parameter space. The main handle to detect uncoloured SUSY particles will be from cascade decays of heavy gluinos and squarks, since in most SUSY scenarios the uncoloured particles are lighter than the coloured ones. Thus, fairly long decay chains giving rise to the production of several supersymmetric particles in the same event and leading to rather complicated final states can be expected to be a typical feature of SUSY production at the LHC. In fact, the main background for measuring SUSY processes at the LHC will be SUSY itself. Many other kinds of SM extensions have been studied for the LHC. For a more complete overview see ref.1.

For the planning of future facilities, it is of particular interest, which kinds of discoveries may be expected from an initial LHC data set that can be collected during the first years of operation [3]. See also contribution [BB2-2.1.23]. Until the year 2010, an integrated luminosity of about 10-30 fb$^{-1}$ appears to be possible. Apart from collecting the integrated luminosity, the first data will have to be used to commission and calibrate the detectors, understand and model the backgrounds and establish the analyses. Possible discoveries during this start-up phase depend also on the complexity of the signal.

For the full Higgs mass range $m_H > 115$ GeV a 5-$\sigma$ discovery can be obtained combining both experiments with 5 fb$^{-1}$ of integrated luminosity. For $m_H < 140$ GeV a combination of three different channels and a very good understanding of the detectors and backgrounds is required. A 95% exclusion over the full mass range can be achieved with 1 fb$^{-1}$. These values of integrated luminosities can be significantly reduced for $m_H > 140$

GeV. For heavier SM Higgs boson, properties like spin-parity can be determined, given sufficient integrated luminosity.

If SUSY particles are not too heavy, they will be produced copiously at the LHC. The inclusive signature of multi-jets accompanied by missing transverse energy is suitable for discovery up to a mass scale of 1.5 (2) TeV for 1 (10) fb$^{-1}$. The understanding and calibration of missing energy requires significant effort. Easier signatures like the presence of kinematic endpoints in di-lepton mass spectra require less calibration effort but the rate is lower and their presence is more model-dependent.

New resonances decaying into lepton pairs like e.g. Z' are expected to be observed relatively fast. A sequential Z' decaying into a muon pair can be detected up to 3 TeV with 10 fb$^{-1}$. On the other hand, signals from new strong interactions replacing the Higgs boson are very difficult to be detected in an initial LHC data set.

In summary, the LHC will provide a very broad sensitivity for detecting a Higgs boson (or several Higgs states) and for discovering high-$p_T$ phenomena at the TeV energy scale. It will perform several precise measurements and provide a first understanding of new physics.

### III-2.1.1 LHC upgrades

#### III-2.1.1.1 LUMINOSITY UPGRADE: THE SLHC[4]

A luminosity upgrade of the LHC, the so-called SuperLHC (SLHC) would allow the maximum exploitation of the existing tunnel, machine and detectors. See also contributions [BB2-2.1.06, 21]. Although the exact physics case is difficult to predict today, since it depends very much on what the LHC will find or not find, in general, the SLHC can extend the LHC mass reach by 20-30%, thereby enhancing and consolidating the discovery potential at the TeV scale. In addition, a tenfold increase in the statistics of the collected data samples should allow more precise measurements of processes that are statistically limited at the LHC.

Experimentation will be difficult already at the LHC design luminosity of $10^{34}$ cm$^{-2}$s$^{-1}$, and even more so at $10^{35}$ cm$^{-2}$s$^{-1}$. The radiation levels in the detectors and the integrated doses will be ten times larger at the SLHC than at the LHC. Other important parameters, such as the particle multiplicity per bunch-crossing, the tracker occupancy, and the pile-up noise in the calorimeters, depend on the machine bunch structure, which is still under study as indicated in Chapter IV. With a bunch spacing of 12.5 ns, as assumed in the studies made by the experiments, and with no changes to the ATLAS and CMS detectors, the tracker occupancy would be a factor of 10 higher at the SLHC than at the LHC, and the pile-up noise in the calorimeters a factor of 3 larger. It is likely that upgrades of the ATLAS and CMS detectors will be necessary, in particular a replacement of the Inner Detectors and the Level-1 Trigger electronics. In order to exploit fully the tenfold increase in luminosity the experiments must be able to reconstruct and identify high-$E_T$ jets, electrons, muons, taus, and b-jets. Some degradation w.r.t. the LHC detectors is expected at L = $10^{35}$ cm$^{-2}$s$^{-1}$. If no detector upgrade and no optimisation of the algorithms for the higher pile-up environment is performed, the rejection against jets faking electrons is reduced by about 30%; the jet energy resolution is degraded from 15% (LHC) to 40% (SLHC) for central jets with $E_T$=50 GeV, and is essentially unaffected for $E_T$ = 1 TeV; for a b-tagging efficiency of 50%, the rejection against light-quark jets decreases by a factor of about 6 (2) for jets with $E_T$ = 80 GeV ($E_T$ = 300 GeV).

The goals of the SLHC will be to extend the discovery reach for physics beyond the SM and to improve the sensitivity for measurements which are rate-limited at the LHC.

Measurements of Triple Gauge Couplings (TGCs), i.e. couplings of the type WWγ and WWZ, probe the non-Abelian structure of the SM gauge group and are also sensitive to new physics. The SLHC can improve the LHC reach by a factor of about two. For λ-type couplings, anomalous couplings that are strongly enhanced at high energy, the accuracy can reach the level of SM EW radiative corrections. The reach for rare decays of the top quark will be improved by a factor of 10, and improvements are expected in the study of its EW couplings.

With increased luminosity, statistical errors on the measurement of ratios of Higgs couplings can be reduced below the level of theoretical and experimental systematic uncertainties; from that point progress would be needed on both experimental systematics and theory. The SLHC should be able to observe for the first time several rare decay modes of a SM Higgs boson, like H → μμ and H → Zγ, which are not accessible at the LHC because their branching ratios are too small. In a narrow Higgs mass range around 160 GeV, SLHC experiments may measure the Higgs self-coupling, which gives direct access to the Higgs potential in the SM Lagrangian. This can be done by looking for the production of a pair of Higgs bosons, in the WWWW final state, which is presently being studied experimentally. If SUSY is realized at the TeV scale the LHC has a sensitivity to squark and gluino masses up to 2.5 TeV. At the SLHC this reach can be extended to up to 3 TeV. SUSY discovery is likely to be based on inclusive signatures, such as events with jets plus missing transverse energy, involving high-$p_T$ calorimetric objects, which suffer very little from the increased pile-up at L = $10^{35}$ cm$^{-2}$ s$^{-1}$. In contrast, precise measurements of the SUSY particles (masses, etc.), which are crucial to constrain the fundamental parameters of the underlying theory, require in most cases the selection of exclusive channels, containing e.g. leptons or b-jets, and therefore the full power of the detectors, including well-performing trackers. Some of these exclusive channels are expected to be rate-limited at the LHC, and would therefore benefit from a luminosity upgrade. The mass reach for discovery of the heavier SUSY Higgs bosons H, A, and H$^\pm$ can be extended by ~100 GeV.

If no Higgs boson will be found at the LHC, one of the most likely scenarios is that electroweak symmetry is broken by a new kind of strong interaction. If this is the case, effects of the strong interaction are expected to manifest themselves in resonant or non-resonant scattering of longitudinally polarized vector bosons at the TeV scale, leading to deviations from the SM expectation. The elementary process is qq → $V_L V_L$qq, where the longitudinal vector bosons $V_L$ are radiated off the incident quarks, and the final state quarks are emitted in the forward regions of the detector (|η| > 2). The latter is a distinctive signature for these processes, and an essential tool, at hadron colliders, to reject the huge backgrounds. A luminosity upgrade to $10^{35}$ cm$^{-2}$s$^{-1}$ offers improved physics prospects w.r.t. the nominal LHC, with some difficulties. The main difficulty is that, because of the higher pile-up of minimum-bias events, the detector tagging performance for forward jets is reduced. Nevertheless, thanks to the larger event statistics, the excess in the non-resonant W$^+$W$^+$ scattering mentioned above should become significant (at the level of 5-8σ, depending on the model). Furthermore, low-rate channels, such as the possible production of resonances in ZZ scattering, could be observed for the first time at the SLHC.

Compositeness is another interesting scenario beyond the SM, motivated in part by the existence of three generations of fermions, which may indicate the presence of more elementary constituents bound together by a force characterized by a scale Λ. If the centre-of-mass energy of the colliding partons is smaller than Λ, compositeness should manifest itself through 4-fermion contact interactions. In particular, 4-quark contact interactions are expected at hadron colliders, which should give rise to an excess of centrally produced high-$p_T$ jets. The LHC should be able to probe compositeness scales Λ up to about 40 TeV, whereas the SLHC should extend this reach to 60 TeV.

More examples and comparisons of the LHC and SLHC physics can be found in ref. [4].

The main general conclusion is that a tenfold increase of the LHC luminosity to $10^{35}$ cm$^{-2}$s$^{-1}$ represents a consolidation and extension of the LHC programme, and the maximum exploitation of the existing infrastructure, machine and experiments. It should allow an extension of the mass reach for singly produced particles by 20-30%, i.e. from about 6.5 TeV to about 8 TeV, to improve precise measurements of standard and new physics, and to enhance sensitivity to rare processes.

### III-2.1.1.2 ENERGY DOUBLING OF THE LHC (DLHC)

There are scenarios for new physics which would benefit from an increase of a factor two in the centre of mass energy at a luminosity of around $10^{34}$ cm$^{-2}$s$^{-1}$. See also contribution [BB2-2.1.06]. Typical examples would be scenarios with new thresholds in the energy range beyond the reach of (S)LHC.

The physics case for an energy doubled LHC is less well studied than that of the SLHC and also requires detailed knowledge from the exploration of the TeV scale. The mass reach of a 28 TeV pp machine is up to 10-11 TeV for singly-produced particles. Supersymmetric particles can be discovered up to 4.5-5 GeV. Compositeness can be probed up to 85 TeV.

A proton-proton collider with 28 TeV centre-of-mass energy would require a new machine and in particular a vigorous R&D effort to develop ~16T magnets if it should be built in the existing LHC tunnel. These aspects will be covered in Chapter IV.

## III-2.1.2 Electron-Proton Collisions in the LHC tunnel (LHeC)

On the occasion of the Orsay Open Symposium a proposal for a 70 GeV electron/positron beam to be collided with one of the 7 TeV LHC proton beams was submitted, the Large Hadron Electron Collider (LHeC). The anticipated luminosity is $10^{33}$cm$^{-2}$s$^{-1}$, and the centre-of-mass energy is 1.4 TeV. The LHeC would make possible deep-inelastic lepton-hadron (ep, eD and eA) scattering for momentum transfers $Q^2$ beyond $10^6$ GeV$^2$ and for Bjorken x down to the $10^{-6}$. New sensitivity to the existence of new states of matter, primarily in the lepton-quark sector would be achieved, much extending the sensitivity of HERA. The aspects concerning QCD are discussed in Chapter IX. For more details we refer to [BB2-2.6.03].

## III-2.2 Physics at the ILC[2,5]

The design of the ILC is being addressed by a world-wide collaboration of physicists in the context of the Global Design Effort [6]. See also contributions [BB2-2.1.08, 09, 12, 13, 20]. The baseline design of the ILC foresees a first phase of operation with a tunable energy of up to about 500 GeV and polarized beams. Possible options include running at the Z-boson pole with high luminosity (GigaZ) and running in the photon-photon, electron-photon and electron-electron collider modes. The physics case of the ILC with centre-of-mass energy of 400-500 GeV rests on high-precision measurements of the properties of the top quark at the top threshold, the unique capability of performing a comprehensive programme of precision measurements in the Higgs sector, which will be indispensable to reveal the nature of possible Higgs candidates, the good prospects for observing the light states of various kinds of new physics in direct searches, and the sensitivity to detect effects of new physics at much higher scales by means of high-precision measurements.

The baseline configuration furthermore foresees the possibility of an upgrade of the ILC to an energy of about 1 TeV. The final choice of the energy and further possible machine and detector upgrades will depend on the results obtained at the LHC and the first phase of the ILC.

The much cleaner experimental environment at the ILC in comparison with the LHC will be well suited for high-precision physics. This is made possible by the collision of point-like objects with exactly defined initial conditions, by the tunable collision energy of the ILC, and by the possibility of polarising the ILC beams. Indeed, the machine running conditions can easily be tailored to the specific physics processes or particles under investigation. The signal-to-background ratios at the ILC are in general much better than at the LHC. In contrast to the LHC, the full knowledge of the momenta of the interacting particles gives rise to kinematic constraints, which allow reconstruction of the final state in detail. The ILC will therefore provide very precise measurements of the properties of all accessible particles.

Direct discoveries at the ILC will be possible up to the kinematic limit of the available energy. Furthermore, the sensitivity to quantum effects of new physics achievable at the ILC will in fact often exceed that of the direct search reach for new particles at both the LHC and the ILC.

The ILC can deliver precision data obtained from running at the top threshold, from fermion and boson pair production at high energies, from measurements in the Higgs sector, etc. Furthermore, running the ILC in the GigaZ mode yields extremely precise information on the effective weak mixing angle and the mass of the W boson (the latter from running at the WW threshold). The GigaZ running can improve the accuracy in the effective weak mixing angle by more than an order of magnitude. The precision of the W mass would improve by at least a factor of two compared to the expected accuracies at the Tevatron and the LHC. However, achieving the accuracy of $10^{-5}$ required for the beam energy calibration needs to be demonstrated. Comparing these measurements with the predictions of different models provides a very sensitive test of the theory, in the same way as many alternatives to the SM have been found to be in conflict with the electroweak precision data in the past.

The ILC is uniquely suited for carrying out high-precision top-quark physics, which plays a crucial role as a window to new physics. Knowing the properties of the top quark with a high accuracy will be essential for identifying quantum effects of new physics. The ILC measurements at the top threshold will reduce the experimental uncertainty on the top-quark mass to the level of 0.1 GeV or below, i.e. more than an order of magnitude better than at the LHC, and would allow a much more accurate study of the electroweak and Higgs couplings of the top quark. A precision of $m_t$ significantly better than 1 GeV will be necessary in order to exploit the prospective precision of the electroweak precision observables. In particular, an experimental error on $m_t$ of 0.1 GeV induces an uncertainty in the theoretical prediction of $M_W$ and the effective weak mixing angle of 0.001 GeV and $0.3 \times 10^{-5}$, respectively, i.e. below the anticipated experimental error of these observables. The impact of the experimental error on $m_t$ is even more pronounced in Higgs physics. In the MSSM, as an example, the uncertainty in the prediction of the lightest Higgs boson mass, $m_h$, induced by an experimental error of $m_t$ of 1 GeV is also about 1 GeV, owing to large top-quark effects scaling with the fourth power of $m_t$. The ILC precision on $m_t$ is mandatory in order to obtain a theoretical prediction for $m_h$ with the same level of accuracy as the anticipated experimental precision on the Higgs-boson mass.

The high-precision information obtainable at the ILC will be crucial for identifying the nature of new physics, and in this way new fundamental laws of nature can be discovered. For instance, once one or more Higgs particles are detected, a comprehensive programme of precision Higgs measurements at the ILC will be necessary to reveal their properties and the underlying physical laws. The mass of the Higgs boson can be determined at the ILC at the permille level or better, Higgs couplings to fermions and gauge bosons can typically be measured at the percent level, and it will be possible to determine unambiguously the quantum numbers in the Higgs sector. Indeed, only the ILC may be able to discern whether the Higgs observed at the LHC is that of the SM or a Higgs-like (possibly composite) scalar tied to a more complex mechanism of mass generation. The verification of small deviations from the SM may be the path to decipher the physics of electroweak symmetry breaking. The experimental information from the ILC will be even more crucial if the mechanism of electroweak symmetry breaking in nature is such that either Higgs detection at the LHC may be difficult or the Higgs signal, while visible, would be hard to interpret. In the example of Higgs - radion mixing mentioned above, the ILC could observe both the Higgs and the radion and measure their properties with sufficient accuracy to establish experimentally the Higgs-radion mixing effects.

If no clear Higgs signal has been established at the LHC, it will be crucial to investigate with the possibilities of the ILC whether the Higgs boson has not been missed at the LHC because of its non-standard properties. This will be even more the case if the gauge sector does not show indications of strong electroweak symmetry breaking dynamics. The particular power of the ILC is its ability to look for $e^+e^- \rightarrow ZH$ in the inclusive $e^+e^- \rightarrow ZX$ missing-mass distribution recoiling against the Z boson. Even if the Higgs boson decays in a way that is experimentally hard to detect or different Higgs signals overlap in a complicated way, the recoil mass distribution will reveal the Higgs boson mass spectrum of the model. The total Higgs-strahlung cross section will be measurable with an accuracy of 2.5% for a Higgs boson with a mass of about 120 GeV.

Should no fundamental Higgs boson be discovered, neither at the LHC nor at the ILC, high-precision ILC measurements will be a direct probe of the underlying dynamics responsible for particle masses. The LHC and the ILC are sensitive to different gauge boson scattering channels and yield complementary information. As mentioned above, this kind of complementarity between lepton and hadron colliders will be similar to the interplay, for instance, of LEP and the Tevatron in exploring the properties of the Z and W bosons with high precision. The combination of LHC and ILC data will considerably increase the LHC resolving power. In the low-energy range it will be possible to measure anomalous triple gauge couplings with a sensitivity comparable to the natural size of EW loop corrections, of order $1/(16\,\pi^2)$. The high-energy region where resonances may appear can be accessed at the LHC only. The ILC, on the other hand, has an indirect sensitivity to the effects of heavy resonances even in excess of the direct search reach of the LHC. Detailed measurements of cross sections and angular distributions at the ILC will be crucial for making full use of the LHC data. In particular, the direct sensitivity of the LHC to resonances in the range above 1 TeV can be fully exploited if ILC data on the cross section rise in the region below 1 TeV are available. In this case the LHC measures the mass of the new resonances and the ILC measures their couplings.

Furthermore, the electroweak precision measurements (in particular from GigaZ running) at the ILC will be crucial to resolve the conspiracy that mimics a light Higgs in the electroweak precision tests. The prospective accuracy on the effective weak mixing angle achievable at GigaZ running of about $1 \times 10^{-5}$ will provide sensitivity to genuine electroweak two-loop and even higher-order effects. Since different kinds of new physics give rise to rather different contributions at this level of accuracy, confronting the theory predictions with the GigaZ data will be a very powerful tool to discriminate between different possible scenarios. The same is true also for Higgs-less models in the context of higher-dimensional theories, where even the current accuracy of the precision observables and the top-quark mass allows to rule out many models. Thus, a thorough understanding of the data of the ILC and the LHC combined will be essential for disentangling the new states and identifying the underlying physics.

For supersymmetric particles, new states arising from extra dimensions of space, and other kinds of new physics, the ILC can provide precise and definite information that will be crucial to unambiguously determine the nature of the new phenomena. The ILC has the capability to run directly at the threshold where a new state is produced. This allows to determine both the spin and the mass of the new state in a model-independent way with high precision. The masses of supersymmetric particles can be measured at the permille level in this way. The precision measurements at the ILC can also give access to further properties of new physics such as couplings, mixing angles and complex phases.

The part of the spectrum of new states accessible at the ILC, for instance of supersymmetric particles, is very likely to be complementary to the LHC. The precise measurements at the ILC will be crucial for revealing the underlying structure, even if only a part of the spectrum is accessible. Since the lightest supersymmetric particle is a promising candidate for cold dark matter in the Universe, studying its properties in detail is of particular importance. The ILC has unique capabilities for performing a high-precision measurement of the mass and further properties of this weakly interacting particle. For instance, the mass of the lightest supersymmetric particle would be measurable at the ILC with an accuracy that is two orders of magnitude better than at the LHC. This precision will be crucial for confronting the properties of dark matter candidates and the nature of their interactions with cosmological observations.

In the scenario of large extra dimensions the ILC with polarised positrons can probe fundamental scales of gravity up to about 20 TeV. The number of extra dimensions can be determined by ILC measurements at different energies. For KK excitations of the gauge bosons the determination of the mass of the first KK excitation at the LHC together with precision measurements at the ILC can be used to distinguish the production of a KK gauge state from a new gauge field in extended gauge sectors.

In summary, the ILC offers a physics programme of precision measurements and discoveries at the TeV scale and beyond that is well motivated and has been studied in great detail. It has been clearly demonstrated that the results from the ILC will lead to definite conclusions about many features of physics at the TeV scale. Thus, an electron-positron collider with the appropriate energy reach has received great attention and is strongly supported world-wide. The physics programme of the ILC is highly complementary to the LHC programme. The synergies arising from the different opportunities at LHC and ILC have been outlined in much detail in ref.[2].

## III-2.3 CLIC[7]

The two-beam acceleration technology being developed for the CLIC electron-positron collider is the most advanced proposal for reaching multi-TeV energies in lepton collisions. See also contribution [BB2-2.1.05]. The CLIC machine parameters are optimized for a collider with a centre-of-mass system energy of 3 TeV, upgradable to 5 TeV. Operations at multi-TeV energies need an elevated luminosity, to compensate for the 1/s dependence of the s-channel annihilation cross sections. In order to have sufficiently high event rates, it is required to operate CLIC at luminosities around $10^{35}$ cm$^{-2}$ s$^{-1}$, leading to data samples of 1 ab$^{-1}$ per year. In the multi-TeV energy range fusion processes mediated by t-channel exchanges, whose cross sections increase logarithmically with the centre-of-mass energy, become comparable in strength and present interesting new physics opportunities.

In order to obtain high energies and luminosities, CLIC will operate in the high-beamstrahlung regime. This leads to large experimental backgrounds, due in particular to coherent and incoherent e$^+$e$^-$ pair production and to hadronic γγ collisions: about four γγ collisions are overlaid per bunch crossing. Whilst the amount of these pile-up events is similar to the number of additional pp collisions per bunch crossing at the LHC during low-luminosity running, pile-up at CLIC is much less problematic, since these are collisions with much lower energy than in e$^+$e$^-$ collisions. The beam-beam interactions also distort the luminosity spectrum at CLIC. Clearly the distorted luminosity spectrum and the backgrounds lead to significant experimental challenges. Nevertheless, detailed simulation studies, which include the machine backgrounds and expected luminosity spectrum, have demonstrated that precision physics is possible at CLIC, provided the detector has sufficient granularity. A solenoidal field of at least 4 Tesla and a minimum distance of the innermost detector from the beam of 3 cm as well as a tungsten mask in the forward region at 120 mrad are required to reduce the backgrounds.

The high energy of a multi-TeV e$^+$e$^-$ collider such as CLIC will extend the reach for heavy Higgs states. While ZH production is suppressed at 3 TeV (and thus the recoil mass technique cannot be exploited), the logarithmic rise of the WW-fusion mechanism, e$^+$e$^-$ → Hνν provides a huge sample of Higgs bosons of O(500k)/ab$^{-1}$. This will add new information on rare Higgs decays, such as the decay H → μμ for $m_H$ in the range 120-140 GeV and the decay H → bb for $m_H$ between 180 and 240 GeV. Double Higgs production in the ννbb and ννWW final states may be exploited in order to measure the trilinear Higgs coupling in the mass range 120-240 GeV to a precision of approximately 10% at 3 TeV energy [BB2-2.1.05]. The large Higgs samples will also be instrumental for increasing the precision on, e.g., the ttH coupling, measurements of possible CP phases, and other Higgs physics.

In case where no Higgs boson has been found at LHC or ILC, CLIC will be better suited than the LHC for the direct production and detection of heavy, broad resonances connected to strong electroweak symmetry breaking, if they are within the kinematic reach. In particular, in contrast to LHC, hadronic final states in νν/ee+VV → νν/ee+4 jets can be observed. Cross sections in the resonant region are typically in the few fb range, yielding a few thousands of events.

The high energy of CLIC will also be instrumental in the search for heavy states of new physics. The spectroscopy of directly produced supersymmetric particles proceeds similarly to that at the ILC, however with reduced mass precision due to the harder beamstrahlung spectrum. With 3 TeV energy, relatively heavy sparticles, in particular squarks and the heavier charginos and neutralinos, will be accessible. Also the sensitivity to new contact interactions will be increased.

## III-2.4 Muon collider [8]

Muon colliders have been proposed as unique tools to study multiple Higgs scenarios and, for the more distant future, as the ultimate probe of energies in the 10 TeV range and possibly above. There are serious technological challenges to building such a collider which will be addressed in Chapter IV. In particular, the potential radiation caused by the resulting beam of high energy neutrinos must seriously be considered.

Muon colliders have the prospect to produce Higgs bosons directly via $\mu^+\mu^-$ annihilation in the s-channel, unaccompanied by spectator particles. For a narrow Higgs resonance a muon collider will have a unique capability for a measurement of the Higgs-boson mass with ultra-high precision. It will furthermore be possible at a muon collider to resolve nearly degenerate Higgs states, for instance the heavy MSSM Higgs bosons H and A that may be very close in mass, and to study CP-violating effects in the Higgs sector.

If the study of an s-channel resonance is to be pursued experimentally, the event rate must be sufficiently large. In the case of a SM Higgs boson this means that the mass must be somewhat less than twice the mass of the W boson, otherwise the large width reduces the peak cross section. This condition need not apply to more complicated Higgs systems, for instance the heavier neutral Higgs bosons of supersymmetry.

## III-3 Detector R&D

Over the next years significant advances in detector technologies need to be achieved and a rigorous R&D programme is needed in order to exploit the physics possibilities mentioned above, see also [BB2-2.1.10]. A large effort has already been invested in detector development for the present LHC programme, with many benefits to other areas in high energy physics and beyond. Nevertheless, there are significant additional and different detector R&D challenges for the ILC and for the SLHC programme. The principal challenges at the (S)LHC are related to the high event rate and the high radiation levels associated with the pp energies and luminosities required to do physics with the parton component of the proton. At the ILC, on the other hand, the reconstruction of the event with the best possible precision poses new and complementary challenges. Backgrounds and radiation levels the detectors have to withstand play a much reduced role due to the collision of point-like particles, the electron and the positron.

The primary new requirements for detectors at the ILC are excellent hermeticity, track-momentum resolution, jet-energy resolution and flavour identification for bottom and charm jets. The most promising method to reconstruct the complex events at the ILC with the required precision, and in particular to measure the jet four-momenta, is the so-called Particle-Flow Algorithm (PFA) [9]. Although not a fundamentally new approach, the ILC for the first time promises to have a detector optimised ab initio for PFA, thereby allowing to reach significantly improved overall performance.

The method of PFA relies on the measurement of momenta of charged particles in jets using the tracking system, the energy of photons and electrons using the electromagnetic calorimeter, and the energy of neutral hadrons from both the electromagnetic and hadronic calorimeters. The first and foremost requirement for a successful application of PFA is the capability to separate close-by particles in a jet inside the calorimeters and the trackers. These requirements call for high granularity, excellent resolution and as small systematic effects as possible for all detector components.

It is important to note here that it has been demonstrated that there is a major advantage in 'luminosity factor' (the factor by which the integrated luminosity would have to be increased to compensate for an inferior detector) if the detector can be built to satisfy the challenging performance criteria. In other words, detector performance is as important as accelerator luminosity for exploring the Terascale. Like increased luminosity, detector performance extends the accelerators' physics reach. Thus novel and far-reaching detector R&D is mandatory which will also be beneficial for other areas of science as already demonstrated in the past. The necessary R&D programmes for ILC detectors have been developed in detail over the last years and coordination on the international level has started.

In order to achieve the physics potential of SLHC mentioned above, the detector performance must be similar to that envisaged for the LHC detectors presently being constructed, however, in an environment with much higher particle multiplicities and radiation levels. Radiation hardness of existing and/or new electronics and detector components such as semiconductor sensors is one of the prime issues. Also higher bunch crossing frequencies and higher readout bandwidth have to be handled. These and other topics need dedicated R&D which needs to be started now. The precise detector needs for the SLHC and its ultimate physics performance will depend on the physics to be studied. Optimisation of luminosity times efficiency (i.e. luminosity versus detector performance) for different physics scenarios has not started yet.

In summary, to achieve the required detector performance for SLHC and ILC, it will be necessary to build up and maintain an effective and efficient programme of detector R&D. It will require a suitable level of funding, matched to the time-scale for the R&D, design and construction phases of the respective accelerator. The recently approved EUDET project for improving the infrastructure for detector R&D in Europe, although primarily aimed at the ILC detector R&D, is a first step towards coordinated detector research in general. A strong and coordinated effort together with improved infrastructures will be of great value also for other detector developments, e.g. for CLIC, as well as for the detector challenges at accelerators in the further future.

# III-4 Open symposium input

The physics opportunities outlined above were discussed at the Open Symposium in Orsay, and further comments from the community were received as written contributions via the web. The written contributions, received until March 15, are collected in Briefing Book 2. In the following a summary of the key points of the written contributions and the discussion following the presentation in the session on 'Physics at the High Energy Frontier' will be given.

## III-4.1 Written contributions

Contributions concerning the session on the High Energy Frontier have been received from several groups and individuals through the web page input to the Strategy Group. They can be found in Briefing Book 2, [BB2-2.1.01] to [BB2-2.1.26]. There are contributions that are mainly focussed on one particular topic. They concern LHC and its upgrades [BB2-2.1.06, 12, 21, 23], including an ep option [BB2-2.6.03], ILC [BB2-2.1.08, 09, 12, 13, 20] and CLIC [BB2-2.1.05]. The main physics aspects of LHC, SLHC, DLHC, ILC, and CLIC described in these contributions are mentioned above and are not repeated here. Theoretical aspects are dealt with in [BB2-2.1.03, 17, 25] and detector R&D is addressed in [BB2-2.1.10]

Several contributions concern aspects of schedule, timeliness and strategy. These are [BB2-2.1.01, 06, 07, 09, 13, 14, 15, 16, 20, 22, 23], as well as contributions in BB2-Chapter 3 and 4.

## III-4.2 Discussion

Besides two remarks about the physics case of SLHC the discussion concentrated on the ILC. The statement concerning the physics case as given in the ECFA document 2001 was discussed and its validity was reaffirmed by all contributions on this subject: The physics case did not change, and the statements made by ECFA and other organisations remain valid.

The need for a timely decision on construction of the ILC was stressed by the majority. The drawbacks of coupling the decision about the construction of the ILC to LHC results were emphasized. The need for the development of a clear strategy on how to react on LHC data was stressed, and it was emphasized that it needs to be avoided that the LHC results / findings that one demands in order to go ahead with the ILC construction become a "moving target". It was stressed that the ILC design, in particular concerning the upgrade after a first phase of running at 400 or 500 GeV, should have enough flexibility to react on results obtained at the LHC and in the first phase of ILC running.

The question was discussed whether the ILC is still interesting in cases where, besides the Higgs boson, either no new states would be found at LHC or only ones beyond the ILC reach. Even in such scenarios the precision studies of Higgs and top quarks would be needed. The case of a Higgs boson with a mass in the range between 160 and 200 GeV which would mainly decay into WW or ZZ pairs and make the determination of fermion couplings more difficult was mentioned. Here, unique measurements could be performed at the ILC, in particular b- and t-couplings and the total decay width, which improve significantly the LHC capabilities.

In conclusion, the discussion revealed a clear majority w.r.t. the validity of the physics case for an ILC, and given present knowledge on physics scenarios, LHC results should not modify this view on the need of an ILC. It became clear also that enough flexibility concerning energy reach and options must be taken into account in the design of such a machine.

# IV HIGH ENERGY FRONTIER: ACCELERATORS

## IV-1 Introduction

A very large fraction of the discoveries and progress accomplished in the field of elementary particle physics over the past decades has been realized thanks to the multiple high-energy frontier accelerator facilities, often operated simultaneously as can be seen in the following diagram. Different types of accelerators were required, based on proton (antiproton) and/or electron (positron) beams, mostly used in collider mode to reach the highest possible energy. Indeed, over the last 3 decades, the c.m. energy has been increased from a few tens of GeV to more than 200 GeV (soon 14 TeV) in $e^+e^-$ ($pp$) collisions.

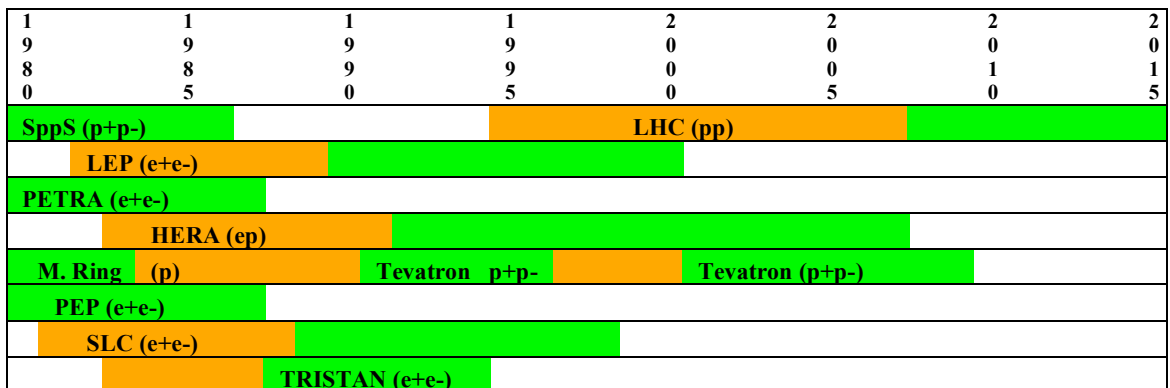

*The construction (orange) and the operation (green) time of the various High Energy Frontier accelerators during the last 25 years.*

The Large Hadron Collider (LHC) is the next major high-energy frontier accelerator. Particle physics will thus enter an extraordinary new era with the likely discovery of the Higgs boson, or whatever takes its place, and the exploration of the physics beyond the Standard Model, including supersymmetry, dark matter and extra dimensions, as well as physics not yet imagined. These discoveries will largely determine the course of particle physics and the launch of the necessary upgrade programmes and/or the construction of new infrastructures. Therefore, it is important to be well prepared to make these strategic decisions and avoid, as far as possible, to be limited by technology.

Past experience shows that the complexity and the size of the recent infrastructures have required an ever longer construction time, following a continuous and vigorous R&D effort. Therefore, should one anticipate that the comprehensive exploration and understanding of the new physics will require several alternative and complementary accelerators, the corresponding R&D program would need to be accomplished in parallel.

Finally, as fewer facilities are available and their exploitation time is becoming longer, reliable operation is mandatory and upgrades are often required.

## IV-2 High-Energy Hadron Colliders

High-energy hadron colliders have been extraordinary sources of discoveries. As illustrations one can quote the Z and W bosons at the $S p\bar{p} S$ or the top quark at the Tevatron. This latter $p\bar{p}$ collider is the only hadron collider in operation to date and reaches a c.m. energy of 1.96 TeV with a peak luminosity of $1.6\times10^{32}$cm$^{-2}$s$^{-1}$ in continuous improvement, thanks to an on-going upgrade programme. The Tevatron is likely to run until the LHC takes over with a large physics reach (see Section 2.1 in Chapter III), possibly around 2009.

## IV-2.1 The Large Hadron Collider (LHC)

The main technical parameters of the LHC are summarized in the following table. It is expected that the luminosity will gradually increase over the first few years of operation, until the injectors reach their limits in terms of bunch intensity and of brightness $N_b/\varepsilon_n$.

| Parameter | Unit | Injection | Collision |
|---|---|---|---|
| Energy | [GeV] | 450 | 7000 |
| Luminosity          nominal ultimate | [cm$^{-2}$s$^{-1}$] | | $10^{34}$ $2.3 \times 10^{34}$ |
| Number of bunches | | | 2808 |
| Bunch spacing | [ns] | | 24.95 |
| $N_b$ intensity per bunch          nominal ultimate | [p/b] | | $1.15 \times 10^{11}$ $1.70 \times 10^{11}$ |
| Beam current nominal ultimate | [A] | | 0.58 0.86 |
| $\varepsilon_n$(transverse emittance, rms, normalised), nominal & ultimate | [µm] | 3.5 | 3.75 |
| Longitudinal emittance, total | [eVs] | 1.0 | 2.5 |
| Bunch length, total (4σ) | [ns] | 1.5 | 1.0 |
| Energy spread, total (4σ) | [10$^{-3}$] | 1.2 | 0.45 |

The operation of the LHC relies strongly on the CERN proton accelerator complex. This is a very important asset and has been instrumental to the decision of building the LHC. The present acceleration chain includes the Linac2 up to 50 MeV, the PS Booster (PSB) up to 1.4 GeV, the PS up to 26 GeV, the SPS up to 450 GeV and the LHC up to 14 TeV. With this complex, a nominal peak luminosity of $10^{34}$ cm$^{-2}$s$^{-1}$ is expected. It could possibly reach an ultimate limit of 2.3 $10^{34}$ cm$^{-2}$s$^{-1}$ by increasing the beam current and brightness after some upgrade of the injector complex.

### IV-2.1.1 Status of the LHC

All efforts are currently made to ensure the timely completion of LHC construction. Despite the fantastic complexity of this frontier infrastructure, the excellent recent progress accomplished concerning the procurements, the tests and the installation of the numerous components allows for reasonable optimism concerning its start up in 2007. The progress can be monitored from the LHC dashboard accessible from the following web site: http://lhc-new-homepage.web.cern.ch/lhc-new-homepage/DashBoard/index.asp

The optimal use of the LHC will require several important steps: the consolidation of the CERN proton accelerator complex, its optimization and improvement to overcome its limitations, and finally the maximization of the physic reach by upgrading the LHC performances [see also BB2-2.1.11].

### IV-2.1.2 Consolidation of the CERN proton accelerator complex

In order to ensure a reliable operation of the LHC, a careful investigation of the limiting items or the components at risk is of prime importance. Indeed, the existing CERN complex is several decades old and suffers age problems, as illustrated by the deterioration of the PS and SPS main magnets. This is due to the combined effects of mechanical fatigue, corrosion and irradiation. Furthermore, it can be anticipated that the damage will be aggravated by the beam loss because of the increased number of protons to be accelerated in the period 2006-2011. Therefore anticipating possible component failures, requiring complicated and lengthy interventions and repairs, would contribute to ensure the reliable operation of the LHC and to maximize the integrated luminosity. Such a consolidation of the existing infrastructures will also allow one to base a possible LHC upgrade programme on solid ground.

### IV-2.1.2.1 *PS AND SPS*

The refurbishment of some 100 PS magnets would ideally be required, of which the 25 more critical ones have already been done, while 25 more will be done by 2010, during scheduled shutdowns. Until the end of this consolidation, it might be wise not to increase the mean cycling rate and the thermal load. Completing this work for the remaining 50 magnets would be important to minimize the risks of disturbing LHC operation if the PS has to be kept operational well beyond 2015. An attractive (though more expensive) alternative would be to rebuild the PS and take this opportunity to increase its energy (PS+) allowing a more robust injection chain for the LHC (see next section).

The SPS magnets also show signs of ageing. Water leaks have shown up in 2004, resulting in a downtime of about one day per event (for a total of 7 in 2004). As of today, non-destructive inspection techniques of the magnet cooling circuits are not available. An adequate and reasonable strategy has recently been proposed to carry out a preventive repair of all magnets during the shutdowns in order to ensure the long term (>10 years) operation of the SPS. Implementing such a programme is essential to ensure reliable operation of the LHC. In the meantime, one should avoid increasing the thermal load and lengthening of the high energy flat tops, which would be an interesting possibility for fixed-target physics.

## IV-2.2 Optimization and improvement of the CERN proton accelerator complex

As mentioned above, the LHC injection scheme based on Linac2, PSB, PS and SPS allows one to reach the "nominal" luminosity of $10^{34}$ cm$^{-2}$s$^{-1}$. Reaching the "ultimate" limit of $2.3 \times 10^{34}$ cm$^{-2}$s$^{-1}$ (see table above) could be possible by increasing the beam current and brightness after an upgrade of the linac injector to the PS booster; also, other improvements concerning the SPS might be necessary. The required modifications would also increase the operation reliability of the LHC and prepare for a further luminosity upgrade.

Injection in the PSB is a well identified bottleneck for the generation of the type of high brightness beams required for reaching the "ultimate" luminosity of the LHC, because of space-charge effects at 50 MeV. A very attractive solution, which overcomes this limitation, is to build a new Linac (Linac4) delivering H$^-$ at 160 MeV, thus halving the space-charge tune shift at injection in the PSB. This new Linac would also lead to a reduced LHC filling time and an increased reliability. It would also help cover the needs of the future LHC luminosity upgrades. Although not critical for the LHC, further robustness in its operation could be obtained by replacing the PSB with a Superconducting Proton Linac (SPL). Such a proton driver could also be used for producing the intense neutrino beams (see Chapter V) or radioactive ion beams.

So far, no particular limitation is expected from the PS, and the nominal LHC intensity is the maximum obtained at 450 GeV in the SPS. However, predictions for ultimate LHC intensity in the SPS are based on scaling and need experimental confirmation. The main difficulty is due to the electron cloud, which generates vertical single bunch instability. The possible SPS magnets consolidation program (discussed above) may also provide the opportunity to improve the impedance and reduce the electron cloud generation by modifying the vacuum chamber. Other sources of instability limiting the LHC intensities could come from the transverse mode-coupling or extraction kickers, which have already been identified as a troublesome source of transverse impedance. Preliminary studies show that significant improvements for these problems can be expected from a higher SPS injection energy (40–60 GeV). This is a strong motivation for preparing the replacement of the PS with a new synchrotron accelerating (PS+) up to ~ 50 GeV, which would both reduce the bottleneck at SPS injection and solve the problem of ageing of numerous equipments in the PS.

## IV-2.3 LHC upgrades

### IV-2.3.1 The Super LHC

Depending on the nature of the discoveries made at the LHC, higher statistics will be necessary, requiring an increase of the LHC luminosity (see Section III-2.2). This LHC upgrade (known as SLHC) might be achieved by increasing the beam current and brightness and modifying the two high-luminosity insertion regions (ATLAS and CMS).

The initial phase concerns the increase of the beam current to the ultimate value, as discussed in the previous section, leading to a peak luminosity of $2.3 \times 10^{34}$ cm$^{-2}$ s$^{-1}$. The baseline luminosity upgrade scenario relies on a new layout of the interaction regions to reduce β* from 0.5 to 0.25 m and increase the crossing angle by a factor √2, to keep the same relative beam separation at the parasitic collision points. The minimum crossing angle depends on the beam intensity and is limited by the triplet aperture. The corresponding peak luminosity is multiplied by a factor 2, provided the bunch length is halved by means of a new RF system. This scheme is the safest option in terms of beam dynamics, machine protection, and radiation risks, but the new IR magnets and the new RF are challenging.

Further increases in luminosity involve major modifications of several LHC subsystems and of the injector chain to exceed the ultimate beam intensity and possibly to inject into the LHC around 1 TeV. It will also require an increased number of bunches and may not be compatible with electron cloud and long range beam–beam effects. Different bunch distances are being considered: 12.5 ns is currently favoured by the experiments and would yield a peak luminosity of $9.2 \times 10^{34}$ cm$^{-2}$ s$^{-1}$.

Dynamic effects due to persistent currents are known to give difficulties at injection energy in all superconducting colliders and are expected to complicate the setting-up of the LHC. Doubling the injection energy would make the magnetic cycle more stable and double the normalized acceptance of the LHC. This would result in a significant simplification and shortening of the setting-up, with a direct benefit for the turn-around time and the integrated luminosity. It would then be possible to increase the luminosity by injecting bunches of larger transverse emittances, provided the experiments could accept the higher event rate per bunch crossing.

### IV-2.3.2 The Double LHC

An increased injection energy into the LHC would also be a natural first step towards a possible LHC energy upgrade by a factor of 2 (known as DLHC), representing a substantial increase of the physics reach. The difficulties in achieving such an improvement are very large, in particular concerning the main dipole magnets, which would require a field of about 16 Tesla.

The construction of such high-field magnets represents a challenge in many ways.

In the first place, one needs to develop cables with new superconducting wires [BB2-2.1.26]. Several candidate materials exist, amongst which $Nb_3Sn$ is a promising one. However, several serious issues have to be solved concerning both the performance of the cables and their utilization. In particular

- High-current density (1500 A/mm² at 15 Tesla) conductor allowing operation with stable current flux is not currently commercially available.
- The superconducting material is brittle and its properties are train-sensitive. The process for winding and impregnating the magnets thus remains a delicate operation, not yet industrialized.

Should the construction of the magnets be solved, the evacuation of the heat produced by the radiation generated by the beams is a difficulty that requires the development of new solutions.

The above issues need to be addressed within a very active R&D programme and reasonable solutions have to be developed for considering the DLHC as a viable option.

## IV-2.4 Summary and required R&D

Linac4 will be essential to improve the injection in the PSB. This will make possible the regular delivery of the ultimate beam to the LHC, reduce its filling time and positively contribute to the overall reliability of the injector complex. To benefit from these improvements already in 2011, Linac4 construction would need to start in 2007. Thanks to the continuous R&D currently carried out on various types of cavity and magnetic components, in particular within CARE [10], such a scenario seems technically realistic.

Further studies in the SPS will help confirming the interest of a new ~ 50 GeV synchrotron (PS+) replacing the PS.

The replacement of the PSB has to be planned in the longer term to get the maximum benefit from a possible PS successor. A superconducting proton linac (SPL) is today a promising accelerator for such purposes in the CERN context. Since its main characteristics might not be critical for the LHC, they would most probably be defined by the needs of other physics facilities concerning, for instance, radioactive ions (EURISOL [11]) and/or neutrinos. As for Linac4, the continuation of the on-going R&D programme on accelerating structures would help developing the critical components.

For the upgrade of the magnets in the LHC interaction regions, and to secure the presence of spare low-β quadrupoles, an intermediate solution would be desirable as soon as possible before 2015. Due to the long lead-time, the Nb-Ti technology is the most practical. However, such magnets would only allow for a moderate luminosity increase, probably up to ~ $3\times10^{34}$ cm$^{-2}$ s$^{-1}$. The development of $Nb_3Sn$ magnets is necessary to get the full benefit of a reduced β* of 0.25 m.

Finally, the experience that will acquired with the commissioning and running-in of the LHC will help determine the difficulty of operating with 450 GeV injection energy and the relative merit of building a new 1 TeV injector for the LHC.

Because of the long lead-time associated with it, critical R&D [BB2-2.1.26] has to begin or be strengthened quickly for

- the superconducting high-field magnets for the LHC interaction regions for the luminosity upgrade and, on a longer term, for the energy upgrade,
- the fast-cycling magnets that may be needed for the superconducting successors of the PS (50 GeV PS+) and/or of the SPS (1 TeV SPS+),
- the superconducting cavities that may be used in a superconducting linac replacing the PSB (SPL).

# IV-3 High-Energy Linear Colliders

The physics motivation for a next-generation $e^+e^-$ collider has been studied in a large number of national and international workshops in Europe, Asia and USA during the past 15 years. Thus a high-energy electron–positron collider is generally considered as an essential facility, which is complementary to the LHC. Besides improved precision measurements of the $Z^0$, W boson and top-quark decays and mass (requiring the collision energy to exceed 400 GeV), it should allow one to study in detail the Higgs boson and the different kinds of particles or new phenomena discovered by the LHC, extending further the domain of exploration for physics beyond the Standard Model (SM). Such a facility would also offer a discovery potential for some phenomena and specific types of new particles that would be very difficult to be observed at the LHC.

The interplay and the complementarities between $p\bar{p}$ and $e^+e^-$ collider have proved to be very efficient in the past. For example, the W and $Z^0$ bosons have been major discoveries made at a hadron collider ($Sp\bar{p}S$). Their precise studies have been carried out at lepton colliders (LEP and SLC), allowing one to test in detail the SM and, in particular, to determine the number of fermion families and the allowed mass range within the SM for the Higgs boson and the top quark. This latter particle was discovered at the Tevatron $p\bar{p}$ collider.

Currently, two types of lepton colliders are under study world-wide: the International Linear Collider (ILC), for which a technical design is worked out, and the Compact Linear Collider (CLIC), for which a design concept is being developed.

## IV-3.1 The International Linear Collider[6] (ILC)

To achieve the main physics goals mentioned above (see section 2.3 in Chapter III), an intense international R&D programme has been set up since many years to develop a high-energy $e^+e^-$ linear collider with the objective of constructing in a timely fashion an International Linear Collider (ILC).

Its main design parameters, driven by physics considerations, are:

- an initial c.m. energy of 500 GeV, upgradable to 1 TeV,
- an integrated luminosity of 500 fb$^{-1}$ in 4 years (100 times that of LEP),
- an energy stability and precision below 0.1%,
- an electron polarization of at least 80%.

After years of parallel R&D on "cold" and "warm" alternative designs at different RF frequencies, a major step forward was made by the International Technology Recommendation Panel (ITRP) in 2004, with the decision to base the ILC on superconducting technology. Indeed the superconducting technology has been demonstrated mature enough to build a linear collider able to achieve the design parameters with an energy of at least 500 GeV.

## IV-3.1.1 Main specificities of the ILC

The choice of the superconducting technology, as expressed by the ITRP, is mainly based on the following reasons:

- The large cavity aperture and long bunch interval simplify operations, reduce the sensitivity to ground motion, permit inter-bunch feedback, and may enable increased beam current.
- The main linac and RF systems, the technical elements of single largest cost, offer comparatively less risk.
- The superconducting European XFEL free electron laser will provide prototypes and test many aspects of the linac.
- The industrialization of most major components of the linac is under-way.
- The use of superconducting cavities significantly reduces the power consumption (the overall power transfer efficiency to the beam is about 20%).

Furthermore, thanks to the work carried out at the TESLA Test Facility (TTF), it is also generally accepted that the cold technology has established the necessary proofs of feasibility allowing us to launch a world-wide effort toward making the final steps up to construction. Therefore, soon after the ITRP decision, a Global Design Effort, involving experts from Asia, America and Europe, started in 2005 with the aim of producing a design for the ILC that includes a detailed design concept, performance assessments, reliable international costing, an industrialization plan and site analysis, as well as detector concepts and scope. A first milestone was achieved at the end of 2005 with the finalization of the baseline configuration, while the Reference Design Report, including cost estimate, is planned for the end of 2006.

### IV-3.1.1.1 REMAINING CHOICES

The choice of the superconducting 9-cell 1.3 GHz niobium TESLA cavities as baseline for the accelerating structure allows for a power-efficient construction, and outstanding progress has been achieved concerning the gradient of the cavities in the past decade. Their performance repeatedly exceeds 35 MV/m with record gradient at the 40 MV/m level. However, the issue of the accelerating gradient remains a very important parameter in the optimization of cost and reliability. A gradient of 35 MV/m is close to the cost optimum, while a lower design value (30 MV/m) would leave a safety margin. As of today, the gradient dispersion of the produced cavities still seems rather large. A major R&D effort is being invested in this area to control the fabrication process better, especially within the CARE program. In particular, the surface treatment and material type (electropolished and baked fine-grain Nb material versus large grain) are investigated in detail as well as the optimization of the cavity shape, for which 3 different geometries are studied.

Important choices have recently been adopted for the baseline configuration with possible alternatives, such as:

- A positron production mechanism based on an undulator scheme, expected to allow positron polarization, with the conventional scheme as alternative.

- A damping ring based on 6 km rings, of dogbone shape as alternative.
- Two separate tunnels for klystrons and accelerator, with a single tunnel layout as alternative.

Meanwhile, the Baseline Configuration Document (which can be found in ref.[3]) has been worked out meanwhile in much detail by the GDE. Some other choices still remain to be made, in particular the optimal crossing angle at the interaction point.

Moreover, since the total cost is considered a key factor in the decision for the ILC construction, cost optimization of all systems is of the highest importance.

### IV-3.1.1.2 REMAINING MAIN SPECIFIC ACCELERATOR ISSUES

Solutions to a few specific issues remain to be consolidated. They include:

- Developing high-gradient superconducting RF systems:
  It requires consolidating and understanding the surface preparation techniques in order to achieve the required cost reduction with respect to using the SC technology as it is today (the XFEL will be constructed using cavities with a gradient of about 25 MV/m).

- Achieving nm-scale beam spots:
  It requires generating high-intensity beams of electrons and positrons; damping the beams to ultra-low emittance in damping rings, including the instability calculations; transporting the beams to the collision point without significant emittance growth or uncontrolled beam jitter; cleanly dumping the used beams.

- Reaching luminosity requirements:
  Present designs satisfy the luminosity goals in simulations. However, a number of challenging issues in accelerator physics and technology must still be solved. In particular the collimation system and the machine–detector interface have to be studied in detail.

None of these points represents a show stopper to the construction of the ILC, but they require ensuring an appropriate and targeted effort. It is worth noting that the last two items apply to linear colliders in general.

## IV-3.1.2 R&D, test facilities and overall schedule

A number of world-wide test facilities were set up to further develop the ILC technology. They include the TESLA test facility linac at DESY, a cryomodule assembly facility at Fermilab, and a test facility at KEK. Europe is active in the ILC R&D via the TESLA Technology Collaboration [12], the European XFEL collaboration, and a dedicated effort within CARE on superconducting RF systems (cavity production, tuners, couplers, diagnostics), while design studies are carried out within EUROTeV [13]. Cryogenic test facilities are also available at DESY and Saclay, but a central major facility would be desirable [BB2-2.1.24].

This world-wide effort represents the inputs to the Global Design Effort. The plan and schedule presented by the GDE are as follows:

- production of a baseline configuration design by December 2005 (done);
- production of a reference design report (incl. cost estimate) by December 2006;
- production of a technical design report (incl. detailed costing) in 2008.

The construction decision could be around 2010, in the context of the first physics results from the LHC and with better knowledge of the status of the R&D for CLIC.

The plan and schedule described above can be summarized as in the following chart:

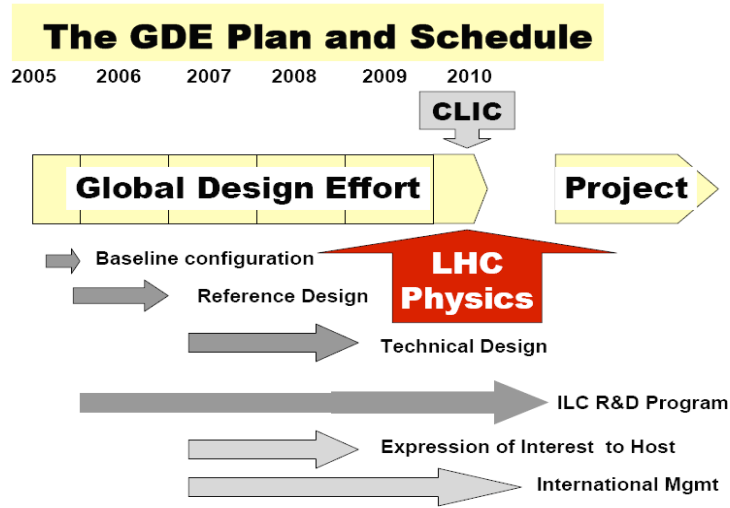

## IV-3.2 Compact Linear collider[14] (CLIC)

Depending on the scale of new physics, a detailed exploration of the energy scale up to several TeV may be required (see section III-2.4). As of today, it is generally accepted that CLIC technology [BB2-2.1.04] is the most promising for realizing high luminosity $e^+e^-$ collisions reaching a c.m. energy of 3 to 5 TeV.

The main features of CLIC are the following:

- An energy range of 0.5–5 TeV c.m., with a luminosity of $10^{34-35}$ cm$^{-2}$ s$^{-1}$;

- An accelerating gradient of 150 MV/m, resulting in a total linac length of 27.5 km for a 3 TeV collider (or 4.8 km for 0.5 TeV) and enabling the energy reach up to 5 TeV;

- A novel design based on the "two-beam scheme" in which the 30 GHz RF power for the main linac acceleration is extracted from a series of low-energy high-current drive beams running parallel to the main linac.

### IV-3.2.1 Main specificities

In order to achieve the above design luminosity, very low emittance beams have to be produced, accelerated and focused down to very small beam sizes at the interaction point (~ 1 nm in the vertical plane). Beam acceleration is obtained using high-frequency (30 GHz) normal-conducting structures, operating at high accelerating fields (150 MV/m). This high gradient significantly reduces the length of the linac. The pulsed RF power (460 MW/m) to feed the accelerating structures is produced by the so-called "two-beam scheme", in which the 30 GHz power is extracted from high-intensity/low-energy drive beams running parallel to the main beam. These drive beams are generated centrally and are then distributed along the main linac. The beams are accelerated using a low-frequency (937 MHz) fully loaded normal-conducting linac. Operating the drive beam linac in the fully loaded condition results in a very high RF-power-to-beam efficiency (~ 96%). The two-beam scheme allows an overall power transfer to the main beam of about 12.5%.

#### IV-3.2.1.1 ATTRACTIVE SPECIFIC FEATURES

The fact that there are no active RF components in the main linac means that CLIC has a single small-diameter (3.8 m) tunnel.

A particularly attractive feature of the CLIC scheme is that to upgrade the energy of the collider, the only change required in the RF power system is in the pulse length of the modulators, which drive the low-frequency (937 MHz) klystrons and not an increase in the number of klystrons (the nominal pulse length for the 3 TeV collider is 100 μs).

Only a relatively small number of klystrons are required for the CLIC scheme – this independently of the final energy. The power for each drive-beam accelerator is supplied by 352 40 MW multibeam klystrons, which are grouped together in the central area of the facility. This central location facilitates power distribution, cooling and maintenance work.

The energy required for acceleration is transported, compressed and distributed using high-power electron beams: conventional systems generate the RF power locally and then transport it over long lossy waveguides; the CLIC energy is only converted into RF power where it is required (typically 60 cm from each CLIC main linac accelerating structure).

The use of a high RF frequency (30 GHz) reduces the peak power that is required to achieve the 150 MV/m accelerating gradient.

### IV-3.2.1.2 SPECIFIC COMPLICATIONS AND DIFFICULTIES

The conditioning of the drive and main linacs with RF power to acceptable breakdown rates is more complicated for a two-beam scheme than for a conventional scheme with conventional RF power sources. This will almost certainly require the provision of some over-capacity in the basic design and the ability to turn the power extraction structures (PETS) on and off.

The higher CLIC RF frequency makes the collider more sensitive to alignment errors and ground stability.

Finally, the drive beam generation system of the CLIC two-beam scheme represents a fixed investment cost, which is independent of the energy. This makes the scheme less cost-effective at low energies.

## IV-3.2.2 Main achievements

Basic designs of all CLIC subsystems and essential equipment have been made, but more work on this design is required. The technical feasibility of two-beam acceleration has been demonstrated in CLIC Test Facility 2 (CTF2). In this test, the energy of a single electron bunch was increased by 60 MeV, using a string of 30 GHz accelerating cavities powered by a high-intensity drive linac.

The nominal CLIC accelerating gradient of 150 MV/m at the nominal pulse length of 70 ns has been obtained during the last CTF3 run of 2005, using molybdenum irises in 30 GHz copper structures. This successful milestone is, however, hampered by the fact that the breakdown rate was found several orders of magnitude above the acceptable rate for a steady operation of the linear collider, and that a recent inspection has shown some damage on the irises. Further modifications and developments are under-way and will be tested within the CTF3 program to demonstrate the feasibility and performance of reliable accelerating structures.

An experimental demonstration of the principle of the bunch combination scheme has been made at low charge using a modified layout of the former LEP Pre-Injector (LPI) complex.

A successful demonstration of full-beam-loading linac operation has been made using the injector of the new CLIC Test Facility 3 (CTF3).

A prototype CLIC quadrupole has been stabilized to the 0.5 nm level in a relatively noisy part of the CERN site, using commercially available state-of-the-art stabilization equipment.

### IV-3.2.3 CLIC-technology-related feasibility issues

Issues common to other linear collider studies are being studied within the framework of the existing world-wide linear collider collaborations. The International Technical Review Committee has indicated a number of crucial items for which the CLIC Collaboration must still provide a feasibility proof (the so-called R1 items) and also a number of issues, which must be investigated in order to arrive at a conceptual design (R2 items).

The three "CLIC-technology-related" R1 issues are:

R1.1 Test of damped accelerating structure at design gradient and pulse length;
R1.2 Validation of the drive-beam generation scheme with a fully loaded linac;
R1.3 Design and test of an adequately damped power-extraction structure, which can be switched on and off.

In addition, two of the "CLIC-technology-related" R2 issues are:

R2.1 Validation of beam stability and losses in the drive-beam decelerator, and design of a machine protection system
R2.2 Test of a relevant linac subunit with beam

CTF3 aims at demonstrating the feasibility of all five of these key issues.

One important CLIC-technology-related feasibility issue is the necessity to synchronize both main and drive beams to avoid excessive luminosity loss due to energy variations. To achieve this, the timing of both main and drive beams have to be measured to a precision of about 10 fs. This problem is being studied within the EUROTeV design studies, as are the effects of coherent synchrotron radiation in bunch compressors and the design of an extraction line for 3 TeV c.m. energy.

### IV-3.2.4 The CTF3 facility

The challenging R&D on CLIC technologies is pursued at the CTF3 facility at CERN: in particular, test of drive beam generation, acceleration and RF multiplication by a factor 10, two-beam RF power generation, and component tests with nominal fields and pulse length. CTF3 is being built in stages by a collaboration of 14 institutes from 9 countries. Several contributions for studying the CLIC concept and critical components for CTF3 are carried out within the EUROTeV and CARE programmes.

The anticipated planning for achieving the proofs of feasibility with CTF3 is shown in the table below. The experience gained with the operation of this facility should therefore lead to the assessment of the CLIC design concept by 2010. It will then take several more years to develop a detailed technical design. In this success-oriented scenario, it would be technically possible to start construction at the end of 2015, at the very earliest.

| Component studied | 2004 | 2005 | 2006 | 2007 | 2008 | 2009 |
|---|---|---|---|---|---|---|
| Drive-beam accelerator | ■ | | | | | |
| 30 GHz high-gradient test stand | ■ | ■ | | | | |
| 30 GHz high-gradient testing (4 months per year) | | ■ | ■ | ■ | ■ | ■ |
| *R1.1 Feasibility test of CLIC accelerating structure* | | | | ▨ | | |
| Delay loop | ■ | ■ | | | | |
| Combiner ring | ■ | ■ | ■ | | | |
| *R1.2 Feasibility test of drive-beam generation* | | | | ▨ | | |
| CLEX | | | ■ | | | |
| *R1.3 Feasibility test of PETS structure* | | | | ▨ | | |
| Probe beam | | | ■ | ■ | | |
| *R2.2 Feasibility test of relevant CLIC linac subunit* | | | | | ▨ | |
| Test beam line | | | ■ | ■ | ■ | |
| *R2.1 Beam stability bench-mark tests* | | | | | ▨ | ▨ |

# IV-4 Very High Energy Frontier Accelerators

## IV-4.1 Muon collider

In general terms, the physics that could be studied with a muon–muon collider would be similar to that with an electron–positron machine (linear collider). However, thanks to the heavier mass of the muon, much less synchrotron radiation is produced, allowing us to reach a much higher energy with smaller radiative correction and associated physics background in circular accelerators.

### IV-4.1.1 Main features

There are several major advantages in using muons instead of electrons:

- Like protons, they can be accelerated and stored in circular rings at energies above 250 GeV up to several TeV, as opposed to electrons, which have to be accelerated in linear machines. Furthermore, since the effective energy in the collision of point particles for carrying detailed measurements is roughly 10 times larger (although this ratio is only about 3 for the direct observation of new particles) than that of protons, a circular muon machine would be a factor of 10 smaller than its proton equivalent. For example, a muon collider of 4 TeV c.m. energy, with a 6 km circumference, would have an effective energy, i.e. physics study potential, similar to that of the 80 km Superconducting Very Large Hadron Collider. Also, it would be smaller than a conventional electron linear collider of the same energy; e.g. a 4 TeV electron collider would be about 50 km long.

- A smaller beamstrahlung, with consequently smaller energy loss and narrow energy spread, opens the possibility of more particles per bunch, yielding greater luminosity.

- The direct s-channel Higgs boson production is greatly enhanced, since the coupling of the Higgs boson to fermions is proportional to their mass, hence leading to a possible "Higgs factory" (see section III-2.5).

### IV-4.1.2 Technical difficulties

There are serious technological challenges to building a muon collider. Amongst those, several major items are well identified. They include issues such as

- the production of an intense muon beam obtained through the decays of pions produced by a multimegawatt proton driver interacting on a high-power target, and captured with a high-efficiency collection system;

- the achievement of very low emittance muon beams essential for the required luminosity of $L = 10^{35}$ cm$^{-2}$s$^{-1}$. Since muons are initially produced with large transverse momentum, a cooling technique is required.

A first step toward a muon collider could be a neutrino factory, whose physics motivation and related R&D activities are discussed in Chapter V.

## IV-4.2 Very Large Hadron Collider

As an alternative to the muon collider, a very large hadron collider is considered for reaching very high energy. To make a significant step forward with respect to the LHC and its upgrades, and to match the possible energy reach of a muon collider, the VLHC should aim at reaching an energy of 100 TeV and a minimum luminosity of $10^{34}$ cm$^{-2}$s$^{-1}$.

### IV-4.2.1 Long tunnel or high field

Two directions are considered:

- With the present LHC technology and its 8-Tesla dipole magnets, the needed tunnel would have a length of about 200 km. However the cost of such a collider would be unacceptably large. Using a 15-Tesla magnet with a new conductor, such as Nb$_3$Sn, as envisaged for the DLHC, would allow the tunnel length to be halved, or the energy to be doubled.

- Ideas exist for realizing low-cost magnets. However the maximum field that can be achieved is about 2 Tesla. The main issues would then reside in the realization of a huge tunnel (about 700 km long) and in the massive manufacturing of the magnets, complete with vacuum and cryogenic system.

### IV-4.2.2 Main R&D needed

The main effort for the R&D should focus on the vacuum system and the magnets, with the objective of drastically reducing the cost, while keeping the highest possible field. The effort needed for the LHC upgrade would be beneficial to such a collider, although it is not clear whether the cost issue can be solved with the currently envisaged technology.

## IV-5 Ultra High Energy Acceleration

The need for ever increasing energy accelerators is likely to continue. Limiting the size of these facilities to practical dimensions calls for developing novel acceleration technologies reaching high gradient, well above GeV/m.

Most of the concepts studied at present rely on the use of laser-driven systems, thanks to the extreme fields achieved in focused short laser pulses. They include: inverse free electron laser (IFEL), inverse Cherenkov effect, and inverse Smith–Purcell effect (called also diffraction-grating acceleration). Other ideas for creating large electric fields are also proposed, such as the use of a dielectric cylinder in conditions of resonantly excited whispering-gallery modes (WGM) [BB2-2.1.18].

However, the recent reports on plasma wakefield accelerators, demonstrating an energy gain equivalent to 100 GeV/m in a few millimetres, renew the hope of realizing particle accelerators.

## IV-5.1 Accelerators with laser-plasma acceleration technique

Present laser-plasma accelerators research relies on the laser wakefield mechanism, which is currently the most promising approach to high-performance compact electron accelerators.

### IV-5.1.1 Resonant laser wakefield acceleration scheme

In the resonant laser wakefield acceleration (LWFA) scheme a short laser pulse, of the order of the plasma period, excites in its wake a plasma wave, or wakefield, that can trap and accelerate electrons to high energy. In the linear or moderately non linear regime of LWFA, or "standard" LWFA, accelerating electric fields are of the order of 1 to 10 GV/m (as measured[1] in experiments and in agreement with simulation), and relativistic electrons injected in the plasma wave are expected to gain an energy of the order of 1 GeV over a few centimetres. The present limitation for the energy of accelerated electrons in standard LWFA is due to the small acceleration distance, limited to a few Rayleigh lengths and typically of the order of 1 mm. Therefore, despite high acceleration gradients (> 1 GV/m), the final energy gain of accelerated electrons, achieved in experiments to date, is rather small (~ 1 MeV) [15]. The extended propagation of a laser pulse over many Rayleigh lengths is necessary to create a long acceleration distance and high-energy electrons, and can be achieved by the use of guiding structures such as plasma channels [16] and capillary tubes [17].

### IV-5.1.2 Non-linear regime of LWFA

In the non-linear regime of LWFA, achieved for laser pulse durations shorter than the plasma period, an intense laser beam drives a highly non-linear wakefield, also called bubble, which traps and accelerates electrons from the plasma. The accelerated electrons observed [18] in this regime emerge from the plasma as a collimated (3–10 mrad divergence), short-duration (sub-50 fs) bunch, with typically a 0.5–1 nC charge in the main peak of the energy spectrum at 170 ± 20 MeV (24 % energy spread) and at 80 ± 1 MeV (2 % energy spread). As the plasma length is of the order of 2 mm, accelerating electric fields are inferred to be larger than 100 GV/m. Recent, as yet unpublished, experimental results of acceleration of electrons up to 500 MeV inside a waveguide in this regime have been reported. In these non-linear regimes, the observed electron energy distribution varies from shot to shot.

### IV-5.1.3 Research and Development

Though higher electric fields can be achieved in strongly non-linear regimes, the standard LWFA regime allows us to control the properties of the accelerating structure and consequently the parameters of the accelerated beam. Recently, the European project EuroLEAP [19] (European Laser Electron controlled Acceleration in Plasmas to GeV energy range) has been launched. The objective is the achievement, in the next 3 years, of a laser-plasma accelerator to test the issues related to the control of the properties of an electron beam accelerated to the GeV energy range in a plasma wave. Short-pulse (10 to 500 fs) electron beams, produced by laser injectors in a plasma or RF photo-injectors, will be accelerated by a linear plasma wave created over a few centimetres. The goal is to produce electron beams in the GeV energy range, with an energy spread less than 1%, in a reproducible way over a distance less than 10 cm. In the frame of this project, it is planned to develop injectors and the plasma medium, and combine these to perform staged and controlled acceleration studies, through the development of advanced fs electron bunches, plasma and laser diagnostics.

This is a crucial step to determine the feasibility of staging in plasma-based accelerators, which seems to be the most viable way to achieve the 10 GeV range in the next decade.

# IV-6 Conclusion

As can be seen from this chapter, many projects addressing the high-energy frontier issues do exist, demonstrating the vitality and the creativity of the community.

However, all proposed future high-energy frontier accelerators need well structured continuous and vigorous R&D efforts. Depending on the level of maturity of the technology required, some of the proposed infrastructures have to focus more on the component reliability developments and industrialization aspects, some others need to establish the proof of feasibility with accelerator test facilities, while for others still one needs to carry out accelerator research and proofs of concept.

The human and financial resources that are required call for setting up large R&D collaborations. This was initiated long ago by the TESLA collaboration for developing the superconducting technology for a linear collider, or by the CTF collaboration for developing the two-beam accelerator technology. More recently, accelerator R&D gained a strong boost, thanks to the EC-funded projects within the $6^{th}$ Framework Programme. We show in the table below the list of recently approved projects, together with their cost.

| Project | Type of programme | Type of beams | Start date | Duration (years) | Total cost (M€) | EC contribution (M€) |
|---|---|---|---|---|---|---|
| CARE | I3 | All | 1/1/2004 | 5 | 55 | 15.2 |
| EUROTeV | DS | $e^+e^-$ (LC) | 1/1/2005 | 3 | 29 | 9 |
| EURISOL | DS | Ion, p (ν β-beam) | 1/1/2005 | 4 | 33 | 9.2 |
| EuroLEAP | NEST | e plasma acceleration | 1/3/2006 | 3 | 4.1 | 2 |
| Total | | | | | > 120 | **35.2** |

Although this has allowed the European effort to increase significantly, it still seems insufficient in regard of the challenges that we are facing for developing the accelerators of the coming decades [BB2-2.1.19]. In particular, the persistent lack of accelerator physicists is worrying and the pressing need for enhancing the education, the training and the recruitment of young accelerator physicists is manifest. Furthermore, the appointment of professors in the field of accelerator physics would need to be strongly encouraged in the universities and supported by the research institutes.

# IV-7 Summary of the Orsay discussion

Following the presentation on High Energy Frontier Accelerators made by P. Raimondi at the Orsay Symposium, a discussion with and within the audience took place, focusing mainly around the linear collider and the LHC upgrades. A brief summary is given by topics, here below.

## IV-7.1 Comparison between ILC and CLIC technologies and status

It was stressed that there is a risk for the linear collider to be very expensive and it was asked whether some extra R&D efforts could drive the price down. The answer was that, as shown in Raimondi's presentation, the optimization of the gradient is not a critical parameter for cost reduction. A more aggressive gradient would not drive the price down significantly, as the optimal gradient from the cost point of view has already been achieved. It was also remarked that the Tesla technology collaboration has been doing R&D for 15 years and they have combined the world know how on superconducting cavities. The cost improved by a factor of 30 in 10 years and there are no more large factors to be gained.

It was stated that when comparing the warm and cold technologies, one should consider also issues like difficulty to reach the nominal luminosity, reliability and cost of operation.

Indeed the general sense was that it is easier to preserve the emittance of the beam when using cold technology, since the higher the frequency the more difficult it is to preserve the emittance. It was recalled that one of the reasons why the cold technology was chosen was that its specific features simplify operations, reduce the sensitivity to ground motion, and may enable increased beam current. The cold technology appears at lower risk thanks to the long-standing R&D and the coming construction of the XFEL, which will provide prototyping and testing of the linac in many ways.

However, on the subject of emittance preservation, it was also pointed out that the tolerances required are tighter for CLIC, but that its elements are also more accessible, because they are not housed in a cryostat. For beam dynamics and tolerance issues, the ILC and CLIC should be simulated with the same tools to compare performances and sensitivities.

It was asked what the relative power consumptions of the ILC and CLIC are. The answer was that, at 1TeV, they have similar operating power.

The question of when CLIC will reach the same degree of maturity as the ILC was also debated. It was said that the CLIC feasibility study will end by 2009, with the completion of CFT3 and the subsequent studies. Then the design phase of the collider will start and it will take several more years to develop a detailed technical design.

It was pointed out that CLIC (3 TeV, 1 ns) and ILC (500 GeV, 300 ns) are two different machines, designed for different physics and therefore they should not be compared as such.

## IV-7.2 Site for a Linear Collider

Concerns were raised on the role of CERN, should the ILC be constructed. CLIC might then be done only on the long term. What would CERN become in the meantime? Would it risk losing its international span? In addition, it was said that it would be difficult for small countries to participate in the ILC programme without the framework of CERN.

Following the concern that the ILC could be in Illinois, somewhere in Asia, or in Europe but NOT in the Geneva region, it was stated that CERN is a candidate for hosting a linear collider and that CERN is participating in the GDE and ILC studies. Indeed, it was also recalled that there are two possible sites considered in Europe in the GDE (CERN and DESY), and that a 50 km long collider can, geographically, be fitted in the Geneva surroundings between Lake Geneva and the Jura mountains.

## IV-7.3 LHC upgrades

To the question "What R&D effort can be put in the LHC luminosity upgrade when LHC is in its construction phase?", it was answered that the large majority of the present effort is put towards the construction of the LHC. However, there are also some efforts devoted to the R&D towards the LHC luminosity upgrade. This R&D is feasible within a reasonable time scale.

It was recalled that the first goal, for any brightness upgrade (SLHC), is the upgrade of the injectors. It was stressed that upgrades of the CERN accelerator complex could also be beneficial to other-than-LHC users and a working group addressing these issues is active at CERN.

Concerning the energy upgrades that would possibly come within a timescale of 10–15 years from now, it was asked what the R&D effort put in this area is. It was answered that the effort must concentrate on the design of magnets that can give twice the field that is achieved today. In order to double the energy dipoles with a field of about 16 Tesla are needed. The $Nb_3Sn$ technology is promising for these field levels. Studies are ongoing, but the support in terms of people and money is low. Efforts are in NED (CARE), INFN, and at Twente University. The minimum incompressible time for this R&D is 10 years from now, but more people working on it are needed, otherwise it will take more than 15 years.

It was pointed out that synergies exist between luminosity and energy upgrades through the development of $Nb_3Sn$ magnets. The effort for developing magnets with $Nb_3Sn$ conductors is also needed for the luminosity upgrade, as the quadrupoles of the IR are likely to use this material for reaching a lower $\beta^*$.

It was stressed that the upgrade of the detectors to follow the luminosity upgrade would probably require a major effort in terms of money. Representatives of CMS and ATLAS stated that this point is very clear and that R&D is going on in the collaborations.

It was asked whether there was some more information about the electron–proton collider that was mentioned by the speaker of the session on physics at the high energy frontier. The answer was that there was a recent paper, submitted to the SG, proposing to collide an electron beam of 70 GeV located in the LHC tunnel with one of the LHC beams. It is claimed that a luminosity of $10^{33}$ cm$^{-2}$ s$^{-1}$ at a c.m. energy of 1.4 TeV could be achieved at such a facility, which could operate simultaneously with the LHC.

# IV-8 Written contributions to High Energy Frontier

Many contributions from individuals and groups have been sent to the Strategy Group. A number of them are addressing the High Energy Frontier. They can be found in Briefing Book 2, references [BB2-2.1.01 to BB2-2.1.27]. Amongst those, some are discussing technical aspects directly relevant to the present chapter. They are [BB2-2.1.11] and [BB2-2.1.26] for the LHC (including the CERN accelerator complex) and its upgrades, [BB2-2.1.24] for R&D infrastructures related to superconducting RF systems, [BB2-2.1.04] for CLIC, [BB2-2.1.18] for novel accelerator techniques and [BB2-2.1.19] on accelerator R&D in general. Other contributions discuss ep colliders [BB2-6.03], linac-ring type colliders (including ep[eA] and γp[γA]) [BB2-2-1-17], neutrino factories [BB2-2-2-03,04 and 05] and flavour factories [BB2-2-3-01] and [BB2-2-3-03], and are discussed in other chapters of this document.

# V OSCILLATIONS OF MASSIVE NEUTRINOS

## V-1 Present status

This is a great surprise of particle physics at the turn of the 21st century: the Standard Model (SM) met triumph with the precision measurements at LEP and SLD, the last missing quark was discovered at Fermilab, the quark-mixing scheme was confirmed in a splendid manner at the B factories. At the same time, the observation of neutrino oscillations, demonstrating that neutrinos have mass and mix, gave the first direct signal of physics beyond the SM.

From the first experimental hints, provided by solar neutrinos already in the early 1970's, to the solid confirmation provided by definitive experiments on atmospheric neutrinos [20] and solar neutrinos [21] natural neutrino sources have provided the initial evidence that neutrinos transform into each other, and therefore are massive and mix. Man-made neutrinos, from reactors [22] or from accelerators [23] together with more precise measurements of solar and atmospheric neutrinos, have confirmed that neutrinos undergo, as expected, a coherent quantum phenomenon called oscillations, which takes place over distances of hundreds to millions of kilometres. Present observations indicate that these oscillations are governed by two distinct sets of mass splittings and mixing angles, one for solar (or reactor) electron-neutrinos with an oscillation length of 17000 km/GeV and a mixing angle ($\theta_{12}$) of about 30º; and the other for atmospheric muon-neutrinos with an oscillation length of 500 km/GeV and a mixing angle ($\theta_{23}$) of about 45º. The present level of precision on these parameters is about 10–20%. Since we know from LEP that there are three families of active light neutrinos, we expect a three-family mixing similar to that of quarks; this should manifest itself by the existence of a third mixing angle $\theta_{13}$, for which a limit of about 10º exists at present, and of a phase $\delta$ responsible for CP violation.

Neutrino masses could in principle be incorporated in a trivial extension of the SM, but this would require i) the addition of a new conservation law that is not now present in the SM, fermion-number conservation, and ii) the introduction of an extraordinarily small Yukawa coupling for neutrinos, of the order of $m_\nu/m_{top} \cong 10^{-12}$. More natural theoretical interpretations, such as the see-saw mechanism, lead to the consequence that neutrinos are their own antiparticles, and that the smallness of the neutrino masses comes from their mixing with very heavy partners at the mass scale of Grand Unification Theories (GUTs). For the first time, solid experimental facts open a possible window of observation on physics at the GUT scale.

There are many experimental and fundamental implications of this discovery. Perhaps the most spectacular one is the possibility that the combination of fermion-number violation and CP violations in the neutrino system could, via leptogenesis, provide an explanation for the baryon asymmetry of the Universe.

The experimental implications are not less exciting. Fermion-number violation, and the absolute mass scale of light neutrinos, should be testable in neutrinoless double beta decay. The direct measurement of the average mass of electron-neutrinos in beta decay could lead to an observable result. The precise values of mass differences, the ordering of masses and the determination of mixing angles is accessible to neutrino-oscillation experiments. Last but not least, the discovery of CP or T violation in neutrino oscillations appears to be feasible, but it requires a new type of experimentation:

precision appearance neutrino-oscillation measurements involving electron-neutrinos. Precision neutrino oscillation experiments, and the CP asymmetry search in particular, require accelerator-based neutrino facilities, on which we concentrate in this chapter. Further discussion of the origin of neutrino masses and of the phenomenology of neutrino oscillations can be found in the physics chapters of Refs. [24] and [25].

## V-2 Neutrino-oscillation facilities

### V-2.1 The present generation

A more complete review can be found in Ref. [26]. Over the next five years, the present generation of oscillation experiments at accelerators with long-baseline $\nu_\mu$ beams (Table V-1), K2K [23] at KEK, MINOS [27] at the NuMI beam at Fermilab, and ICARUS [28] and OPERA [29] at the CNGS beam at CERN, are expected to confirm the atmospheric evidence of oscillations and should improve somewhat the measurements of $\sin^2 2\theta_{23}$ and $|\Delta m^2_{23}|$ if $|\Delta m^2_{23}| > 10^{-3}$ eV$^2$. K2K and MINOS are looking for neutrino disappearance, by measuring the $\nu_\mu$ survival probability as a function of neutrino energy, while ICARUS and OPERA will search for the appearance of $\nu_\tau$ interactions in a $\nu_\mu$ beam by $\nu_\mu \to \nu_\tau$ oscillations, an unescapable, but so far unverified, consequence of the present set of observations in the three-neutrino-family framework. K2K has already completed its data taking at the end of 2004, while MINOS has started taking data at the beginning of 2005. CNGS is expected to start operation in 2006.

**Table V-1 Main parameters for present long-baseline neutrino beams**

| Neutrino facility | Parent-proton momentum | Neutrino baseline | Neutrino beam | pot/yr (10$^{19}$) |
|---|---|---|---|---|
| KEK PS | 12 GeV/c | 250 km | WBB peaked at 1.5 GeV | 2 |
| Fermilab NuMI | 120 GeV/c | 735 km | WBB 3 GeV | 20–34 |
| CERN CNGS | 400 GeV/c | 732 km | WBB 20 GeV | 4.5–7.6 |

These facilities are on-axis, conventional muon-neutrino beams produced through the decay of horn-focused $\pi$ and K mesons. The CNGS $\nu_\mu$ beam has been optimized for the $\nu_\tau$ appearance search. The resulting $\nu_\mu$ beam has a contamination of $\nu_e$ coming from three-body K$^\pm$, K$^0$ and μ decays. The CNGS muon-neutrino flux at Gran Sasso will have an average energy of 17.4 GeV and ~ 0.6% $\nu_e$ contamination for $E_\nu < 40$ GeV.

Although it is not part of the original motivation of these experiments, they will be able to look for the $\nu_\mu \to \nu_e$ transition at the atmospheric wavelength, which results from a non-vanishing value of $\theta_{13}$. MINOS, at NuMI, is expected to reach a sensitivity of $\sin^2 2\theta_{13} = 0.08$, the main limitation being the electron-identification efficiency of the magnetized iron–scintillator detector. The main characteristic of the OPERA detector at CNGS is the emulsion cloud chamber, a lead–emulsion sandwich detector with outstanding angular and space resolution. Although it is designed to be exquisitely sensitive to the detection of tau leptons, this detector is also well suited for the detection of electrons. OPERA can thus reach a 90% CL sensitivity of $\sin^2 2\theta_{13} = 0.06$, a factor of 2 better than Chooz for a five-year exposure to the CNGS beam at nominal intensity, the main limitations being given by i) the mismatch between the beam energy and baseline and the neutrino oscillation length, and ii) the limited product of mass of the detector times neutrino flux.

## V-2.2 The coming generation: searches for $\theta_{13}$

### V-2.2.1 Reactor experiments – Double-Chooz

The best present limit on $\theta_{13}$ comes from the Chooz experiment, a nuclear-reactor experiment. At the low energy of the nuclear-reactor electron-antineutrinos, an appearance measurement is not feasible, and the experiment looks for $\overline{\nu}_e$ disappearance:

$$P(\overline{\nu}_e \to \overline{\nu}_e) = 1 - \sin^2 2\theta_{13} \sin^2(\Delta m_{13}^2 \, L / 4E) + ...$$

The difficulty in this kind of experiment, which looks for a small deficit in the number of observed events, is the flux and cross-section normalization.

The Double-Chooz [30] experiment is set up at the same site at the Chooz reactor, to improve on this limit, mostly by using a near and far gadolinium-loaded liquid-scintillator detectors of improved design. The sensitivity after 5 years of data taking will be $\sin^2 2\theta_{13} = 0.02$ at 90% CL, which could be achieved as early as 2012. It is conceivable to use a second, larger cavern to place a 200 t detector to even improve that bound down to $\sin^2 2\theta_{13} < 0.01$.

A number of other proposals exist in the world (Japan, Brazil, USA and China) for somewhat better optimized or alternate-designed reactor experiments. The advantage of Double-Chooz is that it will use an existing cavern for the far detector, which puts it ahead in time of any other reactor experiment, provided that the final funding decision is made in a timely manner and funding is forthcoming.

Reactor experiments provide a relatively cheap opportunity to search for relatively large values of $\theta_{13}$ in a way that is free of ambiguities stemming from matter effects or from the phase $\delta$. It is clear, however, that the observable $P(\nu_e \to \nu_e)$ is intrinsically time-reversal-symmetric and cannot be used to investigate the sign of $\Delta m_{23}^2$ or CP violation. High-energy neutrino-appearance experiments are necessary to go further.

### V-2.2.2 Off-axis $\nu_\mu$ beams: T2K and NoνA

Conventional neutrino beams can be improved and optimized for the $\nu_\mu \to \nu_e$ searches. An interesting possibility is to tilt the beam axis a few degrees with respect to the position of the far detector (off-axis beams). At a given angle $\theta$ with respect to the direction of the parent pions, the two-body $\pi$-decay kinematics results in a nearly monochromatic muon-neutrino beam. These off-axis neutrino beams have several advantages with respect to the conventional ones: i) the energy of the beam can be tuned to correspond to the baseline by adapting the off-axis angle; ii) since $\nu_e$ mainly come from three-body decays, there is a smaller $\nu_e$ contamination under the off-axis energy peak. The drawback is that the neutrino flux can be significantly smaller.

The T2K (Tokai to Kamioka) experiment [31] will aim neutrinos from the Tokai site to the SuperKamiokande detector, 295 km away. The neutrino beam is produced by pion decay from a horn-focused beam, with a system of three horns and reflectors. The decay tunnel (120 m long) is optimized for the decay of 2–8 GeV pions and short enough to minimize the occurrence of muon decays. The neutrino beam is situated at an angle of 2–3° from the direction of the SuperKamiokande detector, assuring a pion-decay peak energy of 0.6 GeV – precisely tuned to the maximum of oscillation at a distance of 295 km. The beam line is equipped with a set of dedicated on-axis and off-axis detectors, situated at a distance of 280 m. There are significant contributions of European groups to the beam line and to the near detector at 280 m, CERN having donated the UA1/NOMAD magnet, and European groups contributing to various parts of the detector, in particular to the tracker, electromagnetic calorimeter, and to the instrumentation of the magnet.

The T2K experiment is planned to start in 2009, with a beam intensity reaching up to 1.5 MW beam power on target by 2012. The main goals of the experiment are as follows:
1. The highest priority is the search for $\nu_e$ appearance to detect subleading $\nu_\mu \rightarrow \nu_e$ oscillations. It is expected that the sensitivity of the experiment, in a 5-year $\nu_\mu$ run, will be of the order of $\sin^2 2\theta_{13} \leq 0.006$.
2. Precision measurements of $\nu_\mu$ disappearance. This will improve the measurement of $\Delta m^2_{23}$ down to a precision of 0.0001 eV$^2$ or so, and a measurement of $\theta_{23}$ with a precision of a few degrees.
3. Neutral-current disappearance (in events tagged by $\pi^0$ production) will allow for a sensitive search of sterile-neutrino production.

There is an upgrade path for the Japanese programme, featuring: a 2 km near detector station comprising a water Cherenkov detector, a muon monitor, and a fine-grain detector (a liquid argon option has been proposed by European and US groups). The phase II of the experiment, often called T2HK, foresees an increase of beam power up to the maximum feasible with the accelerator and target (perhaps up to 4 MW), antineutrino runs, and a very large water Cherenkov (HyperKamiokande) with a rich physics programme in proton decay, atmospheric and supernova neutrinos, and, perhaps, leptonic CP violation, that could be built around in about 15–20 years from now. An interesting possibility is to install such a large water Cherenkov in Korea, where a suitable off-axis location can be found at a distance from the source corresponding to the second oscillation maximum. The CP asymmetry changes sign when going from one maximum to the other, and the comparison of the effect for the same energy would allow a compensation of systematic errors due to the limited knowledge of the energy dependence of neutrino cross sections.

The NOvA experiment, with an upgraded NuMI off-axis neutrino beam [32] ($E_\nu$ ~2 GeV and a $\nu_e$ contamination lower than 0.5%) and a baseline of 810 km (12 km off-axis), has recently been proposed at Fermilab with the aim to explore the $\nu_\mu \rightarrow \nu_e$ oscillations with a sensitivity 10 times better than MINOS. If approved in 2006, the experiment could start taking data in 2011. The NuMI target will receive a 120 GeV/c proton flux with an expected intensity of $6.5 \times 10^{20}$ pot/year ($2 \times 10^7$ s/year are considered available to NuMI operation, while the other beams are normalized to $10^7$ s/year). The experiment will use a near and a far detector, both using liquid scintillator. In a 5-year $\nu_\mu$ run, with a far detector of 30 kt active mass, a sensitivity on $\sin^2 2\theta_{13}$ slightly better than T2K, as well as a precise measurement of $|\Delta m^2_{23}|$ and $\sin^2 2\theta_{23}$, can be achieved. Because of its relatively long baseline, matter effects are not negligible; hence NOvA can also hope to solve the mass-hierarchy problem for a limited range of $\delta$ and sign($\Delta m^2_{23}$). In a second phase, with the envisaged proton driver of 8 GeV/c and 2 MW, the NuMI beam intensity could increase to $2 \times 10^{21}$ pot/year, allowing an improved sensitivity by a factor of 2, and possibly initiate the search for CP violation.

# V-3 Towards a precision neutrino oscillation facility

Figure V-1 shows the expected sensitivity to $\theta_{13}$, expressed as the 90% C.L. limit that could be achieved in case of a null result, as a function of calendar year. By 2010–12, it should be known whether $\theta_{13}$ is in the 'large range' $\sin^2 2\theta_{13} < 0.01$ or smaller. This knowledge should be sufficient to allow a definition of the parameters (such as baseline, beam energy, detector thresholds, etc.) of the following generation of experiments and to make a definite choice between possible remaining options.

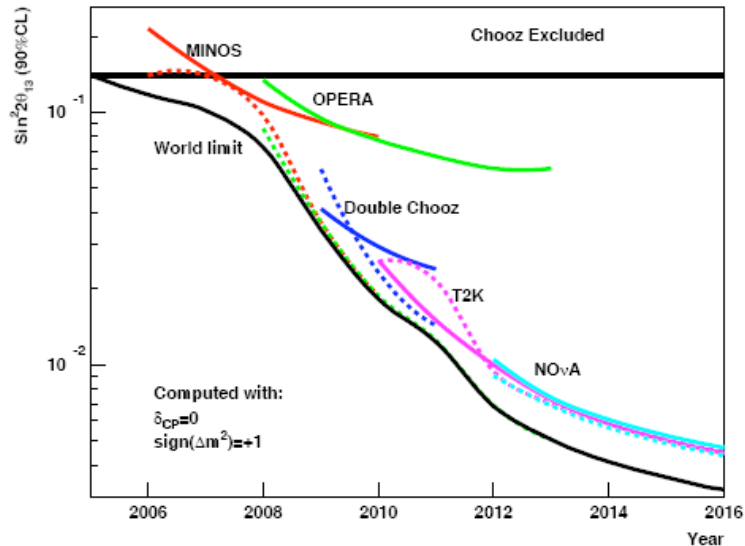

**Figure V-1 Evolution of sensitivities on $\sin^2 2\theta_{13}$ as a function of time. For each experiment we display its sensitivity as a function of time (solid line) and the overall sensitivity computed by combining all experiments but the one under consideration (dashed line). The comparison of the two curves shows the discovery potential of the experiment while it accumulates data. The world overall sensitivity in time is also displayed. The comparison of the over-all world sensitivity with the world sensitivity computed without a single experiment shows the impact of the results of the single experiment. Experiments are assumed to provide results after the first year of data taking.**

At that point in time, the programme of neutrino-oscillation physics will shift emphasis to progressively more challenging measurements, the determination of the mass hierarchy via matter effects, and the study of leptonic CP violation. In addition, basic tests of the general theoretical framework will continue to be performed, such as the unitarity of the leptonic mixing matrix and the precise determination of all mixing angles and mass differences.

The requirements for a precision neutrino facility have been outlined in the studies that took place in the framework of ECFA and CARE.

In order to design a facility, it is important to delineate the main physics objectives that will drive the choice of parameters, while keeping in mind other important physics outcomes and interesting by-products and synergies. Below are a few characteristics of the physics programme of a neutrino facility. Of course, such a hierarchy of physics relevance is a matter of choice and is somewhat subjective. It is not entirely clear that a single facility can do all of this.

1. Main objective: Observe and study CP and T violation, determine mass hierarchy. This can be done, provided neutrino-oscillation probabilities are measured with great precision, in an appearance channel involving electrons, and over a broad range of energies, to decipher the matter effect from the CP violation.

2. Important objectives: unambiguous precision measurements of mixing angles and mass differences, verification of the neutrino mixing framework, unitarity tests.

3. By-products: precision short-baseline neutrino physics and associated nuclear physics, muon-collider preparation.

4. Other physics capabilities: nucleon decay, observation of cosmic events (supernovae, cosmic-ray bursts, etc.), other particle physics (muon–lepton flavour-violating decays, muon EDM).

From a purely European point of view, it is clear that the years 2010–12 will have a strategic importance. Quoting the conclusions of the SPSC workshop in Villars, *'Future neutrino facilities offer great promise for fundamental discoveries (such as leptonic CP violation) in neutrino physics and a post-LHC funding window may exist for a facility to be sited at CERN'*. An ambitious neutrino programme is thus a distinct possibility, but it must be well prepared so that there is a good proposal on time for the decision period, around 2010, when, LHC results being available, the medium-term future of particle physics will be largely decided.

The facilities that have been considered promising for the observation of CP violation are as follows

1. The low-energy (sub-GeV to GeV) avenue: a high-intensity $\nu_\mu$ superbeam combined with a beta beam aiming both at a very large detector (megaton water Cherenkov or liquid-argon detector). We refer to this as the (SB+BB+MD) option.

2. The high-energy avenue: decays of muons $\mu^+ \rightarrow e^+ \nu_e \bar{\nu}_\mu$ contain both flavours of neutrinos, with an energy spectrum reaching all the way up to the parent-muon energy. A neutrino factory based on a muon storage ring aiming at a magnetic detector has been advocated as the ultimate tool to study neutrino oscillations.

It has been argued that the physics abilities of the neutrino factory are superior, but the question is: "What is the realistic time scale?" The timescale is related intimately to the cost of any proposed facility. The (hardware) cost estimate for a neutrino factory is ~1G€ + detectors, but this needs to be verified using a scenario and accounting model specific to a possible location of the facility.

The cost of an (SB+BB+MD) is not very different. The cost driver here (or in the T2HK option) is the very large detector, which is largely site dependent. In addition there will be a hard limit on the size of the largest underground cavern that can be excavated. The issues related to the beta beam are the object of a design study under EURISOL at the moment, and those related to the high power superbeam (4MW on target) are similar to some of those of a neutrino factory.

From this brief discussion it is very clear that a comparative study of cost to physics performance to feasibility is needed; this will be the object of the ongoing 'International scoping study' [33], initiated by the management of the CCLRC Rutherford Appleton Laboratory in the UK. We now describe these two options in turn.

# V-3.1 The beta-beam + superbeam facility

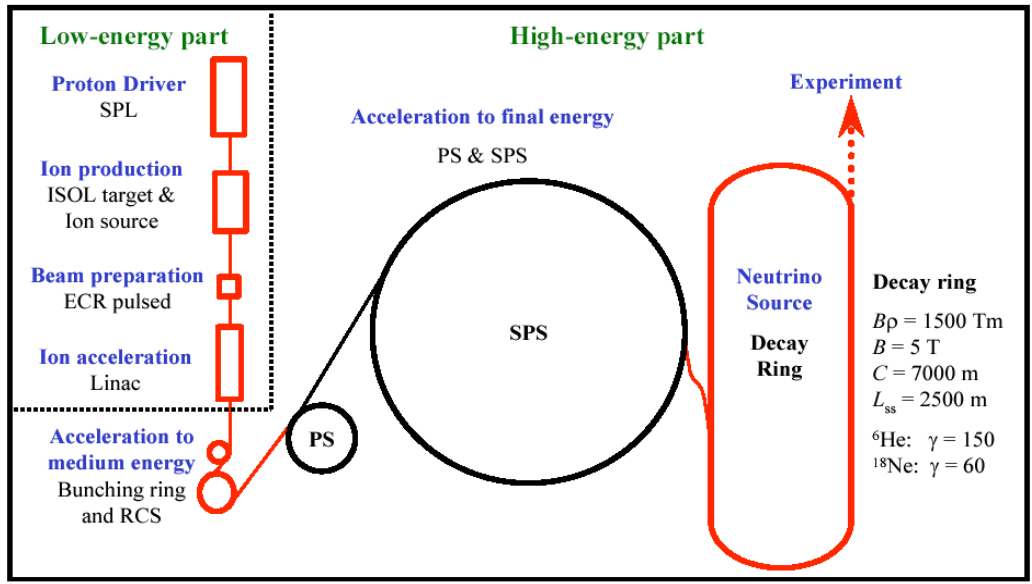

**Figure V-2 Beta-beam base-line design, partially using existing CERN accelerator infrastructure (parts in black).**

The beta-beam [34] concept is based on the acceleration, storage and beta decay of suitable nuclei (see fig.V-2). The preferred ions are

$^{6}He^{++} \rightarrow {}^{6}Li^{+++} \; e^{-} \; \bar{\nu}_e$

$^{18}Ne \rightarrow {}^{18}F \; e^{+} \; \nu_e$

$^{150}Dy + e^{-} \rightarrow {}^{150}Tb \; \nu_e$

The first reaction is normal beta decay and produces a pure wide-band flux of electron-antineutrinos. The second is the beta-plus decay and produces a pure electron-neutrino beam. The third, electron capture on heavier nuclei, is a relatively newer idea, which would allow the production of a pure, monochromatic, $\nu_e$ beam of lower intensity.

The great interest of the beta beam lies in its purity. Its relative practicality is also a strong point: as long as existing proton machines are adequate for the needs of the experiments, the additional required infrastructure is limited to a (challenging) high-intensity ion source and a storage ring. The main drawback is that the facility leads to relatively low energy neutrinos $E_\nu = 2 \gamma E_0$, where $E_0 \sim 3$ MeV is the energy of the neutrino in the decay at rest and $\gamma$ is the Lorentz boost of the accelerated ion. At the CERN SPS, protons can be accelerated to 450 GeV, thus $^{6}$He to 150 GeV/u or $\gamma < 150$. This limits the neutrino energy to about 600 MeV, while already requiring construction of a storage ring with a rigidity equivalent to that of the SPS. The detector of choice for a low-energy beta beam is a large water Cherenkov. For higher energies the detector technology may need to be changed to a fine-grain detector, using scintillator or liquid argon. The higher cross-section and natural focusing at high energy compensates the more difficult realization of massive segmented detectors.

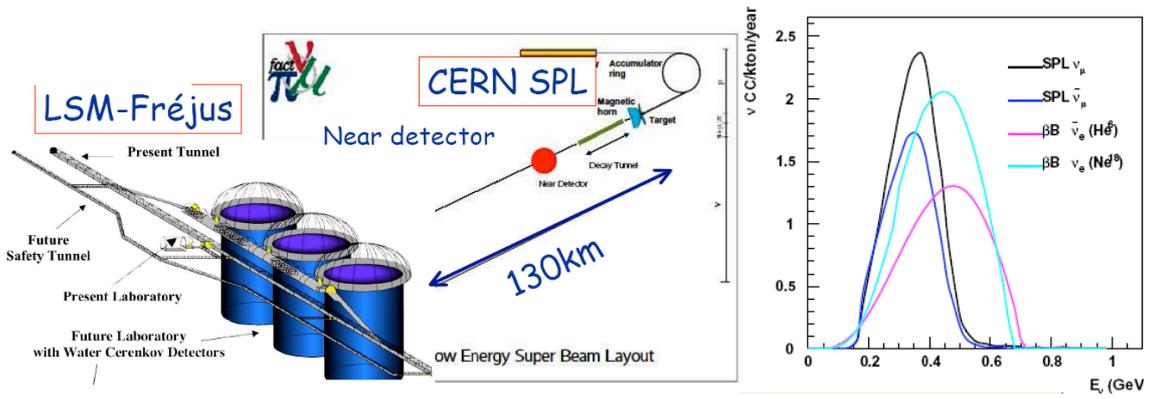

**Figure V-3** The beta beam + superbeam + megaton facility. Left: the schematic layout; right: the number of events without oscillation for a run of 2 years of neutrinos and 8 years of antineutrinos.

The high-intensity flux seems reasonably easy to obtain for antineutrinos with the $^6$He, but it appears to be more difficult with $^{18}$Ne, perhaps smaller by one order of magnitude. The production of $^{150}$Dy seems even more limited; the application may be a wonderful way to measure cross-sections and nuclear effects directly with a monochromatic beam in the near detector.

The superbeam would be a standard horn-focused neutrino beam from pion decay, produced from low-energy protons, with the advantage that the limited kaon production leads to a small and controllable component of electron-neutrinos in the beam, from muon decays. This can be varied and monitored by changing the length of the decay tunnel.

There exists a 'baseline scenario' at CERN for a superbeam + beta-beam facility pointing at a megaton water Cherenkov in the Fréjus laboratory (the MEMPHYS project [35]), with a baseline of 130 km. A preliminary cost estimate yields around 500 M€ for such a detector with a fiducial mass of 440 kt. The on-axis superbeam is optimal if the proton-beam energy is around 3.5 GeV; at this energy, kaon production is very low, and the $\nu_e$ background can be kept at the level of 0.3%. The superbeam and beta-beam neutrino fluxes are shown in Figure V-3. The simultaneous availability of the beta beam and superbeam allows a rather extensive test of symmetry violations (Table V-2).

**Table V-2** Symmetry tests allowed by the simultaneous availability of a beta beam and a superbeam

| Beta beam $^{18}$Ne: $\nu_e \rightarrow \nu_\mu$ | T violation | Superbeam $\pi^+$: $\nu_\mu \rightarrow \nu_e$ |
|---|---|---|
| CP violation | CPT | CP violation |
| Beta beam $^6$He: $\bar{\nu}_e \rightarrow \bar{\nu}_\mu$ | T violation | Superbeam $\pi^+$: $\bar{\nu}_\mu \rightarrow \bar{\nu}_e$ |

A beta beam at higher energies would be more powerful, provided the ion intensity can be kept at a level similar to that for the low-energy scenario described above. The neutrino cross-sections increase linearly with energy, and, as long as the beam energy matches the oscillation length, the ability to separate kinematically the $\nu_e \rightarrow \nu_\mu$ signal from the background generated by pion production improves. It has been suggested that a high-energy beta beam could be run at a possible replacement of the SPS, with a machine of twice the energy. Running at the Tevatron has also been considered. In either case a value of $\gamma \sim 350$ for the helium beam, and 580 for the neon one could be achieved. Clearly the cost of such a facility increases rapidly with energy, since a storage ring of equivalent rigidity would have to be constructed. The performance of such a high-energy beta beam would be similar to that of the neutrino factory from the point of view of CP violation sensitivity, although the number of channels available for oscillation studies is more limited.

## V-3.2 The Neutrino Factory

### V-3.2.1 Description of the facility

In a Neutrino Factory (NF), muons are accelerated from an intense source to energies of several GeV, and injected in a storage ring with long straight sections. The muon decays: $\mu^+ \rightarrow e^+ \nu_e \bar{\nu}_\mu$ and $\mu^- \rightarrow e^- \nu_\mu \bar{\nu}_e$ provide a very well known flux, with energies up to the muon energy itself. The over-all layout is shown in Figure V-4.

Neutrino-factory designs have been proposed in Europe, the US, and Japan. The US design is the most developed, and we will use it here as an example. These studies show that an accelerator complex capable of producing more than $10^{21}$ useful muon decays per year can be built. The NF consists of the following subsystems.

**Proton driver.** It provides 1–4 MW of protons on a pion-production target. For the NF application, the energy of the beam is not critical, in a broad energy range from a few GeV up to 30–50 GeV; it has been shown that the production of pions is roughly proportional to the beam power. The time structure of the proton beam has to be matched with the time spread induced by pion decay (1–2 ns); for a linac driver such as the SPL, this requires an additional accumulator and compressor ring.

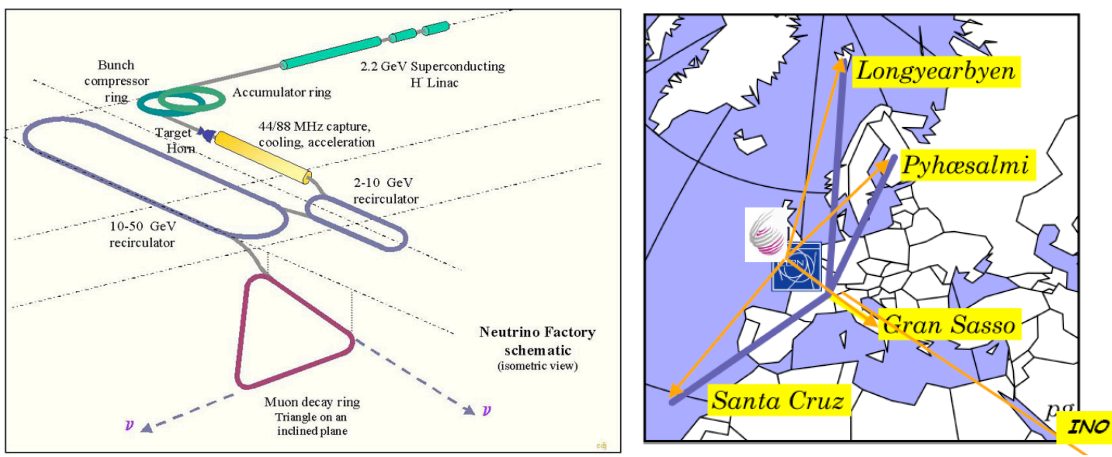

**Figure V-4 Left: Schematic layout of a NF. Right: possible long-baseline scenarios for a European-based facility (INO = Indian Neutrino Observatory)**

**Target, capture and decay.** A high-power target sits within a 20 T superconducting solenoid, which captures the pions and delivers them to a 1.75-T-solenoid decay channel. A design with horn collection has also been proposed. The solenoid scheme offers the advantage that it focuses muons of both signs, which can both be accelerated in the later stages of the machine, thus doubling the available flux.

**Bunching and phase rotation.** A series of warm high-gradient RF cavities (in the frequency range of 88–300 MHz) is used to bunch the muons from the decaying pions and phase-rotate the beam in longitudinal phase-space, reducing their energy spread.

**Cooling.** A solenoidal focusing channel with high-gradient 201 MHz RF cavities, and either liquid-hydrogen or LiH absorbers, is used to reduce the transverse phase space occupied by the beam. The muons lose, by ionization, both longitudinal and transverse momentum as they pass through the absorbers. The longitudinal momentum is restored by re-acceleration in the RF cavities.

**Acceleration.** The central momentum of the muons exiting the cooling channel is 220 MeV/c. A superconducting linac with solenoid focusing is used to raise the energy to 1.5 GeV. Thereafter, a recirculating linear accelerator raises the energy to 5 GeV, and a pair of FFAG[5] rings accelerates the beam to typically 20 GeV or higher.

**Storage ring.** A compact race-track or triangle geometry ring is used, in which 35% of the muons decay in the neutrino beam-forming straight sections. If muons of both signs are accelerated, they can be injected in two superimposed rings or in two parallel straight sections.

Also for a NF, an important R&D effort has been undertaken in Europe, Japan, and the US since a few years. Significant progress has been made towards optimizing the design, developing and testing the required components, and reducing the cost. A rather detailed cost estimate was developed in a study performed in 2001 by the Neutrino Factory and Muon Collider Collaboration in the US[36]. This was based on a significant amount of engineering input, to ensure design feasibility and establish a good cost basis. The hardware cost of the facility from the production target to the muon storage ring was then estimated to 1.65 G$, not including the cost of the proton accelerator. Further optimization has led to a revision published in 2004, with a cost reduction by a factor 0.63, indicating that the total cost of the facility could be of the order of 1 G€. To this should be added the cost of the detectors, which can be evaluated to be in the range of 200–300 M€. This R&D has reached a critical stage in which support is required for two key international experiments: the Muon Ionization Cooling Experiment MICE (at RAL) and the Target experiment MERIT (at CERN), and for a third-generation international design study. If this support is forthcoming, the proponents believe that a NF CDR could be produced by 2010 and that a target date for first beams before 2020 could be realistic.

### V-3.2.2 Oscillations physics at a NF

Considering a NF with beams of positive and negative muons, the 12 oscillation processes shown in Table V-3 can be studied. In addition the neutral-current reactions can be sensitive to the existence of light sterile neutrinos.

Two neutrino flavours are always produced in muon decays. Hence, in addition to providing target mass and identification of the flavour of the lepton produced in charged-current interactions, the detector must provide a measurement of its charge. For muons in the final state (coming from $\nu_\mu$ interactions of from decays of $\tau \rightarrow \mu\nu\nu$), this can be readily done by using a magnetic detector of design extrapolated from that of the MINOS experiment, with an achievable mass assumed to be of the order of 100 kt. Many studies have been performed under this hypothesis, where the main discovery channel is the 'wrong-sign muon', also called 'golden' channel, in which the oscillation produces the appearance of a muon with an 'unexpected' charge. In more challenging detector options, the magnetic field is provided by external coils, surrounding an active volume, such as a fully sensitive segmented scintillator, a liquid-argon TPC, or emulsion cloud chambers, which allow the detection of most of the channels of Table V-3.

---

[5] FFAG: Fixed-field alternating-gradient synchrotron, in which the guiding magnetic field is provided by large-aperture combined-function magnets. The magnetic field has a strong radial dependence, allowing stable orbits and thus acceleration over a momentum range varying by a factor of 2 to 3.

**Table V-3 Oscillation processes accessible to a NF by charged-current interactions**

| $\mu^+ \rightarrow e^+ \nu_e \bar{\nu}_\mu$ | $\mu^- \rightarrow e^- \nu_\mu \bar{\nu}_e$ | |
|---|---|---|
| $\bar{\nu}_\mu \rightarrow \bar{\nu}_\mu$ | $\nu_\mu \rightarrow \nu_\mu$ | Disappearance |
| $\bar{\nu}_\mu \rightarrow \bar{\nu}_e$ | $\nu_\mu \rightarrow \nu_e$ | Appearance ('platinum' channel) |
| $\bar{\nu}_\mu \rightarrow \bar{\nu}_\tau$ | $\nu_\mu \rightarrow \nu_\tau$ | Appearance (atmospheric oscillation) |
| $\nu_e \rightarrow \nu_e$ | $\bar{\nu}_e \rightarrow \bar{\nu}_e$ | Disappearance |
| $\nu_e \rightarrow \nu_\mu$ | $\bar{\nu}_e \rightarrow \bar{\nu}_\mu$ | Appearance: 'golden' channel |
| $\nu_e \rightarrow \nu_\tau$ | $\bar{\nu}_e \rightarrow \bar{\nu}_\tau$ | Appearance: 'silver' channel |

Compared with conventional neutrino beams, NFs yield higher signal rates with lower background fractions and lower systematic uncertainties, especially on the neutrino flux and cross-sections for the initial neutrino flavours, which can be determined in absolute terms with an advocated precision of $10^{-3}$ by using the purely leptonic neutrino interactions in a near detector. These characteristics enable NF experiments to be sensitive to values of $\theta_{13}$, which are beyond the reach of any other proposed facility. Several studies have shown that a non-zero value of $\sin 2\theta_{13}$ could be measured for values as small as $O(10^{-4})$. In addition, both the neutrino mass hierarchy and CP violation in the lepton sector could be measured over this entire range. Even if $\theta_{13}$ is smaller than this value, a $\nu_e \rightarrow \nu_\mu$ oscillation still arises through the same terms as those responsible for the solar-neutrino oscillations; its observation at a NF would allow sufficiently stringent limits to be put on $\theta_{13}$ to suggest perhaps the presence of a new conservation law.

Given the neutrino energies available and the requirement of muon detection and charge identification, the baselines that are optimal for the NF physics are typically 2000 km or longer. At these distances, matter effects become substantial and induce an apparent asymmetry between neutrino and antineutrino oscillations, which can be used to establish the sign of the mass difference $\Delta m^2_{13}$. The matter effects also contribute to the genuine CP asymmetry, with an uncertainty due to the limited knowledge of the material encountered by the beam in its travel from the source to the long-baseline detector.

A common problem to all facilities is that, once a 'golden' signal and/or a CP asymmetry has been observed, the determination of the mixing parameters ($\theta_{13}$, $\delta$) is not free of ambiguities: up to eight different regions of the parameter space can fit the same experimental data. The NF offers several handles against ambiguities, thanks to i) the resolution of high-energy neutrino detectors, which allows the reconstruction of the energy dependence of the oscillation phenomena, ii) the possible availability of two different baselines, and iii) the use of the rich flavour content of the beam with detectors sensitive to electrons, muons and taus.

The study of this latter point was performed, assuming the feasibility of a magnetized liquid-argon detector. By separating the events into several classes, right-sign muon, wrong-sign muon, electron and neutral current, and by performing a fine energy binning down to low energies, it was shown that the matter resonance could be neatly measured, and that the simultaneous observation of the four aforementioned channels allowed resolution of ambiguities to a large extent. Similarly, the tau appearance 'silver' channel, detectable with emulsion detectors, has been advocated as a powerful means of solving ambiguities. This can be readily understood, since this channel has a dependence on δ of a sign opposite to that of the 'golden' one, while having similar dependence on matter effects and $\theta_{13}$. It is even advocated, although a full demonstration is still needed, that, with a NF with two baselines, and with detectors able to measure both the 'golden' and 'silver' channels in addition to the disappearance channels, a fully unambiguous determination of oscillation parameters could be achieved.

## V-3.2.3 Other physics and synergies with other programmes

### V-3.2.3.1 HIGH-FLUX NEUTRINO PHYSICS

The neutrino beams at the end of the straight section of a NF offer an improvement in flux by several orders of magnitude over conventional beams, allowing several times $10^8$ events to be collected per kilogramme and per year. Precision tests of the Standard Model could be carried out in neutrino scattering on nucleon or electron targets, as well as a precise determination of neutrino cross-sections and flux monitoring with per-mille accuracy, thanks to the availability of inverse muon decay $\nu_\mu + e^- \rightarrow \mu^- + \nu_e$. This could also allow a new generation of neutrino experiments, with detailed studies of nucleon structure, nuclear effects, spin-structure functions, and final-state exclusive processes.

### V-3.2.3.2 MUON PHYSICS

As described in Chapter VI, a high-intensity proton source could certainly produce many low-energy muons and thus, provided the beam and experiments can be designed to do so, provide opportunities to explore rare decays such as $\mu \rightarrow e \gamma$, $\mu \rightarrow e e e$, or the muon conversion $\mu N \rightarrow e N$, which are lepton-number-violating processes.

Another fundamental search, as described in Chapter VII, would clearly be the search for a muon electric dipole moment (EDM), which would require modulation of a transverse electric field for muons situated already at the magic velocity where the magnetic precession and the anomalous (g – 2) precession mutually cancel.

### V-3.2.3.3 MUON COLLIDERS

Finally, it is worth keeping in mind that the NF is the first step towards muon colliders [37]. The relevant characteristics of muons are that, with respect to electrons, i) they have a much better defined energy, since they hardly undergo synchrotron radiation or beamstrahlung, ii) their couplings to the Higgs bosons are multiplied by the ratio $(m_\mu/m_e)^2$, thus allowing s-channel production with a useful rate.

These remarkable properties make muon colliders superb tools for the study of Higgs resonances, especially if, as predicted in supersymmetry, there exists a pair H, A of scalars with opposite CP quantum numbers, which are nearly degenerate in mass. The study of this system is extremely difficult with any other machine, and a unique investigation of the possible CP violation in the Higgs system would become possible.

Finally, because muons undergo little synchroton radiation, they can be accelerated in circular machines up to very high energies, providing a possible path to point-like collisions well above 4 TeV centre of mass without the energy spread developed by beamstrahlung, unavoidable at electron colliders. At these energies, however, the potential radiation caused by the resulting beam of high-energy neutrinos must be seriously considered.

## V-3.3 Comparison of facilities

There is a wide consensus in the neutrino community that it will be timely to propose a precision neutrino facility around 2010. As seen previously, several options are open and, for each option, the precise parameters of the design need to be established. By 2010, the improved knowledge of $\theta_{13}$ should allow this process to be finalized, but this must be accompanied by a comparison of options based on performance, feasibility and cost, and by the design and prototyping work necessary to establish a proposal on a firm basis. Meanwhile, the community must remain open to new ideas and technological breakthroughs.

These studies are performed in Europe by ECFA working groups, which are supported by the BENE network (Beams for European Neutrino Experiments) [25], a work package of CARE. There exist several national groups, such as the Groupe de Recherche (GDR) Neutrino in France or the UK NF Collaboration. At the international level, there exist a Neutrino Factory and Muon Collider Collaboration in the US and a Japan Neutrino Factory effort in Japan. These regional entities have joined forces in an International Scoping Study (ISS) [33], which is taking place between June 2005 and August 2006. It will be followed by the preparation of an FP7 funding proposal for design studies that are expected to take place between 2007 and 2010, leading to the Conceptual Design Report (CDR) of a future neutrino facility.

The comparison of performances of the aforementioned facilities cannot be considered concluded at this point, but the following gives a flavour of it. The comparisons have been performed with the neutrino-oscillation-fitting programme GLOBES [38]. A final version of this comparison should be an outcome of the ISS. Several aspects still need to be clarified before a final comparison can be performed:

- Costs, time scales, fluxes of the different accelerator systems are not yet fully worked out.

- Performances and optimization of the detectors are not known or simulated at the same level.

- Systematic errors that strongly influence performances, for instance sensitivity to leptonic CP violation for large values of $\theta_{13}$, are not substantially discussed in the literature. We are confident that facilities where neutrino fluxes can be known a-priori, as the case of beta beams and NFs, will have smaller systematic errors (and smaller backgrounds) than, say, neutrino superbeams, but this difference is not known quantitatively today.

- The concept of the near-detector station(s) and flux monitoring systems has to be proposed together with the facility, in particular for low-energy (few 100 MeV) beta beam and superbeam, where the issues of muon-mass effect, Fermi motion and binding energy uncertainty are highly non-trivial.

- Finally, for the NF, the question of systematics on the prediction of matter effects is essential for the performance at large values of $\theta_{13}$.

Over-all performances will depend on the combination of additional input to the main channels, for instance the information gathered from atmospheric neutrinos observed in the large water Cherenkov or magnetic detectors, or from the various channels available at a given facility. So far, only the dominant channels have been considered.

With these caveats, Figure V-5 compares the reach in $\theta_{13}$ of the Japanese project, T2HK (1-Mt water Cherenkov with 4 MW beam power), of the (BB+SB+MD) project from CERN to Fréjus, and of a NF, while Figure V-6 shows the reach in $\delta$ as a function of $\theta_{13}$, with special attention to systematic errors. The high $\gamma$ option of the $\beta$-beam with a 1-Mt water Cherenkov detector located at 750 Km is also shown in this figure.

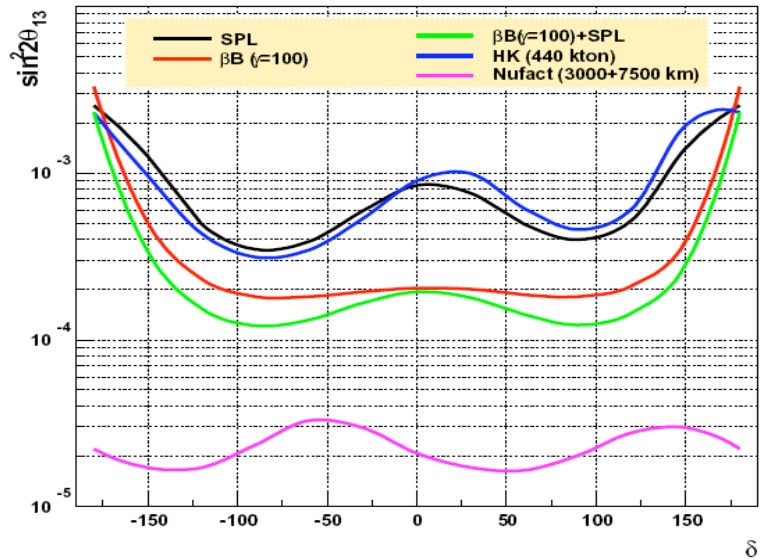

**Figure V-5 Comparison of the sensitivity to $\theta_{13}$ of the beta beam, superbeam, their combination and the neutrino factory. If a value above the limit is found, the limit given here gives an order of magnitude of the precision that would be achieved on its measurement.**

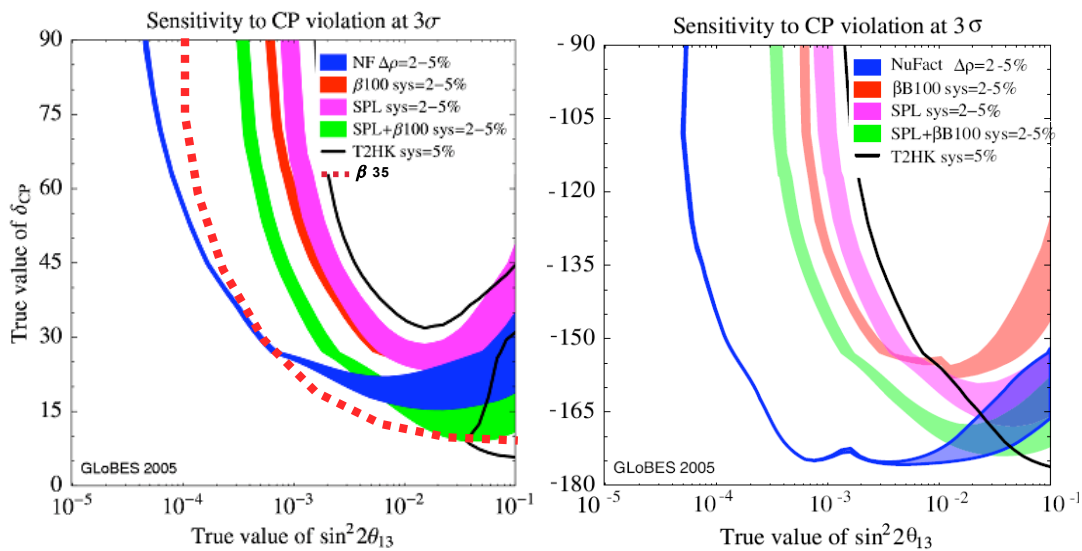

**Figure V-6 Discovery potential on $\delta$ at 3σ, computed for 10 years of running time for the facilities described in the text. These are two of four plots representative of the four possible quadrants of $\delta$ values. The width of the curves reflects the range of systematic errors: 2% and 5% on signal and**

**background errors for SPL-SB and beta beam, 2% and 5% for the matter density Δρ. Other systematic errors are 5% on signal and background at T2HK, and 0.1% for the neutrino-factory signal, 20% for the corresponding backgrounds. The analysis carried out for the beta beam of γ=350 includes systematic errors in a different way. More work is needed to compare under the same assumptions the NF and the high gamma beta-beam.**

# V-4 The design study of the next neutrino facility

Although the contents of the design study will be defined by the on-going scoping study, it is likely to include the investigation and cost estimate of the following accelerator components:

- a high-power proton driver with an energy of 4–5 GeV or more;
- the engineering of the handling, containment and safety aspects of a high-power target and collection station;
- a cost-effective muon phase rotation and cooling channel, involving high-gradient normal-conducting RF operating at a few hundred MHz in magnetic fields of a few teslas;
- non-scaling FFAGs for acceleration of muons (and possibly protons);
- an optimized storage ring for muons;
- higher-gamma and higher-intensity beta beams;

  and a number of technical preparatory (R&D) projects, aiming at demonstrating:

- the existence of at least one adequate choice of target,
- an extended lifetime of the horn prototypes at high rate and radiation,
- muon ionization cooling, by completion of the MICE experimental programme,
- operation of a non-scaling FFAG model and construction of a full-scale FFAG magnet,
- RF cavities and kicker magnets for fast manipulation of muon beams.

At the same time, it will require specific detector R&D and design efforts on a number of topics:

- photodetector development for very large far detectors;
- developments of the liquid-argon technique, including the presence of magnetic field;
- study and tests of a magnetic calorimeter susceptible to be built with a mass of up to 100 kt;
- detectors dedicated for tau detection such as the emulsion cloud chamber;
- last but not least, the necessary near-detector concepts and beam instrumentation that are crucial for the precise flux monitoring needed for CP-violation measurements.

The above structure reproduces that of the existing working groups. Strong support from CERN, other European laboratories and funding agencies will be crucial for the success of this enterprise. The European groups are working in international collaboration for many of these projects, and it will be clear from the start that the study will involve international partners in the US, in Japan, and elsewhere.

# V-5 Conclusions

## V-5.1 The Orsay Symposium

The session on neutrino-oscillation physics at the Orsay Symposium featured the following presentations:

- Theoretical aspects of neutrino physics (P. Huber),

- Experiments and infrastructure (A. Cervera).

These presentations had been circulated beforehand for feedback from the neutrino community, namely: the BENE network (~ 220 people), the OPERA collaboration (~ 100), the liquid-argon community in Europe, the ICARUS collaboration and the GLACIER R&D collaboration (~ 80 people), T2K-Europe (~ 120), the Double-Chooz collaboration (~ 50), MINOS (~ 50), the HARP collaboration (~ 120) and the MICE experiment (~ 120), the members of the ISS (~ 100). Taking into account the large amount of overlap between these collaborations, this represents about 400–500 people in Europe.

A public presentation was organized and intense feedback was given to the speakers, following which the presentations were circulated again for approval. To this extent the conclusions presented at the Orsay Symposium represented the consensus of this community.

In addition, a number of written contributions pertinent to neutrino physics as listed in the list of references below.

During the discussion following the presentations themselves, the conclusions presented by the speakers were endorsed, with special emphasis on two points:

1. The Double-Chooz experiment, which plays an important role in the development of the field in the near future (see Figure V-1), should receive sufficient support if this crucial role is to materialize.

2. The facilities considered are not cheap (> 1 G€) and synergies with other fields of particle and non-particle physics are essential; the following were emphasized:

At the level of the proton driver:

a) synergy with the LHC luminosity upgrade;

b) synergy with the high-intensity physics programme:

- lepton flavour violation searches and other precision muon physics,
- neutrino DIS studies,
- nuclear physics (EURISOL),
- rare kaons decays, depending on proton energy;

c) synergy at the level of detectors, as discussed in the NNN workshops [39]:

- proton-decay searches
- atmospheric neutrinos

- supernovae, solar, and other low-energy neutrinos.

## V-5.2 Overview

Neutrino physics has become one of the most active areas of research in particle physics. This is not surprising, considering that neutrino masses constitute the first clear-cut evidence for physics beyond the Standard Model. It may be more surprising to realize that this new physics may stem from phenomena occurring at a very high energy scale. The physics case for a development of neutrino physics is independent of the arguments for high-energy frontier, and the information gathered on this research front cannot be collected otherwise. The physics questions that are addressed offer the potential for great discoveries, such as a Majorana neutrino mass, and/or leptonic CP violation.

The neutrino (oscillation) community in Europe is very active and developing. Its members are involved in analysing or preparing current experiments (HARP [40], K2K, MINOS, OPERA), and it is working hard on the preparation of experiments for the near future (Double-Chooz, T2K, perhaps NOvA and other reactor experiments). There is a strong interest in this community to prepare actively for the next-generation facilities, and in fact the community is involved in the R&D leading to them, both on the accelerator side (MICE [41], MERIT [42] the beta beam [34]) and on the detectors (liquid-argon TPC [28], [43], water Chrenkov and photosensors [35]) These R&D efforts are, however, severely limited by their budget. There is a general feeling that this emerging field is somewhat under-funded with respect to its physics case, and that CERN should be more actively involved in accelerator-based neutrino physics.

The main wishes expressed by the community can be summarized as follows:

1. Strong support should be made available to make a success of the present and near-future programme. The Double-Chooz experiment should be strongly supported. The involvement of European neutrino physicists in the neutrino physics programme abroad (such as T2K or perhaps NOvA) should be supported in a way that would assure a viable and significant contribution.

2. Europe should get ready to host a major neutrino facility for the precision era, or to play a major role in the preparation and construction of this facility should it be located elsewhere. This would be best achieved if CERN would play a major, perhaps leading, role in the upcoming accelerator-design study and detector R&D, in close collaboration with European laboratories and within an international collaboration.

The European neutrino-oscillation community has high expectations from the CERN strategy to help provide support, priority and resources that it feels the very strong physics case deserves.

In addition, the following written contributions had been submitted to the Symposium

[BB2-2.2.05] Two contributions on the MEMPHYS project.

[BB2-2.2.04] A statement in support of the neutrino factory.

[BB2-2.2.07] A contribution stressing need of R&D programme especially for accelerators.

[BB2-2.2.01] A statement expressing interest in hosting a Neutrino long base line detector in the Pyhäsalmi Mine.

[BB2-2.2.02] A beta-beam contribution.

[BB2-2.2.03] The BENE 2006 report stressing the preparation of a design study proposal as first priority.

[BB2-2.2.06] A statement of interest for a major neutrino oscillation facility in Europe from the neutrino GDR in France.

[BB2-2.1.07] A statement of interest in the CERN-PH contribution

[BB2-2.1.06] From the POFPA report: « discuss how the CERN proton accelerator complex might be upgraded so as to accommodate optimally these two programmes. » (LHC luminosity upgrade and neutrino programme, and high intensity physics, e.g. µ → e γ)

[BB2-2.2.08] A report from the PAF group "Potential for neutrino and radioactive beam physics of the foreseen upgrades of the CERN accelerators"

[BB2-2.2.09] A statement of interest in the Liquid Argon TPC by S. Centre

[BB2-2.2.10] Status report from the International Scoping Study

[BB2-2.2.11] A description of the detector options for future neutrino facilities

# VI FLAVOUR PHYSICS

In the whole of particle physics, the field of flavour is that with the highest complexity and the richest phenomenology. Its phenomena range from strange, charm, bottom and top physics, over mass hierarchy and quark-mixing physics, to CP and T violation, and eventually to the genesis of leptons and baryons that constitute our matter world (cf. Section II-2.1). Although, traditionally, flavour physics is often associated with quark physics, the lepton sector has known a sharp upraise since the discovery of neutrino oscillation. This also boosted the importance of precision measurements in the charged-lepton sector and the search for rare phenomena. We should mention here the precision search for non-V–A currents in $\tau$ and $\mu$ decays (Michel parameters), universality violations in the couplings of charged leptons, and of course the quest for lepton-flavour violation through neutrinoless $\tau$ and $\mu$ decays, as well as T-violating electric dipole moments (EDMs). Flavour physics is both rare and precision physics, and through both windows it is effectively looking beyond the Standard Model (cf. Section II-2.3).

This section covers quark-flavour physics (excluding top physics, which is treated elsewhere) and the search for charged-lepton-flavour violation. We briefly introduce the experimental and theoretical challenges, review the current status of quark-flavour mixing, and finally discuss promising future projects.

# VI-1 Scientific programme

## VI-1.1 Introduction

Throughout the history of particle physics, discoveries and developments in flavour physics have led to spectacular progress in the field. We may recall here the discovery of CP violation through the detection of $K_L \to \pi^+\pi^-$ decays in 1964, the postulation of the charm quark through the GIM mechanism, explaining the smallness of $K_L \to \mu^+\mu^-$ and eventually the tree-level suppression of flavour-changing neutral currents (FCNC) in the SM, Kobayashi–Maskawa (KM) requiring a third quark generation to introduce CP violation into the unitary quark mixing, the mixing frequency between neutral $K$ and $B$ flavour eigenstates indicating the charm- and top-quark mass scales, respectively, and finally the proof of the KM theory by the measurement of $\sin(2\beta)$, in agreement with the KM expectation.

It was the major contribution of the asymmetric-energy *B*-factories at SLAC and KEK to establish that the KM theory represents the dominant source of CP violation at the electroweak scale. However, this success hides the problem that the origin of the observed flavour structure (mass hierarchy, mixing, and CP violation) is *not understood* within the SM. Among the many unanswered questions related to the *known* sector of flavour physics, we may mention the following: Is the CKM phase the only source of CP violation in nature? Why are charged currents left-handed? Why are there no (tree-level) FCNC? What are the relations between neutrinos and charged leptons, and between the quark and lepton sectors?

The central goals of the new generation of flavour-physics experiments must be to uncover physics beyond the Standard Model and to probe the flavour structure of BSM physics that may be discovered elsewhere. High-energy and low-energy precision experiments are thereby complementary. For example, the SUSY flavour structure, studied with flavour experiments, is linked to SUSY breaking measured at the high-energy frontier.

It is also crucial to continue to provide testing grounds to understand better the phenomenology of the Standard Model and its implications. We emphasize in this context the fundamental importance of improved theoretical tools to flavour physics, in particular of the quark sector. Very significant improvements have been achieved in this field. Apart from the powerful phenomenological approaches using flavour symmetries and sum rules, effective theories, such as soft-collinear effective theory (SCET), have emerged in *B* physics out of the observation that soft-gluon exchange between the decay products of *B*-meson decays to two particles is suppressed. Chiral perturbation theory, heavy-quark effective theory, and the heavy-quark expansion are the ingredients to exploit semileptonic *K* and *B* decays to determine accurately CKM matrix elements. Probably the most promising tool for future precision predictions is lattice gauge theory. Modern TeraFLOP computers and algorithmic improvements make it possible to go beyond the quentching approximation that led to large systematic uncertainties in the past (see also [BB2-2.7.01]). The PetaFLOP barrier is expected to be crossed around 2009, which would qualify lattice gauge theory to predictions approaching the per cent level. With such a precision, many measurements in *K* and *B* physics, which are sensitive to BSM physics but currently dominated by theoretical errors in the calculation of the matrix elements, could be revived.

Proposals for future experiments should be examined as a function of their capabilities to derive properly the fundamental parameters of the theory. This requires good control of the hadronic uncertainties involved. The measurements that are aimed at must be competitive with other measurements with a similar focus, and they should provide exploitable sensitivity to BSM physics.

## VI-1.2 Using flavour as a probe of BSM physics

Although the present searches in the quark and charged-lepton flavour sectors have not yet revealed significant signs of BSM physics, this is by no means an unexpected scenario. Indeed, as pointed out in Chapter II, sections 2.2 and 2.3, if low-energy BSM physics existed and possessed *generic*, i.e. unsuppressed, flavour mixing and phases, it would have led to strong deviations from the SM expectation in the mixing and CP-violating observables of the *K* and *B* sectors. Since these have not been seen, either the BSM physics scale is higher than $O(10^4$ TeV), with the fine-tuning problem for the Higgs boson self-energy (discussed also in section II-1.2), or the flavour-mixing structure of the BSM physics is SM-like. This tension between the lower limit on the new-physics scale assuming a generic BSM flavour structure, and the scale needed to resolve the gauge-hierarchy problem is known as the 'flavour problem'. SM phenomenology and the CKM mechanism imply that there is no exact symmetry that protects against flavour mixing. For this reason, a new-physics model, which is completely flavour-blind, with no impact at all on precision measurements in the flavour sector, would be highly unnatural. If the new-physics scale is in the TeV range, some non-SM effects in flavour physics are therefore expected. In the absence of severe fine-tuning, these non-standard effects should at least be competitive in size with the SM higher-order electroweak contributions. Detectable deviations from the SM are also naturally expected in flavour-conserving observables, such as CP-violating electric (and weak) dipole moments of charged leptons and of neutral hadrons or atoms.

It is still possible that the BSM flavour structure is very different from the SM one, with sizable $O(1)$ effects in sectors of flavour physics that have not yet been probed with good precision. For example, GUTs- and neutrino- inspired models can lead to large BSM effects between the 2$^{nd}$ and 3$^{rd}$ generations only. The $B_s$ sector, which will be studied by the LHC experiments (and is currently under study at the Tevatron), is suited to test such scenarios.

Beyond the search for new physics, discussed also in Chapter III, future flavour-physics experiments can play a decisive role when the understanding of the flavour structure of BSM physics that has been discovered, for example at the LHC, will become one of the central challenges of the post-discovery era. We point out, however, that such a task requires cleaner observables than the search for BSM physics alone. While there exist many observables, mainly in the $B$ and charged-lepton sectors, that may establish the existence of BSM physics by deviation from a well controlled SM prediction, not all of them can be exploited as probes of the BSM flavour properties. In particular, if measurements with purely hadronic final states are involved, long-distance QCD effects, which are hard to predict, shadow the interesting physics. An example of such a mode is $B \rightarrow \phi K^0$: while a significant deviation between the value of $\sin(2\beta)$ measured in this mode and the SM value would unmistakably indicate new physics, it is not possible to derive the new CP-violating phase from the measurement without knowing the strong- interaction phases occurring in the decay. Rare semileptonic or purely leptonic decay modes of $K$ and $B$ mesons can lead to cleaner signatures.

## VI-2 *B* Physics

The successful exploitation of the *B*-factories at SLAC and KEK, and of their experiments, BABAR and Belle, led to a quantitative and qualitative reassessment of CKM physics. For the first time it has become possible to determine precisely the CKM phase without hadronic uncertainties, which dominated the previous constraints from indirect (and direct) CP violation in the $K$ system, and the neutral $B$ mixing frequency, as well as semileptonic $B$-meson branching fractions.

### VI-2.1 Quark-flavour mixing and CP violation: the present picture

Traditionally, the constraints on the CP-violating phase of the CKM matrix are represented in the plane of the unitarity triangle (UT). This describes the unitarity relation obtained by multiplying the 1st and the 3rd columns of the CKM matrix: $V_{ud}V_{ub}^* + V_{cd}V_{cb}^* + V_{td}V_{tb}^* = 0$. Dividing the UT relation by $V_{cd}V_{cb}^*$ leads to a phase-invariant form, which allows us to define the observable apex of the UT by $\bar{\rho} + i\bar{\eta} \equiv -V_{ud}V_{ub}^*/V_{cd}V_{cb}^*$. It is the challenge of the CKM flavour-physics programme to over-determine this apex, so as to reveal inconsistencies originating from BSM physics. We stress that it is not the *value* of the apex that is of primary interest (it might well be without fundamental relevance), but its sensitivity to new physics, uncovered by inconsistencies between measurements involving tree-level and loop transitions.

The three angles of the UT, here denoted α, β, γ, and its sides, can all be measured in the $B_d$ system by exploiting time-dependent CP asymmetries in $B_d$ decays to charmonium + $K^0$ (β) and to $hh'$ ($h = \pi,\rho$) (α), direct *CP* asymmetries in $B_d$ decays to open charm (γ), neutral $B_d$ mixing, and semileptonic $B_d$ decays involving $V_{ub}$ and $V_{cb}$ transitions (sides). The $B_s$ system covers the physics represented by a unitarity triangle where the 2nd and the 3rd columns are multiplied by only a small, with respect to the $B_d$ system, weak mixing phase. The primary interest here is the hope that effects from BSM physics may be enhanced in transitions between the 2nd and 3rd quark generations. In addition, the measurement of the neutral $B_s$ mixing frequency reduces the theoretical uncertainty of the UT side measurement obtained from neutral $B_d$ mixing. Since neutral $B_s$ mix roughly 40 times faster than neutral $B_d$, the frequency measurement requires excellent vertexing capabilities, as will be available at LHCb. At present, the combined Tevatron–LEP–SLC lower limit [44] is $\Delta m_s > 16.6$ ps$^{-1}$. [7]

BABAR and Belle have published measurements of all UT angles by now, with however a large variation in precision due to the different statistical reach of the methods: the most precise measurement is the one of $\sin(2\beta)$, with a world average of $0.687 \pm 0.032$ (or $\beta = (21.7 \pm 1.3)°$, when using the SM solution) [44]. Since the measurement of $\alpha$ involves suppressed charmless $B_d$ decays, much less statistics is available. In addition, the extraction of $\alpha$ involves several final states (or a Dalitz plot) to eliminate hadronic uncertainties. The current world average is $\alpha = (99^{+13}_{-8})°$ [45]. The measurement of $\gamma$ requires interference of decay amplitudes that are strongly suppressed. However, the experiments have found a way of enhancing the interference by exploiting the Dalitz plot of the subsequent open charm decay. This leads to the world average $\gamma = (63^{+15}_{-23})°$ [45]. The sum of all angles reads $(184^{+20}_{-15})°$, which is compatible with 180°, as expected from unitarity. The current status of the UT as obtained from the global CKM fit is shown in Fig. VI-1. Agreement between the various measurements is observed.

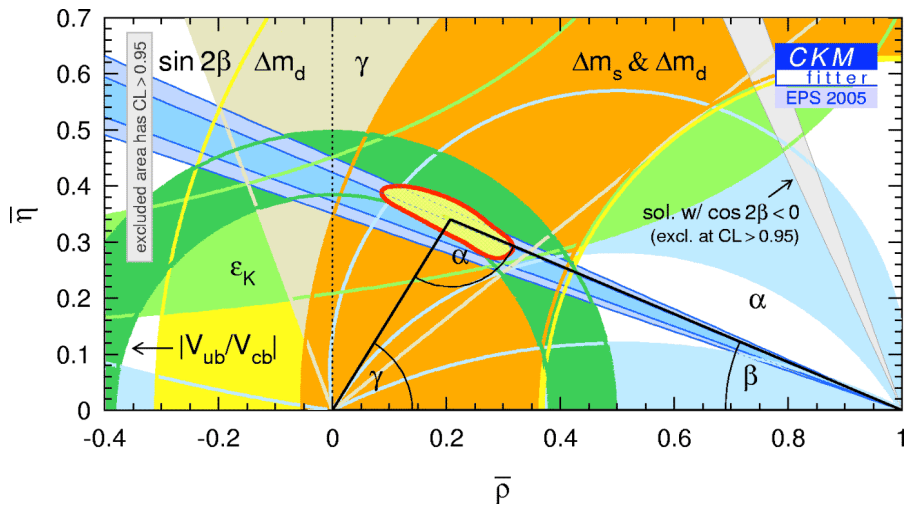

*Fig VI-1:* *The present constraints on the Unitarity Triangle [45]: the 95% confidence level regions for the individual constraints, and the result of the global CKM fit.*

Another important field of activity at the *B*-factories is the measurement of $\sin(2\beta)$ with loop-induced decays (so-called *penguin* decays), e.g., $B_d \to \phi K^0$ (many other such modes exist). Heavy virtual particles from BSM physics may occur in these loops and alter the SM phase. This measurement is complementary to the measurement of the neutral $B_s$ mixing frequency: while the first measurement (approximately) determines the phase of $V_{ts}$, the second one determines its modulus. Using only the theoretically cleaner modes, a discrepancy between the charmonium value for $\sin(2\beta)$ and the penguin average amounts to $2.2\sigma$ (see ref.[44]).

---

[7] Since the completion of the Briefing Book, $16.96$ ps$^{-1} < \Delta m_s < 17.91$ ps$^{-1}$ (95% CL) has been measured by the CDF collaboration.

Since the summer of 2004, direct CP violation (i.e. CP violation in the decay) has been firmly established by both BABAR [46] and Belle [47] in the decay $B^0 \to K^+\pi^-$ with a measured CP asymmetry of $-(11.5 \pm 1.8)\%$, many orders of magnitude larger than in the $K$ system. More exclusive $B$ decays violating CP symmetry are expected to be observed in the coming years.

The new *Zoology*: although it was not part of the initial motivation for the construction of the *B*-factories, charm and charmonium spectroscopy witnessed a renaissance due to the discovery of a number of new states by both BABAR and Belle. Some of these states, such as, the X(3872) and Y(4260), are not yet fully understood (see also section IX-1). The papers reporting on these discoveries are amongst the most cited publications of the *B*-factories.

## VI-2.2 *B* physics at the Tevatron and at the LHC

To a large extent, *B* physics at hadron colliders is complementary to $e^+e^-$ *B*-factories. At $e^+e^-$ machines, the cleaner environment favours measurements of $B_d$ decays to purely hadronic final states, or decays to neutrals and decays with neutrinos, with a large tagging efficiency (also due to quantum coherence of the produced neutral *B* pair) and the reconstruction of the full event; on the other hand, hadron colliders have access to the full spectrum of *B* mesons and baryons, and benefit from a huge *b* production cross section. Moreover, the strong boost of the produced *B* mesons provides a much better proper time resolution than at $e^+e^-$ *B*-factories. Currently both D0 and CDF have an active *B* physics programme at the Tevatron. At the LHC, LHCb is the experiment dedicated to *B* physics, but both ATLAS and CMS are expected to make significant contributions to *B* and heavy-flavour physics as well [BB2-2.3.02]. The *b* production cross sections within the detector acceptances are approximately 100 μb for CMS and ATLAS, and 230 μb for the LHCb forward spectrometer. LHCb has two RICH detectors for kaon-pion separation, giving access to many hadronic decay modes.

Compared to the $B_d$ mesons, the properties of the $B_s$ are currently not well measured. This includes the mass and width differences, $\Delta m_s$ and $\Delta\Gamma_s$, between the two weak eigenstates. A more accurate determination of $\Delta m_s$ will be required to probe the possible existence of physics beyond the SM entering the box diagram for $B_s$ oscillations. LHCb expects to make a 5σ observation of $B_s$ oscillations with one year of data at nominal machine parameters and for any $\Delta m_s$ below 68 ps$^{-1}$. The LHC experiments will also pursue the determination of the $B_s$ mixing phase through the measurement of mixing-induced CP violation in the decay $B_s \to J/\psi\,\phi$ (and similar final states). The precise SM prediction for the CP coefficient is $\sin 2\beta_s = 0.036 \pm 0.003$, so that BSM physics may show up cleanly as an enhancement. Both, LHCb and ATLAS/CMS expect to collect about 100k $B_s \to J/\psi\,\phi$ events per year, leading to $\sigma_{stat}(\sin 2\beta_s) \approx 0.03$ and 0.08, respectively, assuming $\Delta m_s = 20$ ps$^{-1}$.

The class of rare decays that proceed through penguin diagrams such as $B_d \to K^{*0}\gamma$ and $B_d \to K^{*0}\ell^+\ell^-$, and the box-diagram-mediated decay $B_s \to \mu^+\mu^-$, are all sensitive probes of BSM physics. While the branching fractions of $B_d \to K^{*0}\gamma$ and $B_d \to K^{*0}\ell^+\ell^-$ have been measured at the current $e^+e^-$ *B*-factories, significant progress will be limited in this area, because of insufficient statistics, until LHCb starts data taking. The forward-backward asymmetry of the leptons in the $B_d \to K^{*0}\ell^+\ell^-$ decay measured as a function of the di-lepton mass is a sensitive probe for new physics with low systematic uncertainty, and will as such provide a rich area for analysis at the LHC. Assuming an annual yield of 7k reconstructed $B_d \to K^{*0}\ell^+\ell^-$ events, LHCb expects to measure the ratio of effective Wilson coefficients $C_7^{eff}/C_9^{eff}$ with an error of 13% after 5 years, compared with a theoretical uncertainty on this quantity of approximately 5%. The current Tevatron combined limit on BR($B_s \to \mu^+\mu^-$) is $1.5 \times 10^{-7}$ (90% CL), which is still well above the SM expectation of $3.5 \times 10^{-9}$ (here, the uncertainty of the theory prediction is of the order of 30%). Expectations for the LHC experiments are hard to estimate owing to the difficulty to fully simulate sufficient background statistics, but all experiments expect to be able to see a SM signal within 1 year. The decay $B_d \to \mu^+\mu^-$ will be a further challenge, because of its even lower branching fraction. It should have been observed by all experiments after several years of data taking. However, a precision measurement of the left-hand side of the particularly clean relation between the ratios $B_s \to \mu^+\mu^-$ to $B_d \to \mu^+\mu^-$ and $\Delta m_s$ to $\Delta m_d$, will be beyond the scope of the LHC.

Another important task of (mainly) LHCb is to make a precise measurement of the UT angle γ. LHCb has the potential to measure γ with small statistical uncertainty, using a wide range of strategies. The methods that exploit direct CP violation in $B_d \to DK$ decays are dominated by tree-level processes, and hence provide clean measurements of the SM value of γ. The precision that can be achieved is estimated to be around 2.5° after 5 years, depending on the size of the still unobserved colour-suppressed decay amplitude required for the Gronau–London–Wyler and Atwood–Dunietz–Soni methods. Other approaches, such as the combined analysis of $B_d \to \pi^+\pi^-$ and $B_s \to K^+K^-$, involve penguin loops through which new physics effects may be witnessed.

It is anticipated that the defocused luminosity at the LHCb interaction point will gradually rise to a level of around $10^{33}$ cm$^{-2}$ s$^{-1}$, as increased experience is gained with the detector and potential upgrades implemented. This will allow for higher annual statistics, in particular for the leptonic channels. For ATLAS and CMS the main $B$ physics programme will take place in the initial years, before the luminosity reaches the design value of $10^{34}$ cm$^{-2}$ s$^{-1}$ (the search for $B_{d(s)} \to \mu^+\mu^-$ is expected to continue also at high luminosity). For the SLHC it does not seem realistic to anticipate a $B$ physics programme at any of the detectors.

## VI-2.3 Super *B*-factories

Already in the bloom of the data-taking period at the present-generation *B*-factories, their hosts, SLAC and KEK, investigated the potential of a successor project, the *Super B-factory*, with the prospect to collect an integrated luminosity of 50 ab$^{-1}$. A series of common workshops has been organized [48], and comprehensive documents were produced by both collaborations [49,50]. In the meanwhile SLAC has desisted as possible host, but the Super KEK-B project is actively pursued [BB2-2.3.01].

Owing to the complementarity of $e^+e^-$ *B*-factories and *B* physics at hadron colliders, the physics case for a Super *B*-factory is well motivated, even when considering that LHCb will make major contributions to the field. The Super *B*-factory will benefit from a clean environment, allowing for measurements that nobody else can do, such as the leptonic decays $B \to \tau(\mu)\nu$, sensitive to $|V_{ub}|$ and to a BSM-charged Higgs (see Fig. VI-4 for the MSSM), or the rare decay $B \to K\nu\nu$, which is complementary to the corresponding rare-kaon decay and sensitive to many SM extensions. A Super *B*-factory will also outperform LHCb on CKM metrology: a precision measurement of α is only possible at an $e^+e^-$ machine, and also the measurements of β and γ will benefit from a better control of systematic uncertainties. High-precision measurements of time-dependent CP-violating asymmetries in such important hadronic penguin modes as $B_d \to \phi K^0$ and $B_d \to K^*\gamma$ are only possible at a Super *B*-factory. New types of asymmetries, such as the above-mentioned forward–backward asymmetry in various $b \rightarrow s\, \ell^+\ell^-$ decays, can be studied in greater detail. Finally, the full range of interesting τ and charm physics analyses can be exploited with unprecedented statistics. We shall emphasize in particular the search for the lepton-flavour-violating decay $\tau \rightarrow \mu\gamma$, for which sensitivities of the order of $10^{-9}$–$10^{-10}$ can be achieved at a Super B-factory. Such sensitivities are well within the reach of the most prominent BSM physics scenarios.

The KEK scenario has a luminosity goal of approximately $5\times10^{35}$ cm$^{-2}$ s$^{-1}$, which represents an increase over today's peak luminosity of a factor of more than 30, and corresponds to production of $10^{10}$ B mesons and τ leptons per year [50]. KEK-B plans a series of small upgrades in the coming years, including the installation of crab cavities. In 2009, following the finalization of JPARC 1, there is a window opening up for a major upgrade of KEK-B. After 2 years of construction, an upgraded Belle detector would start to collect data again in 2011 and should reach an integrated luminosity of 20 ab$^{-1}$ by 2016.

A new initiative to build a *linear* Super B-factory [51] has emerged recently out of a SLAC/LNF collaboration. It has been discussed at a dedicated workshop at Frascati (see ref. [52]) and a first report discussing the basic design characteristics has been published in ref. [51]. This project benefits from synergy with ILC research, and is attractive in many ways [BB2-2.3.03]. Peak luminosities above $10^{36}$ cm$^{-2}$ s$^{-1}$ with relatively low currents and hence smaller backgrounds are obtained through ultra-small transverse beam-spot sizes ($\sigma_x$ = 4 μm, $\sigma_y$ = 0.028 μm). The necessary small transversal emittance is achieved in 2×3 km or 6 km damping rings with short (< 1.5 ms) damping time. Several alternative designs are still under consideration, one of which is depicted in Fig.VI-2. Here a positron bunch from a 2 GeV damping ring is extracted and accelerated to 7 GeV in a superconducting (SC) linac. Simultaneously, an electron bunch is generated in a gun and accelerated in a separate SC linac to 4 GeV. The two bunches are prepared to collide in a transport line where the bunch lengths are shortened. These bunches are focused to a small spot at the collision points and made to collide. The spent beams are returned to their respective linacs with transport lines where they return their energy to the SC accelerator. The 2 GeV positrons are returned to the damping ring to restore the low emittances. The spent electron beam is dumped.

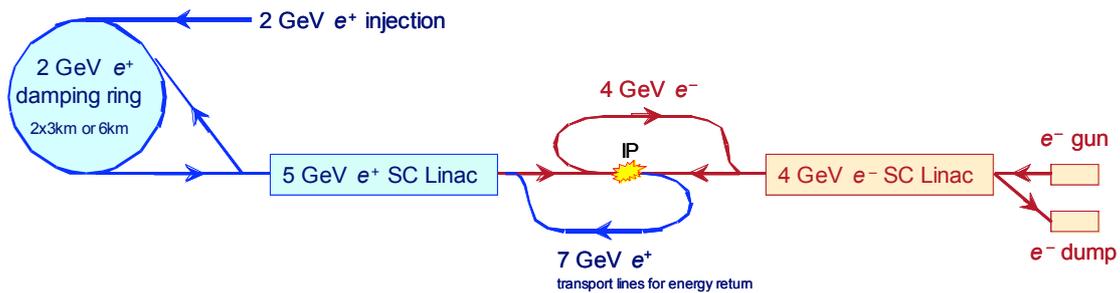

*Fig VI-2: Possible layout of a linear Super B-factory ( from ref.[51]).*

A shortcoming of the small beam-spot size is that it creates an uncertainty in the centre-of-mass (cm) energy, which is proportional to $(\sigma_x\sigma_y)^{-1}$. Because the ϒ(4s) resonance is relatively narrow, this uncertainty leads to an effective reduction of the luminosity. Moreover, since the knowledge of the cm beam energy is one of the primary kinematic constraints exploited in the reconstruction, background levels will increase. One can summarize the qualities and of a linear Super B-factory with respect to the conventional Super B-factory design as follows: although the linear Super B-factory can have smaller currents, it requires smaller damping time and smaller emittance; although smaller machine backgrounds occur in the detector, the increased damping entails significantly higher power consumption; finally, although it has a smaller beam-spot size, better vertex resolution and a better hermeticity, it suffers from a larger beam-energy spread. Different design options are being actively pursued, and more information will be available after the second dedicated Frascati workshop foreseen in March 2006.

To conclude this section we give tentative extrapolations on what can be expected from an integrated luminosity of 50 ab$^{-1}$ collected at a Super *B*-factory. Figure VI-3 shows the individual constraints obtained on the UT, and the constraint from the global CKM fit. Also anticipated for this plot is a measurement of the $B_s$ mixing frequency and progress on lattice QCD calculations. For a better comparison, the expected improvements on α, β, γ from LHCb are not included. As is the case now, the determination of the UT apex will be dominated by the angle measurements. An example for constraints on BSM physics scenarios is given in Figure VI-4: the present (left) and future (right) confidence-level regions obtained in the $m(H^+)$ versus tanβ plane within the MSSM are shown. The white areas are excluded at more than 95% CL.

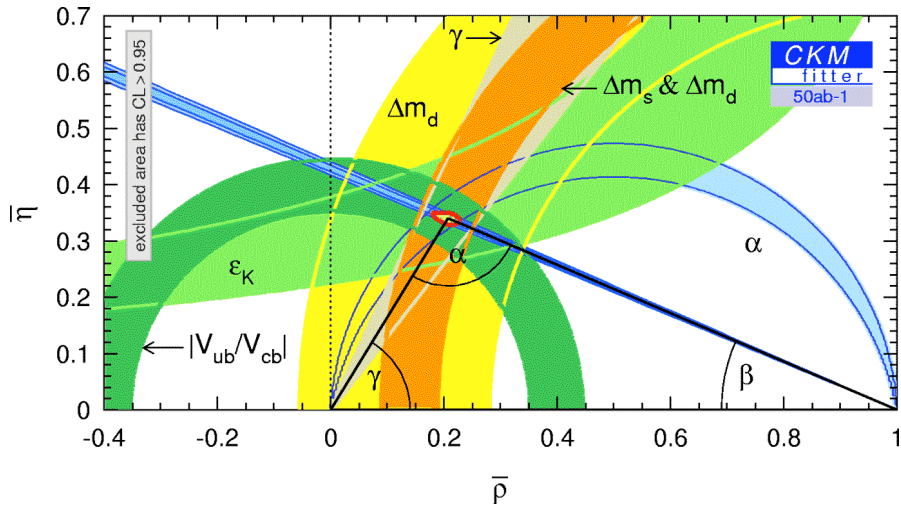

**Fig VI-3:** *Tentative extrapolation of constraints on the UT to an integrated luminosity of 50 ab$^{-1}$ collected at a Super B-factory. Also anticipated for this plot is a measurement of the $B_s$ mixing frequency and progress on lattice QCD calculations. Not included in the plot are the expected improvements on α, β, γ from LHCb. The 95% confidence level regions for the individual constraints, and the result of the global CKM fit [45] are shown.*

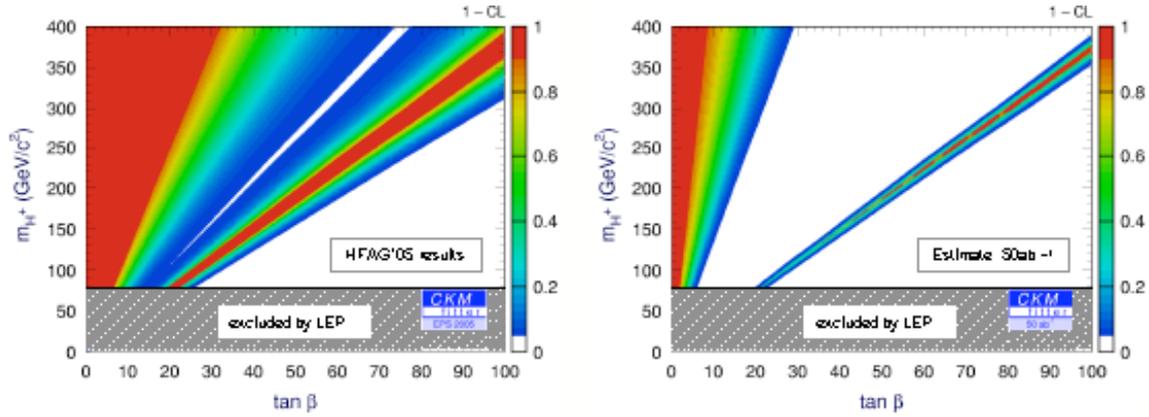

**Fig VI-4:** *Present (left) and future (right) confidence levels obtained in the m(H$^+$) versus tanβ plane within the MSSM. White areas are excluded at more than 95% CL. The shaded areas indicate excluded masses from direct searches at LEP[45].*

## VI-3 Charm physics

Charm physics is naturally performed at the *B*-factories, with spectacular recent successes in discovery spectroscopy. However, essential measurements exist that require a cleaner environment and/or coherent neutral *D*-meson production. Both are available on the ψ(3770) resonance, where coherent neutral *D* pairs are produced almost at rest. Among these measurements is the determination of the neutral and charged *D* decay constants $f_{D(+)}$, which can be accurately predicted by lattice QCD and hence represent a long-awaited precision test of lattice calculations. The experience gained from this comparison in the *D* system can thereafter be extrapolated into the *B* system. It will improve lattice predictions of the neutral *B*-meson mixing frequencies, but also of rare *B* decays governed by annihilation diagrams such as $B \to \tau\nu$, which determines $|V_{ub}|$. A first measurement of $f_{D+}$, based on initial luminosity 281 pb$^{-1}$, has been presented by the CLEO-c collaboration, and the value of (223 ± 17 ± 3) MeV found [53] is in agreement with lattice QCD calculations so far. A total luminosity of up to 1 fb$^{-1}$ is expected to be collected by CLEO-c on the ψ(3770) before the term of the experiment is reached.

At IHEP, Beijing, the τ/charm-factory BEPCII, with its detector BESIII, are currently under construction. BEPCII is designed for a peak luminosity of 10$^{33}$ cm$^{-2}$ s$^{-1}$, implying a number of 30×10$^6$ annual $D\overline{D}$ events when running on the ψ(3770). The commissioning of the storage ring is expected to start early in 2006 and a first physics run is scheduled for February 2007.

Along with the measurement of the decay constants, the study of neutral *D* mixing is a primary challenge of charm physics. Because of the strong CKM suppression, the mixing frequency is much smaller than in the neutral *B* system, and still unobserved. Moreover, CP-violating phases in mixing and decay are also strongly CKM-suppressed so that no CP violation should be observable in the *D* system. Deviations from this null hypothesis would therefore indicate contributions from BSM physics, provided they significantly exceed the size of the experimental systematics and the effects from long-distance strong interactions. Neutral *D* mixing can be searched for, either at a τ/charm-factory through identification of $D^0 \to K^-\pi^+$ double tagged events, or by looking for an apparent lifetime difference between $D^0 \to K^-\pi^+$ and $D^0 \to K^-K^+$ decays. CP violation can be identified in either 2- or 3-body flavour-tagged $D^0$ decays. Measurements of either of these above the 0.1% level would provide strong hints of BSM physics whereas, below this, strong- interaction-related uncertainties are dominant. LHCb will reconstruct several hundred million flavour-tagged two-body decays per year. The search for both mixing and CP violation will almost certainly be limited by systematic errors in the understanding of charge and $K/\pi$ differences in the tracking efficiency. Further studies are required to prove that the limits on searches for CP violation and mixing can go substantially below the 1% level at a τ/charm-factory, a super *B*-factory, or hadron machines.

Another important measurement that can only be performed at a τ/charm-factory with coherent production involves Dalitz-plot fits of CP-tagged $D^0 \to K_S \pi^+ \pi^-$ decays. Such fits could be used as input by the *B*-factories to significantly reduce the model uncertainty of the best method to date to extract the UT angle γ.

In view of the possible future experimental alternatives, it is essential to systematically compare the accuracies on the various measurements that can be obtained in the charm sector, e.g. with a $10^{36}$ cm$^{-2}$ s$^{-1}$ Super *B*-factory and with a $10^{34}$ cm$^{-2}$ s$^{-1}$ τ/charm-factory, respectively.

## VI-4 Rare-kaon decay experiments

Kaon physics has traditionally been studied in Europe, with many important results in particular on CP violation. The chain of CERN proton accelerators provides the possibility of pursuing a very competitive and cost-effective programme at present, and future upgrades of the proton complex would allow the community to plan for next-generation experiments.

The most interesting subject addressed by current kaon experiments is the study of very rare kaon decays. In particular, the highest priorities in kaon decay experiments are studies of the $K \to \pi \nu \nu$ decay modes, both neutral and charged. These flavour-changing neutral decays are particularly interesting, as they are loop processes that can be calculated with good precision in the SM, which may well get significant corrections from extensions such as SUSY, and complement *B* meson measurements, as seen in Figure VI-5.

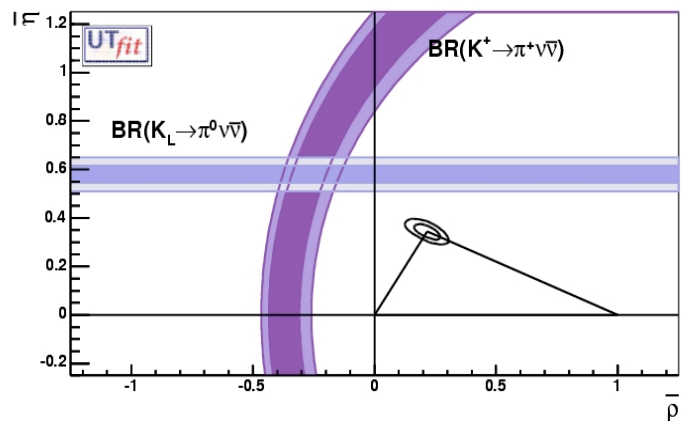

**Fig. VI-5:** *The potential impacts of $K \to \pi \nu \nu$ measurements compared with a present fit to the CKM unitarity triangle, which are largely derived from measurements of B decays [54].*

The current experimental situation is briefly summarized here. The experiment E787 and its upgraded version E949 at the Brookhaven National Laboratory (BNL) observed for the first time the decay $K^+ \to \pi^+ \nu \nu$ by exploiting kaon decays at rest. Three candidate events were published [55], allowing one to quote the following branching ratio: BR($K^+ \to \pi^+ \nu \nu$) = $(15^{+13}_{-9}) \times 10^{-11}$. A lot more data are needed to confront the measurement with the theoretical prediction. The intrinsic theoretical uncertainty of the prediction is mainly due to unsuppressed charm contributions. It has been decreased by recent NNLO calculations [56] of this contribution.

For the theoretically cleaner, due to a negligible charm contribution, but experimentally even more challenging, neutral-kaon decay, progress has been slower. A recent upper limit was presented by the E391a experiment at KEK [57]: BR($K_L \to \pi^0 \nu \nu$) < $2.9 \times 10^{-7}$. It is about four orders of magnitude larger than the SM prediction [58] and therefore a large window of opportunity exists.

Several projects to measure these decays in the United States have been cancelled recently:

- BNL E949 was approved for 60 weeks at the AGS but was run for only 12 weeks before being terminated.
- The CKM proposal to study $K^+ \to \pi^+ \nu \nu$ in flight at the FNAL main injector was not ratified by the P5 Committee.
- The KOPIO proposal at BNL, which had recently completed the R&D, was dropped by the National Science Foundation.

In the rest of the world, there are Letters of Intent in Japan to continue the search for

$K_L \to \pi^0 \nu \nu$ using the E391a technique, and the study of $K^+ \to \pi^+ \nu \nu$ with kaon decays at rest [59] at the new JPARC facility in Japan. A call for proposal was recently announced, and the perspective of these initiatives should soon become clearer. In Europe, the P-326 (NA48/3) proposal [60] was submitted and it is under evaluation by the SPS Committee at CERN. P-326 aims at measuring the charged mode at the SPS, starting taking data in 2009-10, with the objective of obtaining around 80 events by about 2012, assuming the Standard Model branching fraction. The proposal builds on the infrastructure and expertise of the previous CERN kaon experiment (NA48) and needs only a small fraction of the protons that can be delivered by the SPS [BB2-2.3.04]. The key feature of the proposal is to use in-flight kaon decays from a high-energy hadron beam to suppress kinematically the backgrounds originating from the $K^+ \to \pi^+ \pi^0$ decays.

The NA48/3 apparatus could also be modified to serve as a detector of $K_L \to \pi^0 \nu \nu$ if a substantial fraction of the SPS proton intensity is used to produce neutral kaons. In addition, if the apparatus' tracker is retained, it could also be used to measure the $K_L \to \pi^0 e^+ e^-$ and $K_L \to \pi^0 \mu^+ \mu^-$ modes, which are also of interest for BSM physics search.

Follow-up measurements with greater accuracy would be very important should any of these first-generation experiments find a possible discrepancy with the SM prediction based on *B*-physics measurements. This would require Gigahertz kaon rates, for instance a sequel to P-326, using a separate kaon beam originating from a 4 MW proton beam at 50 GeV, such as could be provided by a rapid-cycling replacement for the PS.

It is important to stress that the crucial parameter for the quality of a kaon experiment is the machine duty cycle, which should be as close to 100% as possible.

# VI-5 Charged-lepton-flavour violation

Lepton-flavour violation, recently discovered in the neutral-lepton sector with neutrino oscillation experiments, also actively being searched for in the charged-lepton sector by means of μ and τ rare processes such as μ → eγ, τ → eγ or with μ → e conversion.

In supersymmetric models the amplitude of these decays is derived from the slepton mass matrix and is connected to other observables such as leptonic anomalous magnetic dipole moments (MDMs) and possible electric dipole moments (EDMs). MDM and EDM are related to the real and imaginary part of the smuon diagonal element, while charged- lepton-flavour violation (CLFV) is related to its off-diagonal element.

Relevant non-diagonal terms of the slepton mass matrix are predicted in SUSY GUT models, where these terms arise from radiative corrections from the Planck scale to the weak scale and in SUSY seesaw models, where a suitable scheme of neutrino masses and chiralities is introduced consistently with the existing experimental data of neutrino-oscillation experiments. These models predict CLFV with branching fractions just below to a few orders of magnitude below the current experimental upper limits ($6.8 \times 10^{-8}$ for τ → μγ [61], $1.2 \times 10^{-11}$ for μ → eγ [62] and $8.0 \times 10^{-13}$ for μ → e conversion [63]. The branching fractions predicted for τ → μγ are usually 3-5 orders of magnitude higher than for μ → eγ, which in turn is predicted to have a rate higher by roughly two orders of magnitude than μ → e conversion. CLFV due to neutrino mixing included in the SM frame is suppressed by $(m_\nu/m_W)^4$ and hence unobservable (for instance a BR ≈ $10^{-54}$ is predicted for μ → eγ). The detection of CLFV processes would thus constitute an unambiguous sign of BSM physics.

In μ → eγ searches, a beam of positive muons is stopped in a thin target and the search is made for a back-to-back positron–photon event, with the right momenta and timing coincidence. The main background in present experiments is due to accidental coincidence of independent positrons and photons within the resolutions of the used detectors. The best available detectors for low-energy positrons and photons must therefore be employed. In the MEG experiment at PSI [64] a surface muon beam with an intensity greater than $10^7$ μ/s will be stopped in a thin target. A magnetic spectrometer, composed of a superconducting magnet and drift chambers, will be used for the measurement of the positron trajectory. Positron timing will be measured by an array of scintillators. Photons will be detected by an innovative electromagnetic calorimeter in which a total of about 800 photomultipliers detect the light produced by photon-initiated showers in about 800 l of liquid xenon. The aim of MEG is to reach a sensitivity down to a BR of the order of $10^{-13}$, an improvement of two orders of magnitude with respect to the present experimental bound. The start of the data taking is foreseen in 2006.

Another very promising channel for CLFV investigation, which involves muons but is not limited by accidental background, is muon to electron (μe) conversion in nuclei. Negative muons are brought to a stop in a thin target and are subsequently captured by a nucleus. The energy of a possible converted electron would be equal to the rest muon mass minus the muon binding energy $E_B$. The two main sources of background are: beam-correlated background due to mainly radiative pion capture followed by γ → $e^+e^-$ conversions, and electrons from muon decay in orbit (DIO). The first source of background can be controlled by improving the muon-beam quality, the second one is intrinsic; the DIO electron spectrum extends up to the energy region of electrons from μe conversion, but with a spectrum proportional to $(m_\mu - E_B - E_e)^5$. An excellent electron momentum resolution is fundamental in keeping this background under control. In the PRISM/PRIME [65] project at JPARC, a pulsed proton beam is used to produce low-energy pions that are captured by placing the target inside a superconducting solenoid magnet. The pulsed structure of the beam helps in reducing the beam-correlated background. The beam is then transported in a circular system of magnets and RF cavities (FFAG ring), which acts as a pion-decay section (increasing beam cleaning) and reduces the muon energy spread. The features of this beam would be extremely high intensity ($10^{12}$ s$^{-1}$) of very clean muons of low momentum ($\approx$ 70 MeV/c) and with a narrow energy spread (few % of FWHM). The last feature is essential to stop enough muons in thin targets. If the electron momentum resolution will be kept below 350 keV/c (FWHM) the experiment will be sensitive to μ$e$ conversion down to BR ≤ $10^{-18}$.

Project schedule [66]: PRISM construction and test: 2006-09. Bring PRISM to any high-intensity hadron facility and carry out the μ$e$ conversion experiment (PRIME): after 2010. We note here that the construction of a high-intensity, low-energy proton driver for either neutrino and/or nuclear physics would provide CERN with a scientific opportunity to host a world-leading muon physics facility [BB2-2.1.06].

Concerning τ decays, the luminosity increase foreseen at a future Super $B$-factory scales up by roughly two orders of magnitude with respect to the statistics available to date. One could therefore expect a sensitivity increase reaching $O(10^{-9}–10^{-10})$ for the τ → μγ branching fraction, assuming backgrounds are kept under control. Such a sensitivity is well within the bulk reach of SUSY GUT models.

Also interesting is the search for the CLFV decay τ$^-$ → μ$^-$μ$^+$μ$^-$ (current limit [67]: BR < $1.9\times10^{-7}$), which will be possible at three LHC experiments, and where for instance a limit of $4\times10^{-8}$ is expected to be reached by CMS with 30 fb$^{-1}$.

If CLFV were discovered, the angular distribution of electrons from the CLFV decay of polarized muons could be used to discriminate among the different SUSY GUT SU(5), SUSY GUT SO(10), SUSY seesaw models and others.

# VI-6 Concluding remarks

To reassess the fundamental importance of flavour physics, let us cite from a letter submitted by a large group of theorists working in the field to the Orsay Open Symposium [BB2.3.02]:

> 'It is expected that the experiments at the Large Hadron Collider (LHC) will lead to discoveries of new degrees of freedom at the TeV energy scale. The precise nature of these new phenomena is not known yet, but it is strongly expected that it will answer the crucial question of the origin of electroweak symmetry breaking. This step forward will leave the understanding of the flavour structure of this new physics as a major open question in the field. A deeper understanding of the nature of flavour will most likely be a key element in sorting out the properties of the new phenomena to be uncovered by the LHC. As we shall argue below, neither the LHC nor a possible International Linear Collider (ILC) allow for an exhaustive exploration of the underlying structure of new physics. A diversified and thorough experimental program in flavour physics will therefore continue to be an essential element for the understanding of nature. It should not be endangered, considering in particular the comparatively low cost level of such experiments.
>
> Rare $B$ and $K$ meson decays are highly sensitive probes for new degrees of freedom beyond the standard model (SM); through virtual (loop) contributions of new particles to such observables, one can investigate high energy scales even before such energies are accessible at collider experiments. Today this indirect search for new physics signatures takes place almost in complete darkness, given that we have no direct evidence of new particles beyond the SM. But the day the existence of new degrees of freedom is established by the LHC, the study of anomalous phenomena in the flavour sector will become an important tool for studying their phenomenology. Then, the problem will no longer be to discover new physics, but to measure its (flavour) properties. In this context, the measurement of theoretically clean rare decays will lead to valuable information on the structure of the new-physics models.'

Independently of whether new physics will be discovered at the LHC, flavour physics in all its rich phenomenology represents a privileged window to measure and understand SM *and* BSM phenomena. Once we accept the existence of new physics, the continuous improvement of the experimental sensitivity to new flavour phenomena becomes a necessity. The quantitative assessment of the required accuracies will depend on the specific nature of the BSM physics, which will provide benchmarks for the precision of direct measurements of the BSM parameters, or useful constraints on them. To build a solid physics case for the new facilities in the flavour sector, one must therefore examine whether a possible null result of an experiment can be translated into useful bounds of the BSM parameters, and/or if the available SM physics measurements are themselves of fundamental importance. The lepton-flavour-violation experiments are typical candidates of the first type. LFV discovery would be a conspicuous new physics signature, while a null result reduces the available freedom for model building as is already the case at present (another manifestation of the 'flavour problem'). The rare-kaon experiments and the Super *B*- factories belong to both categories, and a τ/charm-factory mainly falls into the second category. In all cases, it is a valid prerogative to ask whether a given project sets a sufficiently ambitious physics goal and realistically assesses the technical obstacles.

These considerations are met for the projects mentioned in this review. In particular in LFV the ongoing search at PSI, Switzerland, and, if a new proton driver at CERN allows for high-intensity muon beams, a μe conversion experiment. The traditionally strong kaon-physics programme at CERN should be continued by strengthening the P326 project and collaboration, and the potential to measure the more challenging, but also more important, $K_L \to \pi^0 \nu\nu$ decay should be seriously considered when discussing the upgrade of the proton complex [BB2-2.3.05].

The most ambitious of all future flavour-physics projects is the Super *B*-factory. Because of the large investments, a detailed cost/benefit analysis is required. The linear Super B-factory, if realized as proposed, will significantly enhance the benefit by delivering a peak luminosity up to several times larger than the Super KEK-B project. The feasibility of this project would open new prospects for flavour physics, fully justifying an extensive R&D programme spawned by the on-going preliminary studies.

## VI-7 Discussion session

### VI-7.1 Questions

The discussion started with a list of questions presented by the speaker at the end of the overview talk. These questions are quoted below:

### VI-7.1.1 B physics
- Which of the rare modes sensitive to BSM physics will be systematically limited at a Super *B*-factory with 50 ab$^{-1}$?
- Is 50 ab$^{-1}$ enough to do substantially better than the hadron colliders?

- If NP is discovered at the LHC:
    - NP parameters cannot be measured model-independently in hadronic modes
    - Precise measurement of leptonic and rare semileptonic modes are considered
    - What can be done at the LHC? What is the required Super *B* luminosity?
- If no NP is discovered at the LHC: continue indirect search with <u>all</u> modes!
- What is the timescale of the 'proof of principle' of the linear Super *B*-factory?
- Can the linear Super *B*-factory also be τ/charm- and φ-factory?

### VI-7.1.2 K physics
- Should CERN's SPSC-P-326 $K^+ \rightarrow \pi^+ \nu\nu$ proposal get our strong support?
- What are the concrete plans for an ambitious $K_L \rightarrow \pi^0 \nu\nu$ experiment at CERN?

### VI-7.1.3 Charm physics
- In which area is a $10^{34}$ cm$^{-2}$ s$^{-1}$ τ/charm- better than a $10^{36}$ fb$^{-1}$ Super *B*-factory?

### VI-7.1.4 CLFV search
- LFV importance boosted by neutrino discoveries. Can we build a $\mu \rightarrow e$ conversion experiment at a possible new proton driver at CERN?
- What is the required luminosity for a competitive $\tau \rightarrow \mu\gamma$ measurement at a Super *B*-factory?

## VI-7.2 Discussion
During the discussion the following points emerged:

### VI-7.2.1 Comparison of the two proposed Super *B*-factories
It was stated that there is no competition between the two Super *B*-factory proposals. The time scale of the two projects is not the same and it is now too early to anticipate whether a linear *B*-factory can be realized as designed. Therefore, parallel efforts along both lines are needed until the new idea reaches maturity and a cost estimate is available.

Super KEK-*B* is a well advanced project and there is a window of opportunity for this project to go ahead, with construction at the end of 2008, with the finalization of JPARC. No major decision has been made so far to go-ahead, and KEK is open to discussions about other *B*-factory options for the future.

The linear Super *B*-factory has the possibility of polarized beams and may cover an energy range down to the τ/charm region and potentially less. It is designed to reach up to six times higher luminosity than Super KEK-B, although at the expense of a larger energy spread due to strong beam focusing, which will reduce luminosity and increase backgrounds in the analyses.

To set priorities in a more educated way, there is a need to better understand which measurements will be systematics-limited, and why.

### VI-7.2.2 *B*-physics in a hadron environment

The synergy between discoveries at the energy frontier and heavy-flavour measurements should be explored much more, in particular quantitatively – even at the price of strong model dependence. This could take the form of specific BSM scenarios, where the information available from heavy-flavour physics, LHC energy frontier, and the ILC are explored. There are many presentations and publications where such studies are already presented, but it is often difficult to draw conclusions, from the different scenarios considered.

### VI-7.2.3 Very rare kaon decays

We need to put much more emphasis on the importance of the very rare kaon-decay physics. This means that additional effort should be provided to ensure that the trend to cancel kaon-decay experiments is not followed.

### VI-7.2.4 Lepton-flavour violation

Search for lepton-flavour violation is still one of our big opportunities to indirectly discover BSM physics. The PRISM/PRIME experiment to search for μ/e conversions should be strongly supported, including investigations of the proton driver at which the experiment can be sited. Also the LHC experiments should further investigate their reach for the LFV decay $\tau^- \to \mu^-\mu^+\mu^-$.

### VI-7.2.5 Plans at the Frascati Laboratory

Information concerning plans for the short-term future of Daphne and of the Frascati Laboratory were presented. It was stressed how the proposed programme will provide cutting-edge accelerator development and which physics questions can be addressed with a realistic increase of luminosity. More details can be found in the material from the dedicated workshop [68]. The importance and role of the network between laboratories was stressed. It will help not only in the preparation for a specific new machine, but also allows developments of tools generally useful in the future.

# VII PRECISION MEASUREMENTS

## VII-1 Scientific programme

In particle physics the Standard Model (SM) provides a robust framework to describe all processes observed[8]. Despite its success it leaves many questions about the underlying nature of particles, interactions and symmetry principles without satisfactory answers. Among the most intriguing puzzles are the number of particle families, the mass hierarchy of fundamental fermions, and the rather large number of free parameters in the SM. It remains very unsatisfactory that the physical origin of the observed breaking of discrete symmetries in weak interactions, e.g. of parity (P), charge conjugation (C) of time reversal (T) and of the combined CP symmetry, are not revealed, although the experimental facts can be well described within the SM. CP violation plays a particular role through its possible connection to the matter-antimatter asymmetry in the Universe.

A number of speculative models offer answers to these questions. These include approaches that involve left-right symmetry, fundamental fermion compositeness, new particles, leptoquarks, supersymmetry, supergravity and many more possibilities. Their relevance for physics can only be verified in experiments where their (unique) predictions can be measured.

In particle physics, two general complementary approaches exist to test such models:

- direct observations of new particles and processes, and
- precision measurements of observables, which can be calculated to sufficiently high precision within the SM. A significant deviation between a precise measurement and its calculation would undoubtedly indicate new physics.

The direct observation approach is mostly carried out at high energies, such as the searches for supersymmetric particles. Occasionally also at low energies direct searches occur, such as the hunt for axions. Precision work typically (but not exclusively) is performed with low-energy experiments, such as measurements of electron and muon magnetic anomalies or searches for permanent electric dipole moments (EDMs).[9] An experimental verification of SM calculations can be exploited to set limits on parameters in speculative theories. Furthermore it can be utilized to extract most accurate values for fundamental constants and SM parameters.

The landscape of the field of precision measurements is characterized through a number of well motivated, dedicated experiments at the scientific infrastructure that is best suited in each case. Independent of their scale, their potential to discover new physics is very robust and in many cases exceeds the possibilities of direct searches.[10] In the next one to two decades, the progress in precision measurements will depend on four key elements:

---

[8] The recent observations in neutrino physics and their implications for standard theory are discussed in Chapter V.

[9] The prospects of precision experiments at low energies and at typical nuclear physics facilities have been covered recently in the NuPECC Long Range Plan 2004 [69].

[10] There is a plurality of small-scale experiments with a very high potential to influence the development of concepts in particle physics. They do not require larger infrastructures and are therefore not covered here.

(i) Many accelerator-based precision measurements are statistics limited, with their systematic uncertainties well under control. Significantly more intense particle sources, such as a high-power proton driver, are needed for a number of experiments, where successful techniques can be further exploited or where the high particle flux will also allow novel experimental approaches (e.g. rare decays, muon experiments, neutron experiments, nuclear β-decays).

(ii) The low-background, non-accelerator experiments will need an advanced, shielded underground facility (e.g neutrinoless double β-decay).

(iii) Novel ideas exist for accelerator-based, large-scale, dedicated experiments, which require long-term commitments for research and development (e.g. charged-particle EDM, muon magnetic anomaly).

(iv) In all areas of precision experiments a strong interplay between theorists and experimentalists is indispensable for their success.

We will briefly discuss here the physics motivation for the research programmes concerned with precision measurements and indicate the most urgent needs in the field.

## VII-1.1 The nature of the fundamental fermions

The SM has three generations of fundamental fermions, which fall into two groups, leptons and quarks. The latter are the building blocks of hadrons and in particular of baryons, e.g. protons and neutrons, which consist of three quarks each. Forces are mediated by bosons: the photon, the $W^{\pm}$ and $Z^0$-bosons, and eight gluons.

The mass and weak eigenstates of the six quarks (u,d,s,c,b,t) are different and related to each other through the Cabbibo Kobayashi Maskawa (CKM) matrix. Non-unitarity of this matrix would be an indication of physics beyond the SM and could be caused by a variety of possibilities, including the existence of more than three quark generations or could be faked by yet undiscovered muon decay channels. The unitarity of the CKM matrix is therefore a severe check on the validity of the SM and sets bounds on speculative extensions to it.[11] Leptons do not take part in strong interactions. In the SM there are three charged leptons (e$^-$, μ$^-$, τ$^-$) and three electrically neutral neutrinos ($\nu_e$, $\nu_\mu$, $\nu_\tau$) as well as their respective antiparticles. For the neutrinos, eigenstates of mass ($\nu_1$, $\nu_2$, $\nu_3$) and flavour ($\nu_e$, $\nu_\mu$, $\nu_\tau$) are different and connected through a matrix analogous to the CKM mixing in the quark sector.

The reported evidence for neutrino oscillations strongly indicates finite neutrino masses. Among the recent discoveries are the surprisingly large mixing angles $\theta_{12}$ and $\theta_{23}$. The mixing angle $\theta_{13}$, the phases for CP-violation, the question whether neutrinos are Dirac or Majorana particles, the neutrino mass hierarchy, and a direct measurement of a neutrino mass rank among the top issues in (neutrino) physics.

### VII-1.1.1 Dirac versus Majorana neutrino

Neutrinoless double β-decay is only possible for Majorana neutrinos. A confirmed signal would therefore solve one of the most urgent questions in particle physics, i.e., whether the neutrinos are Dirac particles (distinct from antineutrinos) or Majorana particles (identical to antineutrinos). Furthermore, the process violates lepton number by two units and it appears to be the only realistic way at present to discover lepton-number violation.

---

[11] Measurements at highest precision of the largest matrix element ($V_{ud}$) in nuclear and neutron β-decays contribute significantly to the field of flavour physics [69], which is covered separately in this report.

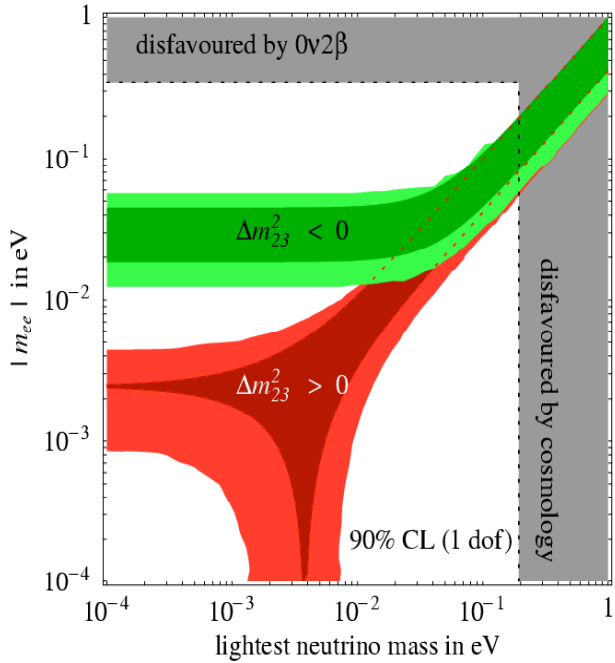

Figure VII-1: *The allowed effective neutrino mass arising in neutrinoless double beta decay versus the lightest neutrino mass. The limits from cosmological and neutrinoless double β-decay searches are indicated [70]. The planned KATRIN experiment will cover about the same region as 0ν2β decays, however, without the restriction to Majorana particles.*

Part of the Heidelberg-Moscow collaboration has reported a positive effect in a $^{76}$Ge experiment [71] at Gran Sasso, with a lifetime of $(0.69\text{-}4.18)\,10^{25}$ years (99.7 % CL). This corresponds to an effective neutrino mass in the few hundred meV range. An independent confirmation of this result is required. The GERDA experiment, also using $^{76}$Ge, aims for a sensitivity beyond $10^{26}$ years, and should thus be able to confirm or rule out this result [72]. This will offer a clarification on the existence of a signal. In case of a successful re-observation one still needs confirmation of the process of neutrinoless double β-decay in at least one other isotope. This could be provided by, e.g., the CUORE experiment [73] In the event that the signal is not confirmed, the next sensitivity goal is set by the effective neutrino mass in the inverted hierarchy scheme. As seen in Fig. VII-1, the effective neutrino mass entering in neutrinoless double beta decay is 10 meV or larger independently of the mass of the lightest neutrino in this hierarchy. Future experiments should therefore aim at this level of sensitivity. This scientific area is characterized by a large number of ongoing and planned experiments, such as NEMO ($^{100}$Mo, $^{82}$Se), CUORE ($^{130}$Te), EXO ($^{136}$Xe), Majorana and GERDA (both $^{76}$Ge). Many different and novel techniques are employed in these experiments, from gas and liquid TPC, to cryogenic bolometers. In themselves they are all well motivated and justified. The experiments will need to be tuned to a background level below $10^{-3}$ events per kg detector mass and per year to reach sensitivities of a few 10 meV. This low level of background can only be achieved by an extremely careful choice of materials. The screening of appropriate materials can only be performed in dedicated facilities, and currently takes a significant amount of time. New, larger scale, facilities for material screening will likely be needed to reach the desired sensitivity levels.

Although the first unambiguous observation of neutrinoless double β-decay will be a breakthrough observation, the full interpretation of the present and future experiments is hampered by the still poor knowledge of the nuclear matrix elements describing the process; yet these are needed to translate a lifetime or lifetime limit into a neutrino Majorana mass or mass limit. The experimental facilities needed to make important measurements (RCNP, Japan, KVI, Netherlands, PSI, Switzerland) will need support. On the theoretical side, conceptually novel ideas are needed to proceed significantly.

Reaching a sensitivity level of 10 meV will require experiments at the 1000kg scale and will require background levels of $10^{-3}$ counts/(kg y). These experiments will be expensive, and choices will have to be made to pick out the most promising isotopes and techniques. The world community should come together once the current round of experiments are underway and more information is available on the different techniques, with the aim to pick out 2-3 experiments which can reach the desired sensitivity level.

### VII-1.1.2 Neutrino Masses

The best directly determined neutrino-mass limits result from measurements of the tritium β-decay spectrum close to its end-point. Since neutrinos are very light particles, a mass measurement can best be performed in this region of the spectrum. In other parts the nonlinear dependences due to the relativistic nature of the kinematic problem cause a significant loss of accuracy, which overwhelms the gain in statistics that could be hoped for. Two groups in Troitzk and Mainz used spectrometers based on magnetic adiabatic collimation, combined with an electrostatic filter (MAC-E technique); they found an effective mass of $m_{\nu e} < 2.2$ eV.

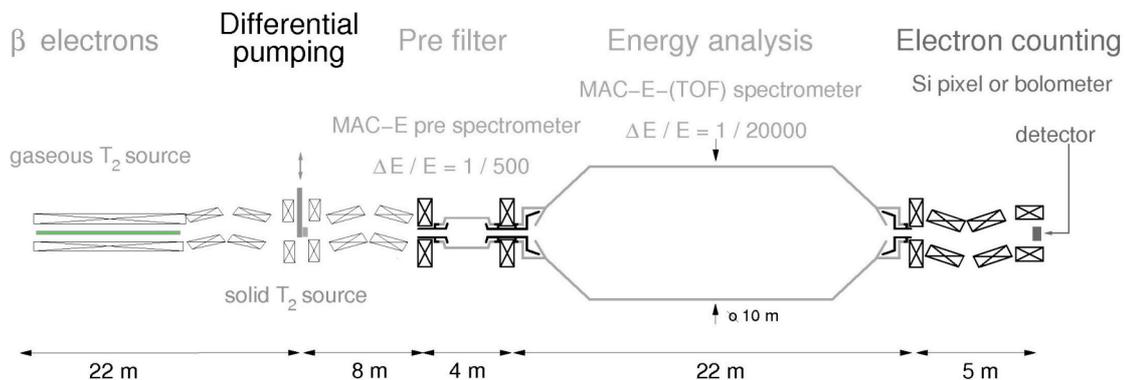

Figure VII-2: The *KATRIN experiment under way at Karlsruhe [74].*

A new experiment, KATRIN [74], is currently being prepared by a world-wide collaboration in Karlsruhe, Germany; it is planned to exploit the same technique. It aims for an improvement by about one order of magnitude. The energy resolution of MAC-E filters is given by the ratio of the maximal magnetic field of 3.5 T in the source region to the minimum field of $3*10^{-4}$ T at the maximum diameter of the apparatus (see Fig. VII-2). The radius of such a device scales inversely with the square of the possible sensitivity to a finite neutrino mass, which will ultimately place a technical limitation on this principle.

The KATRIN experiment will be sensitive to the mass range, where a finite effective neutrino mass was reported from neutrinoless double β-decay in $^{76}$Ge. A direct mass measurement or limitation at that level is well motivated, and KATRIN requires the full support of the community for funding, start running in 2008 and analysis.

There also new calorimetric techniques being developed which will allow sub-eV resolution. There is no fundamental resolution limit for calorimeters. The Genova experiment tries to exploit the Re β-decay with a much smaller Q-value than the tritium decay.

## VII-1.2 The nature of the fundamental interactions

Strong interactions require precise measurements for a refinement of Quantum Chromodynamics (QCD). In this report strong interactions are discussed in Chapter IX. We concentrate here on electromagnetic, weak and gravitational forces, and the major common future needs to conduct this research.

### VII-1.2.1 Electromagnetic interaction

In the electroweak part of the SM, very high precision can be achieved in calculations, in particular within quantum electrodynamics (QED), which is the best tested field theory we know and a key element of the SM. QED allows extracting accurate values of important fundamental constants from high-precision experiments on free particles and light bound systems, where perturbative approaches work very well for their theoretical description. Examples are the fine-structure constant α from measurements of the magnetic anomaly in single electrons or the Rydberg constant $R_\infty$ from atomic hydrogen laser spectroscopy. These results are essential in describing the known interactions precisely. For bound systems containing nuclei with high electric charges, QED resembles a field theory with strong coupling and new theoretical methods are needed.

The muon magnetic anomaly $a_\mu$ has high sensitivity to new physics. Because of its high mass the muon is $(m_\mu/m_e)^2 \approx 40,000$ times more sensitive to heavier particles than the electron. This gives $a_\mu$ sensitivity to a large number of speculative theories, including supersymmetry, compositeness and many others. Any future measurement of $a_\mu$ will be a calibration point for all new models, which they have to satisfy. It has been measured in a series of precision storage-ring experiments at CERN up to the 1970's and recently at the Brookhaven National Laboratory (BNL) for both positive and negative muons with 0.7 ppm accuracy each (Fig. VII-3) [75]. The result is purely statistics limited. The anomaly arises from quantum effects and is mostly due to QED. Further, there is a contribution from strong interactions of 58 ppm, which arises from hadronic vacuum polarization. The influence of weak interactions amounts to 1.3 ppm. Whereas QED and weak effects can be calculated from first principles, the hadronic contributions need to be evaluated through a dispersion relation and experimental input from $e^+e^-$ annihilation into hadrons or hadronic τ-decays. One significant term, the hadronic light-by-light scattering must be taken from calculations only. The values of calculations of the complete hadronic part in $a_\mu$ depend on the choice of currently available experimental hadronic data. The theoretical values for $a_\mu$ differ at present between 0.7 and 3.2 standard deviations from the averaged experimental value, depending both on the chosen theoretical approach and on the experimental input data. This clearly indicates that intense theoretical and experimental efforts are needed to solve the hadronic correction puzzle. More precise experimental hadronic annihilation data up to 10 GeV and quark masses are also required for flavour physics. The community feels that this can be eventually solved. For the muon magnetic anomaly, improvements on the hadronic corrections in both theory and experiment are required, before a definite conclusion can be drawn whether a hint of physics beyond standard theory has been seen. A continuation of the BNL g–2 experiment, with improved equipment and beams, aiming at a factor of 4 to 5 improvement, was scientifically approved in 2004 and is now seeking funding. An experiment with a factor of ten reduction in the error bars has been proposed to J-PARC, where a muon programme is expected to start in 2015 at the earliest.

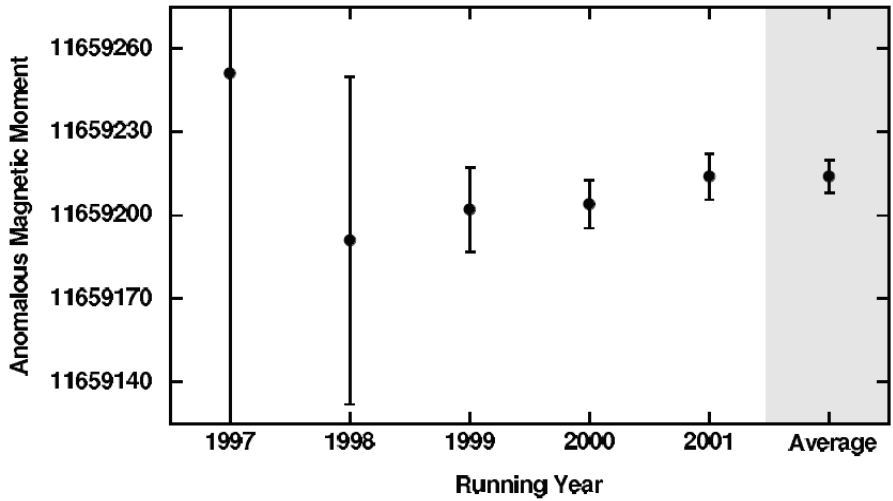

Figure VII-3: *De-velopment of the muon mag-netic anomaly measurements. The results of 2000 and 2001 represent the positive and negative muon final values respectively.*

It should be noted that a novel storage-ring idea with proton beam magnetic field calibration has been suggested recently, which promises significantly higher experimental accuracy. We note, an $a_\mu$ measurement to better than about 0.1 ppm will require a better determination of the muon magnetic moment, which at this point comes from muonium spectroscopy, which in itself is statistics limited and will require a major effort at a new high-flux muon beam (see Table VII-1).

### VII-1.2.2 Weak interactions

The Fermi coupling constant of weak interactions plays an important role in the SM. Its value is at present known to 20 ppm. It can be determined from a precision measurement of the muon lifetime. At this time there are three lifetime experiments on their way, one at the RIKEN-RAL muon facility, and two at the Paul Scherrer Institut (PSI). One can expect an accuracy of order 1 ppm, which will mostly be due to the statistical uncertainty at a chopped continuos muon channel. Recent calculations are sufficiently accurate to allow extracting an improved value for the Fermi coupling constant $G_F$. An intense pulsed facility would be a major advantage here.

In standard theory the structure of weak interactions is V–A, which means that there are vector (V) and axial-vector (A) currents, with opposite relative sign, causing a left-handed structure of the interaction and parity violation. Other possibilities such as scalar, pseudo-scalar, V+A and tensor type interactions would be clear signatures of new physics. So far they have been searched for without positive result. However, the bounds on parameters are not very tight and leave room for various speculative possibilities.

The coefficients describing correlations between observables in the decay products are studied in a number of experiments on selected nuclei, neutrons and muons at this time [76]. These observables are sensitive to non V–A interactions and some of them are T-violating in nature, such as the correlation between neutrino and electron momentum vectors in β-decays of polarized nuclei. From the experimental point of view an efficient direct measurement of the neutrino momentum is not possible. The recoiling nucleus can be detected instead, and the neutrino momentum can be reconstructed using the kinematics of the process. Since the recoil nuclei have typical energies in the tens of eV range, precise measurements can only be performed, if the decaying isotopes are suspended, using extremely shallow potential wells. These exist, for example, in magneto-optical traps, where many atomic species can be stored in traps. In this subfield, progress is achieved in a combination of particle, nuclear and atomic-physics techniques.

Such experiments are carried out at currently at a number of small-scale facilities will in the long run depend on the availability of high flux beams of the (short-lived) radioactive nuclei, for example from a high flux ISOL facility such as EURISOL or in connection with a high-power proton driver.

### VII-1.2.3 Gravity

String and M theories try to find a common description of gravity and quantum mechanics. In their context, there appear predictions of extra dimensions, which could manifest themselves in deviations from the Newtonian laws of gravity at small distances. Therefore a number of searches for such large extra dimensions have been started. At the Institute Laue-Langevin in Grenoble, France, a new limit in parameter space (Fig. VII-4) has been established for extra forces of the type $m_1 m_2 / r^2 [1+\alpha \exp(-\lambda/r)]$ where $\alpha$ determines the strength and $\lambda$ is the Yukawa range of the additional interaction. The experiments with highest sensitivity in the nm range use neutron-nucleus scattering and quantum mechanical interference patterns from ultra-cold neutrons which may be viewed as gravitational matter 'standing' waves [77]. The latter would largely benefit from more intense cold neutron sources. It should be mentioned that searches non-Newtonian forces have also been started using Bose-Einstein condensates, where Casimir-Polder forces were studied in condensate oscillations. These measurements are not yet competitive, but are expected to improve significantly [78].

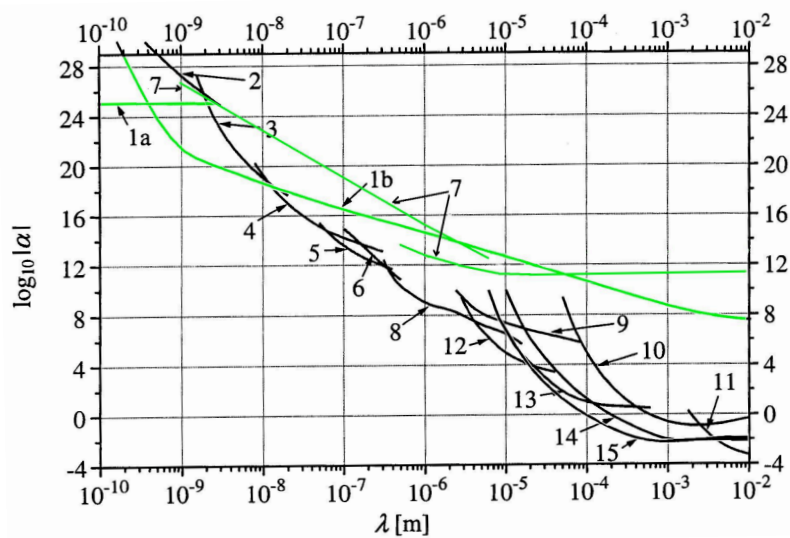

Figure VII-4: *Yukawa constraints for non-Newtonian gravity. In the nanometre region the best limits come from neutron scattering. Data from neutron-nucleus scattering (1a, 1b), neutron bound quantum states (7), Casimir/van der Waals forces (2–9) and torsion pendular/ resonators (10–15).*

### VII-1.2.4 Common future needs for precision fundamental interaction research

The majority of precision experiments on the fundamental interaction properties are significantly statistics limited. Since the systematic uncertainties are well under control for the recently completed and on-going experiments, new and deeper insights as well as improved values of fundamental constants can be expected, from future experiments carried out at higher particle flux-facilities. This requires significantly improved or new accelerator infrastructure, such as a several MW proton driver [79].

## VII-1.3 Symmetries and conservation laws

### VII-1.3.1 Discrete symmetries

#### VII-1.3.1.1 CPT THEOREM

The SM implies exact CPT and Lorentz invariance. As any deviation from it would indicate new physics, CPT conservation has been tested accurately in a number of high-precision experiments. Mostly limits of differences in the properties (such as masses, charges, magnetic moments, lifetimes) of particles and their antiparticles were compared and normalized to the averaged values. To arrive at a dimensionless quantity. The $K^0$–$K^0$bar mass difference had yielded the best test at $10^{-18}$.

Atomic physics experiments as well as the muon storage-ring experiments provide stringent limits on possible CPT-violation when interpreted in terms of a theoretical approach, which allows us to assess experimental results from different fields of physics. Here, additional small terms are introduced into the Lagrangian or Hamiltonian of Dirac particles, and perturbative solutions are searched for [80]. All possible additions violate Lorentz invariance and some of them break CPT. They are associated with the existence of a preferred frame of reference and therefore diurnal variations in physical observables relating to particle spins can be searched for. Here limits have been established at $10^{-30}$ GeV for neutrons, $10^{-27}$ GeV for electrons and protons, and $10^{-24}$ GeV for muons. It remains a controversial theoretical question to know whether the energies associated with CPT-breaking terms should be normalized to the mass of the particles in order to arrive at a dimensionless figure of merit for CPT-violation, which in such case would be most favourable for electrons and neutrons at about $10^{-30}$.

The validity of CPT and Lorentz invariance in atomic systems is currently addressed at CERN/AD. The ALPHA, ATRAP and ASACUSA collaborations are preparing measurements of frequency differences in antihydrogen and comparing them to the hydrogen atom. In the framework of a generic SM extension, they have unique access to parameters in this model. The community is now asking for a well motivated upgrade of the AD facility through the ELENA ring. In the long term, future experiments with larger particle numbers are planned at the FAIR facility at GSI in Darmstadt, Germany [81].

#### VII-1.3.1.2 PARITY

The electroweak theory, which unifies electromagnetism and the weak interaction, is a crown jewel of particle physics and has been confirmed to great precision in high energy experiments. One of the outstanding successful predictions of the theory was the existence of a heavy neutral boson, the $Z^0$, that is mixed with the photon and mediates interactions that do not conserve parity. The $\gamma$-$Z^0$ mixing angle, $\Theta_W$, is a fundamental parameter of the theory,. related to the ratio of the electromagnetic and weak coupling constants by $\sin\theta_W = e/g_w$. Since the electroweak theory is a quantum field theory, these coupling constants vary with scale due to the polarization of the vacuum by article-antiparticle pairs. This "running" of $\sin\theta_W$ from high to low energy is only a poorly tested prediction of the electroweak theory (see Fig. VII-5). This prediction appears not to be in good agreement with all observations. If the value of $\sin^2\theta_W$ is fixed at the $Z^0$- pole, deep inelastic neutrino scattering at several GeV (NuTeV) appears to yield a considerably higher value than predicted [82]. A reported disagreement from atomic parity violation in Cs has recently disappeared, after a revision of atomic theory, but still the agreement is moderate, as it is also for Moeller scattering (E158).

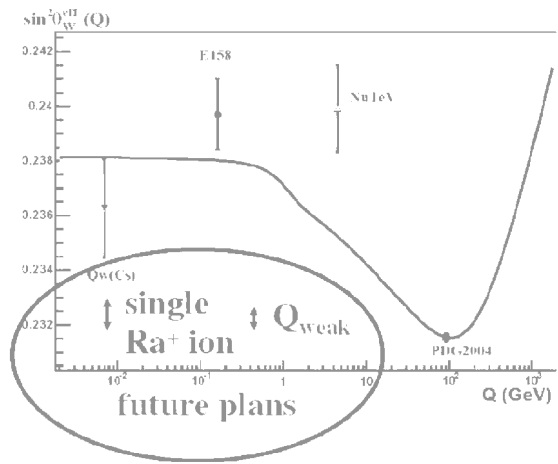

Figure VII-5: *Running of the weak mixing angle due to radiative corrections. Experiments in atomic physics with single ions and electron-proton scattering at few GeV with three times smaller errors are planned to sharpen the situation.*

A new precision measurement (Qweak) is starting at the Jefferson Laboratory in the USA, using parity-violating electron scattering on protons at very low $Q^2$ and forward angles, to challenge predictions of the Standard Model and search for new physics. For atomic-parity violation, higher experimental accuracy will be possible from experiments using Fr isotopes or single or Ra ions in radiofrequency traps. Such experiments have a solid discovery potential for effects of leptoquarks and $Z'$ bosons. The Fr parity experiments in particular will need a most intense source of atoms as it could become available at a high-power proton driver facility.

### VII-1.3.1.3 COMBINED PARITY AND CHARGE CONJUGATION–TIME REVERSAL

An EDM of any fundamental particle violates both P and T symmetries. With the assumption of CPT invariance, a permanent dipole moment also violates CP. EDMs for all particles are caused by T-violation, as is known from the K and B systems through higher-order loops. T-violation has been seen directly in K decays. These are at least 4 orders of magnitude below the present experimentally established limits. Indeed, a large number of speculative models foresee EDMs which could be as large as the present experimental limits just allow. Historically, the non-observation of EDMs has ruled out more speculative models than any other experimental approach in all of particle physics. The field of CP- and T-violation research is a master example of complementarity between low- and high-energy experiments.

EDMs have been searched for in various systems, with different sensitivities. In composed systems such as molecules or atoms, fundamental particle dipole moments of constituents may be significantly enhanced. Particularly in polar molecules large internal fields exist which can cause , e.g. an electron EDM to translate into a much larger (up to several thousand times) observed EDM for the whole molecule.

The T-violating process which underlies an EDM, may arise from the known CP violation in the SM, as described through the CKM mixing (see Fig. VII-6). A variety of models beyond the SM (e.g. supersymmetry, technicolor, or leftright symmetry) and also strong CP-violation could provide additional, new sources of CP-violation. They would translate into particle EDMs, which maybe considered intrinsic properties of leptons or quarks. When these particles are composed into objects that are accessible to experiments, further CP violation maybe introduced through CP-violating forces. Different observable systems (letons, neutrons, nuclei, atoms, molecules) have in general quite significantly different susceptibility to acquire an EDM through a particular mechanism. There is no preferred system to search for an EDM (see Fig. VII-6). In fact, several systems need to be examined in order to unreveal the true nature and the possible contributions from various potential sources. Experimentally directly accessible are particles like the neutrons or leptons. In paramagnetic atoms the EDM of an electron can be seen enhanced as well as in polar molecules. Diamagnetic atoms allow access to a potential nuclear EDM.

This active field of research we has seen recently a plurality of novel developments. They complement the traditional searches for a neutron EDM with stored polarized neutrons, searches for an electron EDM in paramagnetic atoms in atomic beams, and for atomic/nuclear EDMs in diamagnetic atoms in cells. Some experiments exploit the large internal fields in polar molecules such as YbF and PbO, ideas utilizing cryogenic Xe, neutrons in superfluid or solid He, or special solids. Of particular interest is the Ra atom, where significant nuclear and atomic enhancements exist for both a nuclear and an electron EDM. In this area, both novel ideas and up-scaled successful approaches, so far statistics limited, promise progress.[12] Among those experiments are neutron EDM searches underway at ILL, PSI and the Mainz TRIGA reactor. They rely on proven technology in the experiments and improved particle fluxes and improved magnetometry. A major step forward is the high-intensity pulsed ultracold neutron source at Mainz which employs a solid deuterium moderator.

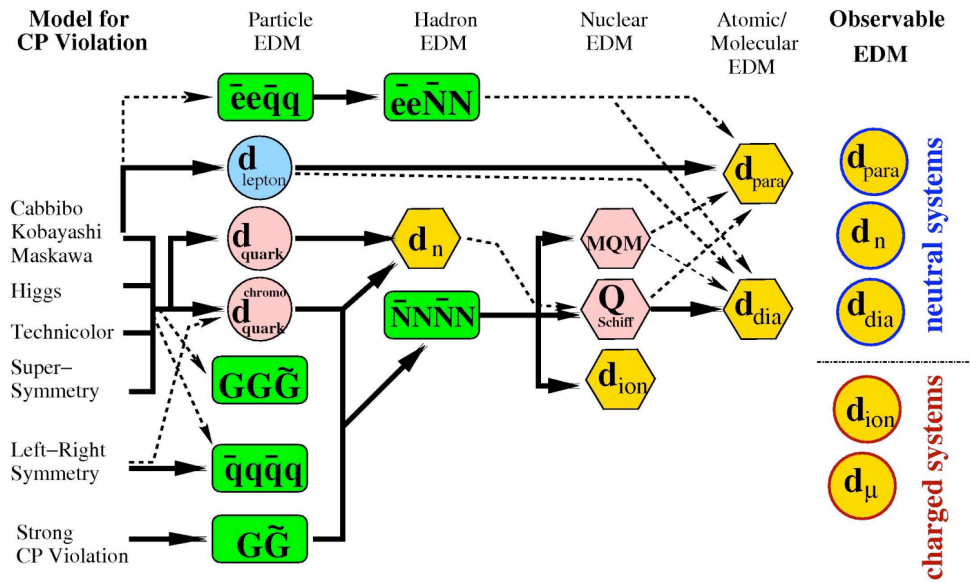

Figure VII-6: *Various possible prosses within the SM and such described in SM extensions could give rise to an experimentally observable EDM .Further there could be CP-odd forces contributing to the binding in composed systems. We need several selected different experiments in order to disentangle fully the underlying physics once a non-SM EDM will have been observed. This will help also to avoid the selection of a sterile system.*

---

[12] Most of these experiments are typically smaller scale and will not be discussed further here despite their enormous discovery potential for new physics.

A very novel idea was introduced for measuring an EDM of charged particles [83]. In this method the high motional electric field is exploited, which charged particles at relativistic speeds experience in a magnetic storage ring (see Fig. VII-7). The method was first considered for muons. For longitudinally polarized muons injected into the ring, an EDM would express itself as a spin rotation out of the orbital plane. This can be observed as a time-dependent (to first order linear in time) change of the ratio of the counting rate on both sides of the orbit plane. For the possible muon beams at the future J-PARC facility in Japan, a sensitivity of $10^{-24}$ ecm is expected. In such an experiment the available muon fluxes are a major limitation. For models with non-linear mass scaling of EDMs, such an experiment would already be more sensitive to certain new physics models than the present limit on the electron EDM. An experiment carried out at a more intense muon source could provide a significantly more sensitive probe to CP-violation in the second generation of particles without strangeness.

The deuteron is the simplest known nucleus. An EDM could arise not only from a proton or a neutron EDM, but also from CP-odd nuclear forces. It was shown very recently that the deuteron [84] can, in certain scenarios, be significantly more sensitive than the neutron, e.g. in the case of quark chromo EDMs. Such an experiment uses the storage-ring technique and polarized particle scattering for spin precession detection. It is considered for a number of research accelerator facilities and would very well fit into the CERN infrastructure. It promises a sensitivity for a deuteron EDM to $10^{-29}$ ecm.

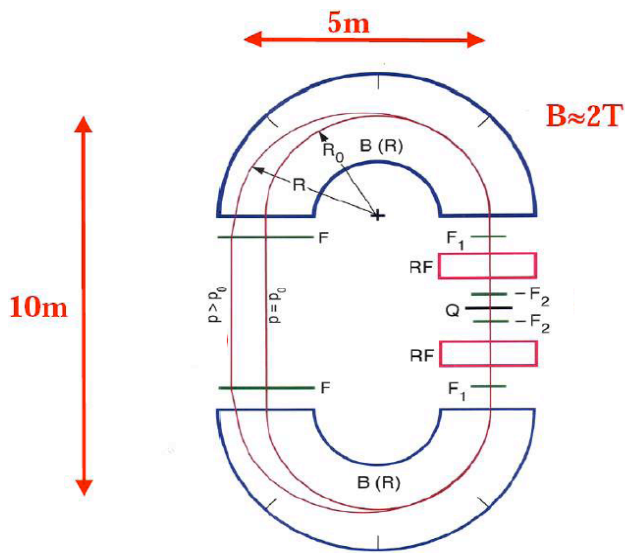

Figure VII-7: *Suggested storage ring for a sensitive search for a charged particle EDM. This method promises significant improvement in particular for deuterons (and muons)*

### VII-1.3.2 Conservation laws

#### VII-1.3.2.1 RARE DECAYS, LEPTON AND CHARGE- LEPTON FLAVOUR NUMBER

In the SM, baryon-number (B) and lepton-number conservation reflects accidental symmetries. There exist a total lepton number (L) and a lepton number for the different flavours, and different conservation laws were experimentally established. Some of these schemes are additive, some obey multiplicative, i.e. parity-like, rules.

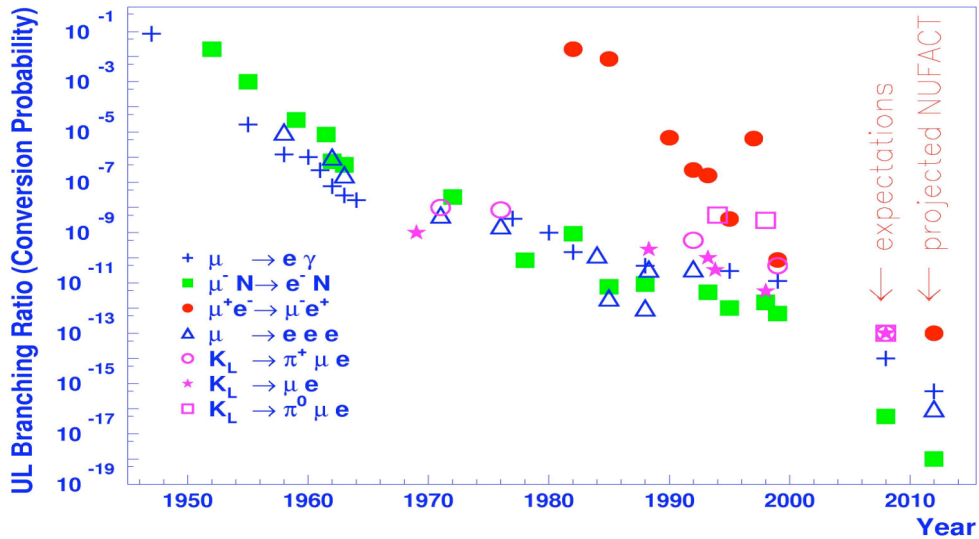

Figure VII-8: *The history of some rare decay experiments. The expectations reflect the proposed goals for on-going activities. The projected NUFACT values have been estimated recently [79]; here in part novel approaches and technologies were assumed, which become possible at a several MW proton facility[85].*

Based on a suggestion by Lee and Yang, in 1955, there is a strong belief in modern physics that a strict conservation of these numbers remains without foundation, unless they can be associated with a local gauge invariance and with new long-distance interactions, which are excluded by experiments. Since no symmetry related to lepton numbers could be revealed in the SM, the observed conservation laws remain without status in physics. However, the conservation of the quantity (B–L) is required in the SM for anomaly cancellation. Baryon-number, lepton-number or lepton-flavour violation appear natural in many of the speculative models beyond the SM. Often they allow probabilities reaching up to the present established limits (see Fig. VII-8). The observations of the neutrino-oscillation experiments have demonstrated that lepton-flavour symmetry is broken and only the total additive lepton number has remained unchallenged. Searches for charged-lepton flavour violation are practically not affected in their discovery potential by these neutrino results. For example, in a SM with massive neutrinos, the induced effect of neutrino oscillation into the branching probability is of order $P_{\mu \to e \gamma} = [(\Delta m_{\nu 1}^2 - \Delta m_{\nu 2}^2)/(2\ eV)^2]^2\ 10^{-47}$ of the ordinary muon decay probability.

This can be completely neglected in view of present experimental possibilities. Therefore we have a clean prospect to search for new physics at mass scales far beyond the reach of present accelerators or of those planned for the future and at which predicted new particles could be produced directly. The rich spectrum of possibilities is summarized in Fig. VII-8. The future projections strongly depend on the availability of a new intense source of particles such as expected from a facility with a high-power (several MW) proton driver.

| Experiment | $q_\mu$ | $\int I_\mu dt$ | $I_0/I_m$ | $\delta T$ [ns] | $\Delta T$ [$\mu$s] | $E_\mu$ [MeV] | $\Delta p_\mu/p_\mu$ [%] |
|---|---|---|---|---|---|---|---|
| $\mu^- N \to e^- N$† | − | $10^{21}$ | $< 10^{-10}$ | $\leq 100$ | $\geq 1$ | $< 20$ | $< 10$ |
| $\mu^- N \to e^- N$‡ | − | $10^{20}$ | n/a | n/a | n/a | $< 20$ | $< 10$ |
| $\mu \to e\gamma$ | + | $10^{17}$ | n/a | n/a | n/a | 1...4 | $< 10$ |
| $\mu \to eee$ | + | $10^{17}$ | n/a | n/a | n/a | 1...4 | $< 10$ |
| $\mu^+ e^- \to \mu^- e^+$ | + | $10^{16}$ | $< 10^{-4}$ | $< 1000$ | $\geq 20$ | 1...4 | 1...2 |
| $\tau_\mu$ | + | $10^{14}$ | $< 10^{-4}$ | $< 100$ | $\geq 20$ | 4 | 1...10 |
| transvers. polariz. | + | $10^{16}$ | $< 10^{-4}$ | $< 0.5$ | $> 0.02$ | 30-40 | 1...3 |
| $g_\mu - 2$ | ± | $10^{15}$ | $< 10^{-7}$ | $\leq 50$ | $\geq 10^3$ | 3100 | $10^{-2}$ |
| $edm_\mu$ | ± | $10^{16}$ | $< 10^{-6}$ | $\leq 50$ | $\geq 10^3$ | $\leq 1000$ | $\leq 10^{-3}$ |
| $M_{HFS}$ | + | $10^{15}$ | $< 10^{-4}$ | $\leq 1000$ | $\geq 20$ | 4 | 1...3 |
| $M_{1s2s}$ | + | $10^{14}$ | $< 10^{-3}$ | $\leq 500$ | $\geq 10^3$ | 1...4 | 1...2 |
| $\mu^-$ atoms | − | $10^{14}$ | $< 10^{-3}$ | $\leq 500$ | $\geq 20$ | 1...4 | 1...5 |
| condensed matter (incl. bio sciences) | ± | $10^{14}$ | $< 10^{-3}$ | $< 50$ | $\geq 20$ | 1...4 | 1...5 |

Table VII-1: *Beam parameters for low-energy muon precision experiments. They have been worked out for as SPL fed neutrino factory complex at CERN and are generically valid for several MW proton driver-based facilities. Most experiments will be possible at a 1-2 GeV machine, only the muon dipole experiments require a beam of several tens of GeV [79].*

### VII-1.3.2.2 BARYON NUMBER.

In most models aiming for the Grand Unification of all forces in nature, the baryon number is not conserved. This has led over the past two decades to extensive searches for proton decays into various channels. Present or planned large neutrino experiments have in part emerged from proton decay searches and such detectors are well suited to perform these searches along with neutrino detection. Up to now numerous decay modes have been investigated and partial lifetime limits could be established up to $10^{33}$ years. These efforts will be continued with existing set-ups over the next decade and the detectors with the largest mass have highest sensitivity. See Chapter VIII for more details.

An oscillation between the neutron and its antiparticle would violate baryon number by two units. Two in principle different approaches have been employed in the most recent experiments. Firstly, such searches were performed in the large neutrino detectors, where an oscillation occurring with neutrons within the nuclei of the detector's material could have been observed as a neutron annihilation signal in which 2 GeV energy is released in the form of pions. Secondly, at ILL a beam of free neutrons was used. A suppression of an oscillation due to the lifting of the energetic degeneracy between n and nbar was avoided by a magnetically well shielded conversion channel. Both methods have established a limit of $1.2 \times 10^8$ s for the oscillation time. Significantly improved limits are expected to emerge from experiments at new intense ultra-cold neutron sources.

# VII-2 Technical status

The detailed (low-energy) precision experiments mentioned in this chapter are at various stages; carried out and funded mostly independently.

The large scale facilities have reached in part the level of conceptual design.

## VII-2.1 Proton driver

In particular, we have studies of a multi-MW proton driver at CERN [79], which is often referred to as SPL (Superconducting Proton LINAC), at FERMILAB [86], i.e. in particular in connection with possible future activities around neutrino and muon physics, and in the framework of the EURISOL[87] design studies. Furthermore, the neutrino factory and muon collider communities are active to identify the optimal intense muon source [88]. See also Chapter V for other possible benefits of an SPL.

## VII-2.2 Underground laboratory

There exists an underground laboratory at Grand Sasso in Italy. A decision will have to be taken about having another underground laboratory before 2010, when the tunnel building machines now at work in the Frejus tunnel will meet. In Finland there is a working mine with road access at Pyhäsalmi. In England a vigorous physics programme is carried out at the Boulby mine. For detailed plans and status, see the Chapters on non-accelerator particle physics and neutrino physics in this report.

## VII-2.3 Large scale dedicated experiments

There are a few dedicated experiments, which run up to a sizable financial volume. Among those are in particular the ring EDM projects, a continuation of the muon g–2 (muon magnetic anomaly) experiment, the search for muon-electron conversion (MECO), and the kaon decay into neutrinos and neutral pions (KOPIO), the direct search for a finite neutrino mass (KATRIN), and several neutrinoless double β-decay experiments. The deuteron/muon EDM search is at the proposal stage, essential details being worked out right now. A technical proposal is expected in 2006. For a continuation of the muon g–2 experiment, scientific approval was obtained at the Brookhaven National Laboratory. This experiment has worked out technical plans for an improvement of a factor of 5 over the present results. Funding is now being sought from American sources. A letter of intent for a new experiment at J-PARC was received positively. The MECO and KOPIO activities have worked out technical proposals, received scientific approval, but funding was cancelled in the United States. These experiments could be technically woven into a new high-energy and high-power facility, e.g. at CERN. The KATRIN experiment has a detailed technical design and almost completed funding, mostly from German sources. The independent searches for neutrinoless double β-decay are at various stages of R&D and construction. They all seek predominatly independent funding.

## VII-2.4 Small scale experiments

The status of the variety of well motivated small-scale experiments at different laboratories can not be discussed here.

# VII-3 Time scale

The typical time scale of small and middle-size precision experiments is 10 to 15 years, where the precision is typically achieved after 2 to 3 iterations.

The time scale for the large-scale facilities needed is not set. The time line of precision accelerator-based experiments, which require high particle fluxes, will depend on the decision of the particle physics laboratories world-wide on the sequencing of future particle accelerators. The experiments could in most cases start with the R&D programme immediately after the site decision and the time schedule for the needed infrastructures are known. Provided a multi-MW proton machine will be proposed and approved, there is little doubt that a flurry of small to medium-scale projects will be proposed. A significant number of them will be ready to take data as soon as the source becomes available.

# VII-4 Required resource scope

## VII-4.1 Megawatt proton driver

Progress in the field of low-energy experiments to verify and test the SM, and to search for extensions to it, would benefit in many cases significantly from new instrumentation and a new generation of particle sources. In particular, a high-power proton driver would boost a large number of possible experiments, which all have a high and robust discovery potential [69].

The availability of such a machine would be desirable for a number of other fields as well, such as neutron scattering, ultra-cold neutron research (e.g. Section 1.2.3), or a new ISOL facility (e.g. EURISOL) for nuclear physics, with nuclei far off the valley of stability. Important synergy effects will result from the collaboration of these research communities with the rare decay searches (Section 1.3.2.1), neutrino physics (Chapter IV) and the next generation of low energy fundamental interaction and symmetry research as it is presently carried out at low energy radioactive beam facilities, e.g. CERN ISOLDE, GANIL, LEGNARO, KVI and GSI. A high-power driver is also of particular interest for high-flux neutrino projects (Chapter 3). The upgrade of the LHC via a replacement of the SPS accelerator could harmonically be included in such a scenario, where the new synchrotron serves also as a several 10 GeV high-power proton machine.

The intercommunity collaboration has been started already between the EURISOL and BETABEAM communities in the framework of an EU-supported design study. Possibilities for a high power machine could arise at CERN [79], Fermilab [86,88], J-PARC [89], EURISOL [87], with either a linac or a rapid-cycling synchrotron.

## VII-4.2 Shielded underground laboratory

The community concerned with non-accelerator precision measurements needs additional screened underground cavities. The existing infrastructures are not sufficient.

## VII-4.3 Continued support for ongoing accelerator and non-accelerator experiments

It will be important to ensure that the physically well motivated smaller-scale experiments with robust discovery potential be continually supported by the community. Beyond the complementary information they offer for direct-observation approaches, most of all at high energy, they provide fundamental constants that are urgently needed. Also, their operation is important in the formation of young scientists and technical personnel.

### VII-4.4 Theory support

The precision experiments require and depend on most accurate calculations of SM observables and the size of possible new physics effects. In particular, sophisticated calculations in the framework of QED and QCD require that young theorists be given at early stages of their career the opportunity to develop the necessary tools and start long-term calculational projects.

## VII-5 Status of organization and the decision process

The field of precision measurement is characterized through a number of significantly different experimental approaches, selecting in each case the best suited experimental facility world-wide. The activities are internally organized mostly in international collaborations.

The decision on a high-power proton machine is coupled to the future facility decisions at CERN, Fermilab and the future of the EURISOL project. In particular, the future of rare decay experiments will depend on this. Existing muon channels at the meson facilities, such as PSI, are not expected to provide sufficient particle flux for the next generation of precision experiments in this sector.

For the muon magnetic anomaly and ring EDM searches on the intermediate time scale, the future of high-energy physics at the Brookhaven National Laboratory will be important. For the long-term, the start of a muon programme at J-PARC or a positive decision by the community of European physicists to join these efforts will have significant impact on the proceedings. For the neutron experiments new ultra-cold sources, such as at the Muenchen research reactor, and the Mainz TRIGA reactor, will provide sufficient particles for the next round of precision experiments. In the USA mostly independent experiments are underway at Los Alamos and the NIST reactor in Gaithersburg. These facilities exist; the experiments are approved and financed. On the long run, improved pulsed sources, as they would be possible at a MW proton driver facility, would be needed. Here no structured approach has been organized yet.

For the large facilities, i.e. a high-power proton driver and a shielded underground laboratory, synergy effects can be expected from the collaboration of different communities. This aspect should be particularly stimulated.

## VII-6 The Open Symposium

Based on a set of summarizing questions concerning the main topics presented in the overview presentation by C.J.G. Onderwater a discussion took place with 17 individual contributions. They covered the broad range of subjects and reflected in part different opinions on the same subjects:

- The general concern was expressed that small-scale (and often low energy) experiments would not be fully appreciated in their potential to advance model building by the particle physics community. In particular it was mentioned that neutron experiments (lifetime, decay correlations, n-nbar oscillations, CKM matrix element $V_{ud}$) and nuclear and muon β-decay experiments (non V–A interactions, T-violation) are important contributions to particle physics.

- The up-scaling of proven techniques, together with the utilization of higher particle fluxes, and the realization of novel experimental ideas, are two equally important ways to proceed.

- The search for neutrinoless double β-decay is one of the most urgent experimental issues in particle physics, because it can clarify the nature of the neutrino, and because it would mean lepton number violation. Different experimental techniques on different candidate nuclei are indispensable. The nuclear matrix elements are important and need theoretical and experimental input. Future experiments will require a well shielded underground laboratory. On the long run the community should collaborate on a world-wide scale.

- The KATRIN experiment on tritium β-decay is very important, as it is the best direct neutrino mass measurement possible at this time. It should get the full community support to receive the remaining funding. Alternate methods using Re or calorimetry may lead to a higher sensitive in the future. On the long run theorists would like to see a mass determination which allows them to solve the mass-hierarchy problem, however, there is no earth-bound experiment conceivable at this point.

- The issue of hadronic vacuum polarization is crucial for the interpretation of the muon g–2 results and also for the running of $\alpha_s$. More reliable and more precise experimental data from, e.g. electron-positron annihilation or τ-decays is needed as well as the necessary theoretical foundation of the cross section extraction.

- The novel suggested technique of measuring EDMs in a magnetic storage-ring would fit well into the framework of the CERN laboratory. In particular, this holds for the well motivated deuteron experiment, which is presently prepared. The European physicists should decide whether they want this. EDM searches in other systems using more standard techniques are independently strongly motivated and should be supported.

- In the field of rare decays, presently used techniques will be ultimately systematically limited. As an example the ongoing μ→eγ activity at PSI was mentioned. This calls for novel techniques, which have been in part discussed in the literature, as in the case of the MECO experiment. Also the same scientific questions can be often addressed using μ→eee instead of μ→eγ.

In private discussions and e-mail exchanges following the meeting some of the issues were later re-addressed and the suggestions were included in this report. The report was sent for comments to members of the community, in particular the colleagues who helped to prepare the session in Paris.

# VII-7 Conclusions

The field of precision measurements offers a variety of possibilities to advance theoretical model building in particle physics. A discovery potential exists, complementary to direct searches for new particles and processes. The field which is characterized through a variety of often small-scale experiments, has made important contributions well beyond providing accurate values for fundamental constants, e.g. by ruling out speculative models. The progress of ongoing activities at typically smaller accelerators and university laboratories are vital for the development of the field and therefore require continuous support.

The future will be characterized by scaled-up experiments utilizing proven techniques at high particle flux facilities and, in particular by a number of novel experimental ideas. The latter have the potential for major steps forward in precision and thereby in the guiding of the model building process.

A high-power proton driver and screened underground laboratory are the most important requests for major future facilities. Both could be built using synergy effects with other particle physics and wider science communities.

# VIII NON-ACCELERATOR AND ASTROPARTICLE PHYSICS

## VIII-1 Introduction

Some of the most fundamental issues of particle physics must be investigated in experiments that do not use particle beams, or that use particles produced in cosmic accelerators. This is because the processes under investigation involve energy scales that cannot be reached otherwise, or no terrestrial accelerator produces (yet) the particles being searched for. Two cases in point are searches for proton decay and Earth-based experiments to detect dark matter (DM) particles.

The use of cosmic accelerators to investigate particle-physics issues is one of the defining aspects of the field of astroparticle physics, which has enjoyed remarkable growth and increasing popularity over the last two decades. An important factor in this growth is the progress both in our knowledge of particle interactions and in detection techniques, which allows carrying out experiments, on Earth and in space, addressing astrophysical questions by methods characteristic of particle physics.

The connection between particle physics and cosmology has also been acquiring increasing conceptual and practical importance. As already mentioned in the opening chapter, one driving issue in this development is the still recent realization that our Universe is predominantly composed of matter and energy fields whose nature is unknown, and cannot be described by the Standard Model of particle physics.

The subject matter of this chapter closely adheres to what was presented and discussed in the non-accelerator particle physics and astrophysics session of the Symposium on European Strategy for Particle Physics, namely searches for dark matter and proton decay, high-energy particle astrophysics, and dark energy. All these research activities, and more, are coordinated in Europe by the Astroparticle Physics European Committee (ApPEC), which under the umbrella of astroparticle physics embraces all physics research that does not use accelerators and has an interface with astrophysics and cosmology. This definition includes, for instance, research on the properties of neutrinos, which were covered in another session, and on gravitation – both on the astronomical and on the table-top-experiment scale – which were not covered in the Symposium.

The purpose of this chapter is to review the state-of-the-art of the research fields addressed in this session of the Symposium, and the outlook for the next decade. A number of large-scale experiments and facilities – some of them involving international collaborations – are at various stages of design or construction. ApPEC is in the process of formulating a comprehensive roadmap for the field, which will include the current cost estimates of future experiments and facilities. For the latter, the reader is referred to the ApPEC submission to the Strategy Group [90].

Several contributions related to this session were received and collected on the CERN Council Strategy Group's web page. These contributions, together with contributions to the discussion session, are mentioned at the appropriate points in the exposition.

# VIII-2 Cosmology and dark matter

Over the last few years, cosmological parameters have been measured with increasing precision, culminating in a unified, quantitative, constrained framework referred to as the 'concordance model'. Measurements of the cosmic microwave background (CMB), of the type Ia supernova luminosity–red shift distributions and of large-scale structures, as well as other observations, all agree in strongly indicating that the Universe is flat ($\Omega = 1$) and that only about 4% of its matter–energy density can be attributed to baryonic matter, which is mostly dark, with only 0.5% of the overall density in stars. The remaining DM accounts for approximately 23% of the overall density, and is composed of particles whose nature is unknown. The evidence is briefly reviewed next.

Since the early 20$^{th}$ century, it has been known that the rotation curves of galaxies indicate the presence of large amounts of non-luminous matter in the galactic halos. Around the end of that century, observations of gravitational microlensing ruled out the possibility that a significant fraction of this matter could consist of macroscopic objects such as planets, brown or white dwarves, or even solar-mass black holes. Also, galaxy surveys indicate that the total matter contents of clusters is about ten times larger than their baryonic matter content. Independently, two separate types of cosmological evidence point to a baryonic matter density of about 4%: primordial-nucleosynthesis calculations produce the observed ratios of light nuclei only for a baryonic density fraction $\Omega_B$ of about 0.04, and the same value of baryon density can be deduced from the angular power spectrum of the CMB. In summary, all the cosmological and astrophysical evidence points to the fact that around and between galaxies there are halos of non-baryonic dark matter, in amounts greatly exceeding baryonic matter.

The prevalent view is that DM consists of stable relics from the Big Bang. Furthermore, considerations of structure formation in the early Universe lead one to expect DM particles to be non-relativistic ('cold' dark matter, or CDM). Then, the observed density $\Omega_{DM} = 0.23$ and the conditions at the time of decoupling suggest CDM particle cross-sections on the weak scale. The lightest SUSY particle (LSP, a neutralino or a gravitino) is one natural candidate. The axion, a non-thermal relic, is another.

## VIII-2.1 WIMP dark matter

The non-detection of the LSP at colliders indicates that neutralinos would be quite massive (M > 50 GeV, roughly). In these experiments, such weakly interacting massive particles (WIMPs) would manifest themselves in events with large missing transverse energies; however, the discovery of a WIMP at a collider would not *per se* prove that it is the cold DM particle needed by cosmology. For that purpose, *direct* detection of WIMPs is necessary – and indeed, many direct CDM searches have been going on for the last two decades [91].

WIMPs may be detected by their elastic scattering on detector nuclei. The detection challenge is considerable, since the signal is small because of the small recoil energy imparted by slow WIMPs, and the interaction rate is very low, always << 1 event/day/ kg of detector. Hence, backgrounds must be reduced by all available means: placing the detectors underground to filter out cosmic rays, using low-radioactivity materials in the detectors and in their immediate environment, catching the irreducible backgrounds in active shielding layers around the detectors, and finally developing powerful event-by-event discrimination techniques in detecting and analysing the signal.

Nuclear recoil signals typically will ionize the medium and release thermal energy (phonons) into it; in addition, scintillation light may be produced. Strong background discrimination can be achieved by detecting two of these signals, and requiring them to be consistent with a signal from nuclear recoil. Thus, experiments such as EDELWEISS (located in the Modane lab, in the Fréjus tunnel) and CDMS (in the Soudan mine, USA) detect the ionization and the phonons produced in germanium crystals (or silicon, for CDMS), ZEPLIN detects ionization and scintillation light in liquid xenon, and CRESST detects phonons and photons in $CaWO_4$ crystals. The most prevalent background is given by γ-rays from natural radioactivity; for instance, in germanium detectors, γ rejection is based on the fact that recoil electrons from these gammas ionize (in proportion to their energy) by a factor of 2.5–3 more than recoil nuclei.

At present, the typical DM experiment has taken or is taking data and is preparing an upgrade, involving greater mass and sensitivity. Thus EDELWEISS is moving from a 1 kg to a 9 kg (eventually, 36 kg) detector; CRESST has tested different detector materials such as BGO and is adding detector modules; ZEPLIN has completed phase I (3 kg) and is planning successive upgrades, spanning until the year 2010 and eventually linking to international xenon experiments.

None of these experiments has claimed a DM detection; however, an annual signal modulation observed over 7 years – a possible WIMP signal, due to the motion of Earth in an essentially stationary galactic particle distribution – has been observed by the DAMA experiment, which uses an array of NaI(Tl) scintillator crystals at the Laboratori Nazionali del Gran Sasso. This result is from a considerable exposure (295 kg·yr) and the statistical significance of the modulation is good; however, the interpretation of the result as evidence for CDM is put into question by the fact that there is only one signature, scintillation light, and that the signal intensity is not compatible with upper limits from other experiments.

All experiments (except DAMA) have published upper limits on the cross-section as a function of WIMP mass (see Fig. VIII-1), where the best sensitivity is around masses of 50–100 GeV, only slowly deteriorating for higher masses. The parameter space allowed by SUSY models is very broad; the mass is limited from below by null results at colliders, and current direct-detection experiments are ruling out cross-section predictions of the most optimistic models. Recent direct searches have set cross-section limits just under $10^{-42}$ cm$^2$; the upgraded versions of current experiments should reach sensitivities around $10^{-44}$ cm$^2$ in about two years.

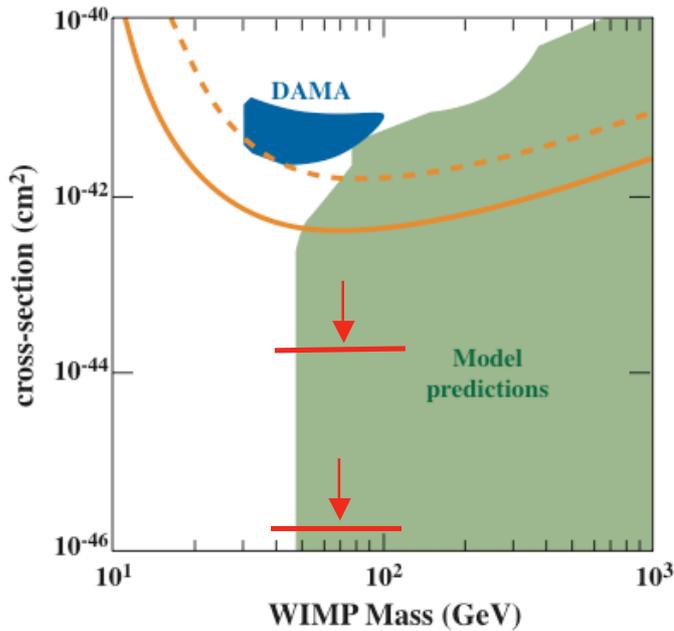

**Fig. VIII-1:** *Wimp mass limits, current and future, as a function of model predictions. The upper dashed line reprsents the current limits from CRESST, EDELWEISS, ZEPLIN; the continuous line is the CDMS limit; the upper arrow points to the sensitivity goal of CDMS-II, EDELWEISS-II, CRESST-II, ZEPLIN-II/III, XENON and XMASS. The lower arrow gives the sensitivity goal of a 1-ton experiment.*

With 1-ton detectors, sensitivities around $10^{-46}$ cm$^2$ should be reachable; there are many proposals for detectors on this scale, both in Europe and the USA, usually calling for international collaboration. Without aiming for completeness, one might mention EURECA, which will combine the detection techniques of EDELWEISS and CRESST in a multitarget approach; Super-CDMS in the USA, using both Ge and Si; XENON in the USA and XMASS in Japan, using liquid xenon, which may be easier to upgrade to large masses than crystal detectors; and WARP and ArDM, designed to use liquid argon, which are in the R&D stage but aim at masses well above 1 ton.

In summary, this research field is experiencing a growth in sensitivity, performance, cost and collaboration size that resembles the evolution of accelerator particle physics over the last decades. There will be a strong synergy between results (positive or not) obtained with WIMP detectors and neutralino searches at colliders.

## VIII-2.2 Axions

As mentioned in the introductory chapter, the axion was originally postulated to solve the strong-CP problem, namely the fact that despite the possibility of a CP-violating term in the QCD Lagrangian, strong CP violation is exceedingly small or absent, as shown by the fact that the EDM of the neutron is at least 10 orders of magnitude smaller than one would naturally expect from QCD. The Peccei–Quinn symmetry was introduced to cure this problem – its prediction is the existence of the axion, which is the extremely light pseudo-Goldstone boson arising from the breaking of this symmetry. The axion acquired its very small mass when the Universe cooled down to a temperature below a few hundred MeV, to the QCD energy scale. The symmetry-breaking mechanism is such that a zero-momentum coherent axion field is created, which fills all space. The relic energy density $\Omega_a$ is related to the axion mass, a free parameter of the theory; for a mass $m_a \approx 10^{-5}$ eV, $\Omega_a \approx 1$. Hence, despite its extremely small mass, the axion is non-relativistic and could constitute CDM.

Several ways of detecting axions make use of a mechanism analogous to the Primakoff effect: in an intense magnetic field, an axion may interact with a virtual photon from the magnetic field, producing a gamma that would be detected. In the lowest mass range, $m_a \leq 10^{-5}$ eV, corresponding to the energy of microwave photons, axion production can be detected in a very low-noise microwave cavity placed in an intense magnetic field. In a frequency scan, a resonance signal would appear when the cavity is tuned precisely to the axion mass. Such experiments have been performed and have almost reached the sensitivity required to constrain the range of axion models. Based on a different technique, CAST, an experiment at CERN, recently placed a strong upper limit on axions that might be produced in the Sun's core by an inverse-Primakoff process, with an energy of the order of keV: here a 9 T magnetic field is provided by an LHC dipole prototype; the magnet was pointed at the Sun, at sunrise and sunset. Photons with $E_\gamma \approx$ keV from axion decay were not detected in this experiment, nor in previous, less-sensitive experiments using such 'helioscopes'.

The recent possible detection of an axion-like particle has caused considerable interest. The PVLAS experiment claims detection of a rotation of the polarization plane of photons in a 6.6 T magnetic field, which might be due to axion-induced birefringence. If confirmed, the effect would indicate the existence of a light scalar particle incompatible with limits on the axion–photon coupling established by CAST. This development has been one of the stimuli behind several currently running or planned axion experiments. This was pointed out in one of the contributions to the Orsay Symposium (see [BB2.2.5.02]) and by a contribution to the discussion by the same author. At a CERN-ILIAS network–CAST workshop [92] that took place at CERN in December 2005 these experiments and several upcoming proposals were discussed.

## VIII-3 Proton Decay

The observation of proton – more generally, nucleon – decay would be an event of enormous importance, because it would be a strong indication of the existence of particles on the mass scale that Grand Unified Theories place at or above $10^{16}$ GeV. The implications for cosmological scenarios in the very early Universe and for the origin of BAU (the baryon asymmetry of the universe, see introductory chapter) would be profound.

Efforts to detect proton decay date from the early 1970's, having begun soon after the first GUTs arose. The main experimental requirements are techniques that allow to detect this extremely rare process anywhere within a large mass, at a rate of at least one-to-a-few decays per year, in an environment highly shielded from cosmic-ray backgrounds. Together, these requirements make large underground laboratories a necessity. The first-generation experiments (Fréjus, IMB, Kamiokande) were on the scale of 1 kton, and ruled out non-supersymmetric SU(5). The second-generation facility SuperKamiokande, with 50 ktons, ruled out minimal supersymmetric SU(5). The facilities of the next generation are designed to test more general supersymmetric models, and aim at being sensitive to mean proton lifetimes of the order of $10^{35}$ years – hence, they must contain $10^{35}$ nucleons, corresponding to a mass of the order of 1 Mton [93].

The main signatures of proton decay are expected to be $p \rightarrow K^+ \bar{\nu}$ and $p \rightarrow e^+ \pi^0$. The former is predicted in several models to occur with a lifetime $\tau \leq$ a few $10^{34}$ yr, but this value is rather model-dependent. The latter is the most model-independent decay channel, and may occur with τ = (a few $10^{34}$ to $10^{35}$) yr. The limits from SuperKamiokande are compared below with the projected sensitivity of MEMPHYS, one of the next-generation's proposed megaton facilities (see [BB2-2.2.05]):

$$\tau\left(p \to K^+ \bar{\nu}\right) > 1.6 \times 10^{33} \text{ yr} \quad \text{vs.} \quad \tau\left(p \to K^+ \bar{\nu}\right) > 2 \times 10^{34} \text{ yr} \quad \text{after 10 years}$$

$$\tau\left(p \to e^+ \pi^0\right) > 5 \times 10^{33} \text{ yr} \quad \text{vs.} \quad \tau\left(p \to e^+ \pi^0\right) > 10^{35} \text{ yr} \quad \text{after 10 years.}$$

Like SuperKamiokande, several of the proposed future detectors are designed to detect Cerenkov radiation in a large volume of water. Other proposals involve very large liquid-argon TPCs or liquid scintillator (see [BB2-2.2.02]). All initiatives share the use of continuously sensitive detectors (no passive absorbers). There are proposals for such facilities in all three regions of the particle-physics world, as shown in the following table:

**Error! Not a valid link.**

There is an interesting complementarity between the water Cerenkov technique on the one hand and liquid argon or liquid scintillator on the other, in that the former allows a larger mass, but is rather inefficient in detecting the $K^+\nu$ channel, whereas the latter two, albeit with smaller masses, can be highly efficient in detecting this channel.

Obviously very large investments are needed for these facilities; however, nucleon decay is not the only exciting physics available. On the astrophysics side, as we learned from SN1987A, these detectors may see neutrinos from type II supernovae; the proposed detectors would have sensitivity all the way to the Andromeda galaxy. In the fortunate case of a supernova in our galaxy (the expectation is one SNII per 50 yr) they would resolve the millisecond time structure of a collapse to a black hole, even for SNe as far as the galactic centre. Furthermore, the diffuse flux of supernova relic neutrinos could be detected by some of the proposed detector types, giving insights into early star formation.

As in the case of SuperKamiokande, solar and atmospheric neutrino studies could be performed, with greater accuracy. However the most important item on the physics menu of such a facility, other than proton decay, would doubtlessly be oscillation and mixing of neutrinos produced at accelerators. For instance, a detector like MEMPHYS could be the far detector of a superbeam and/or beta-beam facility at CERN (see the chapter on neutrino physics).

# VIII-4 Astroparticle physics: the high-energy Universe

It has been known for a long time that the Universe accelerates particles to the highest energies; in fact, it was only 50 or so years ago that Earth-bound accelerators surpassed cosmic accelerators as the most effective tools to discover new particles. The high-energy radiation that hits the Earth may come from sites as relatively near (on the cosmic scale) as galactic supernova remnants (SNRs) or from cosmological distances, such as gamma-ray bursts (GRBs). These messengers may bring us information on novel astrophysical phenomena, but may also shed light on new aspects of particle physics. Hence the recent interdisciplinary field of astroparticle physics, which, as discussed at the Symposium, is a continuum wherein the questions range from being purely astronomical to purely particle physics, as do the observational tools.

The century-old question about the mechanisms that accelerate particles to the highest energies remains at the heart of much of the research in astroparticle physics. The radiation from the cosmos either follows a thermal energy distribution – if the sources are hot bodies such as stars or dust – or has a spectrum extending to much higher energies. Surprisingly, perhaps, the total energy of this non-thermal component is about equal to that of the thermal component of radiation. This fact may be sending us a profound message about the evolution of the Universe, one that we have not yet understood.

The highest energies reached by these cosmic messengers (considered more quantitatively in the following subsection) are particularly interesting to particle physics, and may provide a means of probing the limits of special relativity or even evidence for quantum-gravity phenomena.

These messengers from the Universe span the variety of stable particles. Cosmic rays (charged particles and nuclei) reach the highest-observed energies, but are bent in galactic and intergalactic magnetic fields; therefore, except for the highest part of the spectrum, they carry no information about their sites of origin; on the contrary, photons and neutrinos point to their sources. While high-energy neutrino astrophysics is in its infancy, gamma-ray astrophysics has been through three decades of exciting developments.

## VIII-4.1 The highest-energy cosmic rays

Cosmic rays have been observed over an enormous range of energies, up to and exceeding $10^{20}$ eV, corresponding to a centre-of-mass energy well above that of LHC collisions. The energy spectrum decreases rapidly according to a smooth power law, approximately as $E^{-2.7}$ up to about $10^{15}$ eV (the 'knee' region); around that value, the spectrum gradually changes to an $E^{-3.1}$ behaviour, until it reaches $10^{19}$ eV (10 EeV), where the spectrum hardens again, displaying the 'ankle' feature. Below the knee, it is generally believed that shock-wave acceleration in galactic SNRs explains the observed abundance and the energy spectrum of cosmic rays. Upwards of the knee, cosmic rays are no longer trapped by the galactic magnetic field; above this energy, an extragalactic component is therefore expected to acquire importance. The chemical composition (i.e. the mass-number composition) is also believed to change above the knee. At about 50 EeV, the cross-section for collisions of protons with CMB radiation rises rapidly because the threshold of pion production through the $\Delta(1232)$ resonance is reached. Hence the energy of the projectile is degraded, and an abrupt drop in the spectrum is expected. This effect has been recognized since 1966 and is known as the Greisen–Zatsepin–Kuzmin (GZK) cut-off.

The observation – or not – of the GZK cut-off is of fundamental interest: if it were present, it would imply that there is a maximum distance – a 'horizon' – beyond which we cannot observe protons of more than about 100 EeV; the GZK horizon is 10 to 100 Mpc away, the distance to the nearest galaxy clusters. Should we observe cosmic rays above the cut-off, we would have to conclude either that we do not understand interactions at that energy, or that proton acceleration to such extreme energies occurs within this horizon. However, we do not know of any possible cosmic accelerator within this distance; we would therefore have to assume that rather than being accelerated 'bottom-up', from lower energies, such particles are produced by undiscovered 'top-down' particle-physics production processes, involving unexplored energies. High-statistics observations of EeV cosmic rays might also reveal their sources, because the directions of such extremely energetic particles would not be randomized by intergalactic magnetic fields. Detecting such sources would be of great help in understanding the production or acceleration mechanisms, which are entirely unknown in this energy range.

Because of the power-law drop of the spectrum, the cosmic-ray flux at the highest energies is extremely low: the flux above 10 EeV is of the order of $10^{-16} m^{-2} s^{-1}$, which in more intuitive units is one particle per $km^2$ per century. One reason why such rare events are observable at all is that cosmic rays, interacting high in the atmosphere, develop extended showers that, at these energies, reach sea level with a lateral spread of several km.

Measuring these extended showers relies on detecting the secondaries (mostly low-energy electrons and photons, but also muons from the decay of pions in the hadronic shower) on the ground, or the fluorescence light emitted isotropically by the charged particles as they traverse the atmosphere. Surface detectors typically consist of scintillation counters or water Cerenkov counters, which need to cover only << 1% of the ground, but must be spread out over a huge area. In the case of fluorescence detectors, typically consisting of arrays of PMTs oriented so as to view a large solid angle in the sky, the showers, with a length of several km, are visible from a distance of tens of km, which provides an adequate detection area. In either case, the time structure of the signals is used to reconstruct the direction of the primary.

Calibrating the shower energy from the secondary particle or the fluorescence signal is a delicate process. This very issue may be at the root of the gross disagreement between two past experiments that observed cosmic rays up to the GZK cut-off: the AGASA surface array in Japan, which observed 17 events above 60 EeV, and no spectral feature resembling a cut-off, and the HiRes stereoscopic fluorescence experiment, which observed only two events above this energy (where it should have observed 20, based on the AGASA result), and a spectrum suggesting a down-turn around the GZK cut-off energy.

In a sparsely populated area of Argentina, the Auger collaboration is building a large cosmic-ray observatory that will eventually cover 3000 $km^2$. The facility comprises both ground detectors – of which about 1000, over 60% of the total, have been deployed – and three (eventually, four) fluorescence stations; an appreciable fraction of the events allow the estimation of the primary energies with both types of detectors simultaneously. This allows cross-checking their respective energy calibrations and should avoid the related uncertainties; Auger's current energy calibration has a 20% error margin. Preliminary results were presented in 2005, with limited statistics, at energies above $10^{19}$ eV. These data so far show neither cut-off nor ankle, but they will be superseded by the much larger statistics that are being accumulated as the detector grows. Completion of the array is expected in 2007. No sources (i.e. accumulations of events from particular directions) have been seen, already excluding the excesses observed by AGASA and SUGAR, another large array. Also, events above $10^{19}$ eV have < 26% of primary photons, a result that favours bottom-up over top-down models, because in the former about 50% of the primaries are expected to be photons whereas in the latter at most 10% of the showers would have that origin.

The interest in issues involving the highest-energy cosmic rays has stimulated the development of further proposals. While the Auger project includes from the very beginning an observatory in the Northern hemisphere, which would provide further isotropy tests and could see sources such as the local supercluster of galaxies, novel ways to observe cosmic rays over much larger areas have been proposed. Two of these would record the fluorescence light of the highest-energy showers from orbit: OWL (Orbiting wide-angle light collectors) would observe showers from two satellites, thus obtaining a stereoscopic view, while EUSO (Extreme Universe space observatory), using one satellite, would gather additional information by detecting the diffused image created on the sea or on the ground by the forward-emitted Cerenkov radiation. At present, the time scale of these proposals is unclear, because there are uncertainties in the programmes of space agencies. However one must stress the potential of such experiments for pushing further the limits of fundamental physical laws, such as Lorentz symmetry, as remarked in a contribution to the Symposium (see [BB2.2.5.04]) and in the discussion session.

## VIII-4.2 Gamma-ray astrophysics

The energy range of gamma-ray astrophysics observations is enormous, going from keV to tens of TeV, more than ten orders of magnitude. Two types of 'telescopes' now produce images of the gamma-ray Universe: detectors on satellites, and large ground-based IACTs (imaging atmospheric Cerenkov telescopes). In the current generation of satellites and ground-based instruments, the angular resolution of these images has reached a level of about 0.1º, which permits to connect to phenomena observed in lower energy ranges and greatly expand the scientific reach of gamma-ray astrophysics [94].

Instruments on satellites detect gamma rays ranging from keV to several GeV. At the lower end of the range, gammas are entirely absorbed by the atmosphere, and hence can only be seen from space. Sensitivity at the upper limit of the energy range is limited by the rapidly falling spectrum, which would demand impractically large areas and masses.

The Compton gamma-ray observatory (CGRO), launched in the 1980's, comprised several detectors covering the whole satellite energy range; it was the first of a new generation of instruments, which have identified and characterized hundreds of gamma-ray sources, both galactic and extra-galactic. The EGRET telescope on the CGRO covered the higher-energy range; about half of the 270 or so sources it observed do not correspond to objects known from lower-energy observations. This fact underscores the importance of the emerging multiwavelength approach to gamma-ray astrophysics.

A number of satellites, optimized for different observation programmes, are currently active. Only one of these will be mentioned here, the international gamma-ray astrophysics laboratory (INTEGRAL). In the 20 keV – 10 MeV energy range, the galactic disk is seen at low resolution as a narrow, continuous disk. INTEGRAL has shown that about 90% of this 'gamma fog' is accounted for by 91 sources, of which 47 are X-ray binaries, 3 are pulsars, and 37 are new sources. Pointing at the galactic centre, INTEGRAL has detected a very strong and spatially extended positron annihilation line. There is much debate on the origin of this positron source, with light DM (1–100 MeV) being one of the controversial possibilities.

Until recently it was not possible to do traditional astronomy with TeV gamma rays because the available imaging capability was too poor to clearly show morphological features. Progress with IACTs has now reached the required image quality. IACTs detect the Cerenkov light emitted by showers produced by the interaction of high-energy particles in the upper atmosphere, typically around 10 km above sea level for TeV gammas. The narrow (1º–2º) cone of light intercepts the ground over areas ranging from $10^4$ to $10^5$ m$^2$. The light is collected with large (several m diameter) tessellated mirror systems, and detected in finely-segmented PMT cameras. The optics permit reconstruction of the shower image, a powerful discriminating tool in rejecting the non-gamma background. The use of large light collectors allows to push the gamma detection threshold down to about 100 GeV (tens of GeV in the next generation of IACTs), while the large area of the Cerenkov light 'pool' provides sensitivity up to energies of tens of TeV.

At least two generations of IACTs have been deployed since the Whipple telescope realized the first detection of a TeV gamma source (the Crab) in the late eighties; these more advanced instruments discovered a dozen TeV sources, both galactic such as supernova remnants, and extragalactic, such as active galactic nuclei (AGNs). The crucial improvement of the instrumentation has been realized by H.E.S.S. (High energy stereoscopic system [95]): following in the tracks of the earlier stereoscopic system deployed by the HEGRA collaboration, but with larger light collectors and more finely segmented cameras, H.E.S.S. has reached superior angular resolution and hadronic background rejection. The images thus obtained show detailed features of galactic sources such as SNRs, and for the first time allow the association of the signals to specific morphological features of the sources. It will take several more years to exploit fully the discovery potential of this new tool. However, several important results have already been obtained; on the one hand:

- a survey of the central part the galactic plane, which has revealed 14 new sources, including SNRs, X-ray binaries, and pulsars, but also three with no known counterpart at any wavelength;

- close correlation between X-rays and TeV images from several SNRs, confirming that SNRs are indeed the particle accelerators needed to produce the observed cosmic-ray spectrum up to the knee;

- resolving the SNR expansion wave as the site of cosmic-ray acceleration, which may soon provide the long-sought evidence that part of the gamma-ray spectrum is of hadronic origin (coming from $\pi^o$ decays produced in collisions of protons) rather than originating from electromagnetic processes such as inverse Compton emission.

On the other hand, the search for DM from neutralino pair annihilation in the galactic bulge remains elusive. A hard spectrum from the galactic centre, extending up to 20 TeV, would indicate a very high-mass neutralino or Kaluza–Klein particle, if predominantly attributed to such production mechanisms.

Gamma-ray bursts are also the object of intense investigation, mostly with satellites. Coming from cosmological distances, as was first indicated by their isotropic distribution and later confirmed by measuring large red shifts in optical counterparts, they represent the most energetic events in the Universe. Further insights into the nature of GRBs statistics are expected to come from instruments such as the SWIFT satellite, with observations in the gamma ray, X-ray and optical range. Coincident detection by satellites and by IACTs is being vigorously pursued, particularly by the MAGIC telescope, which has the capability to respond to early GRB alarms from the satellite network and will substantially enhance the range of such multiwavelength observations.

The outlook for this field of research is excellent: the GLAST satellite, to be launched in 2007, is likely to inaugurate a new era (much as the EGRET instrument on the CGRO) by observing thousands of sources. The AGILE satellite, to be launched in 2006, will pursue similar goals. In the IACT arena, the two tendencies represented by HESS (with stereoscopic arrays), and by MAGIC (with very large light collectors), appear to be converging: H.E.S.S. is building a 28 m diameter collector, while MAGIC is building a second 17 m diameter collector, to be used in conjunction with the first. Larger, lower-threshold IACT arrays are already under study; the lower end of their energy range should overlap with the higher end of the energy range of GLAST, thereby allowing useful flux cross-checks. It would be useful to have one such observatory in each hemisphere, overlapping with the operating period of GLAST.

It is worth while to describe a limit of gamma-ray astronomy: photons in the TeV energy range are absorbed because of electron–positron pair production on starlight (recent, or red shifted light from the oldest galaxies). Hence a 'gamma-ray horizon' that makes it hard to observe TeV photons beyond a few hundred Mpc. However this limit may turn into a useful tool, because systematic measurements of the horizon-linked cut-off of gamma-ray spectra from AGNs may permit the measurement of the radiation field from early galaxies. Such a measurement would bear on cosmological issues, such as structure formation in the early Universe.

It is clear from this exposition that despite a still unrealized potential for particle physics discoveries, and a strong relevance to cosmic-ray acceleration issues, gamma-ray astrophysics nowadays is closer to astronomy than to particle physics. This may be one of the reasons why there were almost no gamma-ray astrophysicists at the Symposium.

## VIII-4.3 High-energy neutrino astrophysics

High-energy neutrinos are yet another messenger from the non-thermal Universe. Like photons, they point back to their source; unlike photons and cosmic rays, they are unaffected by interactions with the cosmic-radiation fields that produce the gamma-ray horizon and the GZK cut-off; furthermore, their very low interaction cross-sections with matter makes them ideal probes of dense sources of high-energy radiation. For these reasons, observing high-energy neutrinos should give us unique information about the origin of cosmic rays, and more generally about astrophysical phenomena. The complementarity with gamma-ray astrophysics is an example of the emerging multi-messenger approach.

Cosmic neutrinos must come from the decays of charged pions (and then muons), much like atmospheric neutrinos, except that the pion-producing collisions occur in cosmic accelerators such as AGNs or SNRs. Hence they should be produced at rates similar to the gammas from neutral-pion decay, as discussed earlier. However, TeV gammas may alternatively arise from inverse-Compton processes, whereas there is no such ambiguity for TeV neutrinos; therefore, observing a high-energy neutrino source would be the definitive proof of cosmic hadron acceleration.

Detecting cosmic neutrinos is of course challenging, to put it mildly. The technique is similar to that of proton decay detectors *à la* SuperKamiokande: the detector consists of a large volume of water (or polar ice) in which Cerenkov light radiated by the neutrino-collision products is detected by arrays of large PMTs. Given the expected fluxes and the calculated cross-sections, to reach detection rates of the order of tens of cosmic neutrinos per year [96] requires target/detector volumes of the order of 1 km$^3$. With this technique, neutrinos can be detected over a large range of energies, from roughly 20 GeV to 10 PeV. Atmospheric neutrinos and muons constitute an important background (as well as a useful calibration signal); to filter them out, the detectors must therefore be located at a substantial water (or ice) depth, more than 1 km.

Several major international neutrino telescopes can be seen as prototypes that prepare the next, more ambitious stage; their effective areas are in the $10^4$-$10^5$ m$^2$ range. In order to minimize sensitivity to atmospheric muons, the fields of view of these telescopes are usually oriented downwards, away from the sky. AMANDA, at the South Pole, is complete and has been taking data since 2000; it profits from the extremely small optical absorption of deep Antarctic ice. In the Northern hemisphere, ANTARES, the Baikal telescope and NESTOR have been leading the field. NESTOR, due to its exceptional depth of 4000 m, is the only one looking both at up-going and down-going neutrinos. ANTARES, the largest of the latter three, will be fully deployed in 2007, and is designed to have finer angular resolution due to smaller light scattering in water when compared to ice. AMANDA mostly views sources in the northern hemisphere, while the field of view of the northern hemisphere detectors, located at intermediate latitudes, rotates with the earth; hence they include parts of both hemispheres, albeit with a smaller duty cycle for some source positions. The fields of the northern and southern observatories overlap significantly, which will permit cross-calibration between the detectors. No observatory has announced the observation of a source yet, although AMANDA has an effect, not statistically significant, from the direction of the Crab.

The physics programme of neutrino telescopes is still to be realized, but it is exciting. Besides the already-mentioned potential to elucidate the acceleration mechanisms at play in astrophysical high-energy sources, both the astrophysics and particle-physics potentials are far-reaching:

- Rather straightforward (and hence more credible) considerations, based on the observed cosmic-ray spectrum, and on the absence of horizons, suggest that neutrinos of up to the EeV range might be detected. These extreme energies might have surprises in store.

- Neutrino physics can be studied at energies less extreme but still not reachable in Earth-bound facilities. For instance, the population of τ-neutrinos is expected to be similar to that of the other two species, due to oscillations; 1 PeV $v_\tau$ collisions might be recognized (if the efficiency is high enough) by the spectacular topology of 'double bang' events, in which, thanks to the large Lorentz factor, the bursts of particles corresponding to the τ-lepton production and decay vertices can be separated by a few tens of metres.

- Pairs of neutralinos gravitationally bound to the core of the Earth (or the Sun) may annihilate into neutrino pairs that would be detected in a neutrino telescope.

- Finally, all sorts of exotic phenomena might take place, from breakdown of the equivalence principle to extreme-energy neutrinos from the decay of topological defects, cosmic strings and the like.

The next generation of km$^3$ neutrino detectors is in preparation. At the South Pole, Ice Cube is already being deployed. KM3NeT, a Mediterranean initiative of similar scope, was touched upon in the discussion session.

## VIII-5 Cosmology and dark energy

Over the last decade or two, beginning essentially with the pioneering COBE observation of peaks in the angular power spectrum of the CMB, cosmology has undergone revolutionary developments. Numerous further observations of the CMB, culminating in the WMAP (Wilkinson microwave anisotropy probe) results, have produced precision measurements of several cosmological parameters, thus inaugurating a new era in observational cosmology. The most revolutionary developments have taken place since 1998, when observations of type Ia supernovae showed that at high red shift these standard luminosity candles are fainter than prevalent cosmological models would have predicted. These measurements and independent astronomical evidence point to the fact that the Hubble expansion is accelerating, presumably because the deceleration induced by matter or radiation is less that the acceleration due to a cosmic-energy field that produces a ubiquitous negative pressure [97]. This astonishing entity, dubbed 'dark energy' (DE), accounts for about 70% of the energy density of the Universe.

Naturally, dark energy has attracted enormous attention as well as a healthy dose of scepticism. The phenomenon needs further study, to firm up the evidence and hopefully to understand its nature.

To reduce the systematics tied to the variations in SNe Ia luminosity, and thereby make more accurate the measurements of the related cosmological parameters, the Supernova legacy survey (SLNS) collaboration is observing a very large number of supernovae, using dedicated instruments [98]. Thanks to the 1 square-degree field of view of the Megacam camera, the SNLS observation method involves simultaneous detection of new supernovae and follow up of their light curve, thus greatly improving the efficiency and the quality of the survey; all SNe are monitored with the same instrument (reducing systematics due to the previous use of different telescopes), and their light curve has a much better temporal coverage (reducing uncertainties in the determination of the peak luminosity). The survey started in 2003 and will extend until 2008.

Dark energy appears to be deeply linked to the most fundamental questions of cosmology and particle physics. It is well known that quantum-field-theory (QFT) would 'naturally' predict a vacuum energy density $10^{60}$ to $10^{120}$ times larger than closure density. No symmetry principle that would make it exactly equal to zero has been put forward, and it appears even harder now to find a QFT explanation for a vacuum energy of the order of the closure density. Furthermore, it appears curious that in the current epoch we are in an age of transition in which matter (DM) and energy (DE) appear roughly in balance.

Several theoretical views about the nature of DE have been advanced: Can it be characterized as 'quintessence', a scalar field that would vary with space and time? Or is it the famous cosmological constant introduced and then rejected by Einstein, which would be similar to QFT vacuum energy? Could it indicate that general relativity must be modified? Or do we have to go beyond our current physical frameworks, because the solution is beyond field theories, quantized or not?

One parameter discriminating between theories of DE is the form of the cosmological equation of state. In the relativistic Universe, the relationship between matter density and pressure is $p = w \cdot \rho$, where the parameter $w$ ($w = 0$ for matter, 1/3 for radiation) would be $> -1$, but changing with epochs if it were due to quintessence, exactly $w = -1$ if DE is due to the cosmological constant. By 2008, SNLS may place significant constraints on $w$.

The next-generation DE research instruments are in the making. In the USA, NASA and DOE are planning the Joint dark energy mission (JDEM), which may be implemented either by SNAP (Supernova acceleration probe) or the Destiny dark energy space telescope. Both concepts rely on space telescopes, and will detect and analyse the spectra of thousands of supernovae, to high red shifts.

In addition to high statistics and high red shift observations of type Ia supernovae, several other approaches can shed light on DE, because of its influence on other observables, such as the evolution of structures. Weak gravitational lensing, which induces a shear-like distortion in the images of background galaxies, gives a direct measurement of the distribution of mass in the Universe, which bears information on the evolution of structures. The Dark Universe explorer (DUNE) space mission of ESA will take this approach.

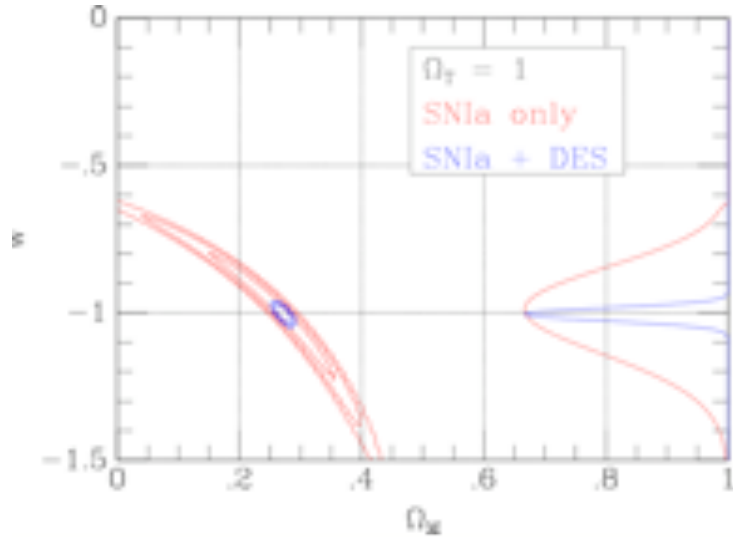

**Fig. VIII-2**: *DES goals in measuring w.*

Ground-based facilities, in addition to being part of space-based observational programs, are an independent path to DE research. Planning to begin observations in 2009, the FNAL-based Dark energy survey (DES) uses four independent methods to study the nature of dark energy: a galaxy cluster survey, a weak lensing study, a galaxy power spectrum measurement and a survey to measure SNe Ia distances. Illustrating the power of the combined approach to DE, Fig. VIII-2 shows how DES plans to combine information on the parameter of the equation of state, *w*, from SNe Ia data with information from weak lensing. The aim is to measure *w* with a precision of ~ 5% and $\Omega_{DE}$ with a precision of ~ 3%.

## VIII-6 Conclusions and outlook, briefly

This brief review should have made it clear that almost all of the activities in non-accelerator particle physics, astroparticle physics and cosmology surveyed here are in a state of ebullient growth, driven by many recent exciting results and/or the intense, often interdisciplinary interest of the issues being addressed. By way of conclusion, and repeating in part what was already said, here are a few of the fundamental questions:

- What is dark matter? WIMPs? Axions? What will be the interplay of colliders, direct searches, and astrophysical evidence in answering this question?

- Do protons decay? What happens at $10^{16}$ GeV? Can we make detectors to answer these questions?

- multimessenger astrophysics: Where and how are cosmic rays accelerated? Can we observe cosmic rays beyond the GZK cut-off? If so, by what process and where are they produced? Can we pinpoint the sources of the highest-energy particles? What can we learn from the highest-energy neutrinos?

- What is dark energy? A quantum field? A different form of gravity? None of these?

Because of intense activity in all of these lines of research, currently envisaged detectors and facilities have reached a new scale of sensitivity, but also of cost. The required investments are not at the level of accelerators at the high-energy frontier; however, for some of the facilities, co-ordination at the world level would be beneficial in order to optimize the overall physics returns. Quoting from ApPEC's Comments on the European Roadmap for Astroparticle Physics (see BB2-4.2.1):

"Cubic-kilometre neutrino telescopes, large gamma ray observatories, Megaton detectors for proton decay, or ultimate low-temperature devices to search for dark matter particles […] are in the 50-500 million Euro range."

## VIII-7 Summary of the discussion session

Participants in the Symposium included several of the main players in non-accelerator and astroparticle physics, but did not represent a comprehensive sample of this research community. To ensure broader coverage of the prospects of the discipline, the discussion session was organized to include very brief presentations illustrating aspects of ApPEC and of two EU networks.

R. Wade, chair of the ApPEC steering committee, emphasized the need for coordination arising from the growth of the field and the cost scale of future large projects. The ApPEC roadmap, to be finalized over the next few months, will set priorities. He also stressed the need for co-ordination with the CERN Council's strategy group, in particular on fields appearing in both strategy papers.

G. Gerbier presented the ongoing ILIAS (Integrated large infrastructures for astroparticle science) network, centred on the R&D common to underground labs and three research themes addressed there: double-beta decay, dark-matter searches, and gravitational-wave detection. Continuing on the theme of underground labs, E. Coccia, director of LNGS, stressed the interdisciplinary potential of these facilities, which includes biology and nuclear physics in addition to the subjects reviewed here. On gravitational-wave detection, he stated that an interferometer at LNGS will be needed in connection with the operation of VIRGO.

On neutrino astrophysics, P. Coyle presented the scope of KM3NeT, wherein three separate neutrino programs (ANTARES, NEMO and NESTOR) are jointly developing a future $\approx 1$ km$^3$ under-sea neutrino detector. On this same theme, it was pointed out that proposals using different detection technologies (radio or acoustic signals) have the potential to discover neutrinos beyond the GZK cut-off and study their interactions, should they exist. It was also pointed out that a source like those discovered by HESS might result in 10 or so neutrino detections per year in a km$^3$ detector. The issue of detecting τ-neutrinos was also discussed, during and after the session.

The discussion about dark energy focused on two issues:

- Could acceleration of the Hubble expansion arise from a misinterpretation of the results? Answers concurred in stating that independent experiments, measuring different phenomena, point to an accelerating expansion; furthermore, several ground-based experiments will soon bring evidence from an even broader variety of observables.

- Does dark energy 'belong' to the field of particle physics, considering that it is observed by astrophysicists? The clear affirmative answer form the floor clearly represented the strong interest of particle physicists in this theme.

Two more general themes were discussed:

- The importance of adequate support for R&D of novel detector techniques and imaginative but daring initiatives. The point of view expressed (see also [BB2.2.5.03]) was that large and expensive astroparticle projects should not monopolize the available resources, but that smaller, novel experiments that might significantly advance the field must be funded.

- Where to practice astroparticle physics? In a submission to the Symposium ([BB2.2.5.01]) it was suggested that CERN should devote a limited amount of human resources to astrophysics research. This is an issue that will certainly receive further attention. In the discussion session, it was also pointed out that not all astroparticle themes are of the same relevance for particle physicists: some use particle physics concepts or techniques to do astrophysics, while others use astrophysics to advance particle physics.

# IX STRONG INTERACTIONS

## IX-1 Overview

It is not easy to sketch a scenario for the developments in Quantum Chromodynamics (QCD), considering the very broad field of applications, which spans many orders of magnitude in the energy scale. At high energies, the running of the QCD coupling $\alpha_S$ ensures that some quantities can be reliably computed perturbatively. On the other hand a broad deployment of techniques, ranging from first principles calculations to ad-hoc models, is needed to describe the behaviour of hadronic matter at low energies.

At the LHC, detailed QCD predictions concerning event rates and characteristics will play a key role in disentangling new physics signals from backgrounds, as well as in the realization of precision measurements (including luminosity). This is a rather paradigmatic case, in which many different aspects of QCD, from perturbative calculations to hadronization models, need to be implemented, in order to unambiguously connect the partonic interactions to the observed particle final states.

In recent years, new ideas have spurred a lot of progress in the development of powerful QCD tools. Techniques imported from string theory provide compact expressions for multi-parton amplitudes, while iterative computational methods allow the leading-order calculation of Standard Model processes of almost any complexity. In addition, Monte Carlo tools provide a description of high-energy interactions of ever increasing accuracy. The availability of high-quality experimental data from HERA and the Tevatron, as well as from the $e^+e^-$ colliders, has assisted enormously in the development and validation of these tools, and more work is underway to attain higher levels of precision.

The study of QCD phase transition in the high temperature domain, where a new state of matter made up of deconfined quarks and gluons might be produced, will be the main focus of the heavy-ion program at the LHC.

New measurements of parton densities at small x, performed at RHIC, provide further support to the HERA measurements, and hint at a saturation of the parton densities, leading to large nonlinear effects, which might be described by the Colour-Glass Condensate (CGC) approach.

The discovery of rapidity-gap events at HERA has revived interest in hard and soft diffraction. Quasi-exclusive diffractive production of Higgs bosons or other new particles at the LHC might prove particularly useful for establishing the charge conjugation and parity quantum numbers of such particles,

Forward physics at the LHC will also help to bridge the gap between ultrahigh energy cosmic rays (UHECRs) and laboratory physics, by providing a more accurate description of hadronic showers, which are a key ingredient in the measurement of the energies of cosmic ray primaries.

Among open questions, the issue of how the nucleon spin is made up by parton spins and orbital angular momenta occupies a prominent role. In contrast to the quark helicity distributions, the distribution of the transverse spin in the nucleon is largely unknown. The recently developed concept of generalised parton distributions (GPD) might lead for the first time to the determination of a three-dimensional 'tomographic' picture of the nucleon and to information about the angular orbital momenta of partons in the nucleon.

The spectroscopy of doubly-charmed baryons, glueballs and hybrids provides detailed insight in the internal dynamics of hadrons. The nonperturbative QCD sector is also challenged by the recent discoveries of new hadrons at $e^+e^-$ facilities. These hadrons appear as very narrow meson resonances evading all standard interpretations as quark-antiquark states. Low-energy colliders also play an important role in exploring effective theories and models based on QCD, and in providing strong-interaction data essential for the interpretation of precision measurements such as $(g-2)_\mu$.

In the following sections we discuss in more detail some of the key issues relevant to future developments in strong interaction physics.

## IX-2 QCD tools for the LHC

Strong interactions will play an essential role in the new physics processes to be hunted at the LHC and in their backgrounds. These processes typically involve large momentum transfer scales (high $Q^2$) and can therefore be treated using perturbation theory and the QCD factorization theorem. Schematically, we have that

$$\frac{d\sigma}{dX} = \sum_X \sum_{j,k} f_j(x_1, Q_i) f_k(x_2, Q_i) \frac{d\sigma^{\hat{X}}_{jk}(Q_i, Q_f)}{d\hat{X}} F(\hat{X} \to X, Q_f)$$

where X represents a given hadronic final state (FS); $\hat{X}$ is an arbitrary partonic FS; $f_j(x,Q_i)$ is the density of partons of type $j$ carrying the momentum fraction $x$ of the nucleon at a scale $Q_i$; $\sigma^{\hat{X}}_{jk}(Q_i, Q_f)$ is the partonic cross section for the transition between the initial partonic state $j,k$ and the final partonic state $\hat{X}$, considered at scales $Q_i$ and $Q_f$ for initial and final state factorization, and $F(\hat{X} \to X; Q_f)$ describes the transition from the partonic final state to the given observable $X$ via fragmentation functions and hadronization effects. It may also include detector response functions, experimental cuts and/or jet algorithms.

Three basic approaches to high-$Q^2$ QCD analyses are available, based on different realizations of the above factorization theorem. They differ mainly in the way in which the initial and final state functions are treated. Ordered in increasing detail of the description of the final state, they are: cross-section "evaluators", parton-level event generators, and shower Monte Carlo event generators.

**The cross section evaluator** is a rapid and effective tool as long as one restricts interest to a limited aspect of the final state, like, for example, the inclusive spectra of leptons produced via the Drell-Yan process. In this case, detector response is directly applied at the parton level, and the inclusiveness of the result allows, via unitarity, the inclusion of higher-order corrections. Next-to-leading order (NLO) results are known for most processes both within and beyond the Standard Model (SM). In addition, next-to-next-to-leading order (NNLO) cross-sections have been calculated for the Drell-Yan process and for Higgs production.

**Parton-level events generators** produce final states consisting of quarks and gluons, with probabilities proportional to the relevant perturbative matrix element (ME). Then a one-to-one mapping of the final-state partons to the observed objects (jet, missing energy, lepton, etc.) is performed, through smart algorithms. Such smart jet algorithms need to assume a detector response that is independent of the jet structure, in order to connect the energy and direction of the measured jets to the originating partons. The advantage over the cross section evaluators is that, with the explicit representation of the kinematics of all hard objects in the event, more refined detector analyses can be performed, implementing complicated cuts and correlations which are otherwise hard to simulate with the inclusive approach. PL event generators are typically used to describe final states with several hard jets. Due to the complexity of the ME evaluation for these many-body configurations, calculations are normally available only for leading-order (LO) cross-sections. In this case, several computational tools (ALPGEN, CompHEP, MadGraph, AMEGIC++, …) have recently become available, covering all of the necessary processes for signal and background LHC studies, with jet multiplicities all the way up to 4, 5 or 6, depending on the specific process. NLO PL event generators are also available for several low-jet-multiplicity final states.

**Shower Monte Carlo generators** provide the most complete description of the final state. Their goal is to generate events consisting of physical, measurable hadrons, with a correct description of their multiplicity, kinematics and flavour composition. These final states can therefore be processed through a complete detector simulation, providing the closest possible emulation of real events.

After the generation of a given PL configuration (typically using a LO ME for $2 \to 1$ or $2 \to 2$ processes), all possible initial- and final-state parton "showers" are generated, with probabilities defined by algorithms that implement the enhanced (collinear and soft) QCD dynamics approximately to all orders. This includes the probabilities for parton radiation (gluon emission, or $g \to qq$ splitting), an infrared cutoff scheme, and a hadronization model. The shower evolution obeys unitarity and therefore it does not alter the overall cross-section, as estimated from the ME evaluation for the initial hard process. This also implies that a shower MC based on LO matrix elements cannot provide an estimate of NLO corrections to the cross section (the so-called K factors). The technique allows also the implementation of quantum-mechanical correlations and coherence providing in this way a more accurate description of the final state.

In the last few years significant progress has been achieved by the inclusion of the NLO correction in the shower MC framework. In order to get to this results one has to develop a procedure that effectively and unambiguously removes double counting of virtual and real effects , which are described in the matrix element calculation as well as in the parton shower.

The inclusion of NLO corrections in the shower MC guarantees that total cross-sections generated by the MC reproduce those of the NLO ME calculation, thereby properly including the K factors and reducing the systematic uncertainties induced by renormalization and factorization scale variations. At the same time the presence of the higher-order corrections generated by the shower improves the description of the NLO distributions, leading to departures from the parton-level NLO result.

The progress made is certainly impressive, but nonetheless a wider effort is needed in order to get the best from the LHC. In this respect, the involvement of young researchers in this field is fundamental. During the open discussion it was pointed out that this might not be trivial to achieve, since this kind of research activity is not highly fashionable and requires long periods of training and program development before results can be produced,  which might discourage young researchers from entering the field.

## IX-3 A new state of matter in heavy-ion collisions

The main motivation behind the heavy–ion experimental program resides in the fact that QCD predicts a phase diagram where quarks and gluons are expected to be deconfined. Whether this new state of matter is a plasma, as initially thought, or a strongly interacting liquid, is still far from being settled.

The SPS experiments at CERN (NA50, NA57, NA60), point at behaviours of the fireball produced in lead-lead collisions, which are not explicable using models of standard hadronic matter. Charmonium suppression and strangeness enhancement in high centrality collisions seem to take place at the onset of a phase transition (or crossover?) from hadronic matter to a new state of matter (deconfined quark-gluon), as the energy density obtained in the collision reaches some critical value (or in other words as the temperature reaches a critical Tc). RHIC at Brookhaven has made considerable progress in the analysis of a possible deconfined phase in Heavy Ion collisions at considerably higher energy densities with respect to the SPS. Since the temperature is expected to scale as the power ¼ of the energy density, SPS experiments and RHIC experiments do not happen to be to far from each other with respect to the phase transition temperature.

Anyway RHIC has found new signals indicating a transition to a new state of matter. Elliptic flow and jet quenching are observables of a different kind with respect to the above mentioned SPS ones, being more connected to the collective properties of a fluid possibly formed in Heavy Ion collisions.

Many apparently uncorrelated experimental signatures point probably at the same physics. ALICE at the LHC has the difficult task to observe the fireball in an energy region never reached before and to make a definitive assessment of this crucial sector of the QCD phase diagram. Understanding this will also be of great help to many crucial problems in cosmology and astrophysics.

At the moment a coherent theoretical framework is missing. Due to the complexity of the system studied and to the variety of phenomena involved, the field is extremely difficult to organize in a unified picture. Some bold speculations aimed at explaining complex experimental phenomena or discrepancies with existing models in terms of very sophisticated theoretical explanations (based on AdS/CFT etc.) have been proposed. While this work is certainly interesting, this field remains the least understood among QCD related topics and the risk of being misled by a plethora of models and conjectures is quite high.

A great experimental/theoretical effort is necessary to be prepared to read the ALICE data. Reflection on the long-term future of this field should be encouraged. Is ALICE the endpoint of this research field?

## IX-4 Nonperturbative QCD and Spectroscopy

BaBar, Belle, CLEO, CDF and D0 agree on the existence of new narrow resonances whose nature evades all standard theoretical assignations. These states named X(3872), X(3940), Y(4260),… resemble charmonium states but behave quite differently from standard charmonium. This situation has triggered the attention of the community and a number of hypothetical assignments such as molecules of D mesons, hybrid states, baryonia, multi-quark states, have been proposed. The most conservative interpretations, the molecular based ones, have the advantage of not predicting other states besides the observed ones. Multiquark interpretations are most fascinating from the physical point of view but predict a number of not yet observed exotic states. Experiments will soon be able to decide between various theoretical proposals.

From the formal standpoint the following question is quite interesting: is QCD allowing hadron body-plans other than quark-antiquark and qqq? This is actually an old question, which may now be close to finding an answer. Lattice QCD could reply to this question with first-principle calculations probing the possibility that multiquark states are formed on the lattice. The interplay between phenomenological models and observations can be very fertile in this field.

We have to stress that the physics case for these particles is different from that of pentaquarks. These objects are mainly observed in clean experimental setups (e+e-).

Sub-GeV scalar mesons can also be considered as non-standard mesons; low-energy colliders, Dafne for example, are in a good position to investigate the nature of these states, whose role in low-energy effective theories of QCD is far from being clear. Chiral perturbation theory has proven to be extremely effective in describing the dynamics of pseudoscalar pseudo-Nambu-Goldstone bosons, but a priori it excludes scalars from the spectrum. A solid theoretical and experimental understanding of scalar hadrons is not yet at hand.

The role of the new mesons quoted above and that of the low-energy scalar mesons is not confined to the interest of spectroscopy. Surely enough, these particles have to be taken into account in the study of several B decay processes, possibly having impact also on CKM physics.

Besides the exploration of hadron spectroscopy, low energy colliders could serve to perform some precision measurements of the e+e- hadronic cross section below the $J/\psi$. Moreover a study of the hadronic cross section in the energy window between 1 and 2 GeV is extremely important for accurate studies of the g-2 of the muon.

## IX-5 Fixed-target hadronic physics at the SPS

A thorough review of the options for fixed-target physics at CERN beyond 2005 was carried out by the SPSC in connection with the Villars meeting on this topic [99]. When compared with facilities at other laboratories globally, the extracted beams from the SPS into the North Hall are a diverse and important resource for physics. However, the SPSC concluded that, for this to remain so, investment in the maintenance and consolidation of the existing infrastructure is required, and that any "cutting-edge" research programme would require a major increase in proton intensity. The options for achieving this were outlined in Chapter III in connection with improvements and upgrades of the LHC, and they would need to be viewed in that context, and also in possible competition with requirements for a neutrino physics programme.

### IX-5.1 Soft and hard hadronic physics

A submission by the COMPASS collaboration [BB2-2.6.02] outlines the hadronic physics that could be studied using muon and hadron beams and an upgrade of the COMPASS spectrometer. Hard exclusive muon scattering processes such as deeply virtual Compton scattering (DVCS) and hard exclusive meson production (HEMP) can be used to measure generalized parton distributions (GPDs), which give insight into the transverse spatial distribution of partons in addition to their longitudinal momentum distribution. With a polarized target, the rich spin structure of GDPs can be explored; this could help to unravel the nucleon spin puzzle since there is sensitivity to the total angular momentum carried by quarks of different flavours. With transverse polarization, the distribution of transversely polarized quarks can be measured using semi-inclusive deep inelastic scattering (SIDIS); this cannot be done with inclusive DIS since the relevant structure function is chiral-odd. Planned measurements at JLAB will cover a more limited range of $x$ and $Q^2$, while proposed studies of transverse spin effects in the Drell-Yan process using polarized beams at FAIR/GSI could not start before 2018.

A wide variety of hadron spectroscopy could be investigated in fixed-target experiments with sufficiently intense proton and pion beams, including doubly-charmed baryons, glueballs and hybrid states. Progress in lattice QCD calculations is expected to provide rather reliable mass predictions for such states within the same timescale.

Long-term projects in this area of physics are under discussion in the USA and Japan, and a small part of the FAIR/GSI programme will be devoted to such topics. However, these facilities will not be operational before around 2020, affording a unique opportunity for interesting physics in the intervening period.

## IX-5.2 Proton-nucleus collisions

A submission by the NA49 collaboration [BB2-2.6.01] proposes a fixed-target programme based on proton-proton, proton-nucleus and nucleus-nucleus collisions in an upgraded version of the NA49 apparatus.

High-precision data on hadron production in hadron-nucleus collisions are needed by long-baseline neutrino oscillation experiments and for the study of ultrahigh-energy cosmic rays (UHECRs). For neutrino beam experiments the predominant muon neutrinos come from pion decay, while the largest background consists of electron neutrinos from kaon decay. Consequently the yields and angular distributions of pions and kaons must be known with high precision. The acceptance and particle identification of the NA49 detector are well suited to this task. For UHECR studies, the energies of primaries interacting in the atmosphere have to be determined from properties of the ensuing extensive air showers, for which the main source of uncertainty is the multiplicity, composition and distribution of the hadronic component. Even for UHECRs the hadronic energy range up to a few hundred GeV, accessible to fixed-target experiments, is very important since it strongly affects the muonic composition and lateral spread of the shower.

The interest in further study of nucleus-nucleus collisions in the SPS energy range arises from the possibility of a critical point in the phase diagram of hadronic matter. Lattice gauge theory and model studies suggest that a line of first-order phase transition extends from high baryon chemical potential and low temperature towards a critical point of second order at a lower but finite chemical potential and a temperature around 180 MeV. Heavy ion collisions at RHIC and LHC explore the region around this temperature but below the critical chemical potential, where the transition to a quark-gluon plasma or liquid phase is a relatively smooth crossover. Collisions at SPS fixed-target energies, on the other hand, probe higher chemical potential and hence could locate the critical point by searching for phenomena characteristic of a second-order transition, such as critical fluctuations.

# IX-6 Deep inelastic scattering

## IX-6.1 Indications from HERA for the LHC

There are many exciting interfaces between physics at HERA and the LHC, which have been explored in a dedicated workshop [100].

Concerning Parton Distribution Functions (PDF), HERA has exposed hints of saturation effects, leading to a breakdown of the simple parton description at small x and large $Q^2$. At small x, there is a large probability that extra gluons are emitted, resulting in a potentially large growth of their number in a limited transverse area. When the transverse density becomes large, partons may start to overlap, and non-linear effects (such as parton annihilation) may appear. The Malthusian growth in the number of gluons seen at HERA is eventually curbed by these annihilation effects when ln(1/x) exceeds some critical x-dependent saturation value of $Q^2$. At larger values of x, the parton evolution with $Q^2$ is described by the usual DGLAP equations, and the evolution with ln(1/x) is described by the BFKL equation. However, at lower values of x and large $Q^2$, a new description is needed for the saturated configuration, for which the most convincing proposal is the Colour-Glass Condensate (CGC).

According to the CGC proposal, the proton wave function participating in interactions at low x and $Q^2$ is to be regarded as a classical colour field that fluctuates more slowly than the collision timescale. This possibility may be probed in Au-Au collisions at RHIC and proton-proton collisions at the LHC: the higher beam energy of the LHC compensates approximately for the higher initial parton density in Au-Au collisions at RHIC. At central rapidities, effects of the CGC are expected to appear only when the parton transverse momentum is less than about 1 GeV. However, CGC effects are expected to appear at larger parton transverse momenta in the forward direction. Pb-Pb collisions at the LHC should reveal even more important saturation effects.

What is the experimental evidence for parton saturation? First evidence came from HERA; at RHIC, in proton-nucleus collisions one expects the suppression of hard particles at large rapidity and small angle compared to proton-proton collisions, whereas one expects an enhancement at small rapidity, the nuclear 'Cronin effect'. The data from the BRAHMS collaboration at RHIC are quite consistent with CGC expectations, but it remains to be seen whether this approach can be made more quantitative than older nuclear shadowing ideas.

## IX-6.2 LHeC

A submission by Dainton et al. [BB2-2.6.03], see also ref. [101], discusses the physics that could be studied by colliding a 70 GeV electron beam with one of the LHC beams (of protons or ions). This would require construction of an electron storage ring in the LHC tunnel – an undertaking comparable to a major upgrade of the LHC. The QCD studies that could be performed with such a machine (the "LHeC") include;

- Physics of high parton densities. Coverage of Bjorken x down to below $10^{-6}$ would allow more detailed investigation of the proton and nuclear parton distributions in the region of high gluon density where saturation effects may occur.
- High-precision parton distributions. Nucleon structure functions over a much wider range of x and $Q^2$ would improve the accuracy of parton distributions and hence the reliability of predictions for virtually all LHC signal and background

- processes. In particular, heavy flavour distributions, which make an important contribution to many new physics processes, would be better constrained.

- Strong coupling constant. Fits to the evolution of DIS structure functions provide one of the best determinations of the strong coupling. If a precision much below the percent level could be achieved, this would not only improve predictions of many signal and background cross sections, but would provide new challenges to models of grand unification.

- Hard diffraction. Diffractive production of new states has been proposed as a possible means of background reduction and quantum number determination. For reliable predictions the diffractive parton distributions need to be studied over a wide range of x and $Q^2$.

- Final state physics. Amongst a variety of interesting possibilities, one could clarify the hadronic structure of real and virtual photons and in particular their gluonic content, and probe the GPDs of the proton through DVCS and HEMP as discussed in the previous section.

- Electron-nucleus scattering. DIS on nuclei at small x would explore a regime of very high parton densities where striking saturation effects could be observed.

## IX-7 Discussion

A lively discussion at the Open Symposium underlined that the field is very active and has a broad and interesting agenda.

While perturbative QCD confirms its solid standing as one of the best-established fields in theoretical physics, issues of technical nature are still at hand, like the extension of the twistor technique to loop calculations.

Conversely, non-perturbative QCD offers a conspicuous number of open questions. like: where are (are there any?) exotic particles with body-plans different from standard hadronic matter? Glueballs? Hybrids? What can lattice say about these objects? There is a general agreement that hadron spectroscopy has a pivotal role in the study of QCD dynamics but it is also clear that the way from fundamental QCD to spectroscopic hadron data is very long and dangerously challenged by the assumptions at the basis of QCD inspired models, thought to reduce this leap. The weakness of theory in this respect is mainly due to the lack of a full theoretical understanding of confinement, which is still the deepest problem in QCD. In the same vein, the dynamics of the soft underlying event and the hadronization process, which play crucial roles at hadron colliders such as the LHC, are even less susceptible to the existing non-perturbative methods than is the problem of static quark confinement, and we have only crude models for these key processes.

Other longstanding problems have also been raised: the spin structure of the nucleon, the transversity, the study of diffractive processes at the LHC. These topics certainly are still triggering a considerable level of attention. Some discussion topics have been raised: HERA has proved that ρ mesons are very effective probes for investigating protons and nuclear matter. How many new insights about proton structure can be gained in other experimental facilities using such indications?

On the side of Heavy Ion collisions one of the points made has been the following: which is more important to ALICE, the fixed target programme at the SPS or the Brookhaven RHIC one? The three experiments are probing different energy density regions. The SPS is closer to the transition region from standard hadronic matter to a possible new deconfined state. RHIC and ALICE explore the new state formed. Is it a fluid? What is the equation of state? Its viscosity? These experiments pose different questions but the hope is that a synthesis of all the signals coming from such diverse experimental situations will describe coherently the same physics. For sure the SPS experiments were the last opportunity to investigate Heavy Ion collisions in proximity of the phase transition. Our knowledge about the character of this phenomenon relies on their findings.

QCD, both in its technical aspects, crucial for the success of the LHC experiments, and in its more physical problems, remains one of the richest sectors in physics.

# X THEORETICAL PHYSICS

## X-1 Introduction

The aim of this chapter is not to review the current status of the available theories of particle physics, i.e. the Standard Model and various alternative scenarios beyond it, but rather to survey the ways in which theoretical particle physics is done in Europe, in order to assist the Strategy Group in

- identifying possible problems or issues that might call for action at a European level;
- identifying and promoting good-practice models that could enhance the progress and impact of European particle physics.

To these ends, we begin with a brief overview of activity as reflected in the citation statistics for different regions and subfields. Comparisons are made here in order to stimulate discussion in view of the above goals, and not to create an impression of aggressive competition in what is in reality a collective world-wide enterprise. Next we consider the different ways in which theoretical research is organized, and how the important connections between theory and experiment can be maintained and strengthened, especially in the forthcoming era of intense experimental activity at the LHC. The special role played by the CERN theory group is discussed. A section then addresses the concerns of lattice field theory, an activity with special equipment needs. The support to theoretical research from EU funding is reviewed, and we end with a summary of the discussion of relevant topics that took place at the Open Symposium.

## X-2 Impact analyses

Theorists working in Europe have contributed many of the key ingredients of the Standard Model and of the leading proposals for physics beyond the SM, such as supersymmetry. However, it is noticeable that the most influential theoretical papers in recent years (on the AdS/CFT correspondence, extra dimensions, braneworlds, …) have come from the United States. This is confirmed by citation analyses of the SPIRES archive [102].

Citation searches must be interpreted with great care: sometimes incorrect papers are cited in order to refute them, and less important papers can become part of a package that is ritually cut-and-pasted into subsequent publications. The success of the SM in explaining most data has meant that more speculative papers can be well cited for reasons of fashion. In addition, the patterns of citation in different sub-fields can be very different and not commensurate with their importance, as for example with long and arduous but essential calculations of higher-order corrections.

That said, the SPIRES analysis shows that the very highest levels of citation are dominated by a few high-impact individuals in the USA (Witten, Randall, Maldacena, …). On the other hand, European theorists make a bigger impact in the range of well cited (up to 100) papers. The speed, volume and quality of the US and European responses to new developments look similar. There are well cited European papers in virtually all fields of particle theory. Relative to the USA, the European impact appears stronger in Standard Model phenomenology, and very competitive in BSM phenomenology, conformal field theory, supergravity, neutrino physics, lattice field theory, astroparticle physics and flavour physics.

Within Europe, the citations show a good level of activity in all countries, the leading role of CERN, the impact of Germany in lattice field theory and increasingly in formal theory (thanks to the Albert Einstein Institute, Potsdam), and the increased UK impact in phenomenology due to the Institute for Particle Physics Phenomenology at Durham.

## X-3 Organization of theory

Theoretical research is organized in many different ways: in established research groups at universities, laboratories and research institutes; in workshops, summer institutes and visitor programmes at research centres, which may or may not themselves have longer-term research staff; or just through collaboration of individuals.

In the case of established research groups, there appears to be no evidence that any particular 'critical mass' is essential for a group to produce good research. There is a clear advantage in having colleagues with whom to discuss and collaborate, but the existence of workshops etc. enables similar results to be achieved by bringing researchers together for limited periods of more intensive work. On the other hand the training of students makes it highly desirable for university groups to have a reasonable range of interests and expertise. Since this is not always possible, it is important for research centres to allow visits by graduate students, either with their supervisors or under the supervision of a staff member with related interests. The CERN theory group has recently set up just such a scheme.

As already remarked, the workshops and visitor schemes of CERN, and other research centres such as DESY, ICTP and the IPPP provide an essential service in promoting collaboration and keeping members of smaller groups in touch. The newly founded Galileo Galilei Institute in Florence will fulfil a similar function. It would be highly beneficial if more funds could be found for smaller institutions to mount their own workshops and visitor programmes on topics of particular interest to local group members.

## X-4 Relations between theory and experiment

Physics is a subject that can only thrive when there is a strong interplay between theory and experiment. New theoretical ideas lead to predictions that can be tested experimentally, and new experimental findings challenge theorists to produce better ideas. In recent years, the great successes of the Standard Model have tended to distort this normal pattern of progress. Experiments have been setting more and more stringent limits on deviations from the SM and measuring its parameters with better precision, while many theorists have shifted their interest to issues that cannot be tested in the laboratory. An exception to this trend is neutrino physics, the only area in which the need for physics beyond the SM is already clearly evident. As discussed in the introductory essay and the relevant chapter, here there are competing theoretical ideas and experiments under way or in preparation that will resolve between them.

In the coming era of LHC physics we expect the situation to become more normal, in the sense that there will be more new phenomena that prompt alternative theoretical explanations, calling in turn for new analyses and/or experiments. Close contact between theorists and experimenters will then be even more important. This is true not only for the development of new theories beyond the SM, but also for the more precise calculation of SM predictions to estimate backgrounds, calibrate signals, and look for small deviations. Theorists need to know what can be achieved experimentally in order to concentrate their efforts on predictions that have the best chance of being tested definitively. They also need the guidance of experimenters in assessing the significance of indications of new phenomena. It is depressing indeed to encounter theorists spending months or years calculating something that an experimenter could have told them could never be measured, or devising explanations of a 'signal' that an expert could tell them is most likely to be a fluctuation or systematic effect.

Close contact between theory and experiment can be achieved in different ways. Probably the most valuable is the daily exposure that can take place at many laboratories and universities. Where this is not possible, workshops and visitor programmes can compensate, provided active measures are taken to ensure that the necessary mixing takes place. Even where contact is possible in principle, some encouragement may be required – joint seminars and coffee arrangements, for example. One can readily recall universities where theorists and experimenters live on different floors and are quite unaware of each other's activities.

Such mixing is especially important at the student level, in order to engender mutual understanding and respect, proper attitudes, and the habit of interaction. Again where it is not possible there should be special measures in the form of visits and combined summer schools.

In the era of LHC data analysis, we anticipate that the distinction between theory and experiment will become less clear, and more like a continuous spectrum of interests and activities (as is the case in astronomy). Some theorists are already playing important roles in the development of analysis tools, event simulation, etc., for the LHC experiments. More thought needs to be given to the career development of those who work in this 'grey area'. Students need to be able to pursue curricula that combine theoretical and experimental work, and proper credit for both aspects should be assigned in hiring and promotions.

In view of the increased collaboration between theorists and experimenters that will undoubtedly occur at the LHC, there are issues that need to be resolved concerning protocols of collaboration: access to data, confidentiality, authorship, freedom to publish, financial contributions, etc. It may be that these can be handled, as in the past, on an informal basis, with a healthy application of common sense. On the other hand, it could be worth while for these matters to be examined alongside those concerning authorship and publication that are already being considered by the LHC collaborations.

# X-5 The CERN theory group

The CERN theory group is the centre of European excellence for all aspects of particle theory. Some 11 papers with more than 500 citations have come out of the group since 1990, about half involving CERN staff members and more than half (because of collaborations) involving visitors. This illustrates the importance of the visitor programme, which supports the whole European theory community and attracts top physicists from outside Europe. Many European theorists rely on visits to CERN as a place to work with collaborators away from the administrative and teaching burdens of their home institutions.

For a statement on the group's objectives and future plans, see their contribution [BB2-2.7.02]. One possible change under consideration is to make the visitor programme more focused by targeting specific fields of interest, as is done at Institutes such as the KITP in Santa Barbara, the GGI in Florence and the INI in Cambridge. This could be a useful addition, provided it does not reduce the opportunity for individual visits to foster and maintain collaboration on a wider and unpredictable range of topics (see Allanach [BB2-2.7.04]).

The CERN Fellowship scheme is also of great value in allowing outstanding young theorists to interact with each other, with world leaders in their field and with experimenters, in a research environment that cannot be matched by national institutes. The success of the scheme can be judged from the very high proportion of Fellows who go on to obtain permanent academic or research positions.

Another essential service of the CERN theory group is the organization of workshops on topics relevant to the experimental programme, resulting in CERN Yellow Reports, which have provided major studies of LEP and LHC physics.

In the coming years it will be vital for the group to support the LHC endeavour with in-house theorists able to help in interpreting the data, while at the same time maintaining the group's tradition of excellence in fundamental physics.

# X-6 Lattice field theory

A primary aim of lattice field theory is to compute non-perturbative physical quantities with sufficient precision to have an impact on experiment, notably, through QCD simulations, to help determine SM parameters, constrain new physics, and study matter at high temperatures and densities. A more general aim is to achieve new insights into the properties of strongly coupled quantum field theories. To meet these objectives, increased computing power and improved algorithms and methods are all needed (see Pène et al. [BB2-2.7.01]).

Increased computing power is needed to increase lattice size, decrease lattice spacing, increase the number of gauge-field configurations, and apply better treatments of fermions. Current European resources amount to a few teraflop/s; the lattice community is seeking to achieve one petaflop/s by 2009. The generation of field configurations is well suited to highly scalable assemblies of units that may even be specifically designed for this purpose, as with the existing QCDOC (QCD-on-a-chip) and ApeNEXT machines. On the other hand the evaluation of field correlators using these configurations calls for the flexibility of more general-purpose machines, and therefore a mix of architectures is needed. Handling and sharing of the huge data sets generated will be made possible by the Europe-wide lattice data-grid system being set up in the framework of the international lattice data grid (ILDG).

Improved algorithms and methods are needed primarily for the inclusion of dynamical fermions (sea quarks), which is important for obtaining correct physical results. Different fermion implementations and actions need to be pursued in order to have confidence in results extrapolated to physical light quark masses and the continuum limit. Overlap or domain wall formulations are the most computationally demanding. Wilson-type fermions are less demanding, but they break chiral symmetry explicitly. Staggered fermions are computationally simple and cheap, but they require more detailed theoretical justification. Improved gauge field and heavy quark actions are also important. In terms of the impact on experiment, a key question is whether lattice calculations can deliver important hadronic parameters reliably enough and soon enough to meet the needs of ambitious future experimental programmes such as super-B factories. Extracting the full value of the data from such programmes depends on adequate funding of the relevant lattice research, which represents only a small fraction of the cost of the experimental facilities.

# X-7 European Union funding of theoretical physics

For many theory groups, especially those in the smaller member and associated countries, EU-funded networks and junior researchers represent an important form of support, not only in finance but also in terms of integration into the research community. In larger countries as well, they provide support for international theory collaborations that are often not adequately funded by national agencies. There are, however, problems with the present EU funding system that have been encountered by many in the particle-theory community:

- The application process is highly time-consuming and has a low success rate, making for a great deal of wasted effort;

- The five-year cycle of EU frameworks, combined with the low success rate, does not promote sustained development or strategic planning;

- Funding is highly project- and goal-oriented, leaving little scope for the flexibility and creativity that is essential in theoretical research;

- The range of programmes for which particle physics is eligible is more limited than that available to other subfields of physics;

- Networks are increasingly oriented towards training rather than research;

- Recruitment is sometimes made difficult by inconvenient timing of grants and the complicated restrictions on nationality and residence.

A more general concern arises from the fact that particle physics is not a recognized sub-field of physics in calls for EU proposals, which engenders doubt that the special characteristics and needs of our field are fully appreciated in the formulation of programmes and the assessment of proposals. It is to be hoped that the existence of a European strategy will open new channels of communication through which these concerns can be addressed.

# X-8 Discussion

Many of the subjects mentioned above were topics of animated discussion at the Open Symposium. We now attempt to summarize the main points made there as follows.

Doubts were expressed about the validity of citation analyses as a way of assessing the quality of theoretical research. Judging by citations is inconsistent with rewarding theorists for doing unglamorous but essential calculations. With the reduction of the flow of new data in recent years, citations reflect more of the fashion and popularity of certain ideas rather than their validity; this will hopefully change in the LHC era. Citation comparisons between countries and between regions are divisive when the object should be to unify. Comparisons should take account of the number of physicists, and the funding per person, in different regions. High citation numbers in certain countries are due to a few people close to retirement, so comparisons could change rapidly.

In connection with comparisons between regions, it was asked whether the USA is doing things right that Europe could learn from? Are the hiring practices of European countries too restrictive? Are salaries too low and/or teaching loads too great? Such questions may need to be addressed to avoid an increased 'brain drain' of the best theorists to the USA.

On the organization of theory, the impact of new centres of excellence such as the IPPP in Durham and AEI in Potsdam is clear and shows the benefit of achieving a certain critical size. However, part of their success may be due to the fact that they are new institutes, staffed by energetic young people. It is essential to maintain excellence in other places, especially in universities, for the proper training of students.

There was much discussion of the relationship between experimental and theoretical research. Should centres of excellence in phenomenology be located in laboratories, or at least in universities with large experimental groups, to facilitate interaction? The IPPP works well without any nearby experimental groups, because experimenters from elsewhere want to interact with the people there. Interaction with a local group is limited to the experiments in which it is engaged, whereas a separate institute can take a broader view. While a theory group should have good contacts with experiment, there are important interactions with other fields, such as mathematics and astronomy, which are best maintained at a university. The important thing is to have sufficient freedom and independence to do the best work that one can. For this a range of different models and styles of theoretical groups needs to be maintained.

On the experimental side, some participants expressed caution about working too closely with theorists, which could lead to bias in the analysis of data. When new results start to flow from the LHC, theorists may be keen to be more involved, while experimenters may be reluctant to give out data for fear of incomplete analyses, misinterpretation of errors, etc. However, others expressed enthusiasm for closer collaboration with theorists, both now in the development of analysis tools and later in their application. Or perhaps the model for interaction between experiment and theory should become more like that in astronomy, where data are released sooner. At the LHC the situation will be different from that at LEP, where the focus was on precision observables that experiment measured and theory calculated.

Concerning lattice field theory, it was emphasized that there is a need for both increased computing power and algorithmic improvements in order to produce results useful for the interpretation of experiments, for example in B physics. Building cost-effective dedicated computers and using general-purpose machines are both required. Europe is well positioned to lead these efforts.

# XI FUTURE ORGANIZATION OF EUROPEAN ELEMENTARY-PARTICLE PHYSICS

## XI-1 Purpose of this text

One of the Zeuthen working groups addressed organizational issues of European elementary-particle physics (EPP). The purpose of this text was to stimulate that discussion.

The text is partitioned in three sections. One describing the European scene with its actors, one illustrating challenges facing us, and a final one giving examples of possible future implementations.

## XI-2 The European Scene

### XI-2.1 Introduction

Europe has an advantageous position for research in elementary-particle physics. The field has pioneered international research collaboration, and indeed shown many other fields what is achievable in terms of collaboration.

Through its collaborations, this field really has during the second half of the $20^{th}$ century created a European Research Area for elementary-particle physics. Many of the research networks created by the European particle physics institutes result in an environment where national boundaries are hardly visible to the individual researcher. The European laboratories play a fundamental role in making this a reality, and CERN is foremost among them. The funding agencies are well integrated actors in this modus operandi. There are good reasons for pride in the excellent established triangular relationship universities – laboratories – funding agencies.

Elementary-particle physics is in one way a victim of its own success because, by being so early in organizing itself, it had fewer incentives to become integrated in the structures that were later established by the European Union. However, the Union is moving towards becoming the centre where, in a general sense, the global European research strategies are being developed; it would be better for particle physics to become a strong actor in this process.

Given the status of CERN and its broad mission for the co-ordination of co-operation among the European states in the field of EPP, it is inevitable that CERN has to take a special place in the discussions to follow.

### XI-2.2 The universities

The universities are the basis for the field. This is the source of rejuvenation and it is from there that the initial steps are mainly taken to engage nationally in international projects. The national funding agencies connect to the international projects after having reviewed proposals made by the national community.

Activities by the EPP teams at the universities span teaching, education of new researchers, research and technical R&D. Indeed, even industrial-scale detector fabrication has been set-up at the Universities to contribute to the construction of the experiments.

The ability and resources for short- and medium-term detachment to the laboratories is important for the university teams. It is made possible in most European countries, but there are exceptions where the resources for the national communities are not at the level with respect to, for instance, the investments in CERN. It would be desirable if this could be strengthened, nationally and/or by European funding. The latter possibility is inhibited by the somewhat disconnected situation that EPP has with respect to the European institutions.

The university teams have the double responsibility to do research and to train new researchers. The latter is successfully done, but it is in the nature of research training that only a fraction of those trained stay in the field; for EPP roughly 50% stay after their PhD. The competences that they take with them to other areas are general problem-solving and modelling, the experience of working in international teams and the experience of a multitude of technical and practical skills; the direct EPP knowledge is of limited applicability. It is therefore reasonable to ask if the EPP PhD's get a training that sufficiently enables them to apply their general capabilities in other fields (the so-called "transferable skills", see the chapter on Knowledge Transfer).

## XI-2.3 The national laboratories

Europe can take pride in having several excellent national laboratories with vibrant research activities. These laboratories are of varying sizes and not all can be explicitly mentioned here. Seven are directly represented in the Strategy Group: DESY (DE), RAL (UK), LNF (IT), LNG (IT), DAPNIA (FR), LAL (FR) and PSI (CH). Most are multidisciplinary, but not all. Their size varies, e.g. DESY has more than 1'500 employees and 3'000 users, of which 50% are from outside Germany. LAL has about 300 employees. There are several other laboratories as well, e.g. NIKHEF (NL), MPI (DE), and LAPP (FR), with well-defined missions, but usually without major accelerator-based facilities or other infrastructures.

These laboratories have several roles. They act as hubs for the national engagement in international projects outside their countries, several host international projects, and they have smaller internal scientific programs. Several laboratories are central to the European accelerator R&D and also contribute to the construction of accelerators at other laboratories, including CERN. The generic accelerator R&D is mostly done at these laboratories, while almost none (apart from CLIC R&D) is pursued at CERN. Several, for example RAL (UK), grew from a purely EPP laboratory into a multidisciplinary one, often built around, but not restricted to, accelerator-based disciplines (e.g. spallation sources, light sources).

The total capacity of these laboratories is significant, but the collaboration between them and with CERN could be strengthened.

Several national laboratories have a strong record in transferring EPP technology to other fields. The most striking example is photon science. This field was not only initiated from EPP, but its influence is continuous, the latest example being the XFEL.

From an EPP perspective, DESY is particularly outstanding. It has a scientific output at the energy frontier, just like CERN and Fermilab, historically the most important being the discovery of gluon radiation. DESY has built a sequence of accelerators at the energy frontier, and operates today Europe's energy-frontier collider, second only to the Tevatron at Fermilab. DESY has also developed the accelerator technology chosen by ICFA as a basis for the International Linear Collider. This success creates a certain competitive tension between DESY and CERN (e.g. DESY was the $e^+e^-$ collider laboratory of Europe when it was decided to build LEP at CERN), and there is also competition for resources.

## XI-2.4 CERN

CERN was established in 1953 as an intergovernmental Organization and plays a special role and has special status on the European particle physics scene:

Under the terms of the CERN Convention, the international treaty that established it as an Intergovernmental Organization financed today by 20 European Member States, CERN's **mission** is to *"provide for collaboration among European States in nuclear research of a pure scientific and fundamental character, and in research essentially related thereto."*

The Convention provides that this mission be implemented through two kinds of **activity**:

1. *"the construction and operation of one or more international laboratories"* with "*one or more particle accelerators"*

2. "*the organization and sponsoring of international co-operation in nuclear research, including co-operation outside the Laboratories*"

Under the CERN Convention, the co-ordination of co-operation in the field of elementary particle physics is explicitly part of its remit. Notwithstanding this broad remit, CERN has hitherto concentrated on successful international co-operation in elementary particle physics **at its Laboratory in Geneva** and has placed less emphasis on co-operation in this field outside the Laboratory or on co-ordination of Europe's position with regard to such international co-operation.

It is evident that CERN has a very strong technical track-record in designing, constructing and operating accelerators, especially strengthened by the ability to apply to such projects a strong engineering base and substantial and stable resources, both human and material.

CERN also has a number of scientific successes, the most striking being the discovery of the intermediate vector bosons. It is the existence of CERN that enables Europe to be a world leader in experimental elementary-particle physics. It would be difficult to create a corresponding situation today from scratch.

In particular, through CERN, there is a significant contribution from all Europe's smaller countries to the common EPP infrastructure. This scale of contributions would never arise through project-based bilateral agreements. It therefore guarantees a voice in large issues to the small countries, and it helps the large countries to get support from them; a win/win situation.

Inspecting the world's citation statistics demonstrates that CERN is Europe's foremost centre of excellence in EPP theoretical physics: a result produced in particular by its visiting scientists programme.

A success of CERN is in international collaboration for projects in Geneva, by direct contacts between the laboratory and national authorities, and through the collegial network of the experimental collaborations. This has evolved from the early days when experiments at CERN were primarily performed by CERN scientists with some external participation, to today's situation with the major experimental contributions coming from outside the laboratory, and with a significant participation from non member states and institutions outside Europe.

Lately, CERN has developed some attractive activities that are more in the area of international policy making than of a research laboratory: the initiative on Open Access publishing, the use of science to counter-act the digital divide, and by addressing the issue of a structure to administrate a multidisciplinary European grid infrastructure.

CERN membership has also been used by new member states to support their case for membership of political organizations, in particular the European Union.

## XI-2.5 The funding agencies

There is no uniform situation of funding-agencies across the European countries. It is therefore difficult to give a general description of their role. Going through each country would result in an excessively extensive discussion. In general these actors provide the peer review and support for the engagement of the national research activities, within the country and in international collaborations. In some cases they also act as the national reference bodies for the membership at CERN, and even have the CERN contribution as part of their budgets. The visibility of the funding agencies in the research collaborations is increasing. They are today signing the MoUs whereas this was previously more at the institutional level.

The national EPP support with respect to the contribution to CERN varies considerably. In the large countries this is of roughly the same size, while it is at the 5–10% level for the smaller countries, and even less for the new CERN member states. Funding agencies have adapted to their role of larger engagement responsibility, especially for the case of the LHC experiments, where these responsibilities have been called upon to address cost increases. These challenges have been addressed despite that the resources provided by the funding agencies and those provided directly by CERN are interlinked.

## XI-2.6 Committees

### XI-2.6.1 European Committee for Future Accelerators[103]

The committee was created in 1963 to address the planning of future accelerator-based infrastructure for European EPP. The construction of PETRA and HERA at DESY, and of SPS, LEP and LHC at CERN, were all preceded by ECFA studies. Through working groups, ECFA helps in building up the engagement of the researcher community in preparations for the next accelerator. It is through ECFA that the European involvement in ILC is being followed. ECFA is also the channel through which the European physicists get represented in ICFA (which itself is constituted through the NGO International Union of Pure and Applied Physics, IUPAP). ECFA also monitors the status of the EPP community in the different CERN member-states.

At the time when the case for PETRA was being prepared, an understanding [104] was reached with the European Science Foundation (ESF, a NGO bringing together the national funding agencies from across Europe) that ECFA would be the scientific reference body for ESF in accelerator-based EPP; ESF has never called on this.

ECFA gets administrative support from the CERN Council secretariat.

All CERN member states are represented in ECFA; their people are nominated through national processes and approved by ECFA. ECFA is therefore the committee that is closest to representing the EPP community.

### XI-2.6.2 The CERN Scientific Policy Committee[105]

This committee is the scientific advisory body of the CERN Council, and is focused on the activities of the Geneva laboratory. The terms of reference of the SPC are:

a) to make recommendations to the Council on the priorities of research programmes and the allocation of research effort both within the Laboratories of the Organization and extramurally;
b) to examine and make recommendations to the Council on the allocation of resources to the various scientific activities of the Organization;
c) to advise the Council from the point of view of scientific policy on the management and staffing of the Organization, including the visitors programme and the nomination of senior staff;
d) to advise the Council on any other matters which affect the scientific activities of the Organization.

The SPC thus provides scientific scrutiny and advice on a very sizable fraction of European EPP and, through the last term, has the right to advice the Council on almost any aspect of particle physics related to CERN's activities.

SPC proposes, by internal election, its future members to the CERN Council for approval. A significant effort is made to have a balanced composition across Europe, together with some members drawn from the Observer states. The members are appointed ad personam.

The chairman of ECFA is ex officio member of the SPC. The chairman of SPC is observer of Plenary ECFA.

### XI-2.6.3 The High Energy Particle Physics Board of the European Physical Society (EPS-HEPP)[106]

This is a rather small committee, mainly focused on arranging the biennial EPS High Energy Physics Conference, and awarding a series of very prestigious scientific prizes. Some of these prizes have preceded Nobel prizes.

The chairman of this board is observer in ECFA, and the ECFA chairman is ex officio in EPS-HEPP.

EPS-HEPP itself appoints its future members when replacing outgoing ones.

### XI-2.6.4 The European Steering Group for Accelerator R&D, ESGARD

The ECFA 2001 roadmap for particle physics recognized the need to strengthen the accelerator R&D in Europe. In the same period it was also recognized that the EU $6^{th}$ framework program opened the possibility to strengthen these activities. Accordingly, the directors of CCLRC, CERN, DAPNIA/CEA, DESY, LNF, Orsay/IN2P3, and PSI in consultation with ECFA decided to form ESGARD.

ESGARD has acted as a coordination body to promote accelerator R&D and ensure a consistent strategy towards proposals to the $6^{th}$ framework, with successful outcomes resulting in several EU-funded activities like CARE and EUROTeV.

### XI-2.6.5 Astroparticle Physics European Coordination, APPEC[107]

ApPEC is a committee set-up by the funding agencies, endorsed by the ESF. It has a steering committee with representatives from the funding agencies, and a peer review committee appointed by the steering committee. ApPEC has worked out a roadmap for particle astrophysics in Europe.

### XI-2.6.6 Nuclear Physics European Collaboration Committee, NUPECC[108]

NuPECC is an expert committee of the ESF. Its members are appointed by the ESF executive Council based on proposals from the NuPECC subscribing institutions (National Funding Agencies). NuPECC has worked out a roadmap for nuclear physics in Europe.

## XI-2.7 The European Science Foundation, ESF[109]

ESF is a NGO, with a secretariat seated in Strasbourg, bringing together the national funding agencies from across Europe. Its chief executive (from 2006) is I. Halliday. ESF influences EU policy matters; for instance ESF was at the origin of the proposal to create a European Research Council. The main purpose of the organization is to promote European research co-operation and to work on policy matters. Its scientific activities are organized in sections, where one is Physical and Engineering Sciences. ESF has expert committees to help its work, NuPECC being one of those. ECFA is the scientific reference body for ESF in accelerator-based EPP, although (as noted above) ESF has never called on this.

## XI-2.8 The European Union[110]

The EU has three decision-making bodies: the Council, the Commission, and the Parliament. In activities, the EU has three pillars: the European Community operated by the Commission, the co-operation in law enforcement, and the common foreign and security matters. For EPP it is relevant to understand the connection with these decision-making bodies and that with the first pillar. It is also relevant to understand that only the Commission can propose initiatives within the first pillar.

The European Strategy Forum on Research Infrastructures (ESFRI[111]) was launched in April 2002 to support a coherent approach to policy making. The Forum brings together representatives, nominated by Research ministers of the EU member states and of European countries associated with the Framework Programme, and a representative of the European Commission. It is chaired by J. Wood, who is also Chief Executive of CCLRC. ESFRI has started to prepare a road map for research infrastructures of pan-European interest in the next 10-20 years, but they will neither decide on priorities nor make scientific peer reviews. They scrutinize if the processes that have led to proposals are sound, i.e. have scientific backing etc. Who then can make proposals to ESFRI? This can be done by countries and members of EIROFORUM [112]. A proposal for infrastructure which is not on the ESFRI list will probably not be considered for support by the EC, including support for R&D and industrialization.

The research support from the EC $6^{th}$ Framework has a certain influence on EPP in Europe through the Marie-Curie Programme and through support for technical R&D and industrialization for accelerators (CARE, EUROTeV and EURISOL), detector development (EUDET) and computing (EDG followed by EGEE, followed by EGEE2).

The $6^{th}$ Framework is coming to its end with the last calls for proposals being launched. What can be expected from the $7^{th}$ Framework (2007–2013 )?

The Commission has requested [113] a seven-year budget of 72.7 G€:

- 44.4 G€ to Collaboration
  - defined in 9 themes, of which ITC (12.7 G€) is the only relevant (Grid) to EPP
  - has to be in the context of across borders and networks
- 11.9 G€ to Ideas
  - here is the European Research Council and nothing else
- 7.1 G€ to Human development
  - here are the Marie-Curie actions and nothing else
- 7.5 G€ to Capacity
  - here are research infrastructure (4 G€), with explicit reference to ESFRI. This also includes technology initiatives (accelerator R&D and industrialization)
  - here are also many other activities
- 1.8 G€ to Common research centres; not relevant to EPP.

The EU budget negotiations led to a reduced level of ambitions. There are some signals that the level for FP7 will end around 48 G€. It is not yet clear how these reductions will affect the different themes in the original Commission proposal, but it is reasonable to expect that what is new since the $6^{th}$ Framework will suffer disproportionately. There are though indications that European Research Council may end around 7 G€.

From the above figures it seems as if major EPP accelerator-construction support cannot be expected from the $7^{th}$ Framework. The drawn up December 2004 "List of Opportunities" [114] from ESFRI has 23 different research infrastructure items on it, which should be compared with the 4 G€ for research infrastructure in the original proposal from the Commission.

It has been speculated that such support could be mobilized from some other area of the 700 G€ EU budget for the same period, e.g. regional support; this would exclude construction at existing laboratory sites.

Except for the regional-support speculation, it is unlikely that there could be EPP accelerator-construction support from the EC comparable with the CERN budget.

However, even if the scope for construction support is limited, it is possible that EU support for specific activities within an overall accelerator project might qualify, e.g. support for R&D and industrialization.

EU support has a larger scope than accelerators. Programmes for training and knowledge transfer, the development of generic infrastructure, support for networks and visitor programmes, support for research groups using the accelerator infrastructure at CERN and elsewhere, and probably many other issues, could be addressed. A lot of these may not directly provide resources to the laboratories, being nevertheless crucial to the field. The European Research Council may. for example, provide opportunities to create new EPP groups.

# XI-3 The new challenges

## XI-3.1 The global projects

One possible scenario for the future is that the major accelerator projects are organized as global collaborations between the three regions Americas/Asia/Europe which would contribute according to some scheme, on top of a special host contribution. In this case, no national authority or existing regional organization such as CERN would take the overall responsibility, but this responsibility would instead be discharged onto a new legal entity.

The International Linear Collider and a neutrino factory are of such a magnitude, as is a far-future µ-collider, that they could all be discussed in that context. In this scenario it is in the interest of European research that participation be under conditions that are advantageous for Europe. This is an issue of negotiation, and will work best if Europe is represented by one fully empowered actor (just as the EU represents all European countries in the WTO negotiations). Several independent actors from Europe will inevitably result in a weakened European position, resulting in turn in less cost-effective conditions, and therefore reduced opportunity for other activities. Such negotiations are not only of initial nature, but on-going over the lifetime of these projects.

The alternative to having a co-ordinated European position on such global projects would be for national funding agencies to make individual commitments, which is very likely to result eventually in a weakened commitment to CERN and consequent loss of European leadership in EPP.

Another issue is to enable European countries to participate in global projects that are not sited in Europe. Given that these facilities will be unique (because of the scale that makes them global projects), it is desirable for all countries to be engaged in order to have a complete involvement in particle physics as a discipline. If countries are not involved from the beginning, then when individual research groups from non-participating countries will want to join at a later stage, it is most likely that they will have to contribute in addition as part of the conditions for participation, and they will have lost opportunities to participate in the R&D and technology benefits from the construction itself.

It should be noted that the global approach is not the only model. Another is that one region takes the lead, and the others join as "foreign participation"; this is how the USA participates in the LHC under very advantageous conditions and without sharing the common risk. This was achieved by a successful negotiation from the American side, demonstrating the virtue of a solid negotiation also for projects organized in this manner. Even under this model, it would probably be advantageous for Europe to be represented as a single entity, rather than as several different and potentially competing entities.

It is also possible that several of these models could be combined; for example, it might be possible for countries to contribute to such global projects both through CERN and directly. Indeed this might well be a mechanism that would be required for a global project constructed in Europe but outside the current CERN laboratory, where the host state might be expected to make a substantial exceptional contribution.

## XI-3.2 Making the case for European EPP

Elementary-particle Physics in Europe is to a large degree a European project, because even activities that are nationally hosted have a significant foreign participation. There is a need to make the case for EPP from this perspective, to show the complementarities of the different programmes and the absence of unnecessary duplication. (In some cases, for example neutrinoless double β decay and dark-matter searches, several experiments studying the same phenomena in different ways is highly desirable.) This approach would strengthen the support both at the European and at the national level, but would also require some framework in which coherence across Europe could be promoted.

One purpose of the Strategy Group activity is to promote this picture and to provide guidance for consistency, thereby helping to realize the full diversity of the European EPP activities. There will be a need to establish mechanisms for continuing this process of promotion and consistency at the European level.

## XI-3.3 Managing the grid

Promoted as a system to make the computing infrastructure for the LHC, and being developed to a production-stable multidisciplinary infrastructure, the time is approaching to create its organizational umbrella; not so much for R&D as for stable operation as a European infrastructure.

This umbrella must be hosted somewhere by a legal entity, since funding, contracts and MoUs must be involved in maintaining and developing the system.

Should this be hosted by an EPP institution and more precisely by CERN? How should this in that case be hosted by CERN without interfering with the core activities of the laboratory? The grid could grow into an activity with a scope surpassing the physics-research activities at CERN.

## XI-3.4 Open-Access publishing

EPP is piloting the change of research-publication culture, by moving over to open access. Whatever the publishing mode may be, there is always a cost to be paid. This cost may be covered by subscribers, authors, or from a third source. In all cases it is tax-funded. The problem with subscription payment is well known. The problem with authors paying is that it becomes an incentive for journals to publish all, and that poorly-funded institutions may not afford publishing by their researchers.

Third-source financing is probably needed. Legal entities taking organizational responsibility, including the administration of collecting contributions and making the payments would be needed.

## XI-3.5 Interactions with the European Union

It is among the activities of the Commission that the reference to what is important with respect to research on the European scale be worked out, and the EU Council is the only forum in Europe where all ministers of Research multilaterally discuss research with a European dimension. The material on which these discussions are based are provided by the Commission. CERN and EPP in general are only weakly connected to the process. This is not a sustainable situation if EPP is to continue to flourish in Europe.

In the immediate future, the issue is the ESFRI document on the future research infrastructure for Europe. The projects needed for European EPP, in Europe and elsewhere, have to enter this compilation. In this round it would be the outcome from the Strategy Group process that is fed to ESFRI through CERN. Since ESFRI will be a continuing process, and the compilation a living document, the issue of how this interaction takes place in the future must be addressed. It is not only construction support that depends on being in that reference list, but also most likely EU R&D and industrialization support for future accelerators.

The Union does have research support, and its positive impact on European EPP could be even greater. When it comes to the difficult discussion on how the $7^{th}$ framework resources are distributed among the various disciplines, it is easy to see that the argument could be made that EPP has CERN, and thus is more than fully covered, compared with other areas; an argument that can be more easily made if European EPP at large has no voice.

Firstly the Marie-Curie Actions and the European Research Council could play an important role in European EPP, maybe not so much for the laboratories as for the research groups using the laboratories, and for theoretical physics. European EPP needs to influence the evolution of these activities to make that happen. Secondly, the Research Infrastructure programme needs to be open to support accelerator R&D and industrialization for EPP; ESFRI is one crucial issue here. Thirdly, one could speculate whether, with additional resources, a structure organizationally under CERN, but also involving other EPP laboratories, could play the more general role of being the European accelerator competence hub for all disciplines. There are many synergies between SPL, EURISOL and ILC on the one hand, and, say, ESS on the other; a dialogue about this could be opened with the Commission.

European EPP needs to be fully connected with the construction of the European Research Area [115], while maintaining its "centre of excellence" at CERN.

## XI-4 Possible Scenarios

A number of scenarios are discussed below. Clearly, many others exist, and also combinations of those mentioned here.

The issues are: Who speaks for Europe?/ Through which channel do the European contributions to global projects pass?/ and In which context is an optimized, balanced and competitive European programme established (if a central process is required for this)?

### XI-4.1 Continue without change

The CERN Council is the only formal Europe-wide decision-making EPP body, in sessions following the agendas being prepared by the CERN management in agreement with the Council President.

Accelerator-based long-term EPP planning is worked out in ECFA, and CERN stays focused on the activities at the Geneva laboratory. The road map would regularly be updated in ECFA and transmitted to the CERN Council through the SPC, to national laboratory managements, and to the national communities through the ECFA members. ECFA has no channel to ESFRI, so this information has to be forwarded by the CERN DG.

ECFA is the body to make the broad case for European EPP, and has to find ways to transmit this in a wide sense relying on support from the laboratories that are ex officio in the committee.

There will be no legal entity in charge of representing Europe in negotiation on European participation in the next accelerator project, and no single institutional structure for the participation of Europe. The policy driven by CERN in such discussions will be significantly influenced by the needs of its own scientific programme, while the independent voices of the national funding agencies will be driven by their respective national motives.

## XI-4.2 Work through new committees

This is essentially the same as the previous scenario, with the same participants meeting under a new umbrella. The only formal Europe-wide decision-making EPP body remains the CERN Council.

## XI-4.3 Create a new European legal entity

It would be politically inappropriate to make this case, since EPP already has an intergovernmental organization with the mission to co-ordinate co-operation among European states in EPP. A new entity would either call together the same people as the CERN Council (if at governmental level) or operate at a lower level than the CERN Council, i.e. at the funding-agency level. Operating at funding-agency level is probably not acceptable when the governance of projects of several G€ are established.

## XI-4.4 Use CERN

Here are only discussed scenarios which can be implemented within the existing Convention.

For CERN to take up the full scope of the two activities in its mission (see 2.4), it has to be organized in a manner so that the responsibility for the Geneva laboratory, does not prevent the impartial co-ordination of European EPP.

### XI-4.4.1 Only use CERN for the policy making

The CERN Council could be responsible for a broad European EPP strategy.

Something like the Strategy Group process could be regularly repeated, but probably not very frequently. A continuous mechanism to develop the Europe-wide EPP strategy would be needed. Today this activity is handled by ECFA, but the ECFA chair is only invited to the CERN Council.

The CERN Director General is responsible for implementing that strategy for the scientific programme at CERN, but would not oversee and co-ordinate the implementation elsewhere.

It has to be clarified how the Strategy is communicated with other institutions, and how Europe negotiates with other organizations.

One issue that has to be solved in this scenario is how the Council's agenda is being prepared to allow the Europe-wide issues to be addressed. The normal operation is that the draft agenda is prepared by the Director General in agreement with the Council President.

The procedures followed by the President of the CERN Council and the Director General would have to be carefully worked out to ensure that Council can take on this wider role of representing the European EPP strategy.

### XI-4.4.2 Use CERN for policy-making and implementation

The European EPP strategy could be decided, as in the previous section, by the CERN Council.

In addition Council could decide to create new programmes of activities within the basic programme, in particular for the implementation of a European EPP strategy. It is useful to mention that the CERN Convention explicitly foresees the possibility for CERN to participate in a national or multinational project, and that this can form a special programme of activities [116]. Such programmes of activities would then exist in parallel, and would be on an equal footing from an organizational point of view. Each programme of activities would have its own Director.

New programmes of activities could be the European participation in a global accelerator project elsewhere, or the organization for the European multidisciplinary Grid infrastructure, etc.

To have different CERN programmes under different Directors is not new; this was the case during the construction of the SPS, when the "CERN II" laboratory was created, geographically close to the original CERN, but with its own Director General responsible for the new construction project.

Each of these programmes of activities would have a defined mission and ring-fenced resources approved by Council according to a Strategy implementation plan proposed by the DG. The funding model of these programmes of activities could be different. The existing programme of activities (including the LHC) would continue to be mainly membership financed as it is today, while the programme of activities for European participation in an accelerator project elsewhere could have mixed funding i.e. an obligatory part and a voluntary part of the member states, and maybe the Grid programme could be mainly funded externally.

The DG would oversee the execution of the different programmes of activities.

This model would give an executive structure representing European EPP in a broad sense.

This model could permit CERN's role to evolve considerably in the future.

## XI-4.5 The European Union taking the lead

Seen from outside our field, many people might think that it would make sense if EPP is handled as just one science out of many whose European perspective is addressed by the European Union. This could mean, for example, that a European participation in a global accelerator project would be negotiated and represented by the European Union (like ITER).

Also issues such as the Grid and publishing could be taken up by the Union. For example, the Grid could be handled in a way similar to GALILEO and publishing through a small agency.

While this approach could initially bring additional resources into European EPP, it would inevitably open the question of the long-term future of CERN. It is likely that it would start a process in which the special position currently occupied globally by European EPP would become less prominent. It would become necessary to address the formal relationship between CERN and the EU, and could weaken the strong connection that currently exists between CERN, the national funding agencies, and the member-state institutions and laboratories.

# XII TECHNOLOGY TRANSFER (TT)

## XII-1 Aim, mission and impact

In the quest to find out what matter is made of and how its different components interact, elementary particle physics (EPP) needs very sophisticated instruments using cutting-edge technologies often requiring considerable inventiveness; it pushes technologies to their limits, often exceeding the available industrial know-how. The technological advancements can find applications useful for society at large, promoting business and general welfare. The innovations created by scientists and engineers working at the frontiers of particle physics can be applied in many fields, such as communication and information technology, medicine, energy, environment and education. Nonetheless, the unaltered reason for doing EPP is the science and not the technology – because there are always 'cheaper ways of developing the non-stick frying pan than putting a man on the moon'.

EPP is supported by our societies, their governments and funding agencies primarily because this prestigious research is an essential part of the culture of our nations or regions. Today, however, other probably equally important sciences exist and request funding. In addition, a number of applied sciences promise a quicker turn-round of the investments into products that can be sold on the world market. Thus, technology transfer (TT) is very attractive to funding agencies and governments. Therefore, in addition to providing knowledge, evidence of economical usefulness and technological relevance are also required from a science, such as EPP.

What distinguishes EPP, as well as some other sciences from, applied areas in industry or commerce is effectively the attitude to risk, more specifically calculated technical risk to meet the technical demands. The advantage is that the intrinsic cutting edge characteristics of resulting technologies from EPP are well established because they are used in EPP laboratories. For scientific machines, such as the Hadron Electron Ring Accelerator (HERA) or the Large Hadron Collider (LHC), the equipment is driven 'at the limit'. In the process of showing that it is possible to operate equipment, such as cavities, couplers, vacuum systems, magnets, detectors etc. reliably, the limit is effectively moved. Additionally, in order to overcome obstacles, that are responsible for a current operational limit, EPP research needs innovative ideas for decisive measures to overcome these obstacles, to finally push the borders of technology even further.

The transfer of technology to society is one of the great benefits of fundamental EPP scientific research. Science research centres contribute to practical benefits in addition to their main activities.

For a successful technology transfer, timing is essential and the dissemination impact strongly depends on investment funds for TT R&D. Therefore in addition to the conventional licensing mode for transferring the technology, a R&D partnership policy is essential to close the gap between technology development for research and commercial products.

However, industry and society are not always ready to adopt new fall-out from the very advanced technologies applied: it may take up to many years to arrive, from a spin-off, at a commercial product. Often this is due to short operation range of the business world and to insufficient support for infrastructure and from public funding. Technology transfer aims at developing interfaces to facilitate the activities and the process within the mission of EPP research centres with the objective to boost the innovation and the benefit to society.

The strategic aims and goals of TT may be defined as

- Maximizing the technological and knowledge return to society without diverting from the scientific EPP mission.
- Supporting technological innovation in industry.
- Encouraging synergies between EPP and TT activities for technological applications.
- Promoting EPP image of a centre of excellence for technology.
- Endeavouring to steadily increase the volume of TT, using pertinent dissemination approaches.
- Securing external resources for TT activities so as to minimize the impact on the resources of the institutions.

The importance and the beneficial fall out from EPP can be categorized into proactive technology transfer actions, procurements and knowledge transfer.

# XII-2 Organization

## XII-2.1 Basic TT process

There are different ways of doing Technology Transfer (TT), in order to fuel innovation to industry. One is through procurements, and this has been the conventional mode used by CERN and other EPP centres since their foundation. TT spurs further: more active methods can be implemented by a technology transfer unit which interacts with all the actors involved. The basic process is executed according to the following sub-processes.

### XII-2.1.1 Technology assessment and evaluation

The assessment and analysis process is a concerted procedure, where the TT unit acts on the advice of a number of people, including, as appropriate, internal and external technical experts, to obtain more targeted information for decision on Intellectual Property protection and for the selection of dissemination approaches.

### XII-2.1.2 Intellectual property evaluation and protection

Intellectual Property (IP) associated with EPP technologies needs to be evaluated, using processes and mechanisms such as prior art search, invention disclosure form, and market survey, and protected, using means such as patents, trademarks, industrial designs, and copyrights. IP protection is a prerequisite to preserve a commercial outcome.

### XII-2.1.3 Technology promotion

Technology promotion may be carried out in a variety of ways, and requires carrying out prior studies of the potential transfer of the technology. The promotional activities currently used include presentations, road shows, conferences, industrial workshops, posters and brochures, and meetings between inventors and industry. Technology promotion facilitates the dissemination and awareness of technologies and their applications.

### XII-2.1.4 Technology dissemination and implementation

The dissemination and implementation process reflects truly successful TT via R&D projects and commercialization of IP. The stages of 'proof of concept', 'prototyping' and 'technology acquiring' will be executed as necessary. These TT activities require a formal framework, such as agreements reflecting the maturity of the technology and the readiness of the acquirers. In order to draft a suitable agreement, close collaboration is needed between the TT unit, the technical experts, the external collaborators, and those involved in the contract-circulation procedure. The agreement tools may encompass pre-competitive collaborative R&D, partnerships, licences and services, and external funding.

## XII-2.2 Institutional set-up

EPP and all other scientific institutions have the choice between two basic options to effectively organize their TT unit. One might be called the in-house solution, while the other could be named the external solution. Both follow the above-mentioned set of measures, the first one does so as an integral part of the institution itself while the latter is either a subsidiary of the scientific institution or an independent entity, which manages and commercializes IP. Both ways have their merits. The actual choice depends on the conditions at hand and the goals to be achieved.

### XII-2.2.1 In-house TT unit

CERN and DESY are good classical models for practising TT via an internal unit aiming at optimizing dissemination. For example, in order to promote TT, CERN introduced a proactive TT policy in 2000 to identify, protect, promote, transfer and disseminate its innovative technologies in the European scientific and industrial environments. Once the technology and IP are properly identified, protected and adequately channelled, they enter into a promotional phase, preparing the ground for their targeted dissemination and implementation.

In order to promote EPP technology more quickly and to further its dissemination outside particle physics, the in-house solution offers interesting financial turnovers as well as the option to back strategic research decisions and requirements by a coordinated IP strategy.

### XII-2.2.2 External TT company

All external solutions for an active TT process require a commercially acting company at some stage of the TT process, this takes over control and responsibility. A key characteristic of these companies is their short-term and often narrowed focus on the commercial aspect of success of technologies rather than a broader strategic research policy. Furthermore their aim is to maximize financial return thus their comparative financial turnovers are normally higher than those from in-house TT units. But this is reached through a much shorter term and often narrowed focus than the options an in-house TT unit can offer for a strategic research policy.

Successful examples for external TT companies are the YEDA Research and Development Company Ltd., which is the commercial arm of the Weizmann Institute of Science, Israel's leading centre of research and graduate education, RAMOT for the University of Tel Aviv, YISSUM for the Hebrew University, and the European Molecular Biology Laboratory (EMBL) Enterprise Management Technology Transfer GmbH (EMBLEM), which is an affiliate and the commercial arm of the EMBL.

EMBLEM identifies, protects and commercializes the IP developed in the EMBL world, from EMBL alumni and from third parties.

These TT firms have as a main role to shield the researcher, as well as the academic institution, from any legal problem related to a given invention, while taking care of the aspects of protection and commercialisation of IP.

The fact that the TT firms mentioned act as independent bodies (unlike the situation in other countries), with their employees (typically 10 - 20) not being part of the academic institution, gives them a high incentive to find the most successful partner. In exchange, some of these firms receive part of the royalties.

### XII-2.2.3 Other approaches

Other approaches between a complete internal and external TT are used when the size of a scientific institution is too small to justify and employ a fully-fledged TT unit. A good example is the TT activity, coordinated by the Helsinki Institute of Physics, a joint institute of the universities of Helsinki and Jyväskylä and the Helsinki University of Technology. The institute has a programme for developing CERN technology for Finnish industry as well as for promoting Finnish technology at CERN. Under this programme, web-based management tools were developed and successfully commercialized. Since 2000, GRID computational technology was developed partly in collaboration with the CERN OPENLAB project and several Finnish IT companies. To assist the Finnish technology industry in bidding for procurement contracts to CERN the Institute collaborates with the Finpro association, which serves the Finnish export industry as an industrial liaison agent in Geneva supported by the Finnish Funding Agency for Technology TEKES. This project has been instrumental in securing major contracts for several Finnish manufacturing enterprises, and has brought the coefficient of return of Finland's contribution to CERN to exceed 1.

## XII-3 Overview of relevant areas

As many different fields of technology are involved in the planning, construction, operation and use of EPP devices, it is no wonder that an impressive list of manifold products has been strongly influenced or even derived from EPP research. Examples of applications of technologies derived from EPP research are provided and span over many fields. [117]

### XII-3.1 Accelerators

Particle accelerators were invented in the 1920s for physics research but have since then developed into a multitude of uses, most of them being quite afar from EPP (see Table XII-1). Some are used for medical diagnosis and care, or to sterilize medical equipment and food. They even appear on production lines for rubber gloves.

| CATEGORY OF ACCELERATORS | NUMBER IN USE (*) |
|---|---|
| High Energy acc. (E > 1 GeV) | ~ 120 |
| Synchrotron radiation sources | > 100 |
| Medical radioisotope production | ~ 200 |
| Radiotherapy accelerators | > 7,500 |
| Research acc. including biomedical research | ~ 1,000 |
| Acc. for industrial processing and research | ~ 1,500 |
| Ion implanters, surface modification | > 7,000 |

| TOTAL | > 17,500 |
|---|---|
| (*) W. Maciszewski and W. Scharf, Int. J. of Radiation Oncology, 2004 | |

Table XII-1. Number of accelerators in use in different technical applications.

Among these, linear accelerators (Linacs) used in radiotherapy represent 40 % of all running accelerators. In France, Germany, Italy, there are 4 units per million inhabitants, while there are, still for 1 million inhabitants, 11 in Switzerland and up to 14 in Finland. Today, accelerators are affordable in cost, small and robust enough to be part of any hospital. [118]

## XII-3.2 Medicine

Many concepts and developments from particle physics find applications in health care. High-quality detectors and accelerators, essential for particle physicists to meet the research request, may be applied as better diagnostic tools and for providing custom radiation treatment of disease.

### XII-3.2.1 Hadron therapy

Hadrons (such as for example the neutron and proton) are the subatomic particles that are influenced by the strong nuclear force and made up of quarks. They were rapidly identified as more appropriate particles than gamma rays for the radiotherapy of deep-seated tumours, because of the dose distribution in tissues. Pioneering studies were carried out at CERN in the late 1960s. Nowadays many centres world-wide are using proton and carbon-ion therapy, from Europe to Japan to Russia and the US. So far some 45,000 patients have been treated with protons and many new centres are under construction. A treatment centre based on an improved version of the PIMMS synchrotron done at CERN, called CNAO (Centro Nazionale di Adroterapia Oncologica) is now being built in the north of Italy. The INFN (Istituto Nazionale di Fisica Nucleare) is co-responsible for the construction of the accelerator.

CERN is also part of ENLIGHT, the European Network for Research in Light-Ion Therapy whose aim is to co-ordinate project at the European level. Measurements of the energy deposition by antiprotons were done at CERN in 1985.

Promising biological investigations for future medical applications with antiprotons have been carried out and proposed for GSI (FAIR, Facility for Antiproton Ion Research machine).

### XII-3.2.2 Isotopes

Many important isotopes were discovered and characterized, and separation techniques developed in the early years of nuclear physics have made them available to society. Now these are used daily for 'in situ' treatment or diagnostics of several million patients each year. Today, most of the isotopes used are produced in nuclear reactors, but many studies on the production of isotopes are made, using particle accelerators, because of the lower production cost expectation and absence of long life radioactive waste production.

The CERN-patented technology (neutron-driven element transmuter) relating to transmutation of elements exposed to an enhanced neutron flux can be used for the production of radioisotopes for medical and industrial applications. The European Isotope Separation On-Line Radioactive Ion Beam Facility (EURISOL) or MW target facilities may also offer opportunities for the production of new isotopes in parasitic mode. These technologies will satisfy the demand for new types of radioisotopes. Some of them may be more interesting for positron emission tomography (PET), others for targeted alpha, or monoclonal antibody therapy, and others to label monoclonal antibodies.

### XII-3.2.3 Detection and imaging

Particle physicists regularly use collisions between electrons and their antiparticles, positrons, to investigate matter and fundamental forces at high energies. At low energies, the electron–positron annihilations can be put to different uses in PET machines. This is a common scanning technique for medical diagnostics as it allows detailed viewing of the chemical processes involved in the functioning of organs in a way previously impossible. Thanks to the improvements of many associated technologies, PET represents a significant step forward in the way clinicians visualize and monitor treatment on-line (i.e. the spatial distribution of radiotherapy treatment). Due to their complementarities, associated with computer tomography (CT) scanners, PET becomes an essential tool for diagnostics. A first image from a PET camera was made at CERN in 1977. Twenty years later, a combined PET/CT scanner has been advocated as the path to true functional and morphological image fusion.

Examples of CERN developments are those of a small animal PET (ClearPET$^{TM}$) for drug discovery actually commercialized, a dedicated brain PET scanner (in collaboration with the Cantonal Hospital of Geneva), which will give more precise resolution, and a Positron Emission Mammography (PEM) prototype using EPP data acquisition, read out techniques and crystals (ClearPEM$^{TM}$). Another example of the use of electrons from radioactive beta decays is in single photon emission computed tomography (SPECT) devices, such as the Compton prostate probe.

The gas electron multiplier (GEM), a device introduced in 1996 at CERN and licensed has opened up for the development of dosimetry for radiotherapy. The hybrid photodetectors (HPD), also called hybrid photodiodes, which surpasses traditional photomultiplier performances make them ideal candidates for instrument to be used for the diagnosis of metabolic disorder.

A single photon counting pixel detector readout chip, the Medipix technology, driven by the requirements to analyse complex interactions of high-energy particle physics, eliminates the background noise associated with more traditional X-ray-imaging approaches and provides energy information that was previously lost. The system has already been transferred to a leading European company in the field of X-ray materials-analysis equipment, and several teams are looking into possible uses of the system in the medical-imaging field.

### XII-3.3 Energy

Energy consumption in the industrialized world tends to increase with economic development. Power engineering and research on heavy-ion fusion, plasma heating, accelerator-driven breeding and fission, radioactive-waste incineration are strongly influenced by EPP technologies. So energy is another crucial domain where EPP can provide new solutions.

An important solution in the nuclear energy field, proposed by Nobel laureate Carlo Rubbia, is the energy amplifier. This concept proposes to produce nuclear energy and/or to eliminate nuclear waste in a sub critical nuclear assembly. In contrast with conventional critical reactors, the nuclear fission reaction chain in the energy amplifier is not self-maintained. The practical implementation of such a device requires some further technological development and a series of experiments and prototypes with increasing power are proposed, expecting to reach a major milestone with an energy amplifier demonstrator of 50 - 80 MW by 2015 - 2020. The final application of energy amplifiers, addressing energy production and waste elimination, will strongly depend on energy demand, on the political decision on the role of nuclear energy, and on the corresponding nuclear fuel cycle of the country in which it is implemented. The energy amplifier opens up the possibility of burning almost any unwanted long-lived radioactive waste, which is a serious environmental issue, and transforming them into exploitable energy without any emissions of $CO_2$, thereby also avoiding the 'green-house effect'.

Solar energy as such has appealing qualities: it is environmentally friendly; it is virtually infinite and may be used to obtain high temperatures for thermal, mechanical, or electric applications, either by light focusing and/or by reducing the thermal losses. Light focusing allows very high temperatures to be reached but, unfortunately, the diffused light (up to 50% in central Europe) cannot be focused and is lost.

Flat solar panels offer many advantages, namely a reduced number of glass-to-metal seals, a larger absorbing area, an easier installation and maintenance, but they have a low efficiency. Evacuated solar collectors are able to reach temperatures of the order of 250°C without focusing, but they have a limited light collection area. Both are commercially available. Flat evacuated solar collectors, combining the advantages of both, will be built commercially thanks to the mastering of ultra-high-vacuum and glass-to-metal seals technologies. This innovative type of solar collector, patented by CERN, is particularly suited for small and medium-sized plants, both for heating, possibly combined with seasonal heat storage, and for cooling or air conditioning. It may also be used for water desalination, agricultural applications (e.g. crop drying), and for the production of heat for industrial processes. Finally, it may produce electricity with efficiency similar to those of photovoltaic cells, with the advantage of a higher combined thermal and electric efficiency. It is currently at the stage of pre-production by industry.

## XII-3.4 Computing and e-science

Information technology, which plays an essential role in scientific research achievements, has undergone a rapid development due to advances in electronics and network technologies. Through the implementation of the World Wide Web (WWW) and now of the Grid, information technology has paved the way to the next generation of computing. The WWW has become part of every-day modern communications. This could thus be considered as one of the most striking examples of TT in the past two decades. It is a worldwide TT that has largely modified both the functioning of modern society and the behaviour of individuals.

The WWW, invented at CERN in order to share information between different physicists, was distributed to the Internet community and became a world-wide phenomenon in the 21$^{st}$ century.

The data flow from the front-end electronics of the LHC experiments is equivalent to the total public telecommunication world-wide. To make sense of such data the latest technologies of computers, communication technology, and software for analysis and simulation are required. To operate large collaborations spread out across the world, essential elements are computer networking including the latest features of tele-cooperation, tele-operation of supervisory tasks in a detector, and exchange of the computer-aided design (CAD) and other digitized data amongst the participants. For example, in international data communication an institute, such as CERN, rates at the level of a medium-sized European country.

As the Web was CERN's response to a new wave of scientific collaboration at the end of the 1980s, the Grid is the answer to the need for data analysis to be performed by the world particle-physics community. The Grid is a very powerful tool tying computing resources distributed around the world into one computing service for all requesting applications. With the LHC, the CERN experiments will have to exploit petabytes of information; this has led the physicists to apply the Grid concept of sharing distributed processing that was originally proposed in the US. At least half of the activities around a big experiment are dealing with informatics. There are therefore a number of successful collaborations with industry, common developments and early equipment tests.

Many developments have been pursued both for the analysis and storage of the LHC data and for developing applications. One example is the EGEE (enabling Grids for e-science) project that builds on recent advances in Grid technology and aims at developing a service infrastructure for Europe available 24 hours a day. Thanks to the Grid, a new way of interaction among scientists and different domains will be made possible with faster dissemination of data, better quality control, and more efficient use and better processing of information sources. These characteristics will allow the rapid spread of the Grid in many different domains of application, from bioinformatics, genomics, astrophysics, epidemiology, pharmacology, biomedical sciences and environmental research.

A rapid and natural consequence of the Grid is its applications in the health field. An EU project for developing a Europe-wide mammograms data-base, called MammoGrid, was led by CERN, for distributing information among doctors and hospitals using Grid technologies. At present a virtual repository, where a total of 30,000 mammograms is stored, is accessible across Europe thanks to the developed prototype in order to provide background reference for follow-up and diagnosis of difficult cases and remote access.

An ongoing transfer is the Network-Emulator technology used to evaluate the performance of applications running over the Grid. The Network Emulator is a configurable network-in-a-box that emulates end-to-end quality degradation likely to appear in wide-area networks. It has a wide range of applications, including Internet telephony, file transfer and web browsing. Network emulation is a technique of reproducing the behaviour of computer networks that enables experiments in real applications and controllable environment to evaluate the effects of protocol choices on overall system performance and cost effectiveness before deployment, to evaluate application behaviour when network conditions are less than ideal, to understand minimal quality requirements for network applications, to emulate interactions between multiple concurrent applications as well as to assess safety-critical failure models.

EPP institutes devote half of their sizeable computing capacities on Monte Carlo simulations and have reached a very high level of competence in this field. In fact, codes, such as EGS (electron gamma showers), developed in international collaborations and distributed by SLAC to more than 1,000 users outside of EPP, are fundamental in all areas of cancer research and treatment by electron accelerators [119]. Geant4, for instance, simulates the passage of particles through matter. Among the various applications are calculations in linear accelerators, shielding for medical use, radiotherapy, brachytherapy, scanners and space satellite. Another important application is the simulation of interaction of radiation with biological systems at cellular and DNA level. FLUKA, a joint INFN-CERN project, was first written for the calculation of radiation-protection shielding; today, it is a multipurpose interaction and transport MonteCarlo code to calculate a variety of particles over a wide energy range in complex geometries. It can be used for microdosimetry calculations, space and accelerator radiation protection as well as for hadron therapy beams.

Tracing the role of EPP is sometimes not easy, as many other sciences are engaged in the field mentioned above. But EPP obviously has played and still plays a leading role in applications related to information technologies. From the pioneering of parallel processing and filtering of vast amounts of data via the full use of information and communication technologies, such as e-communications for individuals, groups and collaborations and the support for the open access to e-information and knowledge to finally the use of grids, e-libraries and persistent digital objects, EPP research nowadays is the leading prototype for e-science or cyber-science.

## XII-3.5 Sensors, diagnostics and micro-electronics

Particle detection is a strong EPP asset which can find applications in many different scientific or commercial areas from data-acquisition systems, to computer controls, calibration of complex apparatus, measuring devices for ionizing particles, on-line pattern recognition, and data selection at very high event rates, controls, instrumentation and survey.

Thin film on application specific integrated circuits (ASIC) detector technology (TFA) offers the possibility to realize a sensor for visible light, X-rays and particles, that is integrated on top of the readout ASIC that performs the readout of the sensor pixel. This technology can be applied in X-ray imaging for intra operative surgical probes devices and gamma detectors for PET/CT. A special pixel electronic, Monopix, is being developed, driven by the requirements of EPP, and has been patented by CERN.

## XII-3.6 Material sciences

Material sciences include condensed matter research as well as most of the research carried out with X-rays from synchrotron radiation sources, ion beams and neutrons from spallation sources.

Other, more applied, examples relate to the use of halogen-free plastic and carbon fibres materials, which have enormously developed in the last decades, as well as to specific surface treatment.

Since the accelerator structures became more complex, CERN decided in the 1980s to use only materials without halogen and/or sulphur agents in order to limit the damage to personnel and material in case of exposure to corrosive and toxic fumes. CERN then encouraged industry to produce halogen-free cables and contributed to this development. The accelerator of the Large Electron Positron (LEP) collider and experiments were equipped with halogen- and sulphur-free cables. The same standards will be applied to the LHC.

CERN has also developed a technology to electro-polish titanium and titanium alloys, which easily reaches a high degree of surface smoothness. The technology was developed for the CERN accelerator cavities, but the process found a number of other commercial applications as well.

High Temperature Superconductors (HTS), including the gas-cooled HTS resistive current leads, used in the LHC, may be useful in the near future to lower electric power consumption for refrigeration of superconducting magnets in medical applications.

# XII-4 Results

Among the standard spin-offs from EPP research are collaborative projects with other sciences or industries, patents and copyrights, revenues from licence contracts and start-up companies based on technology from EPP and the economic utility to industry of high-tech contracts placed by EPP institutes as well as the rare exchange of personnel and the (often informal) consultancy. These results fall in three categories ranging from procurement activity, to proactive TT activity, and to fall out of knowledge transfer.

## XII-4.1 Collaborations

With respect to collaborations EPP needs on subjects such as in-house manufacturing facilities, specialized engineers, technicians and craftsmen as staff members, there are important differences amongst the institutes ranging from complete out-sourcing of the work to companies, such as in Japan, complete in-house production, such as in China and Russia, and intermediate solutions applied in the US and Europe, where prototyping is often done in-house and production by companies.

In Japan, complete collaboration of research with industry is the declared policy of government and industry. So KEK is an organization with relatively few and mostly scientific and limited engineering and technical staff with no craftsmen. Operation of the laboratory, production of components and/or systems and corresponding prototypes are done completely by industry, at the expense of convenience and probably cost.

The institutes in China and Russia have large production capabilities, which allow them to produce in-house a considerable fraction of the parts required for their activities. For example, in Russia, the Budker Institute according to its self-understanding has to earn half of its funds by selling products on the market. The institute may not have always achieved this amount, but it has had some remarkable successes.

The US and European institutes policy is to produce prototypes and to develop new accelerator ideas and detector components in-house. Therefore they still have important production and technological capabilities on site and use them extensively for development work. The same is true, and in this case predominantly in Europe, for the University and research institutes. In the US/European case the combination of scientists, in-house specialized engineers and in-house trained technical development shops are seen to be most effective, convenient and least costly [119].

### XII-4.1.1 CERN collaborations in brief

Funds are needed in order to apply the results of EPP research. These can be obtained in the context of partnerships, which can be of two types: partnerships with institutes and partnerships with industry.

The collaborations established at CERN for applications of EPP developments in the medical field in the framework of technology transfer are the Crystal Clear Collaboration (CCC) and the Medipix2 collaboration.

The collaborations, which are funded by the 6$^{th}$ Framework Programme (Eurotev, Isseg, Etics, Eurons, EU-DET, Dirac, Eurisol, Health-e-Child, BalticGrid, Eela, EUChinaGrid, EUMedGrid) are either collaborations amongst institutes or between institutes and companies, for EPP or TT purposes. The institutionalization of TT has given rise to the establishment of partnerships with industry. With future long-term research projects and participation in prototyping, this will result in a more efficient channel of collaboration and TT.

## XII-4.1.2 EIFast – an industry forum

DESY has a long tradition of participation in joint collaborations and activities between industry and institutes. The official foundation of the Forum has marked a milestone on the road to a close collaboration between European research institutes interested in superconducting radio-frequency (SCRF) technology and companies interested in supplying products for accelerator facilities using that technology. In fact, in October 2005 the European Industry Forum for Accelerators with SCRF Technology (EIFast) has been founded by 61 participants representing more than 34 companies and institutes from nine European countries. They agreed on the Forums' statutes and elected the members of the co-ordination board.

The intention to create the Forum resulted from considerable industrial interest triggered by several large accelerator projects applying SCRF technology, in particular the approved X-ray Free Electron Laser (XFEL) and the planned International Linear Collider (ILC) for particle physics. Both projects use superconducting RF technology, which was substantially advanced during the past decade by the TeV-Energy Superconducting Linear Accelerator (TESLA) collaboration. In addition, the TESLA test facility (TTF) at DESY was built with the strong involvement of European companies and has been in operation since 1996. The TTF has thereby added to the solid base of know-how of European industry in the field of SCRF accelerators.

It was concluded that a Forum is needed to further strengthen the excellent position of European science and industry in SCRF. Moreover, the foundation of forums with an analogue conceptual formulation was arranged in the United States and in Japan. As a strong, common voice of European research and industry, the Forum will promote the realization of the SCRF projects in a coherent way. It aims at bringing together research institutes working in the field of SCRF technology or interested in getting involved, and industrial companies interested in supplying products to projects based on this technology. The main tasks of the Forum include generating strong support of projects at the political level in Europe, ensuring a flow of up-to-date information about projects between institutes and companies, promoting involvement of industry in projects at an early stage, and supporting the members in gaining access to information channels and decision makers otherwise difficult to obtain. [120]

## XII-4.1.3 Collaboration programmes in Israel

The Israeli Government, via the Office of the Chief Scientist of the Ministry of Industry, Labour and Trade, complements activities in the academia by financing a series of initiatives that emphasize common projects between industry and academia. The main programmes are:

i) MAGNET: this programme encourages the creation of consortium of industries and academic institutions in the development of new technologies. These are long-term programmes with a total envelope per project of up to 36 M$ over a 5-year period (up to 8.5 M$/year, with the government covering up to 80% of the budget). The programme exists for more than 10 years and 150 companies have participated.

ii) MAGNETON: this programme supports the TT from the academy to industry and is mainly dedicated to performing feasibility proof. Its overall envelope per project is 800 k$ per project over a period of up to 2 years, 66% financed by Government funding.

iii) NOFAR: It provides assistance to academic research supported by industry for feasibility studies to pass from basic to applied research. Typical Government funds are between 10 and 30 k$ per project.

The above programmes are complemented by a series of 20 - 25 industrial incubators, which range from projects in the phase of applicability to new enterprises and start-ups, supported up to 85% by Government funds.

## XII-4.2 Turn over

Industrial partnership and commercialization of IP represent, at CERN, returns of the order of 1.5 MCHF per year. Taking CERN as an example, it has to be recognized, that its TT unit shows a promising development on the road to a balanced relation between revenue and expenditure.

Turnover from the commercialization of IP will rarely cover the full cost of all technology transfer activities, which includes expenditures for the TT units, payments to inventors, research development groups and institutes, as well as the cost of prototyping, etc. Experience from well-versed TT units of large European institutes, active mainly in applied sciences, shows that it is possible to generate a total income that is nominally above the expenditures for the TT unit itself, but still lower than the over-all financial cost. However, as the WWW example shows, it would be too short-sighted to think only in terms of direct financial return.

The external TT units, YEDA and YISSUM, have been the most successful technology transfer firms in Israel. YEDA, in particular, registers around 80 patents per year and it has obtained royalties that have increased from 98 M$ in 2003 to more than 150 M$ in 2005, 75% of these revenues originating from applied sciences from the biological field (cf. 60% of the scientific activity in biology at the Weizmann Institute).

## XII-4.3 Commercial spin-offs

Commercial spin-offs can take the form of granting contracts to start-ups and companies as a way of exploiting licenses, or can be the result of knowledge transfer. Commercial spin-offs are often incubated in science parks, which are located near the research institutions from which they spring off. In the Geneva area, according to a study carried out by the 'Agence de développement économique du Pays de Gex', it was found that 12 spin-offs have been created 2000 - 2004, when CERN established a proactive TT policy. Of the total inventoried companies for the period 1981 - 2004, 13 are still active by the end of 2004, with a turn over of about 4.5 M€.

## XII-4.4 Procurement & industrial learning

In Europe, some 20 billion € of public money is annually spent on purchasing technology-oriented equipment from industry, of which 2 billion € are for inter-governmental, scientific research projects (EU, 2000). Several studies on CERN, ESA and Fermilab have indicated that there are significant returns on financial investment via 'Big Science' centres. [121–125] Financial multipliers ranging up to 3.7, with a mean value per industrial sector of 3, have been found by CERN, meaning that each currency unit invested from procurements in industry by Big Science generates in average a threefold return for the supplier. These values were obtained from the companies, estimating increase of turnovers due to new products developments, marketing changes, quality improvements, and cost saving in production techniques to be ascribed to the CERN industry relation.

A study carried out at CERN which analysed the technological firms' gain through working in the arena of Big Science by trying to open up the 'black box' of CERN as an environment for innovation and learning [126].

EPP institutes work closely with industry. Only during the time period 1997 - 2001, there were about 7,000 companies involved at CERN, out of which 10% are considered as being high tech companies. More than 50% of these high technology companies declared a gain of technological know-how and skills, and an improvement in their income, related to their collaboration with CERN. More than 35% of these high technology companies developed products as results of their collaboration with CERN.

A series of case studies was carried out, to develop a theoretical framework describing influences on organizational learning. The focus of the survey was CERN-related learning, organizational and other benefits that accrue to supplier companies by virtue of their relationship with CERN.

The learning outcomes documented in the study range from technology development and product development to organizational changes. Technological learning stands out as the main driver (see Figure XII-1 and XII-2). The technological learning impact also varies extensively between supplier projects. Extrapolating the value from the respondent to the total of 629 suppliers suggests that some 500 new products were developed, attracting around 1,900 new customers, essentially from outside high-energy physics. Learning and innovation benefits appear to be regulated by the quality of the relationship. This emphasizes the benefits of a partnership-type approach for technological learning and innovation. Previous studies [127,128] support these findings.

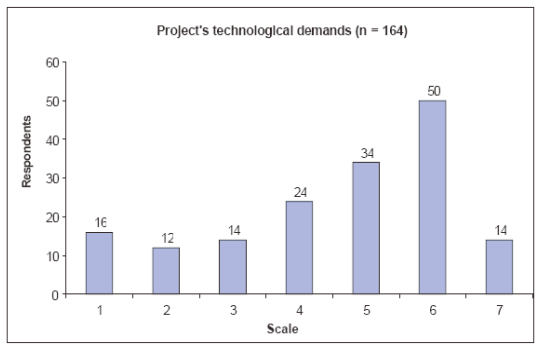
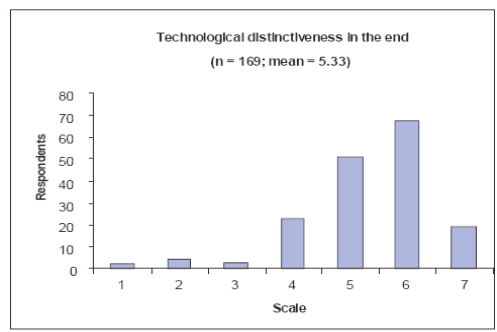

Figure XII.1 Project's estimated technological intensity. [Likert-stile scale 1: 'do not agree', 7: 'agree fully'.]

Figure XII.2 Technological distinctiveness at the end of the project estimated by the companies (reflects the learning effects). [Likert-stile scale 1: 'do not agree', 7: 'agree fully'.]

Additionally, in 2004, a study [129] was carried out, to examine the supply-side effects with a focus on 57 responding companies who supplied components and services to DESY's TESLA Test Facility (now the vacuum ultra violet free electron laser - VUV-FEL). The study showed findings similar to those of the CERN study (see Table XII-2).

|     | CERN study 2003 |     | DESY study 2004 |
| --- | --- | --- | --- |
| 38% | developed new products | 53% | sold new products to other customers |
|     |     | 38% | attained major innovations |
| 13% | started a new R&D team | 46% | made additional investments |
| 14% | started a new business unit |     |     |
| 17% | opened a new market | 60% | noted influences on their overall assortment |
| 42% | increased their international exposure | 82% | see large research infrastructures as important reference for their marketing |
| 44% | indicated technological learning | 23% | noted important learning effects with their employees |
| 36% | indicated market learning |     |     |

Table XII-2. Benefits associated with the procurement activity of CERN and DESY.

Physicists and engineers at CERN also highlighted the benefits that these interactions with industry have for them. This mutual beneficial effect, being independent of the level of in-house experience and participation in high-tech projects, confirms the importance of maintaining an active 'industry-Big Science' interaction, also to motivate highly qualified staff.

## XII-4.5 Socioeconomic Impact

The prospective socio-economic effects of large research infrastructures (RIs) are quite large. As an example from a field related to EPP, the prospects of the European X-Ray Free-Electron Laser (XFEL) facility were investigated in 2003 with a focus on Germany's national economy and analysing the demand-side effects during the building and equipment phase of the XFEL planned in Hamburg. [130]

The calculations in the study were based on the former Technical Design Report (TDR) and cost studies for the joint Linear Collider and XFEL project, where about 615 M€ were foreseen for the construction of the XFEL. Adapted to today's cost estimates (between 900 and 1,000 M€) by an absolute factor of 1.5, the results of this study read as follows: nearly 100% of the personnel cost, 90% of the building cost, and 36% of the equipment cost, added up to 56% of the total cost of 525 M€, are spent in Germany and will unfold supply-side effects. During the 8 years of the building and equipment phase for the European XFEL Facility, will guarantee about 2,000 jobs, an income of 85 M€, and a cumulated turnover of 180 M€ per year in Germany. According to the strong additional commitment of the other European nations, the high technical and industrial skill, and the awareness of European companies in this field, most of the remaining roughly 400 M€ will be spent in the other European countries and unfold likewise effects for a total of more than 3,500 jobs, an income of 150 M€, and a cumulated turnover of 325 M€ per year in Europe The factor of about 3 from project spending per year to cumulated turnover is backed by many other studies in this field.

The socio-economic impact of the international organizations, including CERN, in the Geneva area clearly indicates the positive impact for the region.

A socio-epistemic study analyzing EPP and CERN-LHC experiments was done in the year 1999 comparing the knowledge societies of biology and EPP [131].

The impact for personnel development and individual technological learning, generated by the multicultural and multi-field environment, is well proved. Users are exposed to a great range of knowledge acquisition opportunities, which is used in further professional experiences. It has been assessed that skills are significantly acquired in domains outside the specific work related activity and individual expertise [132,133].

## XII-4.6 Politics

EPP, more than any other area of research has also shown an immense potential for political spin-offs. For instance, it is an obvious fact that EPP scientists from Eastern Europe have collaborated with the West since the sixties. After 1991, the participation in EPP projects, such as the detector R&D and construction for the LHC and HERA, played a major role in the industrial conversion of weapons labs and factories in the former USSR. The existing contacts and mutual trust developed within the EPP community were the key to make this possible and successful. On a regional scale, countries, such as Pakistan and India, Iran and Israel, Greece and Turkey, manage to co-operate under the 'CERN umbrella'. Technology transfer is, together with pure science and economic interests, the driving motive in all these co-operations.

# XII-5 Recommendations

Investments in fundamental research get a higher political visibility because of the trend towards larger, but fewer, projects, increasing the need for visibility of returns. The case for fundamental science would be significantly strengthened if the outcomes given to society are made more visible. As, in general, technological changes originating from EPP have a long-term impact on society, the turnovers take more time to be measurable than those of, say, modern biology. Consequently TT has to be an integral part of the long-term EPP strategy and requires well versed and equipped TT units at EPP institutes.

Technologies that are currently transferred originate from R&D that took place 5 - 10 years ago. The completion of the LHC construction will free staff that can contribute to TT R&D projects and therefore will increase dissemination to society. In the short term, there will thus be an increase of the TT activity. Depending on the launching date for next major EPP R&D programme a possible discontinuity in technology and knowledge transfer potential from EPP resulting in a reduced visibility of the impact of science on society may occur. But in the next years (7 - 12) limited generic R&D funds will lead to a limited renewal of the EPP TT potential.

## XII-5.1 Protect intellectual property

The EPP institutes themselves should have a more active patenting and IP protection policy. Such a policy should preserve the open collaboration and the recognition of the results to the whole field, despite difficulties in some collaborations to identify the IP owners. Even worse from the point of view of IP protection, new ideas and the basic technology are often published or communicated to industry before they are protected. Instead, protection should always be sought for EPP technology with interesting applications or clear origin of IP, while a free licence could be given for EPP research. Common standards for the protection of EPP know-how and use of this IP by EPP and other research areas and commercial enterprises, the gathering of TT information and the dissemination of best TT practice, could be very helpful as today conditions for protecting/licensing of inventions differ greatly among labs.

In the past EPP projects have been fully funded by public funds and the resulting intellectual property was owned by academia. Nowadays interactions with industry are growing. Additionally, in some fields EPP depends on advances from industry, which has pre-existing know-how, being of interest for EPP. An increasing involvement of industry in the pre-procurement phase requires a different mode of interactions with industry in the next major EPP project, including assessment of IP ownership before publication and when establishing partnership with industry. It will become very important to identify win-win situations in the developments of products with industry and to find industry with ideas for technologies of mutual interest. Strategically oriented technology platforms between EPP and industry may form an interesting basic access to tackle these requirements.

The introduction of copyrights or trademark labels, such as 'invented by EPP', emphasizing the origin from institutions of EPP is to be recommended. A patent and licensing service organized across all major institutes participating in EPP could provide the essential clear definitions and the required traceability.

## XII-5.2 Close the application gap

EPP institutes should strive to be more active in technology transfer, in validating their technological developments and their educational impacts. Spin-off, technology transfer and education should be expressed clearly in the mission of each institute, in agreement with the funding agencies, and supported by appropriate funds and human resources.

There is a lack of interest of the pure scientists to develop industrial products or to think about applications. The use of income from know-how transfer to motivate employees is a still underrated incentive mechanism for scientists and technicians working at EPP centres. Contacts at the border lines of sciences, where collaboration between different fields could lead to quantum jumps of development and technology, need to be improved.

Relations between industry and scientific institutes in terms of exchange of personnel are weak or non-existent. The institutes could encourage 'sabbatical' stays in industry of their personnel and invite people from industry to spend some time in EPP institutes to confront and share understanding between these two different environments. Partnerships between EPP institutes and industry should be encouraged. Acknowledgement of each contribution should be recognized.

Special recommendations for the field of e-science aim at the improvement of collaboration and visibility of Information and Communication Technology solutions across EPP, at the encouragement of collaborating institutes to participate in especially critical multidisciplinary e-science projects, at the promotion of applications (e.g. Grids) with other sciences and industry and at the strengthening of the open-access initiatives.

In Europe there is a lack of funding for demonstrating technologies at the pre-competitive level for small productions or building pilot installations. Efforts in this intermediate range of technological developments are mostly improvised, not properly supported, and almost always lack visibility to attract the best people, 'critical mass' for an efficient organisation and at least medium-term efforts (>10 years).

Public funding is needed to assist industry investment to minimize risks. Furthermore, distances from the place where research is done may also play an important role on the low amount of spin-offs observed. The local and the distant infrastructure and the bodies concerned need to interact together.

In the relations to be established for innovative developments it could be envisaged that, in parallel with public funding used for EPP, extra funding is provided to transfer these developments into fields outside EPP.

## XII-5.3 Visualize impacts

The EPP institutes should collect more specified data in a systematic manner on their own usefulness, concerning informal contacts, publications, contracts and personnel. This would also allow for assessing the educational impact of the field. Interdisciplinary contacts and exchanges should be encouraged and supported, for example through active participation in technologically oriented conferences.

In order to maximize the learning benefits and to leverage the knowledge transfer to industry, it is important to ensure that the management of high-tech procurements and pre-procurement projects allows for extensive and frequent interactions of experts, creating an adequate environment for support and access to Big Science resources, the essential element for knowledge sharing. Partnership and consortium approach should be favoured to facilitate exchanges of knowledge within projects and between companies, creating communities of practices. It will be also important in order to apply, whenever needed, corrective measures. Existing purchasing rules will have to be adapted to take into account industry participation in specific types of projects, when high-risk technological development requires consistent investment from the participants, both in term of money and manpower.

Finally, the more general impacts of EPP, which are not measured by classical means, such as patent or turnover statistics, and, although 'felt', are not really understood, quantified or well known to public and politicians, need more attention. They include technical impacts (e.g. extending the limits for technical equipment and applications, strong demand for new technologies) as well as business impacts (e.g. EPP institutes as marketing reference for industry, knowledge build-up and development of new products through institute contracts), and socio-economic impacts (e.g. jobs, taxes, open access to information). These different general impacts should be collected, evaluated and, if possible, quantified. Regular publications on these general impacts are needed, as well as an inclusion in the reports to funding agencies. In the end these supportive measures may substantially aid in sustaining public funding of EPP research.

# XIII KNOWLEDGE TRANSFER

## XIII-1 Introduction

Knowledge transfer (KT) takes place whenever the expertise or discoveries of researchers are disseminated. This can occur in many different ways, for example:

- Through journal publications, preprints, conferences and workshops. This is the usual way of transferring knowledge within the academic and research community;

- Through the provision of training courses;

- When graduate students and post-doctoral researchers move from academic research into industry, business and public bodies;

- When academic researchers make their knowledge and skills available to industry or public bodies, for example through consultancy.

- Through the co-development of research ideas or discoveries with industry, either in co-operation with existing companies or by the formation of spin-out companies;

Since technology transfer, which is involved in some of these, is dealt with elsewhere, we concentrate mainly on other aspects of knowledge transfer here.

## XIII-2 Publications and preprints

We shall not attempt to review in any detail the difficult issues associated with scientific publication [134]. The most important aspects, upon which most of the high-energy physics community is agreed, are:

- Open access. Almost all papers in the field are deposited by the authors on the Cornell University Library electronic archive arXiv.org, which currently provides open access to over 350,000 e-prints in physics, mathematics, computer science and quantitative biology.

- Self-archiving. Authors are encouraged to ensure than the final versions of papers on the archive are those accepted for publication after peer review.

- Peer review. All original research should be subject to peer review, and papers that have passed the peer review process should be distinguished, normally by publication in a recognised scientific journal.

A contentious issue that needs to be resolved is the method of payment for support of the system of peer review and journal publication. In the past, payment has been by readers, or their institutions, through journal subscription. This system has proved very hard on developing countries and increasingly difficult for others as well, in an era of squeezed library budgets. The alternatives are payment by authors and/or subsidy by wealthier institutions or governments.

Another matter of concern is the durability, security and accessibility of archives that are not backed up by paper copies. While the production of paper journals may be regarded by some as obsolete, environmentally damaging, and wasteful of space and resources, electronic storage formats and media have proved short-lived. Wholesale recopying of electronic archives into new formats and media every few years appears inevitable and not without risk, either to the data themselves or to their universal accessibility.

A somewhat different problem concerns the assignment of credit to authors in large collaborations. Traditionally, experimental collaborations have listed all their members as authors on their definitive journal publications, whether or not particular members have played any part in the data analysis being presented. This makes it impossible for readers, including appointment and promotion committees, to assess the significance of any individual author's publication record. A system is needed for highlighting those authors with special responsibility for a given piece of work. At present, this function is served to a limited extent by conference and workshop presentations given by those responsible.

Credit must also be allocated to those collaboration members who make more general but crucial contributions, for example through hardware and software design and construction. There are a few specialist journals dedicated to work of this kind. A recent development is the acceptance by some high energy physics journals of 'physics' notes on technical topics by subgroups within the large collaborations.

## XIII-3 Conferences, workshops and courses

The functions of conferences and workshops within the research community are well known and will not be repeated here. An area in which conferences provide knowledge transfer between particle physics and a wider community is detector development. For example, at the $7^{th}$ International Conference on Position Sensitive Detectors, hosted by the particle physics group at the University of Liverpool in September 2005, out of 124 submitted abstracts, 42 were on technical developments without reference to specific applications, while the rest concerned applications in the fields of medicine (26), particle physics (24), astronomy (11), materials science and engineering (7), accelerator science (7), nuclear physics (6) and biology (1). It is clear that medical imaging in particular has benefited greatly from advances in particle detection, and will undoubtedly continue to do so as long as good KT and TT channels are maintained.

Another field with knowledge transfer to a wide community is accelerator science: out of roughly 17,000 particle accelerators currently operating world-wide, only around 200 are used primarily for particle physics. A crucial role here is played by a number of well-established accelerator schools. The CERN Accelerator School, held twice a year at different locations around Europe, provides courses at various levels, with industrial participation at around 20%, depending on the topic. The Joint Universities Accelerator School (JUAS) in Archamps, France, near Geneva, is organized annually by the European Scientific Institute, with the support of CERN and a group of 11 major European universities. It offers courses on the physics, technology and applications of accelerators, with part-time modular attendance available to staff of organizations using accelerators and companies manufacturing related equipment.

In addition to the direct benefit to industry of staff attending such schools, it should not be forgotten that many of the graduate students attending them will move from pure research into industry later in their careers, forming another channel of KT as discussed in the following section.

## XIII-4 Movement of researchers into industry, business and public bodies

When graduate students or post-doctoral researchers in particle physics move from pure research into other types of employment, they bring with them not only their specific technical skills but also their more general abilities as trained physicists, which, as we shall see, are highly regarded. In this section we survey the objective data that have been collected and draw some conclusions.

Studies of the career paths of students trained in high-energy physics have been carried out by the DELPHI [135] and OPAL [136] Collaborations at LEP and by the UK Particle Physics and Astronomy Research Council (PPARC) [137]. It is relatively easy to track the first employment of students, but more difficult to determine how many move out of particle physics after one or more temporary post-doctoral appointments. Nevertheless all three studies included an element of follow-up to estimate this effect.

## XIII-4.1 Study of DELPHI graduate students

Data were collected on 671 DELPHI students, of whom about half obtained a PhD and the others Master's degrees or (mostly) diplomas. The students were predominantly from Italy (140), Germany (120) and France (80), with Norway and the UK following with about 40 each. Amongst the PhDs, 22% were female. The main data were collected in 1996, with a follow-up in 1999. Only 158 of the ex-students were tracked and included in the follow-up. Based on an extrapolation from this sample, it was estimated that about half of the students eventually migrate out of research into business (7%), management (3%), high-tech industry (24%) and computing (16%). The statistics were similar for men and women. There is some variation by nationality, the migration into industry being significantly smaller for France and higher for Norway.

Interviews with a small sample of former students and/or their employers identified the following skills acquired by the students as being of interest to the private sector:

- Ability to work effectively in large teams;
- Drive and motivation to solve complex problems;
- Exposure to cutting-edge technologies in the electronics and computing domain;
- Familiarity with computing techniques related to handling large quantities of data and performing sophisticated modelling.

## XIII-4.2 Study of OPAL PhD students and Fellows

The OPAL Collaboration kept records of PhD students and CERN Fellows for the period from 1983 to 2004, essentially the whole period of construction, operation and data analysis.

In the case of PhDs the data are for German (93) and UK (90) students only, but they made up two-thirds of the total (183/289 PhD theses). By 2004, about half (47%) of these students were employed in industry (39/93 German, 47/90 UK), 17% in universities, 21% in research centres and 15% elsewhere. Destinations in industry were classified as computing (37%), finance (16%), communications (10%), consulting (11%), chemistry (7%), electronics (3%) and other (15%). Differences between nationalities were not great, with somewhat more German than UK students going into communications and chemistry, and less into computing.

Not surprisingly, far fewer of the 114 CERN Fellows migrated to industry (12%), the majority working in universities (58%) or research centres (25%). About two-thirds of those in universities and one half of those in research centres were employed on fixed-term contracts.

# XIII-4.3 Study of PPARC PhD students

In 2003 the UK Particle Physics and Astronomy Research Council (PPARC) commissioned a study of former students' career paths from DTZ-Pieda Consulting. They surveyed students who finished PPARC PhD awards in 1995-99 (489 eligible, 300-350 traced, 181 responded). Note that respondents included particle physicists (34%), astronomers (48%) and planetary scientists (16%). 20% were women.

Key findings were as follows:

1) 48% worked in the private sector, mostly in IT-related jobs: software design, solutions and management (29% of 48%), financial services (24%), business services (24%); only 13% of 48% = 6% were in manufacturing.
2) 35% were working in universities, 2/3 as postdocs, 17% of 35% = 6% tenured. Historical evidence suggests most will stay in universities, 2/3 getting tenure.
3) 12% were in government and public organizations, half of these in research-related jobs. Others were in the civil service (including intelligence), school teaching, Bank of England, etc.
4) 4% were unemployed (all less than a year: in transition).
5) They were generally in well-paid jobs: 39% earning more than the average UK professional (£32.6k), 80% more than the average worker (£24.5k). In the private sector these figures were 65% and 93%. The lowest-paid were post-docs in universities.
6) 19% were working outside the UK (mainly in Europe & USA), but these were 30% of those in universities (30% of 35% = 11%) and 32% of those in government and public organizations (including CERN: 32% of 12% = 4%).
7) Respondents working in private sector:
    a. Mostly reported the PhD training as essential (10%) or important (78%) in their current job.
    b. Mostly considered the PhD very useful (58%) or quite useful (34%) to their career.
    c. If starting over again, would still choose to take the PhD (92%).
    d. Highlighted the following as the important skills gained from PhD training: problem solving, writing software, quantitative data analysis.
    e. Highlighted the following as skills that could have received more emphasis in PhD training: project management, leadership, entrepreneurial skills, financial management, time management. (Similar responses from other sectors were reported here.)

The conclusions most relevant here would seem to be:

- A large fraction (about half) of UK particle physics and astronomy PhDs end up working in the private sector.
- They are mostly in well-paid IT-related jobs.
- They consider the PhD was important and useful.
- They value most the problem solving, software and data-analysis skills they acquired.
- They would like the PhD to include more management (project, time, financial) and entrepreneurial training.

It is difficult to asses the extent to which these findings can be extrapolated outside the UK, but the need for more management and entrepreneurial training could be a common theme (PPARC is now requiring universities to provide this).

## XIII-4.4 Conclusions

The DELPHI and OPAL studies concerned only experimental high-energy physics students, while the PPARC survey included theorists and astronomers. Furthermore the composition of the three groups by nationality was different. Nevertheless the conclusions to be drawn from the above studies appear broadly similar and are entirely consistent with those of the PPARC study. About half of our PhD students eventually migrate to industry, where they and their employers value the skills acquired during their PhD training, especially their IT and modelling skills and their readiness to tackle complex problems. Not surprisingly, the ability to work effectively in large teams was also highlighted amongst the experimentalists. In return, industry would like to see our PhD students receive more training in project, financial and time management, which could indeed also be beneficial to those who remain in research. Such training would represent knowledge transfer in the opposite direction to that discussed in most of this chapter.

# XIII-5 Transfer of knowledge to industry, business and public bodies

## XIII-5.1 Individual consulting

Knowledge transfer by individual scientists consulting for industry and public bodies is a familiar concept, but one on which it is difficult to gather anything other than anecdotal information. A classic case was the involvement of Richard Feynman in the investigation of the Challenger space-shuttle disaster [138]. This is not such a fanciful example, since it illustrates the fact, often cited by industrialists, that input from physicists is valued, not because of their knowledge of the fundamental laws of nature, but because of their readiness to tackle complex problems with fresh insight.

Informal contacts with colleagues who have spent time in business and financial consultancy also confirm that the general skills a physicist can bring to that arena are:

- Identifying for a complex system a few, relevant degrees of freedom (for example: identifying the key cost drivers in the aggregate telecommunication expenditure of a large company, which results from hundreds of different items and services purchased).
- Devising a simple model to simulate the implications of different scenarios (for example: assessing the financial impact of adopting a new technology, replacing equipment, etc.).
- Communicating the results effectively in a scientific style (soberly, clearly stating key results and model assumptions, etc.).

This agrees well with the findings on graduate students moving into industry discussed in the previous section. In addition, of course, established scientists have their own specific areas of expertise, which can be of value to relevant industries. In that area the primary need is for good channels of communication to make each side aware of the possibilities.

## XIII-5.2 Knowledge transfer institutes

A possible approach to facilitating knowledge transfer is to set up institutions devoted to that objective. As an example of an ambitious scheme along these lines, albeit involving a wide range of research rather than just particle physics, we consider the Cambridge-MIT Institute [139], set up by the UK government in 2000 at Cambridge University in partnership with the Massachusetts Institute of Technology.

CMI was initially set up in 2000 with a commitment of up to £65M public and £16M non-public funding over 5 years. Owing to a slow start-up, this has been extended for a further year, to August 2006, without extra funding.

The aim is 'to undertake education and research designed to improve competitiveness, productivity and entrepreneurship in the UK, and to forge stronger links between academic research and business'. Note that there is no US funding or objective. The role of MIT is to help achieve these aims in the UK, drawing on its successful record in the USA. In return, MIT was able to nominate Cambridge as its 'European sister institution'.

CMI operates through the formation of 'knowledge integration communities' (KICs). These are groups of academics and representatives from industry, business and public bodies, working together on research towards a common objective. They are designed to contain all the elements of the industrial supply chain required to bring about a transition from research to a product or an impact on the economy.

An example of a KIC is the silent aircraft project. This involves academics from Cambridge and MIT with representatives of British Airways, Rolls Royce, the Civil Aviation Authority and National Air Traffic Services. The aim is to reduce aircraft noise to a level unnoticeable outside an airport's perimeter. Other KICs concern connected worlds (linked projects on communications, disruptive technologies and the management of innovation), systems biology (towards personalized medicine) and pervasive computing.

CMI has also developed and launched graduate (MPhil) courses at Cambridge on nanotechnology, computational biology, engineering for sustainable development and technology policy, which include a component on entrepreneurship.

## XIII-6 Conclusions

Open access publishing remains an important issue for the transfer of knowledge within the physics community. Funding agencies need to make provision for the support of publication and peer review in order to optimize the impact of the research they fund.

About half of graduate students trained in high-energy physics move on into careers in industry. They and their employers value the skills they acquired as students, but would also like to see more management and entrepreneurial training. This is a theme that could usefully be promoted at a European level.

An issue that could be considered by the Strategy Group is whether knowledge transfer from the particle physics community could be facilitated by the creation of new organizational structures. For example, research laboratories and institutes could set up KT, as distinct from TT, offices, or combine the two. KT offices could maintain inventories of competences of staff, which would assist other institutions and companies in seeking advice or collaboration, and foster contacts with industry not directly related to technology. Ways of forming knowledge integration communities in areas such as detector development could also be considered.

# XIV EDUCATION

# XIV-1 Introduction

Fundamental research triggers curiosity and fascination in the public, which in its turn creates public support. It is therefore in all our interest to amplify this feedback loop to deliver our excitement to the lives of all people and to induce the science-seed at an early stage in a larger fraction of the population.

A systematic and well planned approach should ideally be made to achieve this goal. Our resources for such activities are rather limited, so they should be focused on giving the broadest maximal impact. Which are then the optimal target-groups to reach out to? Certainly young children, pupils and high school students must be on top of the list. To reach out to teachers is surely crucial, but maybe also to reach out to teacher's teachers could be discussed, especially if primary school pupils are critical.

When communicating our activities to a wide audience, the focus should be on the basic research, it would be a mistake to market fundamental research for its technological and economical aspects. Firstly, because only few of its spin-offs will have a real impact on our daily lives but, more importantly, because it would not do justice to its true motive: curiosity. It is human curiosity that drives fundamental physics: to understand the natural phenomena of our world, the structure of matter and forces, and ultimately the origin and evolution of the Universe.

Research at the frontier of high-energy physics and cosmology has succeeded so well and progressed so far that is has become incomprehensible to all but a few specialists. Obtaining sustained support for this research does not only require a better communication strategy, but it is certain that the communication skills of researchers need to be developed further and used in a much wider context.

## XIV-1.1 Issues in high-school physics education

A worrying phenomenon for physics is the general decline in the number of students that it attracts. For example, in the UK the number of pupils choosing A-level physics has decreased by 35% since 1991, and similar trends are observed in most European countries. Schools are increasingly failing to awaken and to sustain children's interest in science. As a consequence, fewer students consider careers as researchers or young teachers, and future citizens do not obtain the background they need to make intelligent decisions about the future framework for scientific and technological development.

If pupils are to develop an interest in and a desire to study science at higher level, they need to be inspired. And that is not happening at the moment. Students are easily excited by the frontiers of our knowledge, but what they learn in school physics lessons is mainly how to solve formal exercises on physics problems of the 18th and 19th century.

Teachers are on the front line every day. It is largely up to them to excite students about science, prepare them for careers, and give them critical skills necessary to think about the relationship between science and society. But few teachers' training programmes cope with bringing modern physics into the classroom. For teachers, few things are as motivating as direct encounters with front-line research and researchers. Experience shows that this directly transfers into much more exciting teaching. And students can tell whether a teacher is enthusiastic about their subject.

## XIV-1.2 The role of research laboratories in education

Cutting-edge research institutes, an immense source of new scientific information for society, have a unique role to play in providing a direct link between 'science in the making' and teachers. Teachers should be put in the position to grasp the essence of cutting-edge research, which cannot be found in textbooks, and to develop attractive foundation courses in physics for younger students, helping them to master difficult concepts in an intuitive and non-mathematical way.

The vision is that teachers' schools and other programmes can help in creating, identifying, collecting, and stimulating the exchange of successful educational tools and good teaching practices. Working groups of teachers carry out projects they have proposed and follow up on their progress; they explore and further develop existing resources, making them internationally available through translation and networking, thus creating a major, long-term on-line resource.

Teaching teachers needs the active participation of scientists, but we have to provide an easy means to get them involved, and to help them in developing better communication and education skills. It is therefore also important to recruit a larger number of scientists interested in educational matters and to train them in communication skills.

## XIV-1.3 The European dimension

Several training activities (for both students and teachers) are currently coordinated at a national level by the local EPP laboratories and supported by the respective EPP funding agencies, Education ministries and foundations. European Intergovernmental Research Organizations (EIROs) offer, on the other hand, a broader and unique range of education opportunities leading to a fuller integration, on a European scale, of best-practice models.

To start with, the international composition of the scientific staff of EIROs makes it possible to organize educational activities in all the languages of EIROs member states. Tested templates of educational programmes and materials can therefore be proposed in several languages, ensuring that the public and the teachers from all countries get access to the same quality of information. This is particularly valuable for countries where national programmes have not yet been put in place, but which could be seeded by materials made available by the EIROs. Furthermore, national activities can be coordinated across the EIRO member states by ad-hoc bodies, such as, in the case of CERN, the European Particle Physics Outreach Group (EPPOG), to be discussed below.

EIROs can bring together teachers from several countries, and allow them to share experiences and information on their respective syllabuses. Data collected when evaluating these European-scale activities, show that the contact with colleagues from other countries is one of the most valuable components of these experiences. Such exchanges provide the opportunity to establish international contacts between teachers and schools, ultimately leading to Network proposals to be supported by EU grants.

As an example of the activities with European-level impact which are currently taking place, we provide in this chapter an overview of the educational programmes organized or co-organized by CERN.

# XIV-2 CERN educational programmes

CERN runs a large number of educational programmes, aimed at different groups. Since this report refers to the interaction with society at large, it does not contain a description of those programmes, which form part of the professional formation of undergraduate and post-graduate students (summer-student programme, technical- and doctoral-student programme, CERN schools of high-energy physics, accelerators, computing), or academic and technical training programmes for CERN staff and visitors.

## XIV-2.1 School children, general public

The regular visits programme welcomes about 25,000 visitors a year, the majority of which are schools, from all over Europe; this includes about 500—1000 school teachers. These visits last usually for 1/2 day, and include a presentation of CERN, a visit to 1 or 2 CERN sites (e.g. the LHC experiments, the PS/AD accelerator), and a visit to the 'Microcosm', the permanent exhibition.

Open Days are organized every 2 to 3 years. The last event attracted about 33,000 visitors to various itineraries, indicating a very strong interest of the general public.

In the near future the 'Globe of Science and Innovation' will host CERN's new visitors centre, with a permanent exhibition on CERN's physics, accelerators, detectors, derived technologies, and other socio-political aspects. Temporary exhibitions, seminars, public presentations, and many other events for schools and the general public will significantly increase the impact of CERN's communication with the local and regional public.

## XIV-2.2 Teachers' programmes

CERN has organized teachers' schools since 1998, which are now attended by 80-90 school teachers every year.

The in-depth 3-week 'High-School Teacher' (HST) programme, for 30 to 40 teachers, consists of lectures, visits of CERN installations, working groups on educational issues and topical review sessions. The participants come from all the CERN member states, with a quota assigned for non-member states worldwide (for several years, 4 American teachers have also participated, supported by an NSF grant). A large amount of educational material, in several languages, has been developed, assembled and documented by the participants at the website http://teachers.cern.ch/.

A 3-day 'immersion programme' (PhT) gives teachers the opportunity to get a glimpse of the world of CERN, with visits to experiments and some lectures by scientists. The working language of the HST and PhT programmes is English.

The HST lectures have been progressively adapted to the needs of school teachers, and numerous review sessions have given valuable feedback about the most suitable topics for teaching students aged 13 to 18 – and how to present them. The overall cost of these two programmes, which cover all expenses of the participants, is about 100 kCHF per year.

Based on this experience, and in the spirit of establishing even closer links with European schools, a number of 1-week schools for physics teachers from the member states will be organized, starting in 2006. An important development is that these schools, which are intended to be complementary to the HST programme, will be largely delivered in the mother tongue of the participants. The major advantages and objectives of this new approach are

- to better serve the teacher communities in the member states by increasing the number of participants that can follow the training, and by removing the language barrier imposed by the requirement of proficiency in English;
- to have the possibility of organizing courses during the whole year (not only during summer holidays), giving the flexibility to adapt to the time-table of individual countries;
- to record the lectures and to provide the accompanying educational resources in the national language of the participants, making them accessible to many more teachers and directly usable in the classroom;
- to promote the exchange of knowledge and experience among teachers with similar curriculum requirements.

The HST programme has also shown that the location of these schools at CERN offers benefits. It gives the participant a first-hand impression of frontier research in particle physics, promotes direct contact with scientists, and gives the possibility of visiting some of the world's largest scientific installations.

In parallel with the organization of both the HST and the new teacher programmes, and in collaboration with the teachers who attend them, the education group will develop materials and tools that are of direct practical use in the classroom. These new teaching resources – together with the recorded lectures – will be published through a comprehensive web-based library. These materials will be made available in several different languages, since it is more efficient to teach scientific subjects in the mother tongue of students. The feedback from students and teachers will be used to improve and to expand this material over time.

The new programmes shall be organized in collaboration between CERN and the member state of the participants. This will require the appointment of a national contact person that oversees the advertisement, application, selection, and funding procedure in the respective member state.

CERN will provide all scientific, administrative and technical support for the programme, including development of the programme contents and provision of facilitators, lecturers and guides. On the other hand, CERN expects that travel and subsistence funds for teachers will be provided either by the competent national authorities or by other sources, e.g. educational foundations in the member states. This will allow CERN to organize 8 to 10 teacher schools per year, at a cost of about 80 kCHF per year.

## XIV-2.3 Education website

The education website is a single-access point providing classroom-ready material, which is grouped according to age, curriculum topic, and translated in several languages. CERN will provide incentives for translations, since they are an important element for jumping over the language barrier and will allow a more efficient use in many European countries. The website will also give access to a rich archive of lectures, which have been selected for their suitability for school teachers and students.

## XIV-2.4 Other activities establishing links to schools and teachers

The collaboration between scientists and resident teachers, staying at CERN typically for 3 to 6 months, is important for the development of teaching projects which are scientifically correct yet simple enough to be understood by young students.

The use of internet tools in teaching, such as webcasting and live video-conferences, is becoming an important element in the education strategy, since an increasing number of European schools have broadband ($> 512$ kBit s$^{-1}$) connections that are sufficient for audiovisual transmissions. Educational web-casts of 15 – 30 min will be prepared and recorded in collaboration between scientists and teachers. The webcasts are archived and are available on demand, providing first-hand information by scientists on on-going research projects. Similarly, live-video conferences, which are organized by CERN on demand, give the possibility of direct interaction between school classes and scientists.

# XIV-3 CERN collaboration with other HEP institutes

EPPOG consists of representatives from CERN member states. EPPOG has both national and international activities, of which the `master classes' (see below) is the largest common activity.

National activities are organized by universities and research institutes (open days, public lectures), or together with science centres and museums. At various topical occasions (e.g. 1-week celebrations in Berne for Einstein year 2005), institutes also organized activities by visiting schools (`HEP trucks' with portable spark chambers to demonstrate cosmic rays; travelling exhibitions, posters, brochures, video clips, CD-ROMs). Internet-based activities are also provided (tools for analysing selected LEP events), websites.

## XIV-3.1 Master classes

Each year about 3000 high-school students (mainly 17 – 18 years old) in 18 countries all over Europe come to one of about 60 nearby universities or research centres for one day to learn about goals and techniques of particle physics. After following lectures from active scientists about matter and forces, the students perform measurements on real data from particle physics (LEP) experiments, e.g. determining the W/Z mass or the relative Z decay branching ratios. At the end of the day, the participants join a video-conference for discussing and combining their results.

## XIV-3.2 Science ambassadors

Science 'ambassadors' are an excellent way to establish direct contact between the world of science and schools. In many countries, and mostly driven by individual initiatives, young researchers are invited to give lectures at schools. This brings students face to face with potential role models, who are not much older than they are, and which are open to their questions and can present science as an exciting, international endeavour.

# XIV-4 EIROforum activities

## XIV-4.1 Science on Stage

Science on Stage is organized by the seven intergovernmental European research organisations within the EIROforum partnership: CERN, EFDA, EMBL, ESA, ESO, ESRF and ILL. It is an integral part of the EIROforum European Science Teachers' Initiative (ESTI), supported by the European Commission in the context of the new NUCLEUS project for science education.

The Science on Stage programme is an innovative, pan-European science-education activity, aiming to foster a renewal of science teaching in Europe by encouraging the exchange of new concepts and best practices among teachers from all over the continent. The goal is to strengthen the awareness and interest of young people in science and technology, a vital precondition for securing the long-term development of our society. European Science Teacher Awards are prestigious and good incentives to motivate teachers beyond standard duties.

Science on Stage comprises national activities in 29 participating countries, in order to identify the most exceptional teaching projects and most motivating individuals in European science education. These are then invited to a major European Science Teaching Festival, which serves as a showcase and discussion forum. In November 2005, the international festival took place at CERN, and the next edition is jointly organized by ESRF and ILL in Grenoble in April 2007.

Science on Stage events at the national and international level are 'catalyzing' events, attracting attention of the teachers community and stimulating new ideas. The international festival consists of the fair, where teachers exhibit their new ideas for simple, yet stimulating demonstration experiments suitable for reproduction in classrooms. About 40 workshops cover a wide range of topics, with the goal to achieve new insights and inspiration for making science teaching more relevant and appealing. These workshops are complemented by scientific seminars given by front-line researchers from the EIROForum organizations, presenting new results in a pedagogic way. Finally, on-stage performances show how modern science can be presented in an entertaining way.

## XIV-4.2 Science in School journal

Science in School is a new European science-teaching journal, financed jointly by EIROForum and the FP6 programme of the European Commission with the ESTI programme. It has the objective to present the best in science teaching and current research, to bridge the gap between the worlds of research and school. It will be published quarterly and will be freely available from a dedicated website and in 20,000 print copies from spring 2006 onwards. Science in School will cover biology, physics, chemistry, Earth sciences, and mathematics.

The journal will contain teaching material that can be used immediately or adapted, articles on cutting-edge science and important science topics, reports on education projects that teachers can participate in or emulate, announcements of training opportunities, discussion forums and many other topics. It is expected that Science in School will well complement the existing European publication, the European Journal of Physics, whose primary mission is to assist in maintaining and improving the standard of taught physics in universities and other institutes of higher education.

# XV USE OF ACCELERATORS OUTSIDE PARTICLE PHYSICS

There are estimated to be more than 10,000 particle accelerators worldwide (excluding trivial examples such as television sets). Very few of these are used for particle physics experiments – the majority of these are medical accelerators.

The non-particle physics applications can be divided into nine broad categories;

i) nuclear physics (e.g. CEBAF@JLAB, FAIR@GSI, ISAC@TRIUMF, ISOLDE@CERN);

ii) Spallation neutron sources (e.g. ISIS@RAL, J-PARC@Tokai, SINQ@PSI SNS@ORNL);

iii) Synchrotron light sources (there are currently [140] more than 50 light sources world-wide operating or under construction);

iv) Free electron lasers;

v) Archaeology and non-destructive analysis and dating of materials;

vi) Medical applications [141] – radiation therapy with electrons (betatrons), charged hadrons (protons or heavier ions, cyclotrons or synchrotrons), neutrons (e.g. Boron Neutron Capture Therapy) and radio-nuclide production for positron emission tomography and therapies (cyclotrons mainly);

vii) Industrial applications [142] (x-ray lithography, ion implantation, radiation assisted processing such as polymerisation, curing, sterilisation, preservation or micro-engineering);

viii) Energy (as part of fusion programmes, potentially nuclear waste disposal through transmutation, accelerator driven sub-critical systems)

ix) Defence?

These applications involve accelerating particles to energies that range from a few tens of MeV to a few GeV (although the X-ray FEL requires a few tens of GeV). The main lines of development of the non-scientific applications of accelerators are illustrated in fig. XV-1.

The development of the scientific applications of accelerator technology (nuclear physics, spallation sources, light sources) is similar to that for particle physics, and there is significant mutual benefit and cross-fertilization of ideas from the close association of these facilities with the particle physics community, for example in the development of superconducting RF cavities.

The development of accelerators for non-scientific applications (medical and industrial) is always likely to be some way behind the "leading edge" because of the requirements for reliability, stability, ease of maintenance and total cost of ownership. While there are clear benefits, for example, in miniaturising such accelerators (a portable GeV light source could perhaps find many applications, including for the military) the need for extensive shielding is likely to limit such developments.

The key areas where developments in accelerators for particle physics are likely to have a wider impact are:

a) RF technology;

b) Power Converters;

c) Magnet technology;

d) Vacuum technology;

e) Diagnostics and Instrumentation;

f) Controls.

In each of these areas, the development of new (usually "more efficient") technologies will be of more general benefit through improvements in price/performance, once industrialisation of the process has been completed, in particular the introduction of suitable quality control procedures for the advanced technologies that are involved.

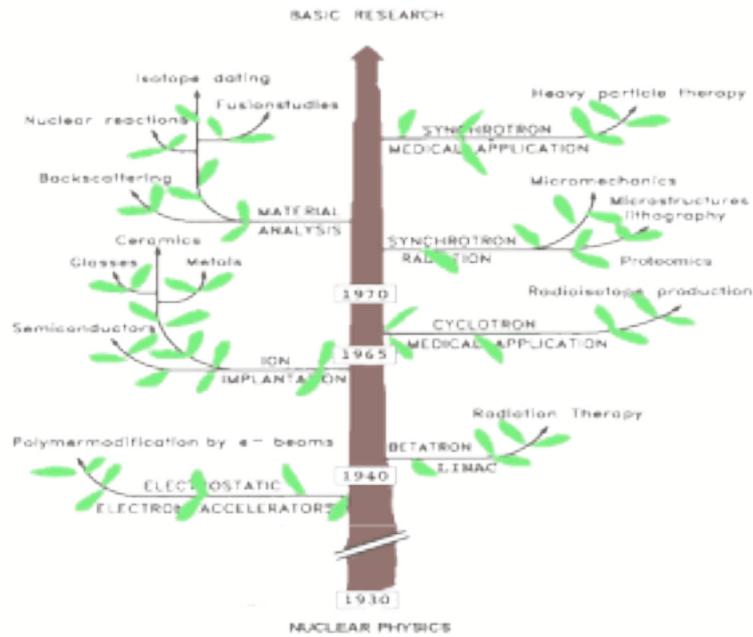

**Fig XV-1**: The Time Tree gives a pictorial view of the development of accelerators applications in both modification processes and sample analyses [141].

# XVI INDUSTRIAL INVOLVEMENT IN RESEARCH PROJECTS

The major particle physics construction projects, like the LHC, depend upon industry to provide much of the volume production, for both the accelerators and the detectors. Much of this work involves high technology or engineering close to the leading edge of what is achievable with contemporary technology. Often, the development showing that required performance can be achieved took place in a particle physics laboratory or institute, and the relevant technologies have then to be transferred into industry for prototyping and production.

There are very many examples, in both the accelerator and detector, where this process has been very successful. However, in a small number of cases, some with a fairly high profile, this process has encountered difficulties. It is essential that the causes of these difficulties are understood, and action taken to minimise such risks in future.

It is inappropriate here to analyse any particular problem, but it is possible to extract some general issues that contribute to such difficulties. However, it is important in discussing this topic that the differences between the industrial and the research environments are understood *by the research side*. In essence, the motivation for industry is to return value to the shareholders who have invested in the company (indeed, in many jurisdictions, this is a legal requirement on company directors); it does this by manufacturing and selling products. The consequence is that it is in the interests of the company to sell the product for the highest possible price while manufacturing it for the lowest possible cost. While this may seem obvious, it is not in general the way in which research institutes approach a similar problem. This difference in approach can be illustrated by a simple (fictitious) example. Suppose that there is a requirement on a certain parameter that 80% should be within 100 microns, and 100% within 120 microns. In a research environment, an outcome where 100% were within 100 microns, and 80% were within 50 microns would be considered an exceptional result, and the team responsible would receive high praise for excellent performance. A similar result in industry is likely to cause adverse comment, because the manufacturing standards used to achieve this result would have been more stringent (which means in the end, more costly) than was needed to meet the customer requirements, and hence the profit margins would have been lower than they could have been.

Given this difference in approach, there are at least five reasons why contracts may fail to deliver satisfaction.

1. The company failed to understand the full implications of the specification, or the specification was incomplete, so that manufacturing costs turn out to be higher than was anticipated when the tender was submitted. [Changes to the specification after the contract has been agreed, even if apparently minor, are a serious error on the research side.]

2. Because the company (in general) sees only the component and not the rest of the system in which the component is embedded, it is possible that manufacturing techniques could be used which lower the cost of production and which do not impact upon the performance *of the component* but which adversely affect the performance *of the system*.

3. Sometimes, especially for products that are very innovative or particularly challenging, individuals or a small group within the company may have a particular, essentially *personal*, interest in the project. While from the research point of view this might be considered a very *good* thing, from a company point of view it can, in some circumstances, be undesirable. Since this is essentially a personal commitment of individuals within the company, the commitment of the company may diminish if that individual or group leaves, or is transferred to other tasks.

4. The long time (more than 5 years in many cases) between the negotiation of the contract and the completion of the order means that changes in the industrial environment can affect the attitudes of the company to the contract. Companies can become insolvent, or be subject to takeover or divestment, any of which can lead the new owners or administrators taking a fresh look at the profitability of the contract, and seeking to restore margins either through re-negotiation or redesign of the manufacturing process.

5. Particularly for complex products, it is very likely that the contracting company will subcontract substantial parts of the manufacture to other companies. The subcontractors will be subject to the same pressures as the prime contractor, leading to the secondary risks outlined above. This is more serious for the customer, since there is usually no direct link with the subcontractor and therefore less chance of detecting potential problems early.

Given that many high technology contracts deliver successfully, for both the machine and the detectors, it seems likely that there are no fundamental obstacles to success. It would nevertheless be interesting, once the LHC construction is complete, to examine critically some contracts (some which were entirely satisfactory and some where there were serious difficulties) to look for examples of good and poor practice, and to see what lessons can be learnt from the experience.

This is a difficult area to investigate. Much of the information is necessarily anecdotal. When things work out well, there is rarely any real enthusiasm for a critical analysis of what *might* have gone wrong but did not. When things have gone wrong, the process of settlement often includes clauses that restrict the amount of information about the settlement that can be made public. The result is that much of the "evidence" is anecdotal, and rather biased towards the "problems" at the expense of the "successes". However, without a proper understanding of the factors that can influence both the probability of success and the likelihood of failure, it is difficult to see how mitigating strategies can be developed.

Issues that need to be examined include whether there is any correlation between the degree of technical risk and the probability that a contract fails to deliver satisfactorily, and whether, for commodities, Commercial Off The Shelf (COTS) systems and services, the experience in particle physics is significantly different from that in any other sector. Once this information is known, it might be useful to examine whether, for contracts with a significant degree of risk, there should be additional quality control, procurement or contract management procedures to take these risk factors into account, and so reduce the risks of similar problems arising with major new construction projects in particle physics.

# XVII APPENDIX

# XVII-1 Table of contents and references for Briefing Book 2

All documents quoted below, and referred to in the text of this paper as [BB2-…], can be accessed from the Strategy Group web page, at the URL http://council-strategygroup.web.cern.ch/council-strategygroup/BB2/Inputs.html .

**[2.1] THE HIGH-ENERGY FRONTIER**

- [2.1.1] B. Allanach High-energy frontier
- [2.1.2] F. Zarnecki High-energy frontier
- [2.1.3] R. Hofmann EWSB
- [2.1.4] . Wilson et al, CLIC Study Group The CLIC study of a Multi-TeV Linear Collider
- [2.1.5] A. de Roeck and J. Ellis Physics prospects with CLIC
- [2.1.6] J. Ellis et al, POFPA Working Group Physics Opportunities for Future Proton Accelerators
- [2.1.7] J-J Blaising et al, CERN PH staff physicists High-energy physics options for the next decades
- [2.1.8] B. Foster et al, LCUK International Linear Collider, UK
- [2.1.9] B. Foster et al, GDE Executive Committee Global Design Effort for the International Linear Collider
- [2.1.10] C. Da Via et al Detector and instrumentation development beyond the LHC
- [2.1.11] R. Garoby et al, PAF Working Group Preliminary accelerator plans for maximizing the integrated LHC luminosity
- [2.1.12] G. Weiglein et al, LHC/ILC study group The EPP questions: response from the LHC-ILC study group
- [2.1.13] C. Damerell Contribution to the ILC debate
- [2.1.14] E. Arik and S. Sultansoy Future perspectives in HEP
- [2.1.15] G. Moortgat-Pick Thoughts about the future structure in HEP
- [2.1.16] A.K. Ciftci and S. Sultansoy Linac-Ring type colliders
- [2.1.17] M.A. Sanchis-Lozano Non-minimal Higgs scenarios
- [2.1.18] W. Zakowicz Extreme high-energy accelerators
- [2.1.19] R. Aleksan et al, ESGARD European Strategy Group for Accelerator R&D
- [2.1.20] F. Richard et al, the LC World Wide Study Organizing Committee World Wide Study for a Linear Collider
- [2.1.21] F. Gianotti The physics potential of the LHC upgrades

[2.1.22] R. Klanner The european strategy

[2.1.23] J-J Blaising et al, CERN-PH Potential LHC contributions to Europe's future Strategy at the high-energy frontier

[2.1.24] A. Wagner et al, TESLA Technology Collaboration LoI about a European SC RF facility

[2.1.25] S. Sultansoy SUSY or not

[2.1.26] A. Devred et al, European Superconducting Accelerator Magnet R&D aimed at the LHC Luminosity Upgrade

[2.1.27] A.F. Zarnecki et al, Polish ILC community Contribution on the ILC

**[2.2] NEUTRINO PHYSICS**

[2.2.1] J. Peltoniemi et al, LENA Neutrino physics, underground laboratories

[2.2.2] M. Benedikt, M. Lindroos et al, EURISOL Beta-beam task Neutrino physics: beta beams

[2.2.3] V. Palladino et al, BENE Beams for European Neutrino Experiments

[2.2.4] P. Migliozzi et al Neutrino physics: neutrino factory

[2.2.5] J-E Campagne et al, MEMPHYS A large scale water Cerenkov detector at Frejus

[2.2.6] M. Dracos et al, GDR Neutrino Neutrinos groupement de recherche

[2.2.7] K. Long et al Precision measurements of neutrino oscillation parameters

[2.2.8] R. Garoby et al, PAF WG Potential for neutrino and radioactive beam physics of the foreseen upgrades of the CERN accelerators

[2.2.9] S. Centro et al, ICARUS Liquid Argon detectors for underground experiments

[2.2.10] A. Blondel et al, ISS programme committee International Scoping Study interim report

[2.2.11] V. Palladino et al, BENE Statement of interest in a European SC RF facility

**[2.3] FLAVOUR PHYSICS**

[2.3.1] M. Yamauchi et al Super-B factory

[2.3.2] T. Hurth et al Flavour physics

[2.3.3] M. Giorgi, et al Linear Super-B factory

[2.3.4] A. Ceccucci et al, P326 Search for rare K decays

[2.3.5] F. Cervelli Flavour physics with high-intensity hadron beams

**[2.4] PRECISION MEASUREMENTS**

[2.4.1] O. Naviliat et al Neutron EDM at PSI

[2.4.2] K. Kirch et al, nEDM collaboration EDM searches

**[2.5] NON-ACCELERATOR AND ASTROPARTICLE PHYSICS**

[2.5.1] J. Steinberger Astro-particle and cosmology

[2.5.2] K. Zioutas, et al, CAST Axion searches